\documentstyle[12pt,a4,psfig,amssymb]{article}

\def\bbbn{{\rm I\!N}}
\def\bbbr{{\rm I\!R}}
\def\bbbq{{\mathchoice {\setbox0=\hbox{$\displaystyle\rm Q$}\hbox{\raise
0.15\ht0\hbox to0pt{\kern0.4\wd0\vrule height0.8\ht0\hss}\box0}}
{\setbox0=\hbox{$\textstyle\rm Q$}\hbox{\raise
0.15\ht0\hbox to0pt{\kern0.4\wd0\vrule height0.8\ht0\hss}\box0}}
{\setbox0=\hbox{$\scriptstyle\rm Q$}\hbox{\raise
0.15\ht0\hbox to0pt{\kern0.4\wd0\vrule height0.7\ht0\hss}\box0}}
{\setbox0=\hbox{$\scriptscriptstyle\rm Q$}\hbox{\raise
0.15\ht0\hbox to0pt{\kern0.4\wd0\vrule height0.7\ht0\hss}\box0}}}}
\def\bbbz{{\mathchoice {\hbox{$\sf\textstyle Z\kern-0.4em Z$}}
{\hbox{$\sf\textstyle Z\kern-0.4em Z$}}
{\hbox{$\sf\scriptstyle Z\kern-0.3em Z$}}
{\hbox{$\sf\scriptscriptstyle Z\kern-0.2em Z$}}}}

\title{The Quantum Three-Dimensional Sinai Billiard --- 
       a Semiclassical Analysis}

\author{
  Harel Primack 
  \\
  {\it Fakult\"at f\"ur Physik, 
    Albert--Ludwigs Universit\"at Freiburg,}
  \\
  {\it Hermann--Herder--Str.\ 3, D--79104 Freiburg, Germany}
  \\
  email: harel@physik.uni-freiburg.de
  \\ \\
  and 
  \\ \\
  Uzy Smilansky
  \\
  {\it Department of Physics of Complex Systems,}
  \\
  {\it The Weizmann Institute, Rehovot 76100, Israel} 
  \\
  email: fnsmila1@weizmann.weizmann.ac.il
  \\ \\}

\date{Submitted to Physics Reports, 1.6.99}

\begin{document}

\maketitle
\begin{abstract}

  We present a comprehensive semiclassical investigation of the
  three-dimensional Sinai billiard, addressing a few
  outstanding problems in ``quantum chaos''. We were mainly concerned
  with the accuracy of the semiclassical trace formula in two and
  higher dimensions and its ability to explain the universal spectral
  statistics observed in quantized chaotic systems. For this purpose
  we developed an efficient KKR algorithm to compute an extensive and
  accurate set of quantal eigenvalues. We also constructed a
  systematic method to compute millions of periodic orbits in a
  reasonable time. Introducing a proper measure for the semiclassical
  error and using the quantum and the classical databases for the
  Sinai billiards in two and three dimensions, we concluded that the
  semiclassical error (measured in units of the mean level spacing) is
  independent of the dimensionality, and diverges at most as $\log
  \hbar$. This is in contrast with previous estimates. The classical
  spectrum of lengths of periodic orbits was studied and shown to be
  correlated in a way which induces the expected (random matrix)
  correlations in the quantal spectrum, corroborating previous results
  obtained in systems in two dimensions. These and other subjects
  discussed in the report open the way to extending the semiclassical
  study to chaotic systems with more than two freedoms.

\end{abstract}

\\


\tableofcontents

\section{Introduction}
\label{sec:intro}

%
The main goal of ``quantum chaos'' is to unravel the special features
which characterize the quantum description of classically chaotic
systems \cite{LH89,Gut90}. The simplest time independent systems which
display classical chaos are two dimensional, and therefore most of the
research in the field focused on systems in 2D. However, there are
very good and fundamental reasons for extending the research to higher
number of dimensions. The present paper reports on our study of a
paradigmatic three-dimensional system: The 3D Sinai billiard. It is
the first analysis of a system in 3D which was carried out in depth
and detail comparable to the previous work on systems in 2D.

%
The most compelling motivation for the study of systems in 3D is the
lurking suspicion that the semiclassical trace formula \cite{Gut90}
--- the main tool for the theoretical investigations of quantum chaos
--- fails for $d > 2$, where $d$ is the number of freedoms. The
grounds for this suspicion are the following \cite{Gut90}. The
semiclassical approximation for the propagator does not exactly
satisfy the time-dependent Schr\"odinger equation, and the error is of
order $\hbar^2$ {\em independently of the dimensionality}. The
semiclassical energy spectrum, which is derived from the semiclassical
propagator by a Fourier transform, is therefore expected to deviate by
${\cal O}(\hbar^2)$ from the exact spectrum. On the other hand, the
mean spacing between adjacent energy levels is proportional to
$\hbar^d$ \cite{LLqm} for systems in $d$ dimensions. Hence, the figure
of merit of the semiclassical approximation, which is the expected
error expressed in units of the mean spacing, is ${\cal
  O}(\hbar^{2-d})$, which diverges in the semiclassical limit $\hbar
\rightarrow 0$ when $d > 2$! If this argument were true, it would have
negated our ability to generalize the large corpus of results obtained
semiclassically, and checked for systems in 2D, to systems of higher
dimensions. Amongst the primary victims would be the semiclassical
theory of spectral statistics, which attempts to explain the universal
features of spectral statistics in chaotic systems and its relation to
random matrix theory (RMT) \cite{Ber85,BK96}. RMT predicts spectral
correlations on the range of a single spacing, and it is not likely
that a semiclassical theory which provides the spectrum with an
uncertainty which exceeds this range, can be applicable or relevant.
The available term by term generic corrections to the semiclassical
trace formula \cite{GA93,AG93,VWR94} are not sufficient to provide a
better estimate of the error in the semiclassically calculated energy
spectrum. To assess the error, one should substitute the term by term
corrections in the trace formula or the spectral $\zeta$ function
which do not converge in the absolute sense on the real energy axis.
Therefore, to this date, this approach did not provide an analytic
estimate of the accuracy of the semiclassical spectrum.

Under these circumstances, we initiated the present work which
addressed the problem of the semiclassical accuracy using the approach
to be described in the sequel. Our main result is that in contrast
with the estimate given above, the semiclassical error (measured in
units of the mean spacing) is {\em independent}\/ of the
dimensionality. Moreover, a conservative estimate of the upper bound
for its possible divergence in the semiclassical limit is ${\cal O}(|
\log \hbar |)$. This is a very important conclusion. It allows one to
extend many of the results obtained in the study of quantum chaos in
2D to higher dimensions, and justifies the use of the semiclassical
approximation to investigate special features which appear only in
higher dimensions. We list a few examples of such effects:
%
\begin{itemize}

\item The dual correspondence between the spectrum of quantum energies
  and the spectrum of actions of periodic orbits
  \cite{ADDKK93,Coh98,CPS98} was never checked for systems in more
  than 2D. However, if the universality of the {\em quantum}\/
  spectral correlations is independent of the number of freedoms, the
  corresponding range of correlations in the spectrum of {\em
    classical\/} actions is expected to depend on the dimensionality.
  Testing the validity of this prediction, which is derived by using
  the trace formula, is of great importance and interest. It will be
  discussed at length in this work.

\item The full range of types of stabilities of classical periodic
  orbits that includes also the loxodromic stability \cite{Gut90} can
  be manifest only for $d > 2$.

\item Arnold's diffusion in the KAM regime is possible only for $d >
  2$ (even though we do not encountered it in this work).

\end{itemize}
Having stated the motivations and background for the present study, we
shall describe the strategy we chose to address the problem, and the
logic behind the way we present the results in this report.

%
The method we pursued in this first exploration of quantum chaos in
3D, was to perform a comprehensive semiclassical analysis of a {\em
  particular}\/ yet typical system in 3D, which has a well studied
counterpart in 2D. By comparing the exact quantum results with the
semiclassical theory, we tried to identify possible deviations which
could be attributed to particular failures of the semiclassical
approximation in 3D. The observed deviations, and their dependence on
$\hbar$ and on the dimensionality, were used to assess the
semiclassical error and its dependence on $\hbar$. Such an approach
requires the assembly of an accurate and complete databases for the
quantum energies and for the classical periodic orbits. This is a very
demanding task for chaotic systems in 3D, and it is the main reason
why such studies were not performed before.

%
When we searched for a convenient system for our study, we turned
immediately to billiards. They are natural paradigms in the study of
classical and quantum chaos. The classical mechanics of billiards is
simpler than for systems with potentials: The energy dependence can be
scaled out, and the system can be characterized in terms of purely
geometric data. The dynamics of billiards reduces to a mapping through
the natural Poincar\'e section which is the billiard's boundary. Much
is known about classical billiards in the mathematical literature
(e.g.\ \cite{Tab95}), and this information is crucial for the
semiclassical application. Billiards are also very convenient from the
quantal point of view. There are specialized methods to quantize them
which are considerably simpler than those for potential systems
\cite{KS84}. Some of them are based on the Boundary Integral Method
(BIM) \cite{BW84}, the KKR method \cite{Ber81}, the scattering
approach \cite{DS92a,SS95} and various improvements thereof
\cite{VS95,Pro97,Pro97a}. The classical scaling property is manifest
also quantum mechanically. While for potential systems the energy
levels depend in a complicated way on $\hbar$ and the classical
actions are non-trivial functions of $E$, in billiards, both the
quantum energies and the classical actions scale trivially in $\hbar$
and $\sqrt{E}$, respectively, which simplifies the analysis
considerably.

\begin{figure}[p]
  \begin{center}
    \leavevmode

    \begin{tabular}{c}
      \psfig{figure=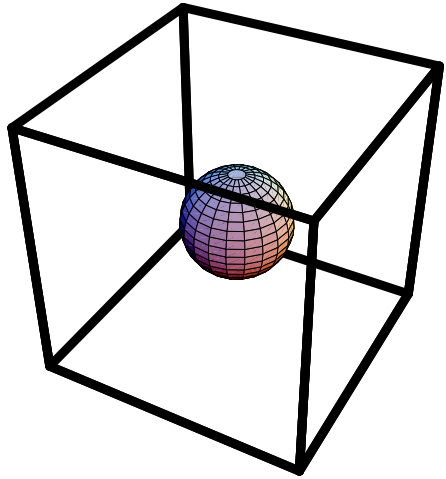,width=6.5cm} \\
      (a) \\
      \mbox{} \vspace{1cm} \\
      \begin{tabular}{cc}
        \psfig{figure=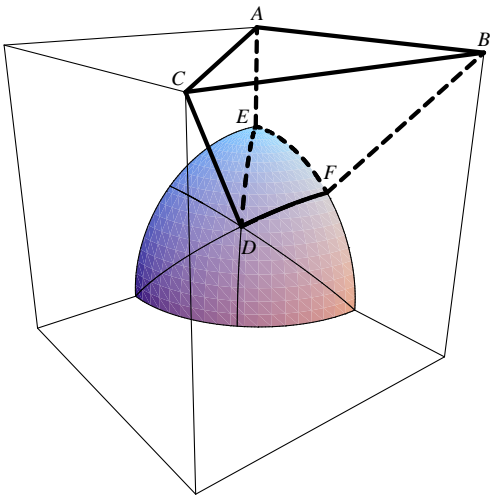,width=6.5cm} &
        \psfig{figure=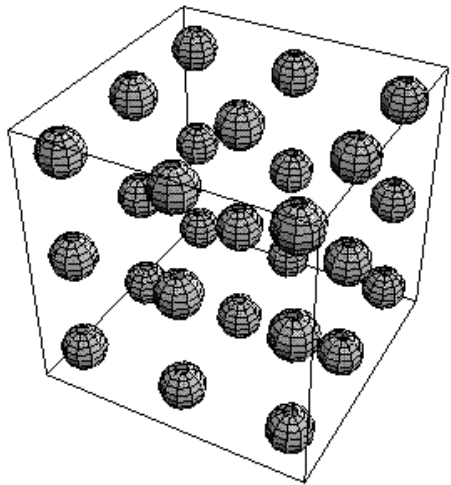,width=6.5cm} \\
        (b) & (c)
      \end{tabular}
    \end{tabular}

  \end{center}

  \caption{Three representations of the 3D Sinai billiard:
    (a) original, (b) 48-fold desymmetrized (maximal desymmetrization)
    into the fundamental domain, (c) unfolded to $\bbbr^3$.}

  \label{fig:sb}

\end{figure}

%
The particular billiard we studied is the {\em 3D Sinai billiard}. It
consists of the free space between a 3-torus of side $S$ and an
inscribed sphere of radius $R$, where $2R < S$. It is the natural
extension of the familiar 2D Sinai billiard, and it is shown in figure
\ref{fig:sb} using three complementary representations.
The classical dynamics consists of specular reflections from the
sphere. If the billiard is desymmetrized, specular reflections from
the symmetry planes exist as well. The 3D Sinai billiard has several
advantages. It is one of the very few systems in 3D which are
rigorously known to be ergodic and mixing \cite{Sin70,Nak93,BR97}.
Moreover, since its introduction by Sinai and his proof of its
ergodicity \cite{Sin70}, the 2D Sinai billiard was subject to thorough
classical, quantal and semiclassical investigations
\cite{Sin70,Ber81,BGS84,Bun85,DS92,Nak93,SS95,Bun95}. Therefore, much
is known about the 2D Sinai billiard and this serves us as an
excellent background for the study of the 3D counterpart. The
symmetries of the 3D Sinai billiard greatly facilitate the quantal
treatment of the billiard. Due to the spherical symmetry of the
inscribed obstacle and the cubic-lattice symmetry of the billiard (see
figure \ref{fig:sb}(c)) we are able to use the KKR method
\cite{KR54,Kor47,HS61,Ber81} to numerically compute the energy levels.
This method is superior to the standard methods of computing generic
billiard's levels. In fact, had we used the standard methods with our
present computing resources, it would have been possible to obtain
only a limited number of energy levels with the required precision.
The KKR method enabled us to compute many thousands of energy levels
of the 3D Sinai billiard. The fact that the billiard is symmetric
means that the Hamiltonian is block-diagonalized with respect to the
irreducible representations of the symmetry group \cite{Tin64}. Each
block is an independent Hamiltonian which corresponds to the
desymmetrized billiard (see figure \ref{fig:sb}(b)) for which the
boundary conditions are determined by the irreducible representations.
Hence, with minor changes one is able to compute a few independent
spectra that correspond to the same 3D desymmetrized Sinai billiard
but with different boundary conditions --- thus one can easily
accumulate data for spectral statistics. On the classical level, the
3D Sinai billiard has the great advantage of having a symbolic
dynamics. Using the centers of spheres which are positioned on the
infinite $\bbbz^3$ lattice as the building blocks of this symbolic
dynamics, it is possible to uniquely encode the periodic orbits of the
billiard \cite{Bun95,Sch96}. This construction, together with the
property that periodic orbits are the single minima of the length
(action) function \cite{Bun95,Sch96}, enables us to systematically
find all of the periodic orbits of the billiard, which is crucial for
the application of the semiclassical periodic orbit theory. We
emphasize that performing a systematic search of periodic orbits of a
given billiard is far from being trivial (e.g.\
\cite{Gut90,BK92,BD97,Han95,HC95}) and there is no general method of
doing so. The existence of such a method for the 3D Sinai billiard was
a major factor in favour of this system.

\begin{figure}[p]
  \begin{center}
    \leavevmode

    \begin{tabular}{c}
      \psfig{figure=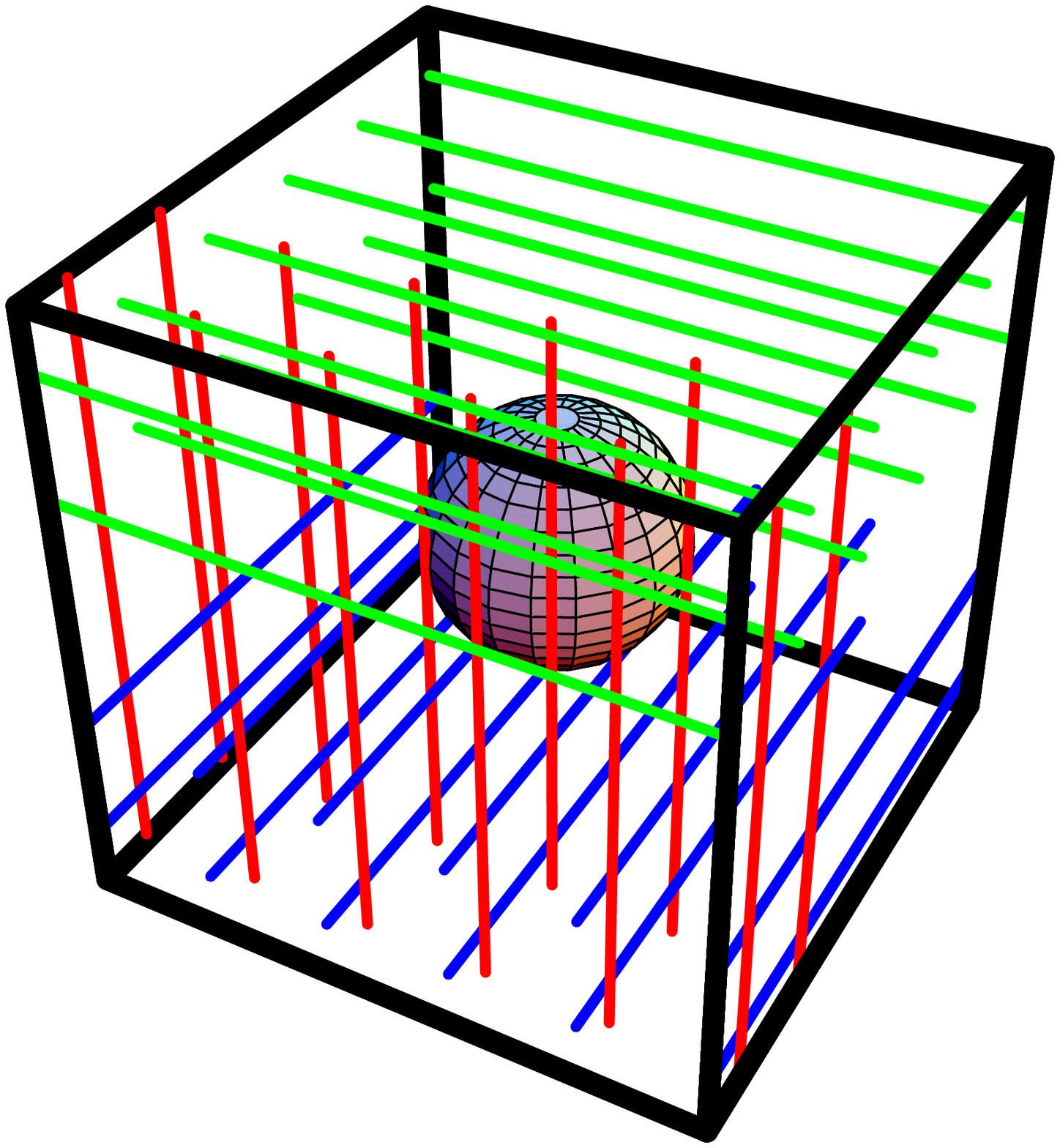,height=8cm} \\
      \mbox{} \vspace{1cm} \\
      \psfig{figure=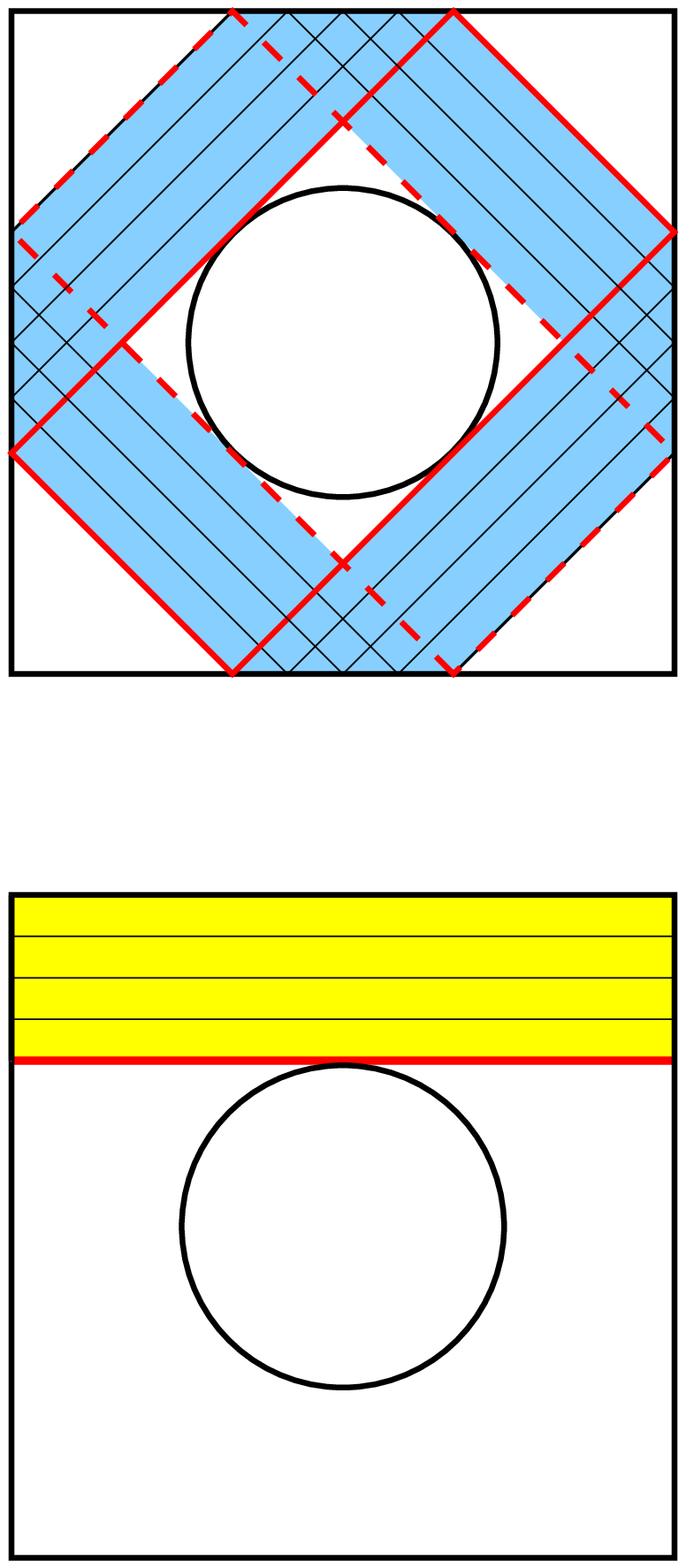,height=6cm,angle=270} \\
      \mbox{}
    \end{tabular}

    \caption{Some bouncing-ball families in the 3D Sinai billiard.
      Upper figure: Three families parallel to the $x, y$ and $z$
      axes. Lower figure: Top view of two families.}

    \label{fig:bb}

  \end{center}
\end{figure}

%
The advantages of the 3D Sinai billiard listed above are gained at the
expense of some problematic features which emerge from the cubic
symmetry of the billiard. In the billiard there exist families of
periodic, neutrally stable orbits, the so called ``bouncing-ball''
families that are illustrated in figure \ref{fig:bb}. The
bouncing-ball families are well-known from studies of, e.g., the 2D
Sinai and the stadium billiards \cite{Ber81,SS95,SSCL93,PSSU97}. These
periodic manifolds have zero measure in phase space (both in 2D and in
3D), but nevertheless strongly influence the dynamics. They are
responsible for the long (power-law) tails of some classical
distributions \cite{DA96,FS95}. They are also responsible for
non-generic effects in the quantum spectral statistics, e.g., large
saturation values of the number variance in the 2D Sinai and stadium
billiards \cite{SSCL93}. The most dramatic visualization of the effect
of the bouncing-ball families appears in the function $D(l) \equiv
\sum_n \cos(k_n l)$ --- the ``quantal length spectrum''. The lengths
$l$ that correspond to the bouncing-ball families are characterized by
large peaks that overwhelm the generic contributions of unstable
periodic orbits \cite{PSSU97} (as is exemplified by figure
\ref{fig:bare-ls}). In the 3D Sinai billiard the undesirable effects
are even more apparent than for the 2D billiard. This is because, in
general, they occupy 3D volumes rather than 2D areas in configuration
space and consequently their amplitudes grow as $k^1$ (to be
contrasted with $k^0$ for unstable periodic orbits). Moreover, for $R
< S/2$ there is always an infinite number of families present in the
3D Sinai billiard compared to the finite number which exists in the 2D
Sinai and the stadium billiards. The bouncing balls are thoroughly
discussed in the present work, and a large effort was invested in
devising methods by which their effects could be filtered out.

%
After introducing the system to be studied, we shall explain now the
way by which we present the results. The semiclassical analysis is
based on the exact quantum spectrum, and on the classical periodic
orbits. Hence, the first sections are dedicated to the discussion of
the exact quantum and classical dynamics in the 3D Sinai billiard, and
the semiclassical analysis is deferred to the last sections. The
sections are grouped as follows:
\begin{itemize}

\item Quantum mechanics and spectral statistics (sections
  \ref{sec:quantization} and \ref{sec:quantal-spectral-statistics}).

\item Classical periodic orbits (section \ref{sec:classical-pos}).

\item Semiclassical analysis (sections \ref{sec:sc-analysis},
  \ref{sec:sc-accuracy}, \ref{sec:sc-specstat}).

\end{itemize}

%
In section \ref{sec:quantization} we describe the KKR method which was
used to numerically compute the quantum spectrum. Even though it is a
rather technical section, it gives a clear idea of the difficulties
encountered in the quantization of this system, and how we used
symmetry considerations and number-theoretical arguments to reduce the
numerical effort considerably. The desymmetrization of the billiard
according to the symmetry group is worked out in detail. This section
ends with a short explanation of the methods used to ensure the
completeness and the accuracy of the spectrum.

The study of spectral statistics, section
\ref{sec:quantal-spectral-statistics}, starts with the analysis of the
integrable billiard ($R = 0$) case. This spectrum is completely
determined by the underlying classical bouncing-ball manifolds which
are classified according to their dimensionality. The two-point form
factor in this case is {\em not}\/ Poissonian, even though the system
is integrable. Rather, it reflects the number-theoretical degeneracies
of the $\bbbz^3$ lattice resulting in non-generic correlations.
Turning to the chaotic ($R > 0$) cases, we investigate some standard
statistics (nearest-neighbor, number variance) as well as the
auto-correlations of the spectral determinant, and compare them to the
predictions of RMT. The main conclusion of this section is that the
spectral fluctuations in the 3D Sinai billiard belong to the same
universality class as in the 2D analogue.

Section \ref{sec:classical-pos} is devoted to the systematic search of
the periodic orbits of the 3D Sinai billiard. We rely heavily on a
theorem that guarantees the uniqueness of the coding and the
variational minimality of the periodic orbit lengths. The necessary
generalizations for the desymmetrized billiard are also explained.
Once the algorithm for the computation of periodic orbits is outlined,
we turn to the definition of the spectrum of lengths of periodic
orbits and to the study of its statistics. The number of periodic
orbits with lengths smaller than $L$ is shown to proliferate
exponentially. We check also classical sum rules which originate from
ergodic coverage, and observe appreciable corrections to the leading
term due to the infinite horizon of the Sinai billiard. Turning our
attention to the two-point statistics of the classical spectrum, we
show that it is not Poissonian. Rather, there exist correlations which
appear on a scale larger than the nearest spacing. This has very
important consequences for the semiclassical analysis of the spectral
statistics. We study these correlations and offer a dynamical
explanation for their origin.

The semiclassical analysis of the billiard is the subject of section
\ref{sec:sc-analysis}. As a prelude, we propose and use a new method
to verify the completeness and accuracy of the quantal spectrum, which
is based on a ``universal'' feature of the classical length spectrum
of the 3D Sinai billiard. The main purpose of this section is to
compare the quantal computations to the semiclassical predictions
according to the Gutzwiller trace formula, as a first step in our
study of its accuracy. Since we are interested in the generic unstable
periodic orbits rather than the non-generic bouncing balls, special
effort is made to eliminate the the effects of the latter. This is
accomplished using a method that consists of taking the derivative
with respect to a continuous parameterization of the boundary
conditions on the sphere.

In section \ref{sec:sc-accuracy} we embark on the task of estimating
the semiclassical error of energy levels. We first define the measures
with which we quantify the semiclassical error, and demonstrate some
useful statistical connections between them. We then show how these
measures can be evaluated for a given system using its quantal and
semiclassical length spectra. We use the databases of the 2D and 3D
Sinai billiards to derive the estimate of the semiclassical error
which was already quoted above: The semiclassical error (measured in
units of the mean spacing) is {\em independent}\/ of the
dimensionality, and a conservative estimate of the upper bound for its
possible divergence in the semiclassical limit is ${\cal O}(| \log
\hbar |)$.

Once we are reassured of the reliability of the trace formula in 3D,
we return in section \ref{sec:sc-specstat} to the spectral statistics
of the quantized billiard. The semiclassical trace formula is
interpreted as an expression of the duality between the quantum
spectrum and the classical spectrum of lengths. We show how the length
correlations in the classical spectrum induce correlations in the
quantum spectrum, which reproduce rather well the RMT predictions.

The work is summarized in section \ref{sec:summary}.

%
To end the introductory notes, a review of the existing literature is
in order. Only very few systems in 3D were studied in the past. We
should first mention the measurements of 3D acoustic cavities
\cite{Wea89,BLSS91,DSW94,EGLLNO95,EGLNO} and electromagnetic
(microwaves) cavities \cite{DKS95,AGHRRRSW96,ADGHR97,DSBK97}. The
measured frequency spectra were analyzed and for irregular shapes
(notably the 3D Sinai billiard) the level statistics conformed with
the predictions of RMT. Moreover, the length spectra showed peaks at
the lengths of periodic manifolds, but no further quantitative
comparison with the semiclassical theory was attempted. However, none
of the experiments is directly relevant to the quantal (scalar)
problem since the acoustic and electromagnetic vector equations cannot
be reduced to a scalar equation in the configurations chosen.
Therefore, these experiments do not constitute a direct analogue of
quantum chaos in 3D. This is in contrast with flat and thin microwave
cavities which are equivalent (up to some maximal energy) to 2D
quantal billiards.

%
A few 3D billiards were discussed theoretically in the context of
quantum chaos. Polyhedral billiards in the 3D hyperbolic space with
constant negative curvature were investigated by Aurich and Marklof
\cite{AM96}. The trace formula in this case is exact rather than
semiclassical, and thus the issue of the semiclassical accuracy is not
relevant. Moreover, the tetrahedral that was treated had exponentially
growing multiplicities of lengths of classical periodic orbits, and
hence cannot be considered as generic.
Prosen considered a 3D billiard with smooth boundaries and 48-fold
symmetry \cite{Pro97,Pro97a} whose classical motion was almost
completely (but not fully) chaotic. He computed many levels and found
that level statistics reproduce the RMT predictions with some
deviations. He also found agreement with Weyl's law (smooth density of
states) and identified peaks of the length spectrum with lengths of
periodic orbits. The majority of high-lying eigenstates were found to
be uniformly extended over the energy shell, with notable exceptions
that were ``scarred'' either on a classical periodic orbit or on a
symmetry plane. Henseler, Wirzba and Guhr treated the $N$-sphere
scattering systems in 3D \cite{HWG97} in which the quantum mechanical
resonances were compared to the predictions according to the
Gutzwiller trace formula. A good agreement was observed for the
uppermost band of resonances and no agreement for other bands which
are dominated by diffraction effects. Unfortunately, conclusive
results were given only for non-generic configurations of two and
three spheres for which all the periodic orbits are planar. In
addition, it is not clear whether one can infer from the accuracy of
complex scattering resonances to the accuracy of real energy levels in
bound systems. Recently, Sieber \cite{Sie98} calculated the $4 \times
4$ stability (monodromy) matrices and the Maslov indices for general
3D billiards and gave a practical method to compute them, which
extended our previous results for the 3D Sinai billiard
\cite{PS95,Pri97}.


\section{Quantization of the 3D Sinai billiard}
\label{sec:quantization}
%
In the present section we describe the KKR determinant method
\cite{KR54,Kor47,Ewa21,HS61} to compute the energy spectrum of the 3D
Sinai billiard, and the results of the numerical computations. The KKR
method, which was used by Berry for the 2D Sinai billiard case
\cite{Ber81}, is most suitable for our purpose since it allows to
exploit the symmetries of the billiard to reduce the numerical effort
considerably. The essence of the method is to convert the
Schr\"odinger equation and the boundary conditions into a {\em
  single}\/ integral equation. The spectrum is then the set of real
wavenumbers $k_n$ where the corresponding secular determinant
vanishes. As a matter of fact, we believe that only with the KKR
method could we obtain a sufficiently accurate and extended spectrum
for the quantum 3D Sinai billiard. We present in this section also
some numerical aspects and verify the accuracy and completeness of the
computed levels.

We go into the technical details of the quantal computation because we
wish to show the high reduction factor which is gained by the KKR
method. Without this significant reduction the numerical computation
would have resulted in only a very limited number of levels
\cite{DKS95,ADGHR97}. The reader who is not interested in these
technical details should proceed to subsection \ref{subsec:low-lying}.
To avoid ambiguities, we strictly adhere to the conventions of
\cite{VMK88}.

\subsection{The KKR determinant}
\label{subsec:kkr-det}
%
We first consider the 3D ``Sinai torus'', which is the free space
outside of a sphere of radius $R$ embedded in a 3--torus of side
length $S$ (see fig.\ \ref{fig:sb}). The Schr\"odinger equation of an
electron of mass $m$ and energy $E$ is reduced to the Helmholtz
equation:
\begin{equation}
  \nabla^2 \psi + k^2 \psi = 0 \: , 
  \; \; \;
  k \equiv \sqrt{2 m E} / \hbar \: .
\end{equation}
The boundary conditions on the sphere are taken to be the general
linear (self-adjoint) conditions:
\begin{equation}
    \kappa \cos \alpha \cdot \psi + 
    \sin \alpha \cdot \partial_{\hat{n}} \psi = 0 
    \: ,
  \label{eq:mbc-alpha}
\end{equation}
where $\hat{n}$ is the normal pointing outside the billiard, $\kappa$
is a parameter with dimensions of $k$, and $\alpha \in [0, \pi / 2]$
is an angle that interpolates between Dirichlet ($\alpha = 0$) and
Neumann ($\alpha = \pi / 2$) conditions. These ``mixed'' boundary
conditions will be needed in section \ref{sec:sc-analysis} when
dealing with the semiclassical analysis. Applying the KKR method, we
obtain the following quantization condition (see \cite{Pri97} for a
derivation and for details):
\begin{equation}
  \det \left[ A_{lm, l'm'}(k) +
    k P_l(kR; \kappa, \alpha) \delta_{ll'} \delta_{mm'} \right] = 0 \; ,
   \label{eq:kkr-det}
\end{equation}
\[
\; \; \; 
l, l' = 0, 1, 2, \ldots , 
\; \; 0 \leq m \leq l 
\: , 0 \leq m' \leq l'
\: ,
\]
where $k$ is the wavenumber under consideration and:
\begin{eqnarray}
  A_{lm, l'm'}(k)
  & \equiv &
  4 \pi i^{l-l'} \sum_{LM} i^{-L} C_{LM, lm, l'm'} D_{LM}(k) \; ,
  \nonumber \\
  & & 
  L = 0, 1, 2, \ldots \ , \; M = 0, \ldots , L \ ,
  \label{eq:almlm}
  \\
  D_{LM}(k)
  & \equiv &
  (-i k) \left[ \sum_{\vec{\rho} \in \bbbz^3 / \{\vec{0}\}}
  h^+_L(k S \rho) Y^*_{LM}(\Omega_{\vec{\rho}}) +
  \frac{1}{\sqrt{4 \pi}} \delta_{L0} \right] \ ,
  \label{eq:dlm}
  \\
  C_{LM, lm, l'm'}
  & \equiv &
   \int_{0}^{\pi} {\rm d}\theta \int_{0}^{2 \pi} {\rm d}\phi \,
     Y_{LM}(\theta, \phi) Y^*_{lm}(\theta, \phi) Y_{l'm'}(\theta, \phi) \ ,
  \label{eq:clm}
  \\
  P_l(kR; \kappa, \alpha)
  & \equiv &
  \frac{\kappa R \cos \alpha \cdot n_l(kR) {-} 
    k R \sin \alpha \cdot n_l'(kR)}
  {\kappa R \cos \alpha \cdot j_l(kR) {-} 
    k R \sin \alpha \cdot j_l'(kR)}
  \\
  & = &
  \cot [ \eta_l (kR; \kappa, \alpha) ] \: .
  \label{eq:pl}
\end{eqnarray}
In the above $j_l$, $n_l$, $h^+_l$ are the spherical Bessel, Neumann
and Hankel functions, respectively \cite{VMK88}, $Y_{lm}$ are the
spherical harmonics \cite{VMK88} with argument $\Omega_{\vec{\rho}}$
in the direction of $\vec{\rho}$, and $\eta_l$ are the scattering
phase shifts from the sphere, subject to the boundary conditions
(\ref{eq:mbc-alpha}).

The physical input to the KKR determinant is distributed in a
systematic way: The terms $A_{lm, l'm'}(k)$ contain information only
about the structure of the underlying $\bbbz^3$ lattice, and are
independent of the radius $R$ of the inscribed sphere. Hence they are
called the ``structure functions'' \cite{KR54,HS61}. Moreover, they
depend on a smaller number of ``building block'' functions $D_{LM}(k)$
which contain the infinite lattice summations. The diagonal term $k
P_l(kR) \delta_{ll'} \delta_{mm'}$ contains the information about the
inscribed sphere, and is expressed in terms of the scattering phase
shifts from the sphere. This elegant structure of the KKR determinant
(\ref{eq:kkr-det}) prevails in more general situations and remains
intact even if the $\bbbz^3$ lattice is replaced by a more general
one, or if the ``hard'' sphere is replaced by a ``soft'' spherical
potential with a finite range (``muffin-tin'' potential)
\cite{KR54,HS61,Kor47}. This renders the KKR a powerful quantization
method. In all these cases the structure functions $A_{lm, l'm'}$
depend only on the underlying lattice, and the relation (\ref{eq:pl})
holds with the appropriate scattering matrix. Thus, in principle, the
structure functions (or rather $D_{LM}$) can be tabulated once for a
given lattice (e.g.\ cubic) as functions of $k$, and only $P_l$ need
to be re-calculated for every realization of the potential (e.g.\ 
changing $R$). This makes the KKR method very attractive also for a
large class of generalizations of the 3D Sinai billiard.

The determinant (\ref{eq:kkr-det}) is not yet suitable for numerical
computations. This is because the lattice summations in $D_{LM}$ are
only {\it conditionally convergent}\/ and have to be resummed in order
to give absolutely and rapidly convergent sums. This is done using the
Ewald summation technique, which is described in appendices
\ref{app:ewald}--\ref{app:dlm3}. The further symmetry reductions of
the KKR determinant, which are one of the most important advantages of
this method, are discussed in the following.

\subsection{Symmetry considerations}
\label{subsec:kkr-symmtery}

As can be seen from equations (\ref{eq:almlm}--\ref{eq:pl}) and from
appendix \ref{app:ewald}, the main computational effort involved in
computing the KKR determinant is consumed in the lattice sums
$D_{LM}(k)$ which need to be evaluated separately for every $k$.
Therefore, it is imperative to use every possible means to economize
the computational effort invested in calculating these functions. For
this purpose, we shall exploit the cubic symmetry of the 3D Sinai
billiard as well as other relations that drastically reduce the
computational effort.

\subsubsection{Group--theoretical resummations}

For the practical (rapidly convergent) computation, the functions
$D_{LM}$ are decomposed into three terms which are given in appendix
\ref{app:ewald} (see also appendix \ref{app:ewald-physical}).
Equations (\ref{eq:dlmewa1})--(\ref{eq:dlm3ewa1}) express
$D_{LM}^{(2)}$ as a sum over the direct cubic lattice, whereas,
$D_{LM}^{(1)}$ is a sum over the reciprocal cubic lattice, which is
also a cubic lattice. Thus, both sums can be represented as:
\begin{equation}
  D_{LM}^{(j)}(k) 
  =
  \sum_{\vec{\rho} \in \bbbz^3}
    f^{(j)}(\rho; k) Y^{*}_{LM}(\Omega_{\vec{\rho}}) \; ,
    \; \; \; j = 1, 2 \: .
  \label{eq:dlm-sr1}
\end{equation}
We show in appendix \ref{app:lrho} that lattice sums of this kind can
be rewritten as:
\begin{equation}
  D_{LM}^{(j)}(k) =
    \sum_{\vec{\rho}_p} \frac{f^{(j)}(\rho_p; k)}{l(\vec{\rho}_p)}
    \sum_{\hat{g} \in O_h} Y^{*}_{LM}(\Omega_{\hat{g} \vec{\rho}_p}) \; ,
  \label{eq:dlm-sr2}
\end{equation}
where $O_h$ is the cubic symmetry group \cite{Tin64}, and
$\vec{\rho}_p \equiv (i_1, i_2, i_3)$ resides in the {\em fundamental
  section} $0 \leq i_1 \leq i_2 \leq i_3$. The terms $l(\vec{\rho}_p)$
are integers which are explicitly given in appendix \ref{app:lrho}.
The inner sums are independent of $k$, and can thus be tabulated once
for all. Hence the computation of the $k$ dependent part becomes 48
times more efficient (for large, finite lattices) when compared to
(\ref{eq:dlm-sr1}) due to the restriction of $\vec{\rho}_p$ to the
fundamental section.

A further reduction can be achieved by a unitary transformation from
the $\{Y_{LM}\}$ basis to the more natural basis of the irreducible
representations (irreps) of $O_h$:
\begin{equation}
  Y^{(\gamma)}_{LJK} (\Omega) \equiv
    \sum_{M} a^{(L)*}_{\gamma J K, M} Y_{LM}(\Omega) \ ,
  \label{eq:ylj-def}
\end{equation}
where $\gamma \in [1, \ldots, 10]$ denotes the irrep under
consideration, $J$ counts the number of the inequivalent irreps
$\gamma$ contained in $L$, and $K = 1, \ldots, \dim (\gamma)$ is the
row index within the irrep. The functions $Y^{(\gamma)}_{LJK}$ are
known as the ``cubic harmonics'' \cite{LB47}. Combining
(\ref{eq:dlm-sr2}) and (\ref{eq:ylj-def}), and using the unitarity of
the transformation as well as the ``great orthonormality theorem'' of
group theory \cite{Tin64} we arrive at:
\begin{eqnarray}
  D_{LM}^{(j)}(k)
  & = &
  \sum_{J} a^{(L)*}_{s J, M} D_{LJ}^{(j)}(k)
  \\
  D_{LJ}^{(j)}(k) \: ,
  & = &
  48 \sum_{\vec{\rho_p}} \frac{f^{(j)}(\rho_p; k)}{l(\vec{\rho_p})}
    Y^{(s)*}_{LJ}(\Omega_{\vec{\rho_p}}) \ .
  \label{eq:dlj}
\end{eqnarray}
The superscript $(s)$ denotes the totally symmetric irrep. The
constant coefficients $a^{(L)*}_{s J, M}$ can be taken into the
(constant) coefficients $C_{LM, lm, l'm'}$ resulting in:
\begin{eqnarray}
  A_{lm, l'm'}(k)
  & = &
  4 \pi i^{l-l'} \sum_{LJ} i^{-L} D_{LJ}(k) C_{LJ, lm, l'm'}
  \\
  D_{LJ}(k)
  & = &
  D_{LJ}^{(1)}(k) + D_{LJ}^{(2)}(k) + D_{00}^{(3)}(k)\delta_{L0}
  \\
  C_{LJ, lm, l'm'}
  & = &
  \sum_{M} a^{(L)*}_{s J, M} C_{LM, lm, l'm'} \ .
\end{eqnarray}

We show in appendix \ref{app:ylj} that for large $L$ the number of
$D_{LJ}(k)$'s is smaller by a factor $\approx 1/48$ than the number of
$D_{LM}(k)$'s. This means that the entries of the KKR determinant are
now computed using a substantially smaller number of building blocks
for which lattice summations are required. Thus, in total, we gain a
saving factor of $48^2 = 2304$ over the more naive scheme
(\ref{eq:almlm}--\ref{eq:clm}).

\subsubsection{Number--theoretical resummations}
\label{subsubsec:nt-deg-kkr}
%
In the above we grouped together lattice vectors with the same
magnitude, using the geometrical symmetries of the cubic lattice. One
can gain yet another reduction factor in the computational effort by
taking advantage of a phenomenon which is particular to the cubic
lattice and stems from {\em number theory}. The lengths of lattice
vectors in the fundamental sector show an appreciable degeneracy,
which is not connected with the $O_h$ symmetry. For example, the
lattice vectors $(5, 6, 7)$ and $(1, 3,10)$ have the same magnitude,
$\sqrt{110}$, and are not geometrically conjugate by $O_h$. This
number--theoretical degeneracy is both frequent and significant, and
we use it in the following way. Since the square of the magnitude is
an integer we can write:
\begin{equation}
  D_{LJ}^{(j)}(k) =
  \sum_{n=1}^{\infty} f^{(j)}(\rho_p{=}\sqrt{n}; k)
  \left[ \sum_{\rho_p^2=n} \frac{48}{l(\vec{\rho_p})}
    Y^{(s)*}_{LJ}(\Omega_{\vec{\rho_p}}) \right] \ .
  \label{eq:dlj-sr3}
\end{equation}
The inner sums incorporate the number theoretical degeneracies.  They
are $k$ independent, and therfore can be tabulated once for all.

To show the efficiency of (\ref{eq:dlj-sr3}) let us restrict our
lattice summation to $\rho_p \leq \rho_{\rm max}$ (which we always do
in practice). For large $\rho_{\rm max}$ the number of lattice vectors
in the fundamental domain is $\pi \rho_{\rm max}^3 / 36$, and the
number of summands in (\ref{eq:dlj-sr3}) is at most $\rho_{\rm
  max}^2$. Thus, the saving factor is at least $\pi \rho_{\rm max} /
36$. In fact, as shown in appendix \ref{app:nt-deg}, there are only
(asymptotically) $(5/6) \rho_{\rm max}^2$ terms in (\ref{eq:dlj-sr3}),
which sets the saving factor due to number--theoretical degeneracy to
be $\pi \rho_{\rm max} / 30$. In practice, $\rho_{\rm max} = {\cal
  O}(100)$ and this results in a reduction factor of about 10, which
is very significant.

\subsubsection{Desymmetrization}
\label{subsec:kkr-desym}

The symmetry of the 3D Sinai torus implies that the wavefunctions can
be classified according to the irreps of $O_h$ \cite{Tin64}.
Geometrically, each such irrep corresponds to specific boundary
conditions on the symmetry planes that define the desymmetrized 3D
Sinai billiard (see figure \ref{fig:sb}). This allows us to
``desymmetrize'' the billiard, that is to restrict ourselves to the
fundamental domain with specific boundary conditions instead of
considering the whole 3--torus. We recall that the boundary conditions
on the sphere are determined by $P_l(k)$ and are independent of the
irrep under consideration. For simplicity, we shall restrict ourselves
to the two simplest irreps which are both one--dimensional:
\begin{description}
  
  \item{$\gamma = a$:} This is the totally antisymmetric irrep, which
    corresponds to Dirichlet boundary conditions on the planes.
  
  \item{$\gamma = s$:} This is the totally symmetric irrep, which
    corresponds to Neumann boundary conditions on the planes.

\end{description}

The implementation of this desymmetrization is straightforward (see
\cite{Pri97} for details) and results in a new secular equation:
\begin{equation}
  \det \left[ A_{lj, l'j'}^{(\gamma)}(k) +
    k P_l(kR) \delta_{ll'} \delta_{jj'} \right] = 0
  \label{eq:kkr-desym}
\end{equation}
where $\gamma$ is the chosen irrep and:
\begin{eqnarray}
  A_{lj, l'j'}^{(\gamma)}(k)
  & = &
  4 \pi i^{l-l'} \sum_{LJ} i^{-L} D_{LJ}(k) 
    C_{LJ, lj, l'j'}^{(\gamma)}
  \\
  C_{LJ, lj, l'j'}^{(\gamma)}
  & = &
  \sum_{mm'} a^{(l)}_{\gamma j, m}
    a^{(l')*}_{\gamma j', m'} C_{LJ, lm, l'm'} \\
  & = &
  \sum_{Mmm'} a^{(L)*}_{s J, M} a^{(l)}_{\gamma j, m}
    a^{(l')*}_{\gamma j', m'} C_{LM, lm, l'm'} \, .
\end{eqnarray}

The desymmetrization of the problem has a few advantages:
\begin{description}
  
\item[Computational efficiency:] In appendix \ref{app:ylj} we show
  that for large $L$'s the number of cubic harmonics
  $Y^{(\gamma)}_{LJK}$ that belong to a one--dimensional irrep is
  $1/48$ of the number of the spherical harmonics $Y_{LM}$.
  Correspondingly, if we truncate our secular determinant such that $L
  \leq L_{\rm max}$, then the dimension of the new determinant
  (\ref{eq:kkr-desym}) is only $1/48$ of the original one
  (\ref{eq:kkr-det}) for the fully symmetric billiard.  Indeed, the
  desymmetrized billiard has only $1/48$ of the volume of the
  symmetric one, and hence the density of states is reduced by $48$
  (for large $k$). However, due to the high cost of computing a
  determinant (or performing a Singular Value Decomposition)
  \cite{NAG90} the reduction in the density of
  states is over-compensated by the reduction of the matrix size,
  resulting in a saving factor of $48$. This is proven in appendix
  \ref{app:symm-reduction}, where it is shown in general that levels
  of desymmetrized billiards are computationally cheaper than those of
  billiards which possess symmetries. Applied to our case, the
  computational effort to compute a given number $N$ of energy levels
  of the desymmetrized billiard is $48$ times cheaper than computing
  $N$ levels of the fully symmetric billiard.
    
\item[Statistical independence of spectra:] For each irrep the
  spectrum is statistically independent of the others, since it
  corresponds to different boundary conditions. Thus, if the fully
  symmetric billiard is quantized, the resulting spectrum is the union
  of 10 independent spectra (there are 10 irreps of $O_h$
  \cite{Tin64}), and significant features such as level rigidity will
  be severely blurred \cite{Boh89}. To observe generic statistical
  properties and to compare with the results of RMT, one should
  consider each spectrum separately, which is equivalent to
  desymmetrizing the billiard.
    
\item[Rigidity:] The statistical independence has important practical
  consequences. Spectral rigidity implies that it is unlikely to find
  levels in close vicinity of each other. Moreover, the fluctuations
  in the spectral counting functions are bounded. Both features of
  rigidity are used in the numerical algorithm which computes the
  spectrum, and is described in more detail in section
  \ref{subsec:numerical-aspects}.

\end{description}

To summarize this subsection, we have demonstrated that the high
symmetry features of the 3D Sinai billiard are naturally incorporated
in the KKR method. This renders the computation of its spectrum much
more efficient than in the case of other, less symmetric 3D billiards.
Thus, we expect to get many more levels than the few tens that can be
typically obtained for generic billiards \cite{DKS95,ADGHR97}. In
fact, this feature is the key element which brought this project to a
successful conclusion. We note that other specialized computation
methods, which were applied to highly symmetric 3D billiards, also
resulted in many levels \cite{Pro97,Pro97a}.

This completes the theoretical framework established for the efficient
numerical computation of the energy levels. In the following we
discuss the outcome of the actual computations.

\subsection{Numerical aspects}
\label{subsec:numerical-aspects}

We computed various energy spectra, defined by different combinations
of the physically important parameters:
\begin{enumerate}
  
\item The radius $R$ of the inscribed sphere (the side $S$ was always
  taken to be $1$).
    
\item The boundary conditions on the sphere: Dirichlet / Neumann /
  mixed: $0 \leq \alpha \leq \pi / 2$.
    
\item The boundary conditions on the symmetry planes of the cube:
  Dirichlet / Neumann. These boundary conditions correspond to the
  antisymmetric / symmetric irrep of $O_h$, respectively. Due to the
  lattice periodicity, Dirichlet (Neumann) boundary conditions on the
  symmetry planes induce Dirichlet (Neumann) also on the planes
  between neighbouring cells.

\end{enumerate}
The largest spectral stretch that was obtained numerically
corresponded to $R=0.2$ and Dirichlet boundary condition everywhere.
It consisted of 6697 levels in the interval $ 0 < k \leq 281.078$. We
denote this spectrum in the following as the ``longest spectrum''.

\begin{figure}[p]
  \begin{center}
    \leavevmode 

    \psfig{figure=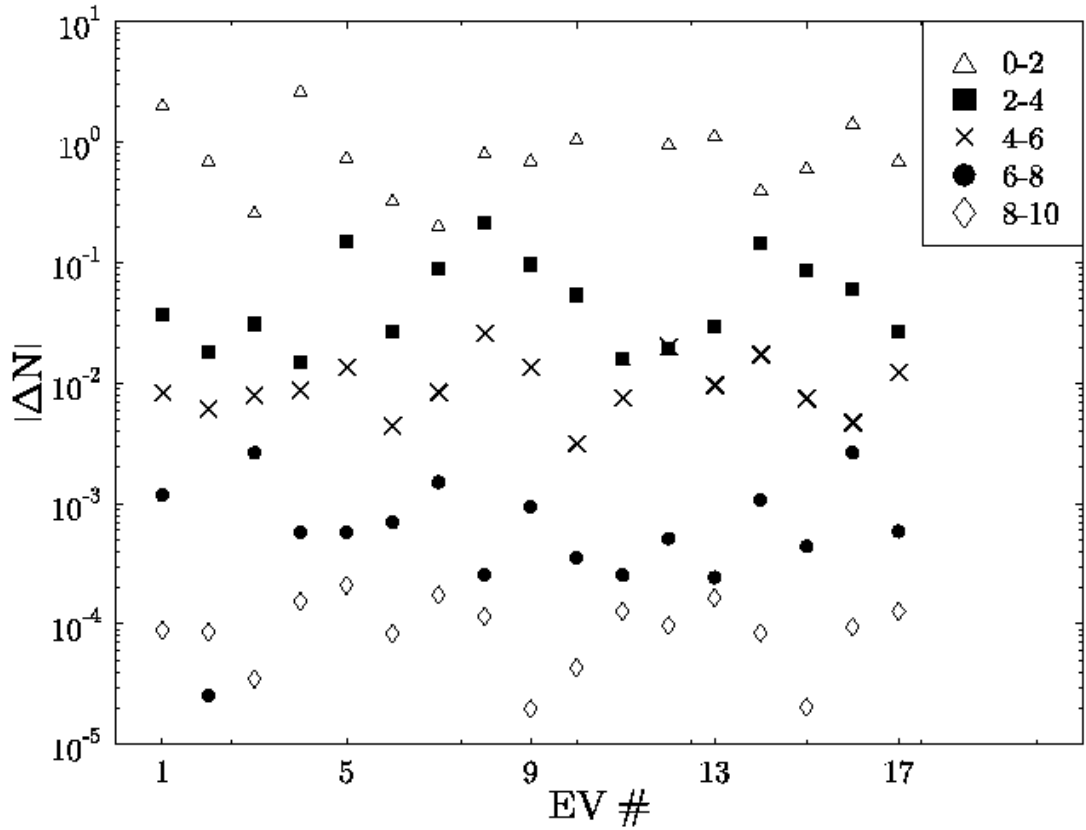,width=16cm}

    \caption{Accuracy of eigenvalues as a function of the number of 
      evanescent modes $l_{\rm evan}$. The case considered was $R =
      0.2$ and Dirichlet boundary conditions everywhere. The figure
      shows the absolute differences of the eigenvalues between two
      successive values of $l_{\rm evan}$, multiplied by the smooth
      level density. That is, ``0-2'' means $\bar{d}(k_n) \left|
        k_n(l_{\rm evan}{=}2) - k_n(l_{\rm evan}{=}0) \right| \equiv |
      \Delta N_n |$. We show 17 eigenvalues in the interval $199.5 < k
      < 200$.}

  \label{fig:evanescent-modes}

  \end{center}
\end{figure}

The practical application of (\ref{eq:kkr-desym}) brings about many
potential sources of divergence: The KKR matrix is infinite
dimensional in principle, and each of the elements is given as an
infinite sum over the cubic lattice. To regulate the infinite
dimension of the matrix we use a physical guideline, namely, the fact
that for $l > kR$ the phase shifts decrease very rapidly toward zero,
and the matrix becomes essentially diagonal. Therefore, a natural
cutoff is $l_{\rm max} = kR$, which is commonly used (e.g.\ 
\cite{SS95}).  In practice one has to go slightly beyond this limit,
and to allow a few evanescent modes: $l_{\rm max} = kR + l_{\rm
  evan}$. To find a suitable value of $l_{\rm evan}$ we used the
parameters of the longest spectrum and computed the 17 eigenvalues in
the interval $199.5 < k < 200$ with $l_{\rm evan} = 0, 2, 4, 6, 8, 10$
($l_{\rm max}$ has to be odd). We show in figure
\ref{fig:evanescent-modes} the successive accuracy of the computed
eigenvalues between consecutive values of $l_{\rm evan}$. The results
clearly indicate a 10-fold increase in accuracy with each increase of
$l_{\rm evan}$ by 2. A moderately high accuracy of $O(10^{-4})$
relative to level spacing requires $l_{\rm evan} = 8$ which was the
value we used in our computations.

To regulate the infinite lattice summations in $D_{LJ}$ we used
successively larger subsets of the lattice. The increase was such that
at least twice as many lattice points were used. Our criterion of
convergence was that the maximal absolute value of the difference
between successive computations of $D_{LJ}$ was smaller than a
prescribed threshold:
\begin{equation}
  \max_{LJ} \left| D_{LJ}^{i} - D_{LJ}^{i+1} \right| < \epsilon \, .
\end{equation}
The threshold $\epsilon = 10^{-6}$ was found to be satisfactory, and
we needed to use a sub-lattice with maximal radius of 161.

The KKR program is essentially a loop over $k$ which sweeps the
$k$-axis in a given interval. At each step the KKR matrix $M(k)$ is
computed, and then its determinant is evaluated. In principle,
eigenvalues are obtained whenever the determinant vanishes. In
practice, however, the direct evaluation of the determinant suffers
from two drawbacks:
\begin{itemize}
  
\item The numerical algorithms that are used to compute $\det M(k)$
  are frequently unstable. Hence, it is impossible to use them beyond
  some critical $k$ which is not very large.

\item For moderately large $k$'s, the absolute values of $\det M(k)$
  are very small numbers that result in computer underflows (in double
  precision mode), even for $k$-values which are not eigenvalues.
  
\item Due to finite precision and rounding errors, $\det M(k)$ never
  really vanishes for eigenvalues.

\end{itemize}
A superior alternative to the direct calculation of the determinant is
to use the Singular Value Decomposition (SVD) algorithm \cite{NAG90},
which is stable under any circumstances. In our case, $M$ is real and
symmetric, and the output are the ``singular values'' $\sigma_i$ which
are the absolute values of the eigenvalues of $M$. The product of all
of the singular values is equal to $|\det M|$, which solves the
stability problem. To cure the other two problems consider the
following ``conditioning measure'':
\begin{equation}
  r(k) \equiv \sum_{i=1}^{\dim M(k)} \ln \sigma_i (k) \, .
\end{equation}
The use of the logarithm circumvents the underflow problem. Moreover,
we always expect some of the smallest singular values to reflect the
numerical noise, and the larger ones to be physically relevant. Near
an eigenvalue, however, one of the ``relevant'' singular values must
approach zero, resulting in a ``dip'' in the graph of $r(k)$. Hence,
by tracking $r$ as a function of $k$, we locate its dips and take as
the eigenvalues the $k$ values for which the local minima of $r$ are
obtained. Frequently one encounters very shallow dips (typically $\ll
1$) which are due to numerical noise and should be discarded.

To ensure the location of all of the eigenvalues in a certain $k$
interval, the $k$-axis has to be sampled densely. However,
oversampling should be avoided to save computer resources. In order to
choose the sampling interval $\Delta k$ in a reasonable way, we
suggest the following. If the system is known to be classically
chaotic, then we expect the quantal nearest--neighbour distribution to
follow the prediction of Random Matrix Theory (RMT) \cite{Gut90}. In
particular, for systems with time reversal symmetry:
\begin{equation}
  P(s) \approx \frac{\pi}{2} s \; , 
  \; \; \; s \ll 1 \; ,
  \; \; \; s \equiv (k_{n+1}-k_{n}) \, \bar{d}((k_{n}+k_{n+1})/2)
\end{equation}
where $\bar{d}(k)$ is the smooth density of states. The chance of
finding a pair of energy levels in the interval $[s, s+{\rm d}s]$ is
$P(s)\,{\rm d}s $. The cumulative probability of finding a pair in
$[0, s]$ is therefore crudely given by:
\begin{equation}
  I(s) \approx 
  \int_{0}^{s} P(s') \, {\rm d}s' \approx \frac{\pi}{4} s^2 \; ,
  \; \; \; s \ll 1 \, .
\end{equation}
A more refined calculation, taking into account all the possible
relative configurations of the pair in the interval $[0, s]$ gives:
\begin{equation}
  Q(s) \approx \frac{\pi}{6} s^2 \; ,
  \; \; \; s \ll 1 \, .
\end{equation}
If we trace the $k$-axis with steps $\Delta k$ and find an eigenvalue,
then the chance that there is {\em another} one in the same interval
$\Delta k$ is $Q(\Delta k \bar{d}(k))$. If we prescribe our tolerance
$Q$ to lose eigenvalues, then we should choose:
\begin{equation}
  \Delta k = \frac{s(Q)}{\bar{d}(k)} \approx \frac{1}{\bar{d}(k)}
  \sqrt{\frac{6 Q}{\pi}} \, .
\end{equation}
In the above, we assumed that the dips in $r(k)$ are wide enough, such
that they can be detected over a range of several $\Delta k$'s. If
this is not the case and the dips are very sharp, we must refine
$\Delta k$. In our case dips were quite sharp, and in practice we
needed to take $Q$ of the order $10^{-5} \div 10^{-6}$.

\subsection{Verifications of low--lying eigenvalues}
\label{subsec:low-lying}

After describing some numerical aspects of the computation, we turn to
various tests of the integrity and completeness of the computed
spectra. In this subsection we compare the computed low--lying
eigenvalues for $R>0$ with those of the $R=0$ case. In the next one we
compare the computed stair--case function to Weyl's law.

\begin{figure}[p]
  \begin{center}
    \leavevmode

    \psfig{figure=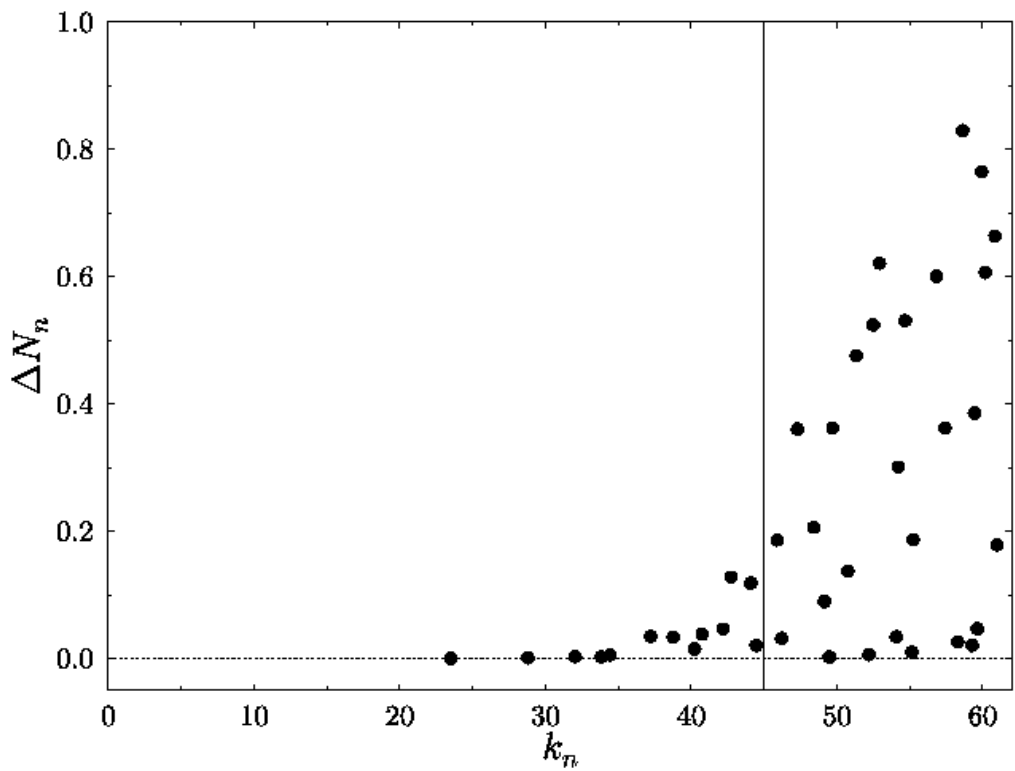,width=15cm}

    \caption{The unfolded differences $\Delta N_n$ for the low--lying 
      levels of the 3D Sinai billiard with $R=0.2$ and Dirichlet
      everywhere. We indicated by the vertical line $k = 45$ the
      theoretical expectation for transition from small to large
      $\Delta N$. The line $\Delta N = 0$ was slightly shifted upwards
      for clarity.}

  \label{fig:lowevs}

  \end{center}
\end{figure}

The theoretical background for the comparison between low--lying
eigenvalues to those of the $R=0$ case is as follows. The lowest $l$
value, for which there exist antisymmetric cubic harmonics, is $l=9$
\cite{LB47}. Consequently, for cases with Dirichlet conditions on the
symmetry planes, the lowest $l$-values in the KKR matrix is $l=9$.
Thus, for $kR<9$ the terms $P_l(kR)$ in equation (\ref{eq:kkr-desym})
are very small, and the matrix is essentially as if the inscribed
sphere was not present. In that case of the ``empty tetrahedron'' the
eigenvalues can be calculated analytically:
\begin{equation}
  k_n^{R=0} = \frac{2 \pi}{S} \sqrt{l^2 + m^2 + n^2} \; ,
  \; \; \; 0 < l < m < n \, .
\end{equation}
We hence expect:
\begin{equation}
  k_n \approx k_n^{R=0} \; \; \; \; 
  \mbox{for} \; \; k_n \lesssim 9 / R \, .
\end{equation}
Similar considerations were used by Berry \cite{Ber81} for the 2D
Sinai billiard, where he also calculated the corrections to the
low--lying eigenvalues. In figure \ref{fig:lowevs} we plot the
unfolded difference $\Delta N_n \equiv \bar{d}(k_n) \left| k_n -
  k_n^{R=0} \right|$ for the longest spectrum ($R=0.2$, Dirichlet
everywhere). One clearly observes that indeed the differences are very
small up to $k = 9 / 0.2 = 45$, and they become of order 1 afterwards,
as expected. This confirms the accuracy and completeness of the
low--lying levels. Moreover, it verifies the correctness of the rather
complicated computations of the terms $A_{lj, l'j'}$ which are due to
the cubic lattice.

\subsection{Comparing the exact counting function with Weyl's law}
\label{subsec:verify-weyl}
\begin{figure}[p]
  \begin{center}
    \leavevmode

    \psfig{figure=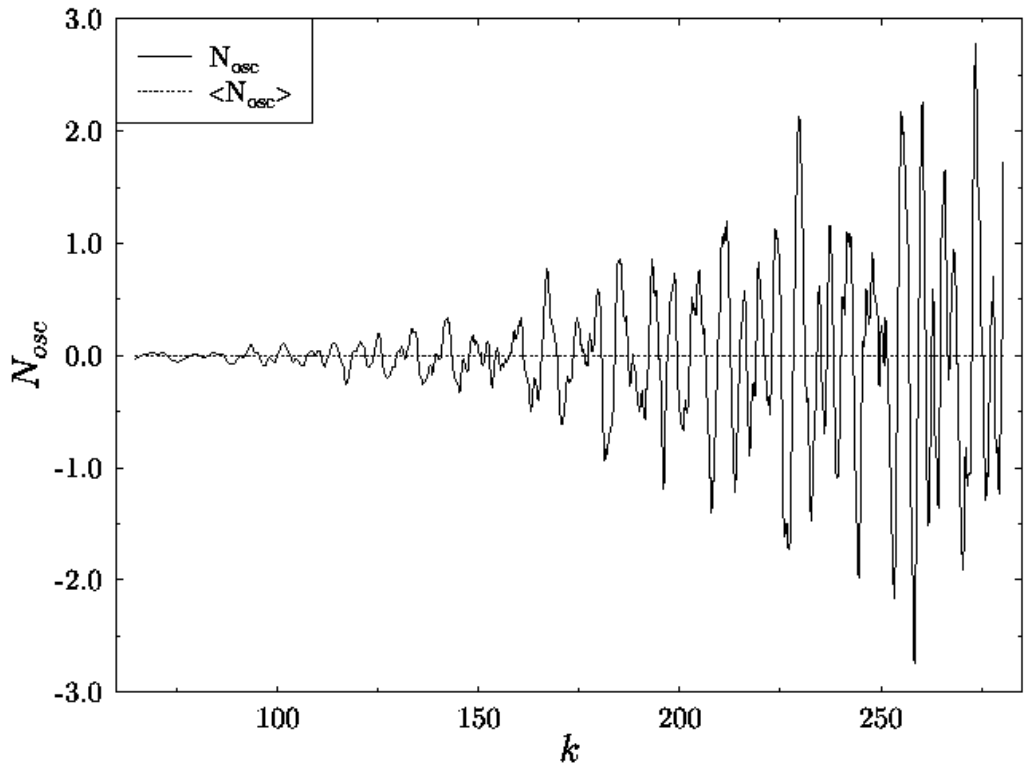,width=15cm}

    \caption{$N_{osc}(k)$ for the longest spectrum of the 
      3D billiard. The data are smoothed over 50 level intervals.}

  \label{fig:exact-weyl}

  \end{center}
\end{figure}

%
It is by now a standard practice (see e.g.\ \cite{SS95}) to verify the
completeness of a spectrum by comparing the resulting stair--case
function $N(k) \equiv \# \{ k_n \leq k \}$ to its smooth approximation
$\bar{N}(k)$, known as ``Weyl's law''. In appendix \ref{app:weyl} we
derived Weyl's law for the 3D Sinai billiard (equation
(\ref{eq:weyl})), and now consider the difference $N_{osc}(k) \equiv
N(k) - \bar{N}(k)$. Any jump of $N_{\rm osc}$ by $\pm 1$ indicates a
redundant or missing eigenvalue. In fact, this tool is of great help
to locating missing eigenvalues. In figure \ref{fig:exact-weyl} we
plot $N_{osc}$ for the longest spectrum. It is evident that the curve
fluctuates around 0 with no systematic increase/decrease trends, which
verifies the completeness of the spectrum. The average of $N_{osc}$
over the available $k$-interval is $(-4) \cdot 10^{-4}$ which is
remarkably smaller than any single contribution to $\bar{N}$ (note
that we had no parameters to fit). This is a very convincing
verification both of the completeness of the spectrum as well as the
accuracy of the Weyl's law (\ref{eq:weyl}). We also note that the
typical fluctuations grow quite strongly with $k$. This is due to the
effects of the bouncing--ball families (see section \ref{sec:intro})
and will be discussed further in section \ref{subsec:two-point}.


\section{Quantal spectral statistics}
\label{sec:quantal-spectral-statistics}
%
Weyl's law predicts the smooth behaviour of the quantal density of
states. There is a wealth of information also in the fluctuations, and
their investigation is usually referred to as ``spectral statistics''.
Results of spectral statistics that comply with the predictions of
Random Matrix Theory (RMT) are generally considered as a hallmark of
the underlying classical chaos \cite{BGS84,Boh89,Gut90,Sie91,SS95}.

In the case of the Sinai billiard we are plagued with the existence of
the non-generic bouncing--ball manifolds. They influence the spectral
statistics of the 3D Sinai billiard. It is therefore desirable to
study the bouncing balls in some detail. This is done in the first
subsection, where we discuss the integrable case ($R=0$) that contains
only bouncing--ball manifolds.

For the chaotic cases $R>0$ we consider the two simplest spectral
statistics, namely, the nearest--neighbour distribution and two--point
correlations. We compute these statistics for the levels of the 3D
Sinai billiard, and compare them to RMT predictions. In addition, we
discuss the two--point statistics of spectral determinants that was
recently suggested by Kettemann, Klakow and Smilansky \cite{KKS97} as
a characterization of quantum chaos.

\subsection{The integrable $R=0$ case}
\label{subsec:integrable}
\begin{figure}[p]
  \vspace*{-2cm}
  \begin{center}
    \leavevmode

    \begin{tabular}{c}
      \psfig{figure=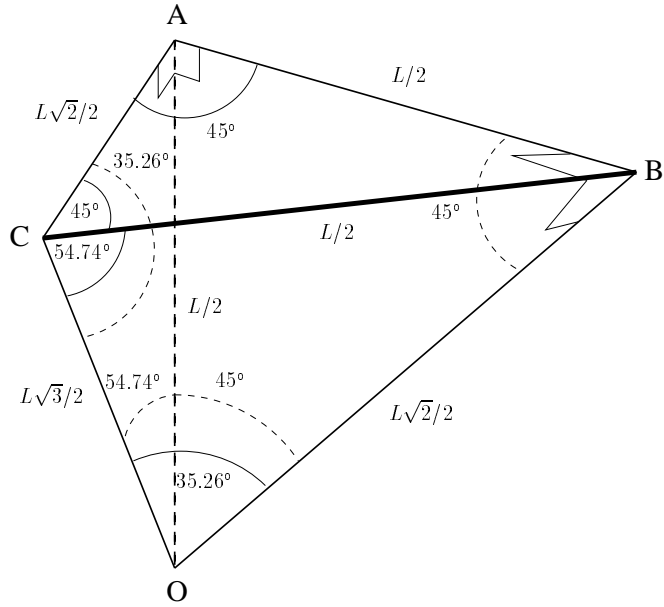,height=8cm} \\
      \mbox{} \\
      \psfig{figure=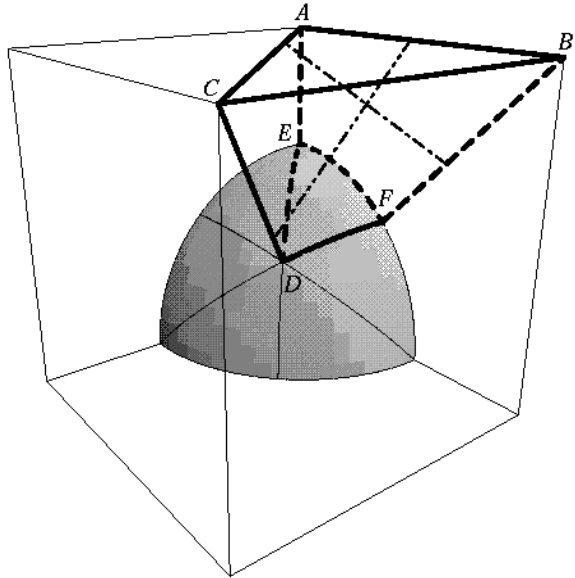,height=8cm}
    \end{tabular}

    \caption{Upper: Geometry of the tetrahedron $(R=0)$ billiard.  Lower:
      Neutral periodic orbits in the desymmetrized 3D Sinai.  The
      billiard is indicated by boldface edges. Dot--dash line: The
      shortest neutral periodic orbit of length $S/\protect\sqrt{3}$.
      Double dot--dash line: Neutral periodic orbit of length
      $S/\protect\sqrt{2}$.}

    \label{fig:pyramid}

  \end{center}
\end{figure}

%
If the radius of the inscribed sphere is set to 0, we obtain an
integrable billiard which is the irreducible domain whose volume is
1/48 of the cube. It is plotted in figure \ref{fig:pyramid}. This
tetrahedron billiard is a convenient starting point for analyzing the
bouncing--ball families, since it contains no unstable periodic orbits
but only bouncing balls. Quantum mechanically, the eigenvalues of the
tetrahedron are given explicitly as:
\begin{equation}
  k_{(nml)} = \frac{2 \pi}{S} \sqrt{n^2+m^2+l^2} \; ,
  \; \; \; 0 < n < m < l \in \bbbn \, .
\end{equation}
The spectral density $d_{R=0}(k) = \sum_{0<n<m<l}^{\infty} \delta ( k
- k_{(nml)} )$ can be Poisson resummed to get:
\begin{eqnarray}
  d_{R=0}(k) 
  & = & 
  \frac{S^3 k^2}{96 \pi^2} \sum_{pqr \in \bbbz} 
    {\rm sinc} \left( k S \sqrt{p^2+q^2+r^2} \right)
  \nonumber \\
  & - & 
  \frac{S^2 k}{32 \pi} \sum_{pq \in \bbbz} 
    J_0 \left( k S \sqrt{p^2+q^2} \right) -
  \frac{S^2 k}{16 \sqrt{2} \pi} \sum_{pq \in \bbbz} 
    J_0 \left( k S \sqrt{p^2+\frac{q^2}{2}} \right)
  \nonumber \\
  & + &
  \frac{3 S}{16 \pi} \sum_{p \in \bbbz} 
    \cos \left( k S p \right) + 
  \frac{S}{8 \sqrt{2} \pi} \sum_{p \in \bbbz} 
    \cos \left( k \frac{S}{\sqrt{2}} p \right)
  \nonumber \\
  & + &
  \frac{S}{6 \sqrt{3} \pi} \sum_{p \in \bbbz} 
    \cos \left( k \frac{S}{\sqrt{3}} p \right)
    - \frac{5}{16} \delta \left( k-0 \right) \, .
  \label{eq:dkint}
\end{eqnarray}
In the above $\mbox{sinc}(x) \equiv \sin(x)/x$, $\mbox{sinc}(0) \equiv
1$, and $J_0$ is the zeroth order Bessel function. Let us analyze this
expression in some detail.  Terms which have all summation indices
equal to $0$ give the {\em smooth} part of the density, and all the
remaining terms constitute the {\em oscillatory} part.  Collecting the
smooth terms together we get:
\begin{equation}
  \bar{d}_{R=0}(k) = \frac{S^3 k^2}{96 \pi^2} 
                   - \frac{S^2 k}{32 \pi}(1+\sqrt{2})
                   + \frac{S}{144 \pi}(27+9\sqrt{2}+8\sqrt{3})
                   - \frac{5}{16} \delta(k-0) \, .
\end{equation}
This is Weyl's law for the tetrahedron, which exactly corresponds to
(\ref{eq:weyl}) with $R=0$ (except the last term for which the limit
$R \rightarrow 0$ is different).

As for the oscillatory terms, it is first useful to replace $J_0(x)$
by its asymptotic approximation \cite{AS65} which is justified in the
semiclassical limit $k \rightarrow \infty$:
\begin{equation}
  J_0(x) \approx \sqrt{\frac{2}{\pi x}} 
    \cos \left( x - \frac{\pi}{4} \right) \; ,
    \; \; \; x \rightarrow \infty \, .
\end{equation}
Using this approximation we observe that all of the oscillatory terms
have phases which are of the form $(k \times \mbox{length} +
\mbox{phase})$. This is the standard form of a semiclassical
expression for the density of states of a billiard. To go a step
further we notice that the leading--order terms, which are
proportional to $k^1$ (first line of (\ref{eq:dkint})), have lengths
of $S\sqrt{p^2+q^2+r^2}$ which are the lengths of the periodic orbits
of the 3-torus, and therefore of its desymmetrization into the
tetrahedron. This conforms with the expressions derived by Berry and
Tabor \cite{BT76,BT77} for integrable systems. The other, sub-leading,
oscillatory contributions to (\ref{eq:dkint}) correspond to
``improper'' periodic manifolds, in the sense that their dynamics
involves non-trivial limits. Some of these periodic orbits are
restricted to symmetry plane or go along the edges. Of special
interest are the periodic orbits that are shown in figure
\ref{fig:pyramid}.  They are isolated, but are neutrally stable and
hence are non-generic.  Their contributions are contained in the last
two terms of (\ref{eq:dkint}), and the one with length $S/\sqrt{3}$ is
the shortest neutral periodic orbit. Other sub-leading oscillatory
contributions are discussed in \cite{Pri97}. We therefore established
an interpretation in terms of (proper or improper) classical periodic
orbits of the various terms of (\ref{eq:dkint}).

\subsubsection{Two-point statistics of the integrable case}
\label{subsubsec:two-point-int}
%
We continue by investigating the two--point statistics of the
tetrahedron, which will be shown to provide some nontrivial and
interesting results. Since we are interested in the limiting
statistics as $k \rightarrow \infty$ we shall consider only the
leading term of (\ref{eq:dkint}), which is the first term. Up to a
factor of 48, this is exactly the density of states $d_{T^3}$ of the
cubic 3-torus, and thus for simplicity we shall dwell on the 3-torus
rather than on the tetrahedron:
\begin{equation}
  d_{T^3}(k) = \sum_{\vec{\rho} \in \bbbz^3} 
  \delta \left( k - \frac{2 \pi}{S} \rho \right) = 
  \frac{S^3 k^2}{2 \pi^2} \sum_{\vec{\rho} \in \bbbz^3} 
  \mbox{sinc}(k S \rho) \, .
  \label{eq:dkt3}
\end{equation}
We observe that both the quantal spectrum and the classical spectrum
(the set of lengths of periodic orbits) are supported on the cubic
lattice $\bbbz^3$, and this strong duality will be used below.

The object of our study is the spectral form factor, which is the
Fourier transform of the two--point correlation function of the energy
levels \cite{Boh89}. For billiards it is more convenient to work with
the eigenwavenumbers $k_n$ rather than with the eigenenergies $E_n$.
Here the form factor is given by:
\begin{equation}
  K(\tau; k) = 
  \frac{1}{N} \left| \sum_{n=n_1}^{n_2} 
    \exp \left[ 2 \pi i \bar{d}(k) k_n \tau \right] \right|^2
  \label{eq:ktau}
\end{equation}
In the above $N \equiv n_2-n_1+1$, and $k_n$ are the eigenvalues in
the interval $[k_{n_1}, k_{n_2}]$ centered around $k = (k_{n_1} +
k_{n_2})/2$. It is understood that the interval contains many levels
but is small enough such that the average density is almost a constant
and is well approximated by $\bar{d}(k)$.

In the limit $\tau \rightarrow \infty$ the phases in (\ref{eq:ktau})
become random in the generic case, and therefore $K(\tau) \rightarrow
1$. However, if the levels are degenerate, more care should be
exercised, and one obtains:
\begin{equation}
  K(\tau; k) = 
  \frac{1}{N}{\sum_n} g_k(k_n) =
  \frac{1}{N}{\sum_i}^{\prime} g_k^2(k_i) \; , 
    \; \; \; \tau \rightarrow \infty \; ,
\end{equation}
where $g_k(k_n)$ is the degeneracy of $k_n$ and the primed sum is only
over {\rm distinct} values of $k_i$. Since $N = \sum_i^{\prime}
g_k(k_i)$ we obtain:
\begin{equation}
  K(\tau; k) = 
  \frac{{\sum_i}^{\prime} g_k^2(k_i)}{{\sum_i}^{\prime} g_k(k_i)} =
  \frac{\langle g_k^2(k) \rangle}{\langle g_k(k) \rangle} \; ,
  \; \; \; \tau \rightarrow \infty \; ,
  \label{eq:ktau-large}
\end{equation}
where $\langle \cdot \rangle$ denotes an averaging over $k_i$'s near
$k$. In the case of a constant $g$ the above expression reduces to
$K(\tau \rightarrow \infty) = g$, but it is important to note that
$K(\tau \rightarrow \infty) \neq \langle g \rangle$ for non-constant
degeneracies. Using the relation $\rho = kS/(2\pi)$ (see equation
(\ref{eq:dkt3})) and equations (\ref{eq:g-rho}), (\ref{eq:g2-rho}) in
appendix \ref{app:nt-deg} we get:
\begin{equation}
  K_{T^3}(\tau; k) = 
  \frac{\langle g_{\rho}^2(kS/(2 \pi)) \rangle}
       {\langle g_{\rho}(kS/(2 \pi)) \rangle} =
  \frac{\beta S}{2 \pi} \, k \; ,
  \; \; \; \tau \rightarrow \infty \, ,
\end{equation}
where $\beta \approx 9.8264$ is a constant. That is, contrary to the
generic case, the saturation value of the form factor grows linearly
with $k$ due to number--theoretical degeneracies.

Turning to the form factor in the limit $\tau \rightarrow 0$, we first
rewrite (\ref{eq:dkt3}) as $d_{T^3}(k) = \bar{d}(k) + \sum_j A_j
\sin(k L_j)$. Then, using the diagonal approximation as suggested by
Berry \cite{Ber85,Ber89}, and taking into account the degeneracies
$g_{\ell}(L_j)$ of the lengths we have:
\begin{equation}
  K(\tau; k) = 
  \frac{1}{4 \bar{d}^2(k)} 
    {\sum_j}^{\prime} g_{\ell}^2(L_j) |A_j|^2 
    \delta (\tau - L_j/L_{\rm H}) \; ,
  \; \; \; \tau \ll 1 \; .
\end{equation}
In the above the prime denotes summation only over {\em distinct}\/
classical lengths, and $L_{\rm H} \equiv 2 \pi \bar{d}(k)$ is called
the Heisenberg length. The coefficients $A_j$ are functions of $L_j$
and therefore can be replaced by the function $A(\tau)$. For $\tau$
large enough such that the periodic manifolds have a well-defined
classical density $\bar{d}_{\rm cl}(\ell)$, the summation over delta
functions can be replaced by multiplication with $L_{\rm H}
\bar{d}_{\rm cl}(\ell)/\langle g_{\ell}(\ell) \rangle$ with $\ell =
L_{\rm H} \tau$ such that:
\begin{equation}
  K(\tau; k) = 
  \left( \frac{\pi |A^2(\tau)| \bar{d}_{\rm cl}(\ell)}
              {2 \bar{d}(k)} \right)
    \frac{\langle g_{\ell}^2(\ell) \rangle}
         {\langle g_{\ell}(\ell) \rangle} \; ,
  \; \; \; \tau \ll 1 \, .
\end{equation}
A straightforward calculation shows that the term in brackets is
simply $1$, which is the generic situation for the integrable case
(Poisson statistics) \cite{Ber85,Dit96}. Hence, we obtain:
\begin{equation}
  K(\tau; k) = 
  \frac{\langle g_{\ell}^2(\ell) \rangle}
       {\langle g_{\ell}(\ell) \rangle} \; ,
  \; \; \; \tau \rightarrow 0 \, .
\end{equation}
Since, as we noted above, the lengths of the classical periodic orbits
are supported on the $\bbbz^3$ lattice, we can write using $\ell = S
\rho$:
\begin{equation}
  K(\tau; k) = 
  \frac{\langle g_{\rho}^2(\ell/S) \rangle}
       {\langle g_{\rho}(\ell/S) \rangle} =
  \frac{\beta k^2 S^2}{\pi} \ \tau \; ,
  \; \; \; \tau \rightarrow 0 \, .
\end{equation}
where we used again equations (\ref{eq:g-rho}), (\ref{eq:g2-rho}).
This is a very surprising result, since it implies that contrary to
the generic integrable systems, which display Poisson level statistics
with $K = 1$, here $K \propto \tau$ which is typical to chaotic
systems! This peculiarity is manifestly due to the number theoretical
degeneracies of $\bbbz^3$.

\begin{figure}[p]
  \begin{center}
    \leavevmode

    \psfig{figure=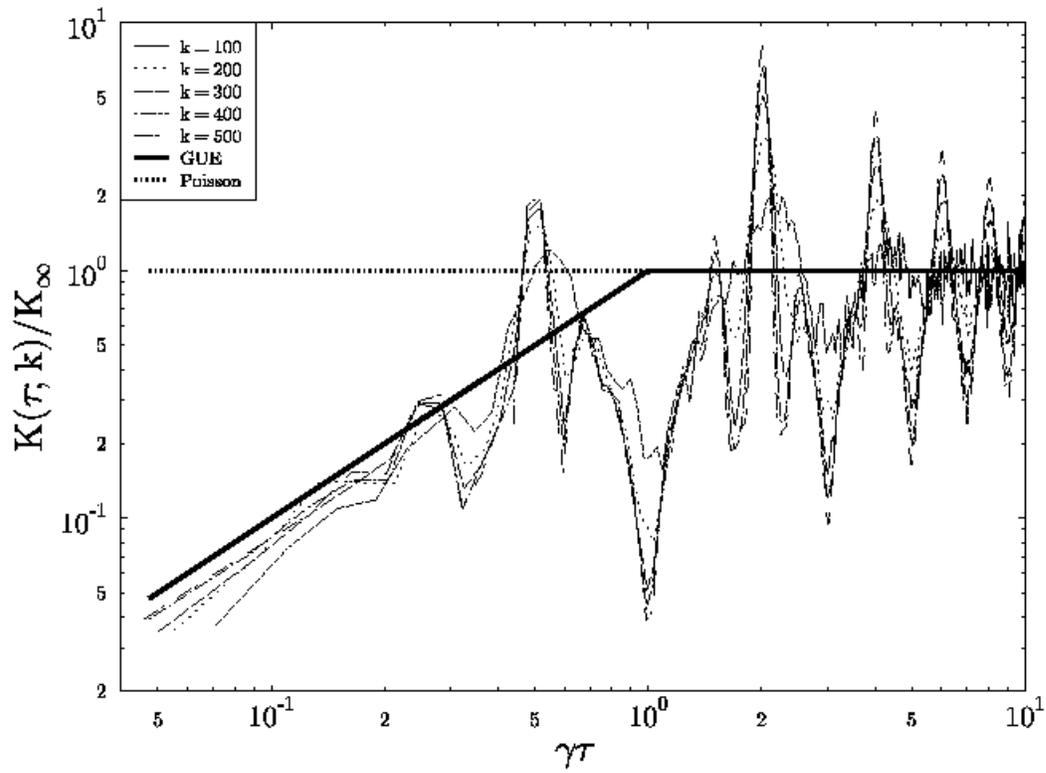,width=15cm}

    \caption{The scaled quantal form factor of the tetrahedron for 
      various $k$-values compared with GUE and Poisson. Note the
      log-log scales.}

    \label{fig:kt-pyr}

  \end{center}
\end{figure}

If we now combine the two limiting behaviours of the form factor in
the simplest way, we can express it as a scaled RMT-GUE form factor:
\begin{equation}
  K_{T^3}(\tau; k) \approx 
  K_{\infty} \cdot K_{\mbox{\tiny GUE}} (\gamma \tau)
 \label{eq:ktt3}
\end{equation}
where $K_{\infty} = S \beta k / (2 \pi)$ and $\gamma = 2 S k$. For the
tetrahedron we have the same result with $K_{\infty} \rightarrow
K_{\infty}/48$ and $\gamma \rightarrow \gamma/48$. This prediction is
checked and verified numerically in figure \ref{fig:kt-pyr} where we
computed the quantal form factor of the tetrahedron around various
$k$-values. The agreement of the two asymptotes to the theoretical
prediction (\ref{eq:ktt3}) is evident and the difference from Poisson
is well beyond the numerical fluctuations.

\subsection{Nearest--neighbour spacing distribution}
\label{subsec:nearest-neighbour}
\begin{figure}[p]
  \begin{center}
    \leavevmode

    \begin{tabular}{c}
      \vspace*{-0.5cm}
      \psfig{figure=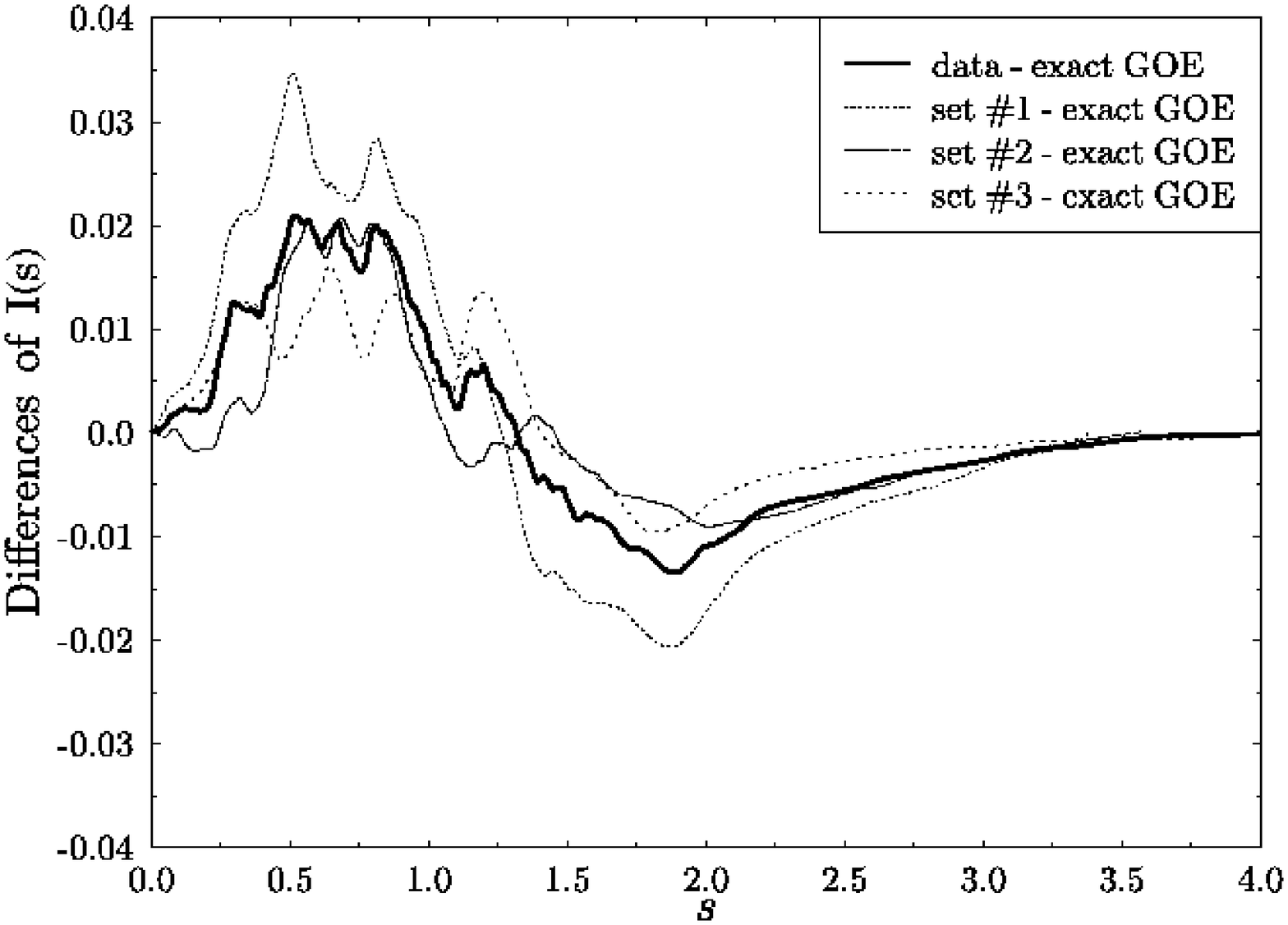,height=8cm} \\
      \psfig{figure=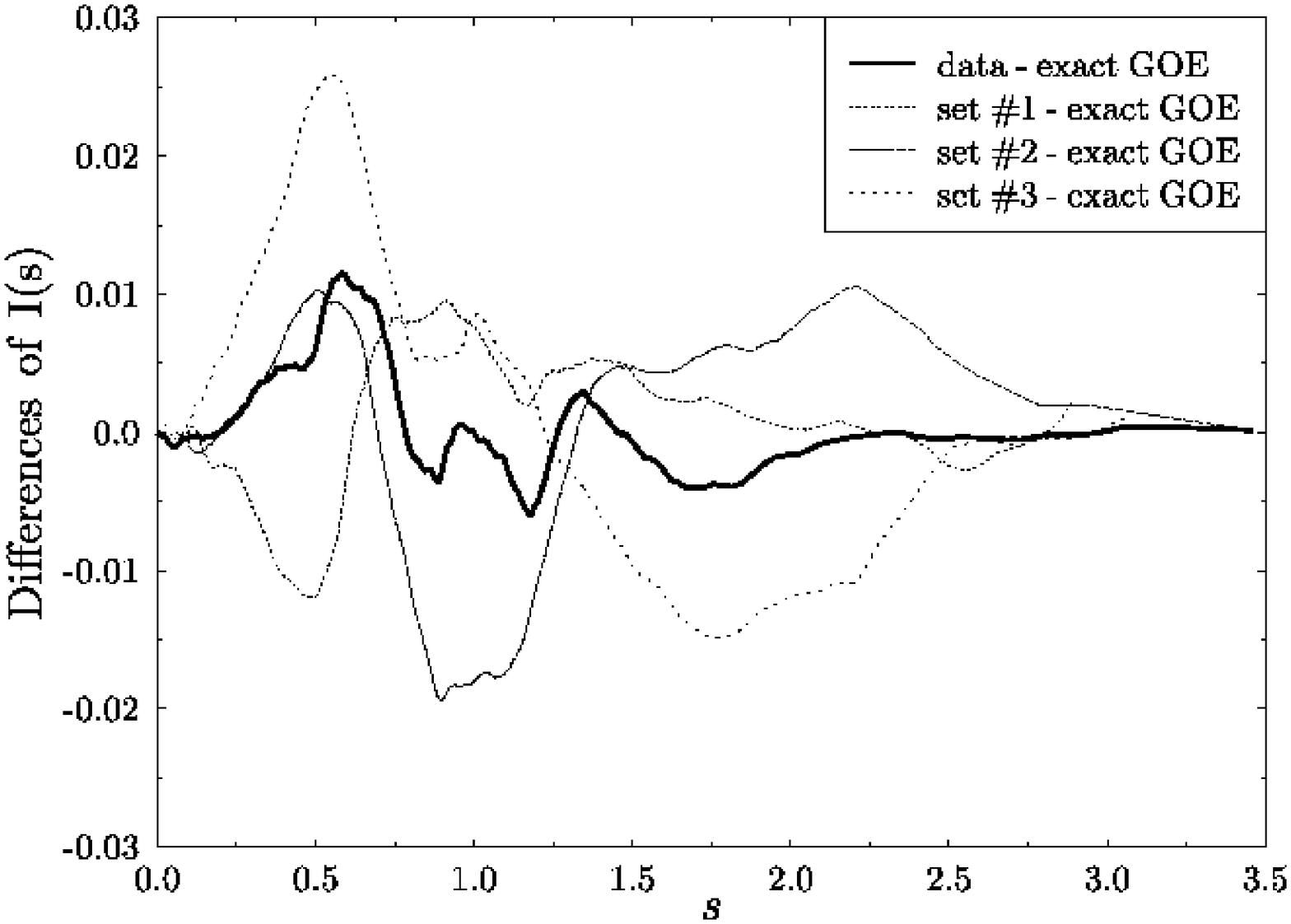,height=8cm}
    \end{tabular}

    \caption{Differences of integrated nearest--neighbour distribution
      for $R=0.2$ (up) and $R=0.3$ (down). Set \#1, 2, 3 refer to the
      division of the spectrum into 3 domains. Data are slightly
      smoothed for clarity.}

  \label{fig:is-diff}

  \end{center}
\end{figure}

%
We now turn to the chaotic case $R > 0$. One of the most common
statistical measures of a quantum spectrum is the nearest--neighbour
distribution $P(s)$. If fact, it is the simplest statistics to compute
from the numerical data. We need only to consider the distribution of
the scaled (unfolded) spacings between neighbouring levels:
\begin{equation}
  s_n \equiv \bar{N}(k_{n+1}) - \bar{N}(k_n) \approx
  \bar{d}(k_n) ( k_{n+1} - k_{n} ) \, .
\end{equation}
It is customary to plot a histogram of $P(s)$, but it requires an
arbitrary choice of the bin size. To avoid this arbitrariness, we
consider the cumulant distribution:
\begin{equation}
  I(s) \equiv \int_{0}^{s} {\rm d}s' \, P(s')
\end{equation}
for which no bins are needed. Usually, the numerical data are compared
not to the exact $P_{\rm RMT}(s)$ but to Wigner's surmise
\cite{Gut90}, which provides an accurate approximation to the exact
$P_{\rm RMT}(s)$ in a simple closed form. In our case, since we found
a general agreement between the numerical data and Wigner's surmise,
we choose to present the differences from the exact expression for
$I_{\rm GOE}(s)$ taken from Dietz and Haake \cite{DH90}. In figure
\ref{fig:is-diff} we show these differences for $R=0.2, 0.3$ and
Dirichlet boundary conditions (6697 and 1994 levels, respectively).
The overall result is an agreement between the numerical data and RMT
to better than $4\%$.  This is consistent with the general wisdom for
classically chaotic systems in lower dimensions, and thus shows the
robustness of the RMT conjecture \cite{BGS84} for higher--dimensional
systems (3D in our case).

Beyond this general good agreement it is interesting to notice that
the differences between the data and the exact GOE for $R=0.2$ seem to
indicate a systematic modulation rather than a statistical fluctuation
about the value zero. The same qualitative result is obtained for
other boundary conditions with $R=0.2$, substantiating the conjecture
that the deviations are systematic and not random. For $R=0.3$ the
differences look random and show no particular pattern. However, for
the upper third of the spectrum one observes structures which are
similar to the $R=0.2$ case (see figure \ref{fig:is-diff}, lower
part).

Currently, we have no theoretical explanation of the above mentioned
systematic deviations. They might be due to the non-generic bouncing
balls. To assess this conjecture we computed $P(s)$ for $R=0.2, 0.3$
with Dirichlet boundary conditions in the spectral interval $150 < k <
200$. The results (not shown) indicate that the deviations are smaller
for the larger radius. This is consistent with the expected weakening
of the bouncing--ball contributions as the radius grows, due to larger
shadowing and smaller volumes occupied by the bouncing--ball families.
Hence, we can conclude that the bouncing balls are indeed prime
candidates for causing the systematic deviations of $P(s)$. It is
worth mentioning that a detailed analysis of the $P(s)$ of spectra of
quantum graphs show similar deviations from $P_{\rm RMT}(s)$
\cite{KS99}.

\subsection{Two--point correlations}
\label{subsec:two-point}
%
Two--point statistics also play a major role in quantum chaos. This is
mainly due to their analytical accessibility through the Gutzwiller
trace formula as demonstrated by Berry \cite{Ber85,Ber89}. There is a
variety of two--point statistical measures which are all related to
the pair--correlation function \cite{Boh89}. We chose to focus on
$\Sigma^2(l)$ which is the local variance of the number of levels in
an energy interval that has the size of $l$ mean spacings. The general
expectation for generic systems, according to the theory of Berry
\cite{Ber85,Ber89}, is that $\Sigma^2$ should comply with the
predictions of RMT for small values of $l$ (universal regime) and
saturate to a non-universal value for large $l$'s due to the
semiclassical contributions of short periodic orbits. The saturation
value in the case of generic billiards is purely classical
($k$-independent). The effect of the non-generic bouncing--ball
manifolds on two--point spectral statistics was discussed in the
context of 2D billiards by Sieber et al.\,\cite{SSCL93} (for the case
of the stadium billiard). They found that $\Sigma^2$ can be decomposed
into two parts: A generic contribution due to unstable periodic orbits
and a non-generic contribution due to bouncing balls:
\begin{equation}
  \Sigma^2(l) \approx 
  \Sigma^2_{\rm UPO}(l) + \Sigma^2_{\rm bb}(l) \, .
\end{equation}
The term $\Sigma^2_{\rm bb}$ has the structure:
\begin{equation}
  \Sigma^2_{\rm bb}(l) = 
  k F_{\rm stadium}(l / \bar{d}(k)) \, ,
\end{equation}
where $F_{\rm stadium}$ is a function which is determined by the
bouncing balls of the stadium billiard, and is given explicitly in
\cite{SSCL93}. In particular, for large values of $l$ the term
$\Sigma^2_{\rm bb}$ fluctuated around an asymptotic value:
\begin{equation}
  \Sigma^2_{\rm bb}(l) \approx  
  k F_{\rm stadium}(\infty) \; ,
  \; \; \; l \rightarrow \infty \, .
\end{equation}
One can apply the arguments of Sieber et al.\,\cite{SSCL93} to the
case of the 3D Sinai billiard and obtain for the leading order
bouncing balls (see (\ref{eq:dkt3})):
\begin{equation}
  \Sigma^2_{\rm bb}(l) \approx 
  k^2 F_{\rm 3Dsb}(l / \bar{d}(k)) \, ,
  \label{eq:s2bb3d}
\end{equation}
with $F_{\rm 3Dsb}$ characteristic to the 3D Sinai billiard.
Asymptotically, we expect:
\begin{equation}
  \Sigma^2_{\rm bb}(l) \approx  k^2 F_{\rm 3Dsb}(\infty) \; ,
  \; \; \; l \rightarrow \infty \, .
  \label{eq:s2bb3d-saturate}
\end{equation}
The function $F_{\rm 3Dsb}$ can be written down, albeit it contains
the areas of the cross--sections of the various bouncing--ball
manifolds, for which we have no explicit expressions. Therefore, we
shall investigate the scaling features of $\Sigma^2_{\rm bb}$ without
insisting on its explicit from.

\begin{figure}[p]

  \begin{center}
    \leavevmode
    \psfig{figure=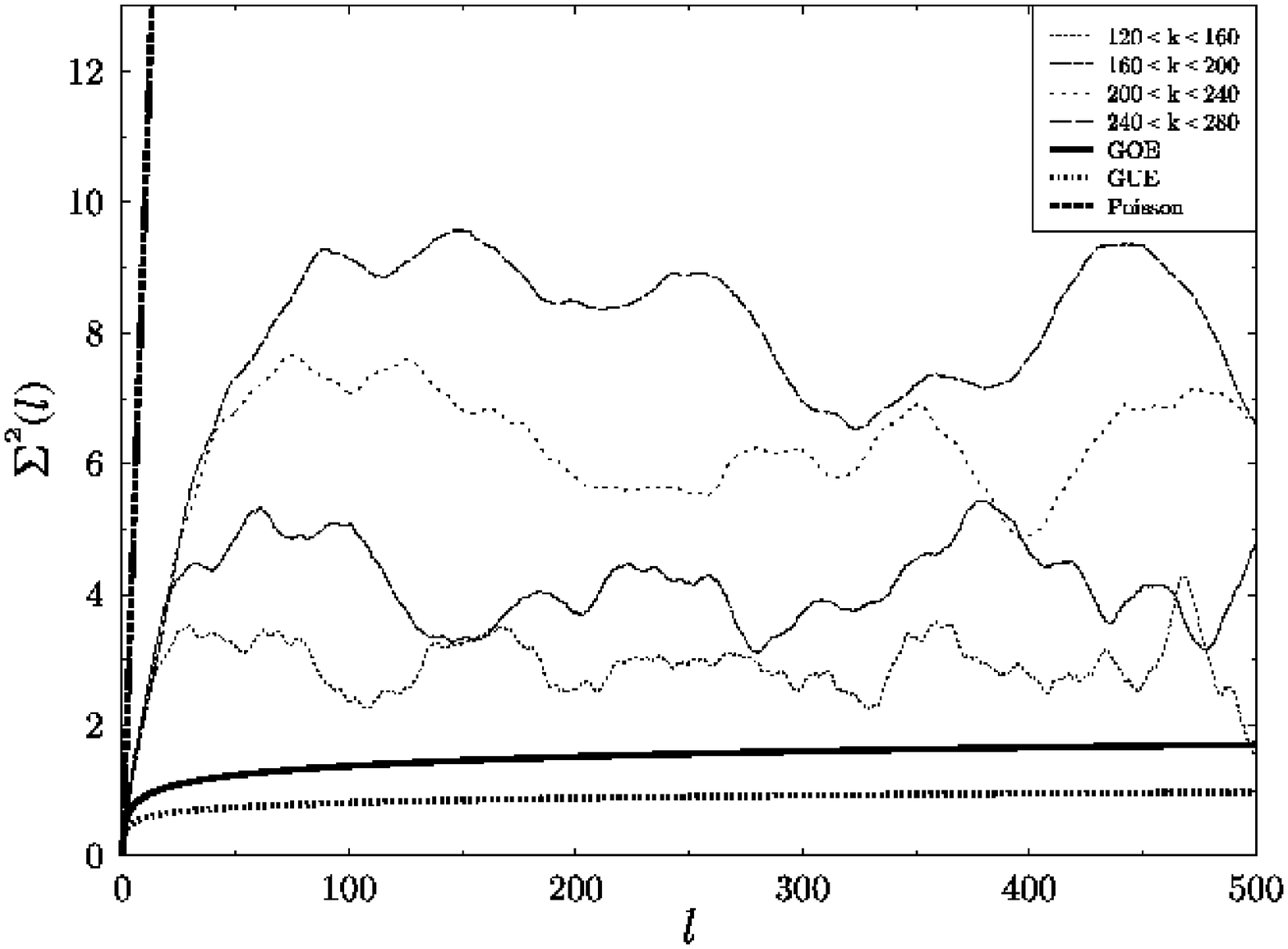,height=8cm}
    \psfig{figure=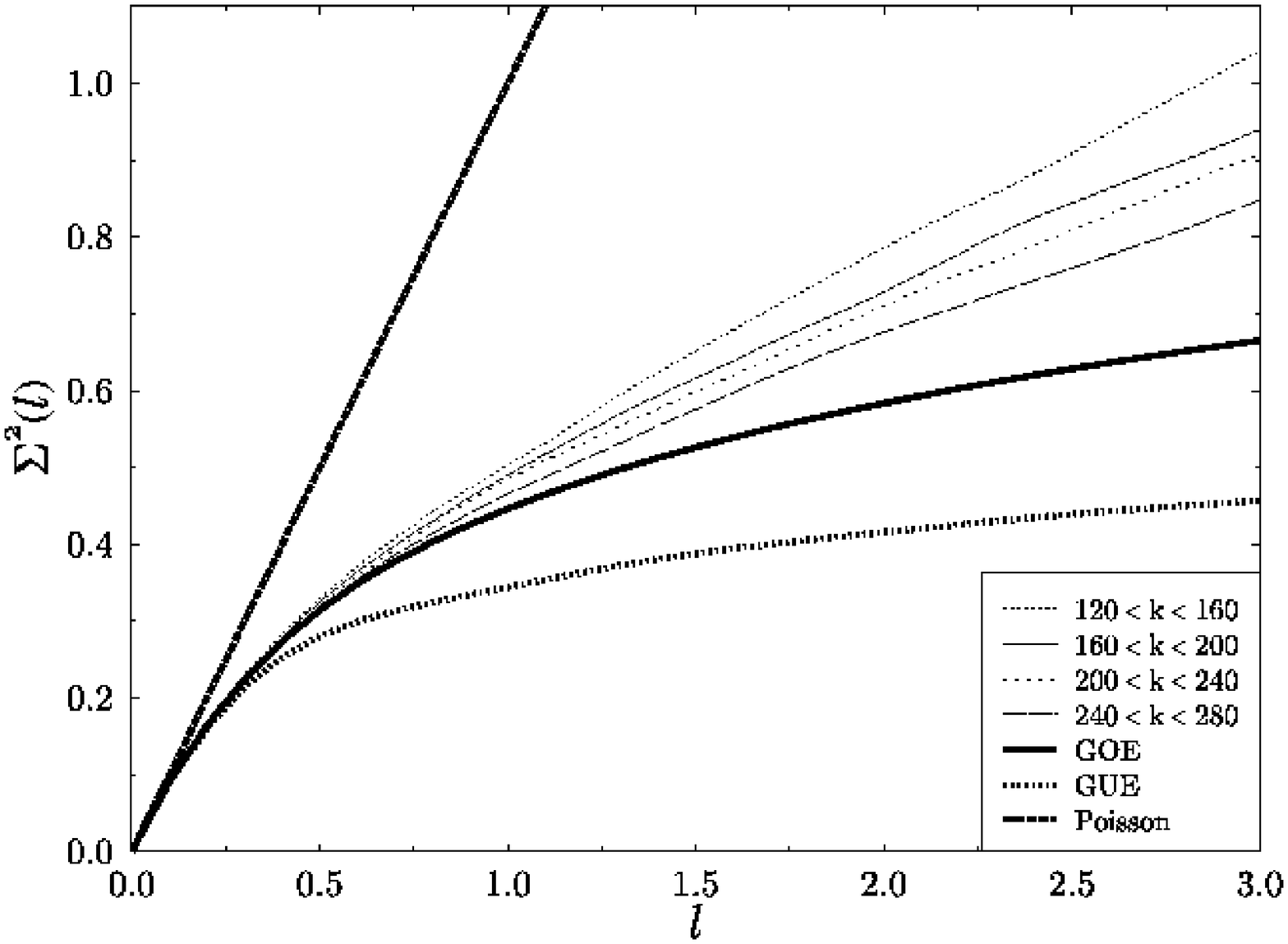,height=8cm}
  \end{center}

  \caption{The number variance $\Sigma^2(l)$ for the longest spectrum.
    Upper plot: Full $l$-range, lower plot: A magnification of small
    $l$ range.}

  \label{fig:s2-20}
\end{figure}

\begin{figure}[p]

  \begin{center}
    \leavevmode    
    \psfig{figure=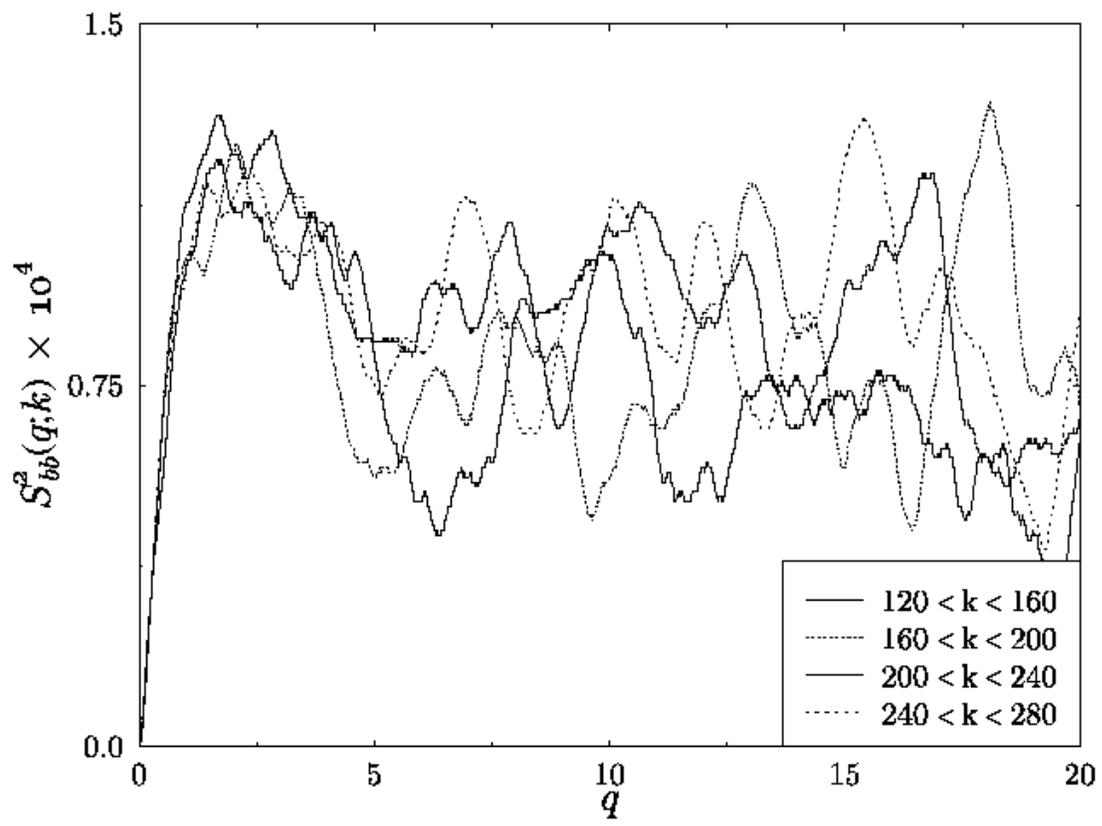,width=16cm}
  \end{center}

  \caption{Rescaled number variance (\protect\ref{eq:rescaled-s2}) for
    the longest spectrum.}

  \label{fig:s2-sdif}
\end{figure}

The numerical computations of $\Sigma^2$ for the longest spectrum
($R=0.2$, Dirichlet everywhere) are shown in figure \ref{fig:s2-20}.
We divided the spectrum into 4 intervals such that $\bar{d}$ did not
vary much within each interval. This is a pre-requisite for a
meaningful semiclassical analysis.
It is evident from the figure that for small values of $l$ (up to
$\approx 1$) there is an agreement with GOE. Moreover, the agreement
with GOE is much better than with either GUE or Poisson, as expected.
This is in agreement with the common knowledge in quantum chaos
\cite{Boh89}, and again, substantiates the RMT conjecture also for
chaotic systems in 3D. For larger $l$ values there are marked
deviations which saturate into oscillations around a $k$-dependent
asymptotic values. It is clearly seen that the saturation values grow
faster than $k$, which is consistent with (\ref{eq:s2bb3d-saturate}).
To test (\ref{eq:s2bb3d}) quantitatively, we plotted in figure
\ref{fig:s2-sdif} the rescaled function:
\begin{equation}
  S^2_{\rm bb}(q; k) \equiv
  \frac{1}{k^2} \left[ \Sigma^2 (q \bar{d}(k)) - 
    \Sigma^2_{\rm GOE} (q \bar{d}(k)) \right]
  \label{eq:rescaled-s2}
\end{equation}
which according to (\ref{eq:s2bb3d}) is the $k$-independent function
$F_{\rm 3Dsb}(q)$. Indeed, there is a clear data collapse for $q
\lesssim 5$, and the saturation values of $S^2_{\rm bb}$ are of the
same magnitude for all values of $k$.  This verifies (\ref{eq:s2bb3d})
and demonstrates the important part which is played by the bouncing
balls in the two--point (long range) statistics.

\begin{figure}[p]

  \begin{center}
    \leavevmode
    \psfig{figure=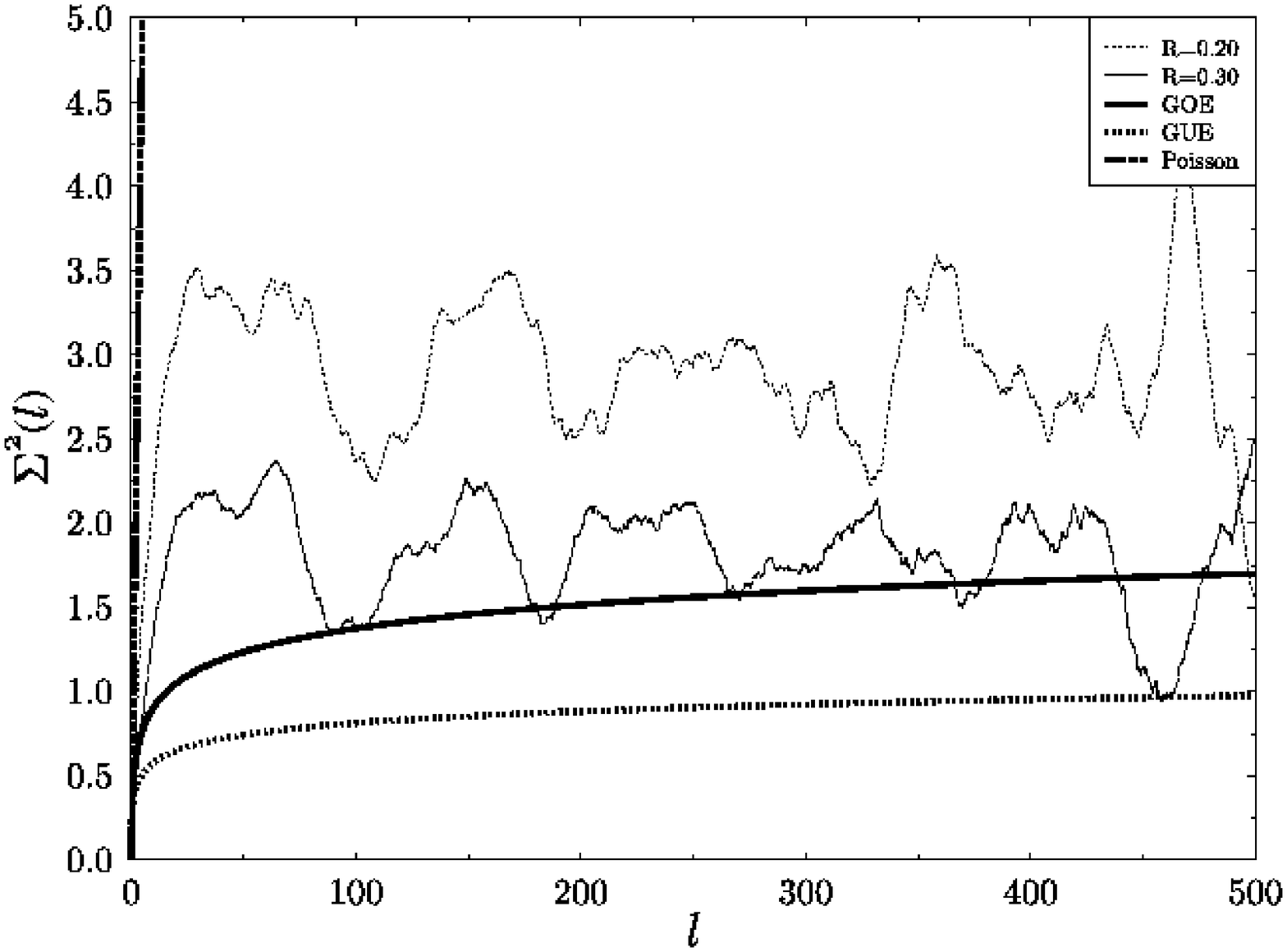,height=8cm}
    \psfig{figure=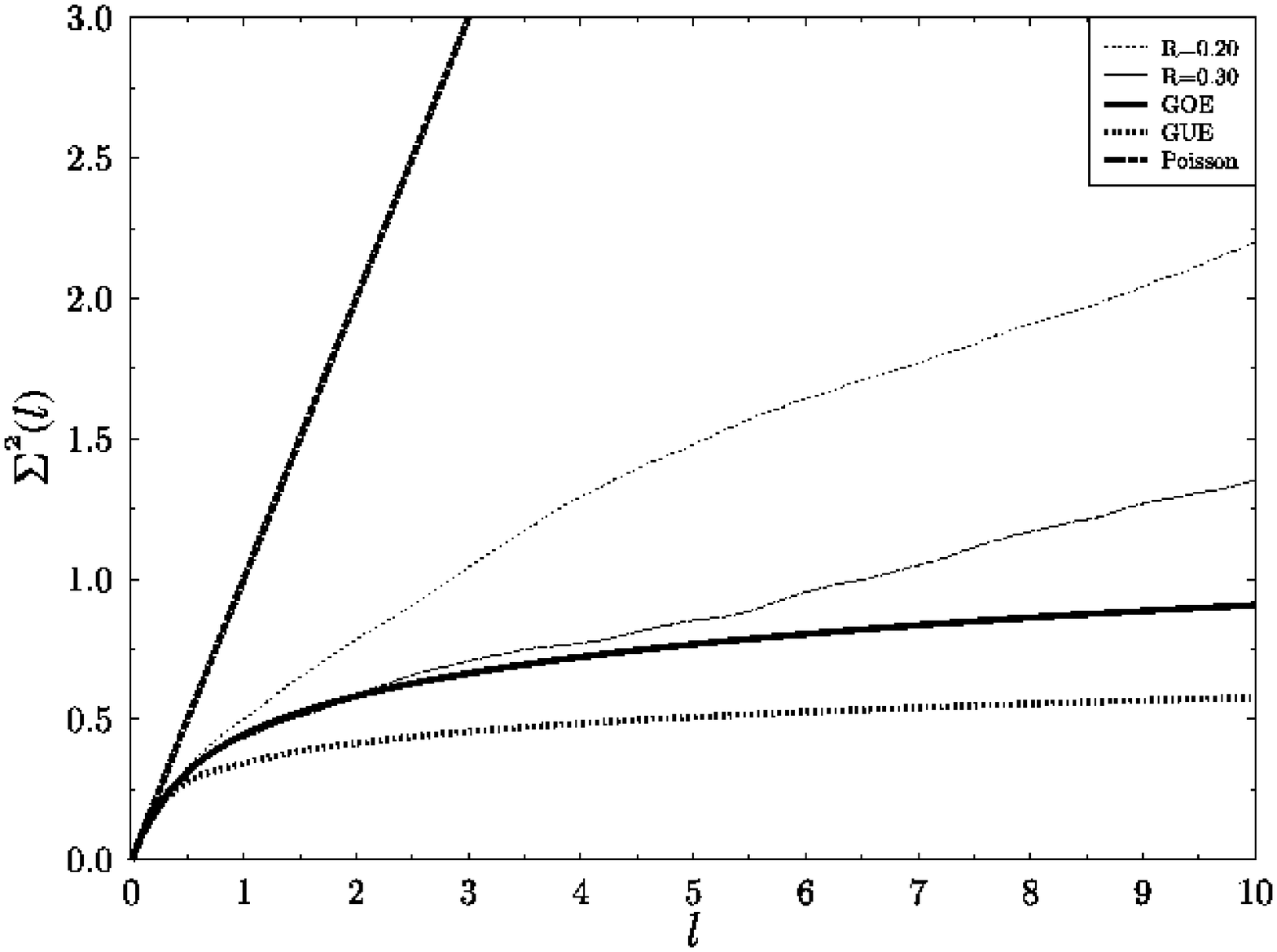,height=8cm}
  \end{center}

  \caption{Comparison between the number variances for two 
    different radii $R=0.2, 0.3$ of the inscribed sphere of the 3D
    Sinai billiard. In both cases we considered the spectral interval
    $120 < k < 160$ and used Dirichlet boundary conditions.}

  \label{fig:s2-2030}

\end{figure}

For generic systems the agreement between $\Sigma^2$ and RMT should
prevail up to $l^*$, where:
\begin{equation}
  l^{*} = 
  \frac{L_H(k)}{L_{\rm min}} = 
  \frac{2 \pi \bar{d}(k)}{L_{\rm min}} \, .
  \label{eq:lstar}
\end{equation}
In the above $L_{\rm H}$ is Heisenberg length and $L_{\rm min}$ is the
length of the shortest periodic orbit. For the cases shown in figure
\ref{fig:s2-20} the value of $l^*$ is of the order of $100$.
Nevertheless, the deviations from the universal predictions start much
earlier. This is again a clear sign of the strong effect of the
bouncing-balls. To substantiate this claim, we compare in figure
\ref{fig:s2-2030} the number variances for $R=0.2$ and $R=0.3$ in the
same $k$ interval and with the same boundary conditions (Dirichlet).
The influence of the bouncing-balls is expected to be less dominant in
the $R=0.3$ case, since there are fewer of them with smaller cross
sections. This is indeed verified in the figure: The agreement with
GOE predictions lasts much longer (up to $l \approx 6$) in the $R=0.3$
case, and the saturation value is smaller, as expected.

\subsection{Auto-correlations of spectral determinants}
\label{subsec:two-point-det}
%
The two-point correlations discussed above are based on the quantal
{\em spectral densities}. Kettemann, Klakow and Smilansky \cite{KKS97}
introduced the auto-correlations of quantal {\em spectral
  determinants}\/ as a tool for the characterization of quantum chaos.
Spectral determinants are defined as:
\begin{equation}
  Z(E) = 0 \; \; \; \Longleftrightarrow \; \; \; E=E_n \; ,
\end{equation}
that is, they are $0$ iff $E$ is an eigenenergy. The (unnormalized)
correlation function of a spectral determinant is defined as:
\begin{equation}
  C(\omega; E) 
  \equiv 
  \frac{1}{\Delta E} \int_{E - \Delta E / 2}^{E + \Delta E / 2} {\rm d}E'
    Z     \left( E' + \frac{\omega}{2 \bar{d}} \right)
    Z^{*} \left( E' - \frac{\omega}{2 \bar{d}} \right) \; ,
    \; \; \; \omega \ll \Delta E \ .
  \label{eq:cw}
\end{equation}
There are various motivations to study the function $C(\omega)$
\cite{KKS97}:
\begin{enumerate}
  
\item There is a marked difference in the behaviour of $C(\omega)$ for
  rigid and non-rigid spectra. For completely rigid spectra the
  function $C(\omega)$ is oscillatory, while for Poissonian spectra it
  rapidly decays as a Gaussian. For the RMT ensembles it shows damped
  oscillations which are due to rigidity.

\item The function $C(\omega)$ contains information about all
  $n$-point correlations of the spectral densities. Thus, it is
  qualitatively distinct from the two-point correlations of spectral
  densities and contains new information.

\item The Fourier transform of $C(\omega)$ exhibit in an explicit and
  simple way symmetry properties which are due to the reality of the
  energy levels.
  
\item In contrast to spectral densities, the semiclassical expressions
  for spectral determinants can be regularized using the method of
  Berry and Keating \cite{Kea93}. Regularized semiclassical spectral
  determinants contain a finite number of terms, and are manifestly
  real for real energies.

\item The semiclassical expression for $C(\omega)$ is closely related
  to the classical Ruelle zeta function.

\end{enumerate}

\begin{figure}[p]
  \begin{center}
    \leavevmode

    \psfig{figure=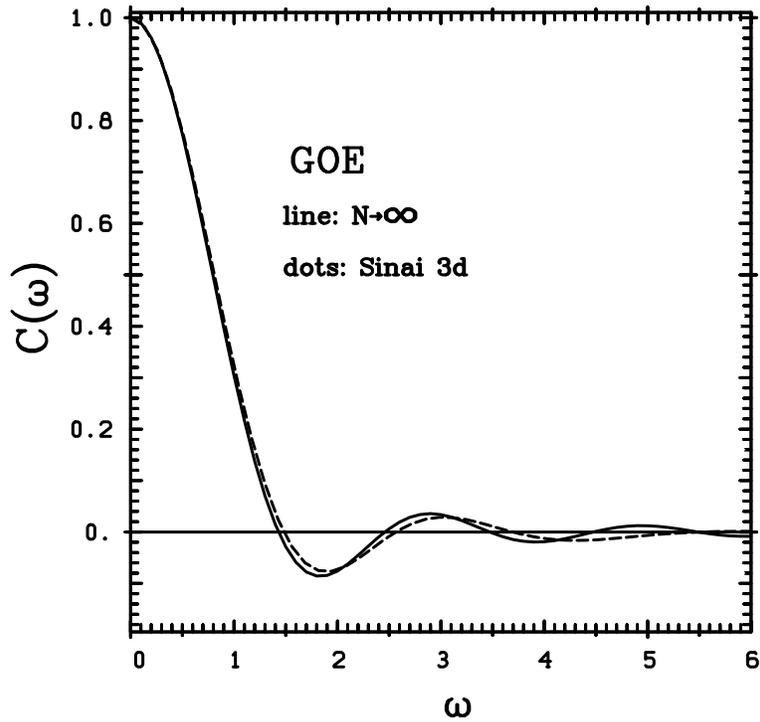,width=10cm}

    \caption{The two-point correlation function of spectral 
      determinants $C(\omega)$ for the 3D Sinai billiard (longest
      spectrum). The spectrum was divided into 167 intervals of 40
      levels each and the average correlation function is shown. The
      continuous line is the RMT-GOE theoretical curve, and the dashed
      line is the numerical correlation. The correlation function is
      normalized to 1 for $\omega = 0$. With kind permission from the
      authors of \protect\cite{KKS97}.}

  \label{fig:zeta-corr}

  \end{center}
\end{figure}

To study $C(\omega)$ numerically, regularizations are needed. For the
3D Sinai billiard the longest spectrum was divided into an ensemble of
167 intervals of $N = 40$ levels, and each interval was unfolded to
have mean spacing $1$ and was centered around $E = 0$. For each
unfolded interval $I_j$ the function $C_j(\omega)$ was computed using
equation (69) of \cite{KKS97}, with $\Delta E = \sqrt{N}$. The
ensemble average function $C(\omega)$ was normalized such that $C(0) =
1$. The results of the computation are shown in figure
\ref{fig:zeta-corr}. The agreement with RMT is quite good up to
$\omega \approx 3$, that is for short energy scales for which we
indeed expect universality to hold.


\section{Classical periodic orbits}
\label{sec:classical-pos}

In this section we present a comprehensive study of the periodic
orbits of the 3D Sinai billiard. By ``periodic orbits'' we mean
throughout this section generic, isolated and unstable periodic orbits
which involve at least one bounce from the sphere. Thus,
bouncing--ball orbits are not treated in this section. The classical
periodic orbits are the building blocks for the semiclassical
Gutzwiller trace formula, and are therefore needed for the
semiclassical analysis to be presented in the next section.

\subsection{Periodic orbits of the 3D Sinai torus}
\label{subsec:pos-torus}

We found it necessary and convenient to first identify the periodic
orbits of the {\bf symmetric} 3D Sinai billiard on the torus, and to
compute their lengths and stabilities. The periodic orbits of the {\bf
  desymmetrized} 3D Sinai billiard could then be derived by an
appropriate classical desymmetrization procedure.

\begin{figure}[p]
  \begin{center}
    \leavevmode
    
    \psfig{figure=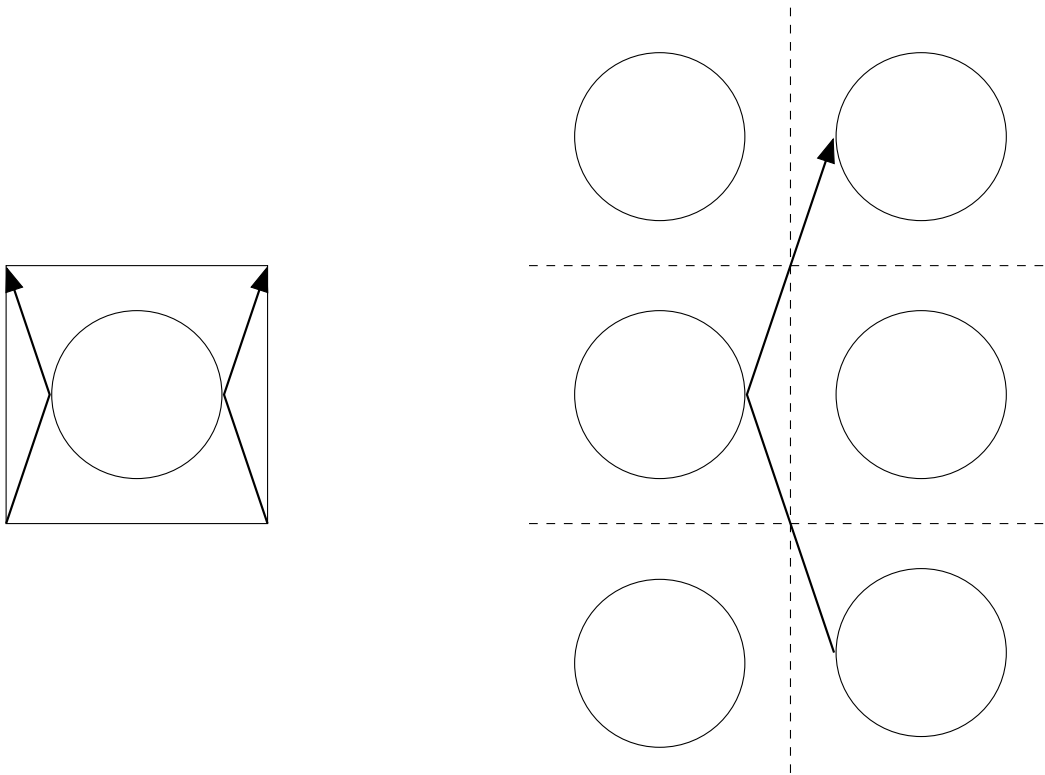,width=14cm}

    \caption{Representation of a periodic orbit of the Sinai 2-torus.
      Left: One cell representation, Right: Unfolded representation.}

    \label{fig:stpo}

  \end{center}

\end{figure}

The basic problem is how to find in a systematic (and efficient) way
all the periodic orbits of the 3D Sinai billiard up to a given length
$L_{\rm max}$. In dealing with periodic orbits of the Sinai billiard
it is very helpful to consider its unfolded representation that
tessellates $\bbbr^3$ --- as is shown in figure \ref{fig:sb}. We start
by considering the periodic orbits of the fully symmetric 3D Sinai
billiard on the torus (ST). This case is simpler than the
desymmetrized billiard, since it contains no boundaries and the tiling
of the $\bbbr^3$ space is achieved by simple translations along the
cubic lattice $\bbbz^3$. In the unfolded representation every orbit is
described by a collection of straight segments which connect spheres.
At a sphere, the incident segment reflects specularly. A {\em
  periodic}\/ orbit of period $n$ is not necessarily periodic in the
unfolded representation, but rather, it obeys the restriction that the
segments repeat themselves after $n$ steps {\em modulo}\/ a
translation by a lattice vector (see figure \ref{fig:stpo}).
If we fix an origin for the lattice, we can assign to every orbit (not
necessarily periodic) a ``code word'' by concatenating the
``addresses'' (locations of the centers on the $\bbbz^3$ lattice) of
the spheres from which it reflects. The code word can consist of
either the {\em absolute}\/ addresses of the spheres or alternatively,
the address of the sphere {\em relative}\/ to the previous one. We
shall adopt the latter convention and use the relative addresses as
the ``letters'' from which the code word is composed. This relative
coding has the advantage that a periodic orbit is represented by a
periodic code word. The number of possible letters (``alphabet'') is
obviously infinite and the letter $(0, 0, 0)$ is excluded. A periodic
orbit can be represented by any cyclic permutation of its code. To
lift this ambiguity, we choose a convenient (but otherwise arbitrary)
lexical ordering of the letters and use the code word which is {\em
  lexically maximal}\/ as the unique representative of the periodic
orbit:
\begin{equation}
  \begin{array}{l}
    \mbox{(periodic orbit of ST)} \longmapsto
    W = (w_1, w_2, \ldots, w_n) \; ,
    w_i \in \bbbz^3 \backslash (0, 0, 0)
    \\ \\
    W = \max \{ W, \hat{P}W, \hat{P}^2 W, \ldots, \hat{P}^{n-1} W \} \: ,
  \end{array}
  \label{eq:st-coding}
\end{equation}
where $\hat{P}W = (w_2, w_3, \ldots, w_n, w_1)$ is the operation of a
cyclic permutation of the code word.

Let us consider the code word $W$ with $n$ letters:
\begin{equation}
  W = (w_1, w_2, \ldots, w_n) \: ,
  \; \; \; w_i = (w_{ix}, w_{iy}, w_{iz}) \: .
\end{equation}
It relates to the $n+1$ spheres centered at $c_1 = (0, 0, 0)$, $c_2 =
w_1$, $c_3 = w_1+w_2, \ldots, c_{n+1} = w_1 + \cdots + w_n$. Let us
choose arbitrary points on each of the spheres, and connect them by
straight segments. We get a piecewise straight line which leads from
the first to the last sphere, which, in general, is not a classical
orbit because the specular reflection conditions are not satisfied.
To find a periodic orbit, we specify the positions of the points on
each sphere by two angles $\theta_i$ , $\varphi_i$. The length of the
line is a function of $\{ (\theta_i, \varphi_i)|_{i=1,\cdots,n}
\}$. Periodic orbits on the ST must have identical
coordinates for the first and the last points (modulo a lattice
translation), hence
$\theta_{n+1} = \theta_1$, $\varphi_{n+1} = \varphi_1$ and we have
only $2n$ independent variables to completely specify a periodic set
of segments, with length:
\begin{equation}
  L_{W}(\theta_1, \ldots, \theta_n, \varphi_1, \ldots, \varphi_n)
  =
  \sum_{i=1}^{n} L_{i} (\theta_i, \theta_{i+1},
                        \varphi_i, \varphi_{i+1}) \: ,
\label{eq:length-of-line}
\end{equation}
where $L_i$ are the lengths of the segments that correspond to the
letter $w_i$. To satisfy the condition of specular reflection we
require that the length $L_W$ is extremal with respect to any
variation of its variables.

\begin{figure}[p]
  \begin{center}
    \leavevmode

    \psfig{figure=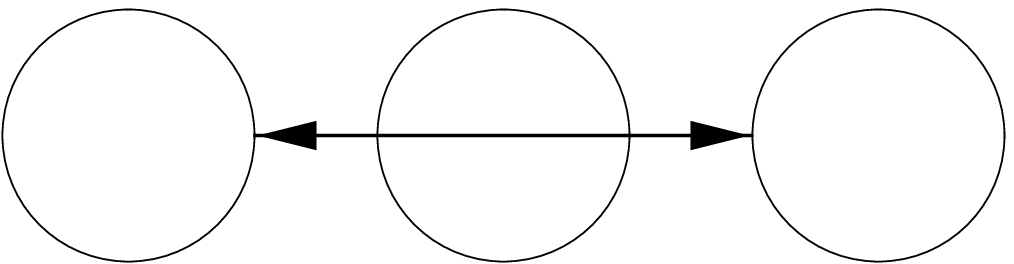,width=10cm}

    \caption{A shadowed (classically forbidden) periodic orbit of
      the Sinai 3-torus.}

    \label{fig:st-shadow}

  \end{center}
\end{figure}

The following theorem guarantees two essential properties of the
coding and of the periodic orbits which are identified as the extrema
of (\ref{eq:length-of-line}) \cite{Bun95,Sch96}:
\begin{description}
\item[{\bf Theorem:}] To each code word $W$ of the 3D ST there
  corresponds {\em at most}\/ one periodic orbit which is the only
  minimum of $L_W$.
\end{description}
The theorem contains two statements: First, that periodic orbits are
necessarily minima of the length, and not saddles or maxima. Second,
that there are no local minima besides the global one. The phrase ``at
most'' in the theorem above needs clarification: For each code word
$W$ the length function $L_W$ is a continuous function in all of its
variables over the compact domain which is the union of the spheres.
Therefore $L_W$ must have a global minimum within this domain. This
minimum can be, however, classically forbidden, meaning that at least
one of its segments cuts through one or more spheres in the lattice
(that might or might not be a part of the code) rather than reflecting
from the outside. This is called ``shadowing''. An example is shown in
figure \ref{fig:st-shadow}.
The forbidden periodic orbits are excluded from the set of classical
periodic orbits. (They also do not contribute to the leading order of
the trace formula \cite{BB72,Ber81} and therefore are of no interest
in our semiclassical analysis.) If all the segments are classically
allowed, then we have a valid classical periodic orbit. Finally we
would like to mention that the minimality property was already implied
in the work of Sieber \cite{Sie91}, and the explicit versions of the
theorem were proved simultaneously by Bunimovich \cite{Bun95} (general
formulation, applies in particular to the 3D case) and Schanz
\cite{Sch96} (restricted to the 2D Sinai billiard).

The number of letters in the codes of periodic orbits of length less
than $L_{\rm max}$ can be bounded from above by the following
argument. To each letter $w$ there corresponds a minimal segment
length $L_{\rm min}(w) > 0$ which is the minimum distance between the
spheres centered at $(0,0,0)$ and at $w = (w_x, w_y, w_z)$:
\begin{equation}
  L_{\rm min}(w) 
  = 
  S \sqrt{w_x^2 + w_y^2 + w_z^2} - 2R \: .
\end{equation}
In the above, $S$ is the lattice constant (torus's side) and $R$ is
the radius of the sphere. The smallest possible $L_{\rm min}(w)$ is
obtained for $w = (1, 0, 0)$ and equals $S - 2R \equiv L_{\rm min}$.
We readily conclude that the code word cannot contain more letters
than the integer part of $L_{\rm max} / L_{\rm min}$.

We are now in a position to formulate an algorithm for a systematic
search of {\em all}\/ the periodic orbits of length up to $L_{\rm
  max}$ of the 3D Sinai torus:
\begin{enumerate}

\item Collect all of the admissible letters into an alphabet. An
  admissible letter $w$ satisfies:
  \begin{enumerate}
  \item $w \neq (0, 0, 0)$.
  \item $w$ is not trivially impossible due to complete shadowing,
    e.g., like $(2, 0, 0) = 2 \times (1, 0, 0)$.
  \item $L_{\rm min}(w) \leq L_{\rm max}$.
  \end{enumerate}

\item Define an arbitrary lexical order of the letters.

\item From the admissible alphabet construct the set of admissible
  code words $W = (w_1, \ldots, w_n)$, such that:
  \begin{enumerate}
  \item $L_{\rm min}(W) \equiv \sum_{i=1}^{n} L_{\rm min}(w_i) \leq
    L_{\rm max}$.
  \item $w_i \neq w_{i+1}$ --- no a-priori complete shadowing.
  \item $W$ is lexically maximal with respect to cyclic permutations:
    $W = \max \{ \hat{P}^{i}W, i = 0, \ldots, n-1 \}$.
  \end{enumerate}

\item For each candidate code word $W$ minimize numerically the
  function $L_W$. According to the theorem, there should be exactly
  one minimum, which is the global one.

\item Check whether the resulting periodic orbit is shaded. Accept
  only periodic orbits which are not shaded.
 
\end{enumerate}

 Once the periodic orbit is identified,  its monodromy
(stability) matrix  is computed according to the recipe given in appendix
  \ref{app:3d-monodromy}.

\subsection{Periodic Orbits of the 3D Sinai billiard ---
            Classical desymmetrization}
\label{subsec:cl-desym}

\begin{figure}[p]
  \begin{center}
    \leavevmode

    \psfig{figure=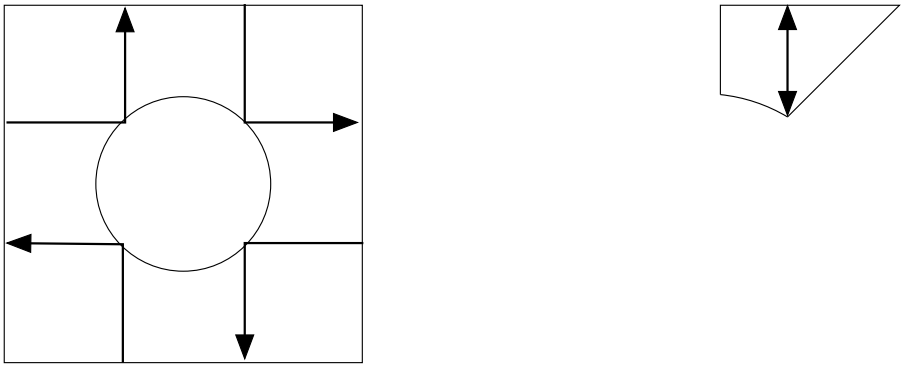,width=12cm}

    \caption{Desymmetrization of orbits from the Sinai torus to
      the Sinai billiard. For clarity we show an example in 2D. Left:
      A primitive periodic orbit in the ST. Right: The corresponding
      periodic orbit in the SB. We observe that the latter is 4 times
      shorter than the former.}

    \label{fig:st-sb}

  \end{center}
\end{figure}

If we desymmetrize the ST into the Sinai billiard (SB), we still find
that the SB tessellates the $\bbbr^3$ space. Hence, each periodic
orbit of the ST is necessarily also a periodic orbit of the SB. The
converse is not true, i.e., periodic orbits of the SB are not
necessarily periodic in ST. However, it is easy to be convinced that
if a periodic orbit of SB is repeated sufficiently many times, it
becomes also periodic in ST. An example is shown in figure
\ref{fig:st-sb}.
From a more abstract point of view, this is because the cubic group
$O_h$ is finite. Thus in principle one could use the algorithm given
above to systematically find all the periodic orbits of the SB. This
is, however, highly inefficient because by analyzing the group $O_h$
we find that in order to find all the periodic orbits of the SB up to
$L_{\rm max}$ we must find all of the periodic orbits of ST up to $6
L_{\rm max}$. Due to the exponential proliferation of periodic orbits
this would be a colossal waste of resources which would diminish our
ability to compute periodic orbits almost completely. To circumvent
this difficulty, without losing the useful uniqueness and minimality
properties which apply to the ST, we make use of the property that
periodicity in the SB is synonymous to periodicity in ST {\em
  modulo}\/ an element $\hat{g} \in O_h$. This simple geometrical
observation is a manifestation of the fact that the tiling of
$\bbbr^3$ by the SB is generated by the group $O_h \otimes \bbbz^3$.
Thus, we can represent the periodic orbits of the SB by using their
unfolded representation, augmented by the symmetry element $\hat{g}$
according to which the periodic orbits closes:
\begin{equation}
  \mbox{Periodic orbit of SB}
  \longmapsto
  \hat{W} \equiv (W; \hat{g})
  =
  (w_1, w_2, \ldots, w_n; \hat{g}) \: .
\end{equation}
The coding is not yet well-defined since a given periodic orbit can in
general be represented by several codes. Similarly to the case of the
ST, there is a degeneracy with respect to the starting point. However,
in the case of the SB this is not simply related to cyclic
permutations. Rather, if a periodic orbit is described by $(w_1, w_2,
\ldots, w_n ; \hat{g})$ then it is also described by:
\begin{equation}
  \begin{array}{l}
    (w_2, w_3, \ldots, w_n, \hat{g} w_1; \hat{g}),
    (w_3, w_4, \ldots, \hat{g}w_1, \hat{g} w_2; \hat{g}), \ldots, \\
    (\hat{g} w_1, \hat{g} w_2, \ldots, \hat{g} w_n; \hat{g}),
    (\hat{g} w_2, \hat{g} w_3, \ldots, \hat{g}^2 w_1; \hat{g}), \ldots \\
    \vdots \\
    (\hat{g}^{\phi(\hat{g})-1} w_1,
    \hat{g}^{\phi(\hat{g})-1} w_2, \ldots,
    \hat{g}^{\phi(\hat{g})-1} w_n; \hat{g}), \ldots,
    (\hat{g}^{\phi(\hat{g})-1} w_n, w_1, w_2, \ldots, 
    w_{n-1}; \hat{g}) \: .
  \end{array}
\end{equation}
In the above $\phi(\hat{g})$ is the {\em period}\/ of $\hat{g}$, which
is defined as the smallest natural number for which
$\hat{g}^{\phi(\hat{g})} = \hat{e}$, where $\hat{e}$ is the identity
operation. For $O_h$ in particular $\phi(\hat{g}) \in \{ 1, 2, 3, 4, 6
\}$. The above generalized cyclic permutation invariance is due to the
periodicity modulo $\hat{g}$ of the periodic orbits of the SB in the
unfolded representation. In addition to the generalized cyclic
invariance there is also a geometrical invariance of orbits of the SB
in the unfolded representation. Indeed, if we operate on an orbit in
the unfolded representation with any $\hat{h} \in O_h$ we obtain the
same orbit in the SB. This symmetry is carried over also to the codes.
If a periodic orbit is described by $(w_1, w_2, \ldots, w_n; \hat{g})$
then it is also described by:
\begin{equation}
  (\hat{h} w_1, \hat{h} w_2, \ldots, \hat{h} w_n;
   \hat{h} \hat{g} \hat{h}^{-1}) \; \; \;
   \forall \hat{h} \in O_h \, .
\end{equation}
To summarize, a periodic orbit of the SB can be encoded into a code
word up to degeneracies due to generalized cyclic permutations and
geometrical operations. The set of operations which relate the various
codes for a given periodic orbit is a group to which we refer as the
invariance group.

In order to lift this degeneracy and to obtain a unique mapping of
periodic orbits of the SB to code words we need to specify a criterion
for choosing exactly one representative. There are many ways of doing
this, but we found it convenient to apply the natural mapping of
periodic orbits of the SB to those of the ST, and there, to choose the
maximal code. More specifically:
\begin{enumerate}

\item Select the alphabet according to the rules prescribed in the
  preceding subsection, and define ordering of letters.

\item Extend the word $\hat{W}$ into $\tilde{W}$:
  \begin{eqnarray}
    \tilde{W}
    & \equiv &
    (w_1, w_2, \ldots, w_n,
    \hat{g} w_1, \hat{g} w_2, \ldots, \hat{g} w_n, \nonumber \\
    & &
    \hat{g}^2 w_1, \ldots,
    \hat{g}^{\phi(\hat{g})-1} w_1, \ldots,
    \hat{g}^{\phi(\hat{g})-1} w_n) \, .
  \end{eqnarray}
  The code $\tilde{W}$ describes the periodic orbit of the SB which is
  continued $\phi(\hat{g})$ times to become periodic in the ST.
  Applying a generalized cyclic permutation on $\hat{W}$ is equivalent
  to applying the standard cyclic permutation on $\tilde{W}$. Applying
  a geometrical operation $\hat{h}$ on $\hat{W}$ is equivalent to
  operating letter by letter with $\hat{h}$ on $\tilde{W}$. The
  invariance group corresponding to $\tilde{W}$ is ${\cal H} = {\cal
    C} \otimes O_h$, where ${\cal C}$ is the group of cyclic
  permutations of order $n \cdot \phi(\hat{g})$. The simple
  decomposition of ${\cal H}$ is due to the commutativity of ${\cal
    C}$ and $O_h$, and it greatly facilitates the computations.

\item If $\tilde{W}$ is maximal with respect to the invariance group
  ${\cal H}$, then the corresponding $\hat{W}$ is the representative
  of the periodic orbit.

\end{enumerate}
A comment on the uniqueness of this selection process is appropriate
at this point. For any $\hat{W}$ we can uniquely construct the
corresponding $\tilde{W}$ and the invariance group and check the
maximality of $\tilde{W}$. Hence, we are able to uniquely decide
whether $\hat{W}$ is a valid representative code or not. However,
there are cases in which more than one $\hat{W}$ correspond to the
{\em same maximal}\/ $\tilde{W}$. It is straightforward to show that
in these cases the basic code word $W$ is symmetric under some
operation(s): $W = \hat{k}W$, $\hat{k} \in O_h$. To such symmetric
codes must correspond symmetric periodic orbits, which is necessitated
by the uniqueness theorem for the ST. But for the SB the symmetry of
the orbit means that it is wholly contained in a symmetry plane, and
therefore is not a proper classical orbit. Such orbits are
nevertheless required for the semiclassical analysis and will be
treated in the next section when dealing with semiclassical
desymmetrization. In summary, we have shown so far that the mapping of
a given {\em proper}\/ periodic orbit to a code is well-defined and
unique.

In order for the coding to be useful and powerful, we need to
establish uniqueness in the opposite direction, that is to show that
for a given (unsymmetrical) $\hat{W}$ there corresponds at most one
(proper) classical periodic orbit. The mapping $\hat{W} \mapsto
\tilde{W}$ is very useful in that respect. Indeed, if there were two
distinct periodic orbits of the SB with the same coding $\hat{W}$,
then we could repeat them $\phi(\hat{g})$ times to get two distinct
periodic orbits of the ST with the same code $\tilde{W}$, which is in
contradiction with the theorem above. This proves the uniqueness of
the relation between codes and periodic orbits.

To facilitate the actual computation of periodic orbits of the SB, we
have to establish their minimality property, similarly to the ST case.
We need to prove that the length of a periodic orbit is a minimum, and
that it is the only minimum. The minimality of a periodic orbit of the
SB is proven by using again the unfolding to periodic orbits of ST,
and noting that a minimum of $L_{\tilde{W}}$ is necessarily also a
minimum of $L_{\hat{W}}$, since the latter is a constrained version of
the former. Thus, periodic orbits of the SB are minima of
$L_{\hat{W}}$. We finally have to show that there exists only a single
minimum of $L_{\hat{W}}$. The complication here is that, in principle,
a minimum of $L_{\hat{W}}$ does not necessarily correspond to a
minimum of $L_{\tilde{W}}$, since there are, in general, more
variables in the latter. We resolve this difficulty by using arguments
from the proof of Schanz \cite{Sch96} as follows. A necessary
condition for minimality is that orbits are either externally
reflected from the scatters or cut through them in straight segments.
Internal reflections are not allowed for a minimum. Thus, if we extend
a minimum of SB to ST, we necessarily get an orbit with no internal
reflections. According to Schanz \cite{Sch96}, there is exactly one
such orbit, which is the minimum in ST. This proves the
uniqueness of the (global) minimum of $L_{\hat{W}}$ in SB.

These results allow us to use essentially the same algorithm as for
the ST for the systematic search of periodic orbits of the SB. We need
to extend the codes and the length functions to include a group
element $\hat{g}$, and to modify the rules according to which we
choose an admissible and lexically maximal code word $\hat{W}$. One
also has to modify the computations of the monodromy matrix, as
described in appendix \ref{app:3d-monodromy}.

\subsection{The properties and statistics of
            the set of periodic orbits}
\label{subsec:pos-figs+stat}

\begin{figure}[p]
  \begin{center}
    \leavevmode
    \begin{tabular}{ll}
      \psfig{figure=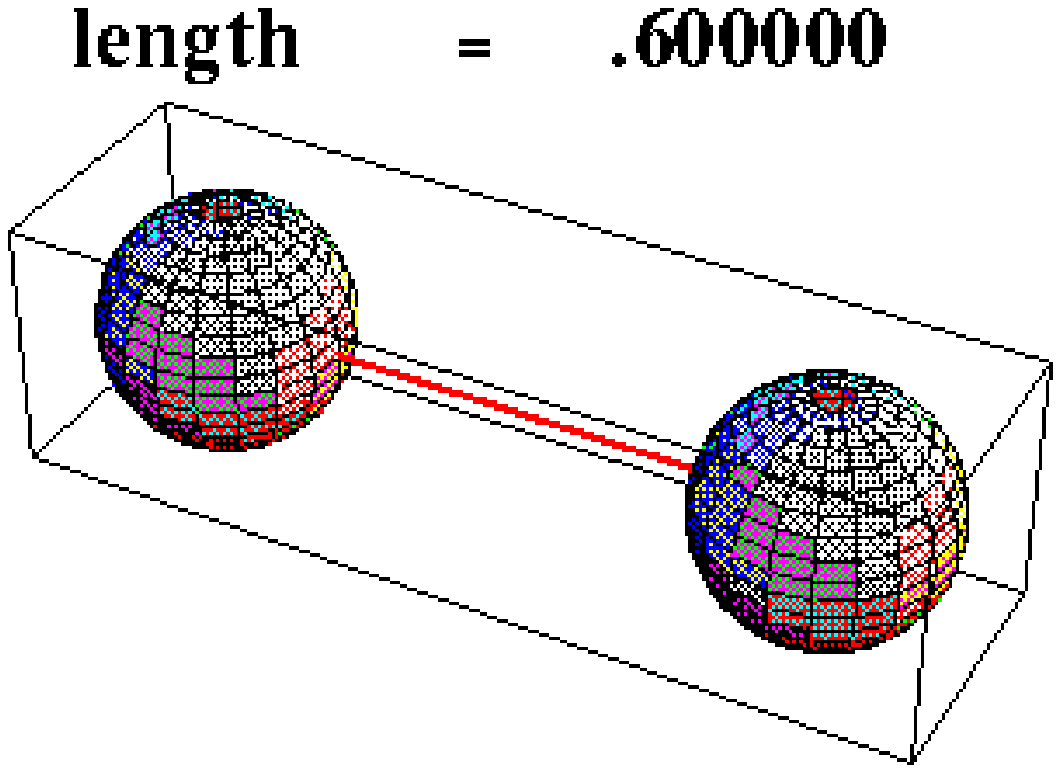,height=5cm} &
      \psfig{figure=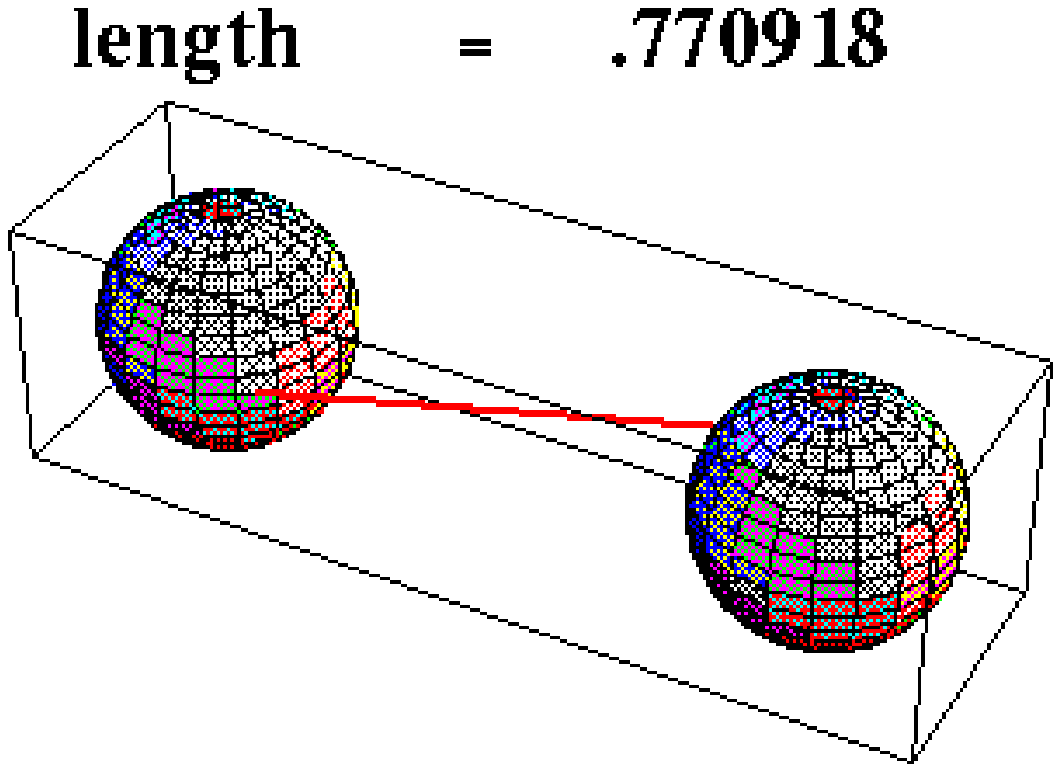,height=5cm} \\
      \psfig{figure=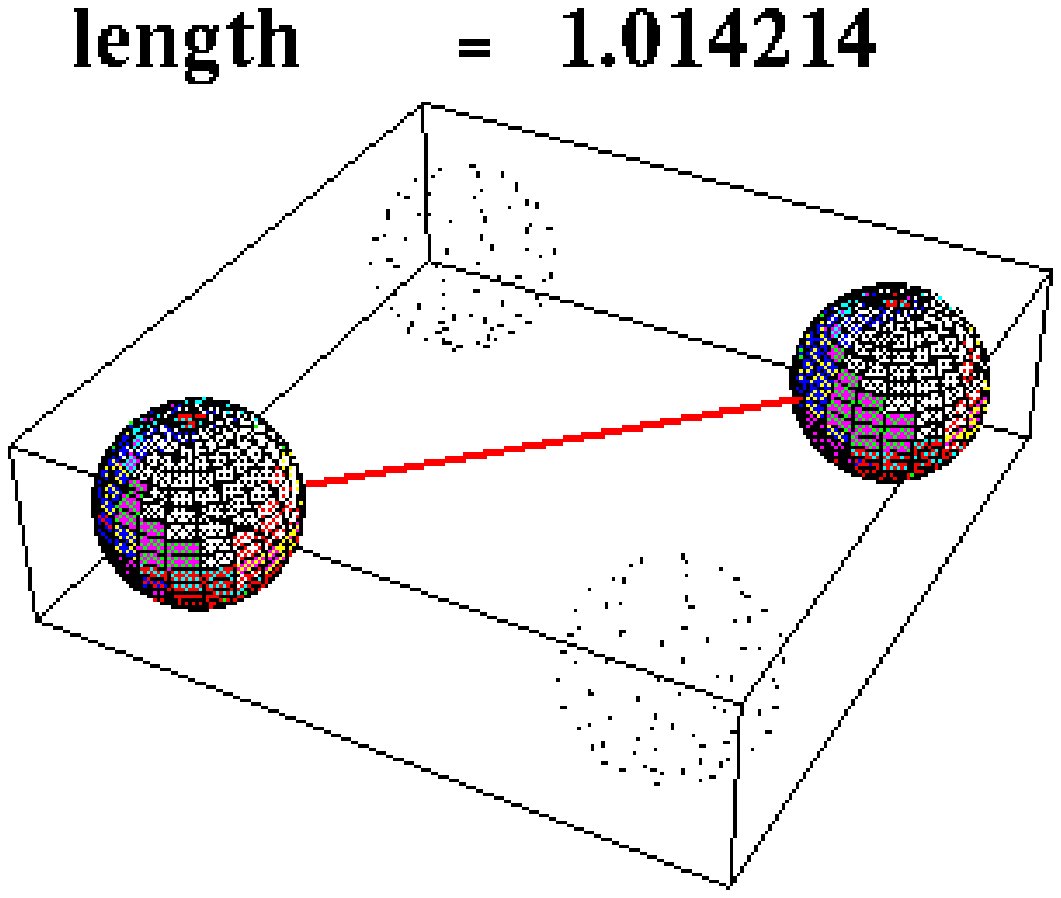,height=5cm} &
      \psfig{figure=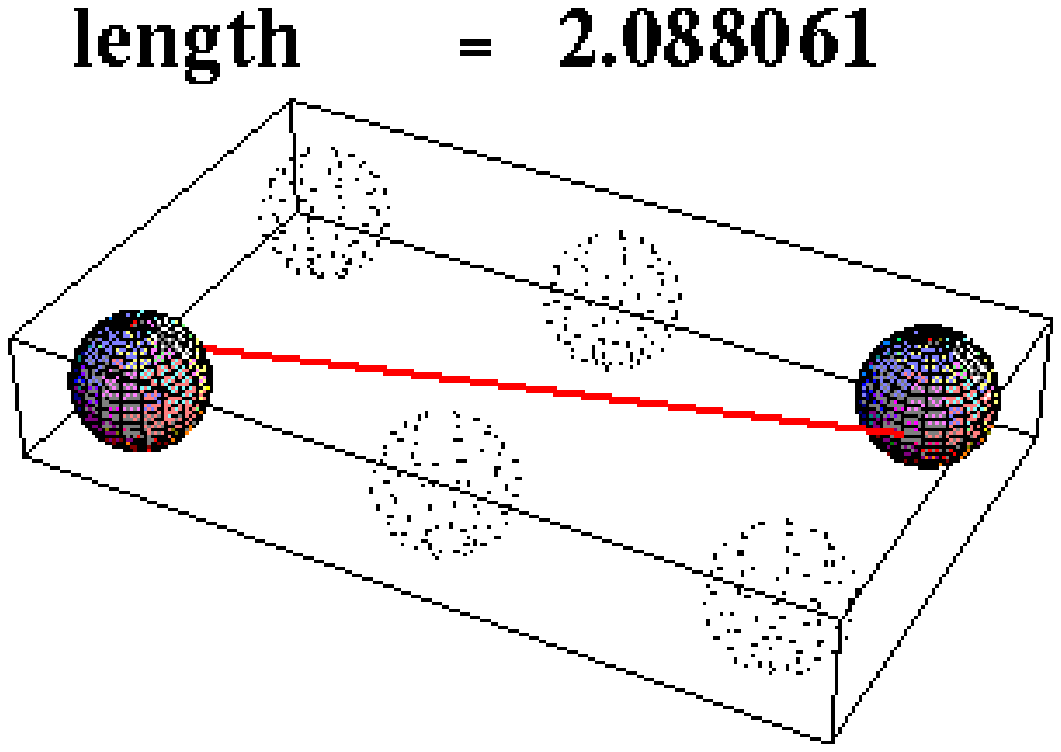,height=5cm} \\
      \psfig{figure=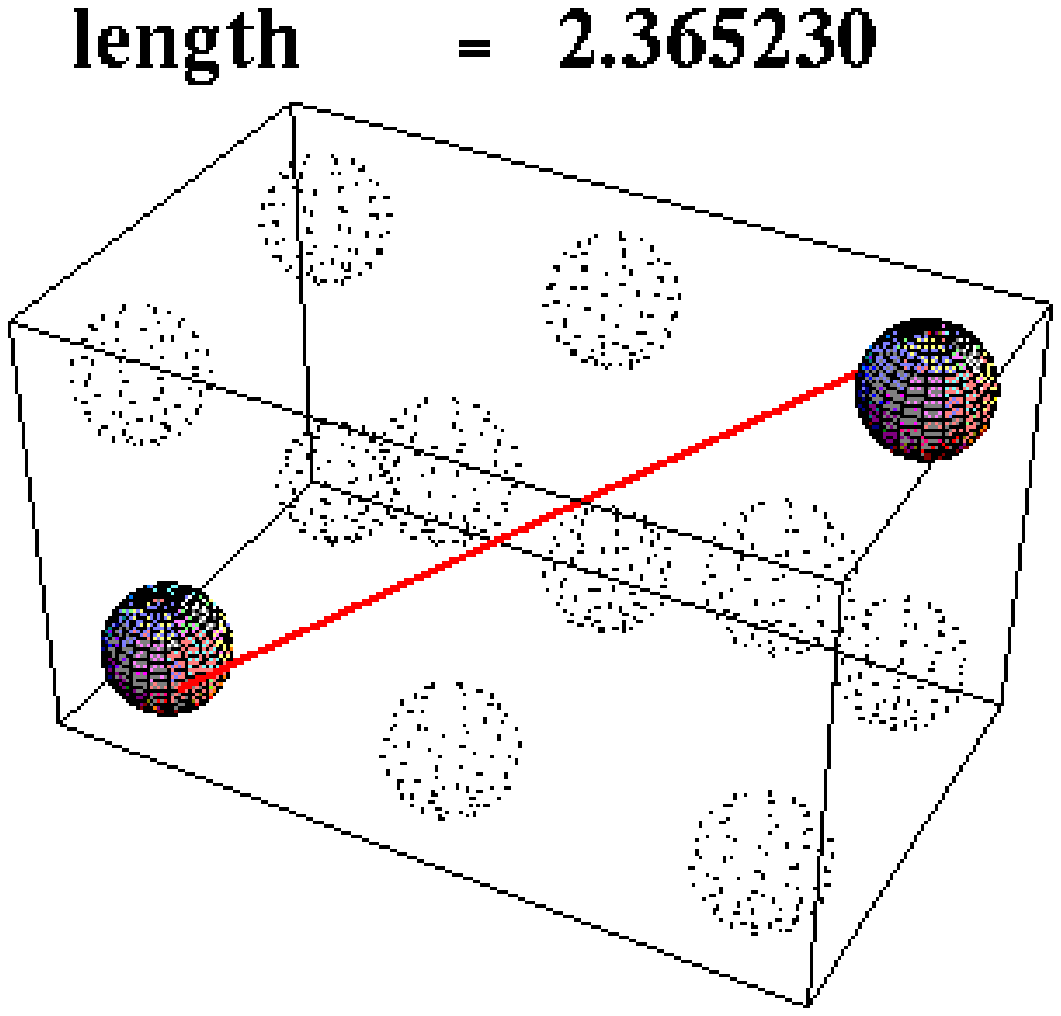,height=5cm} &
      \psfig{figure=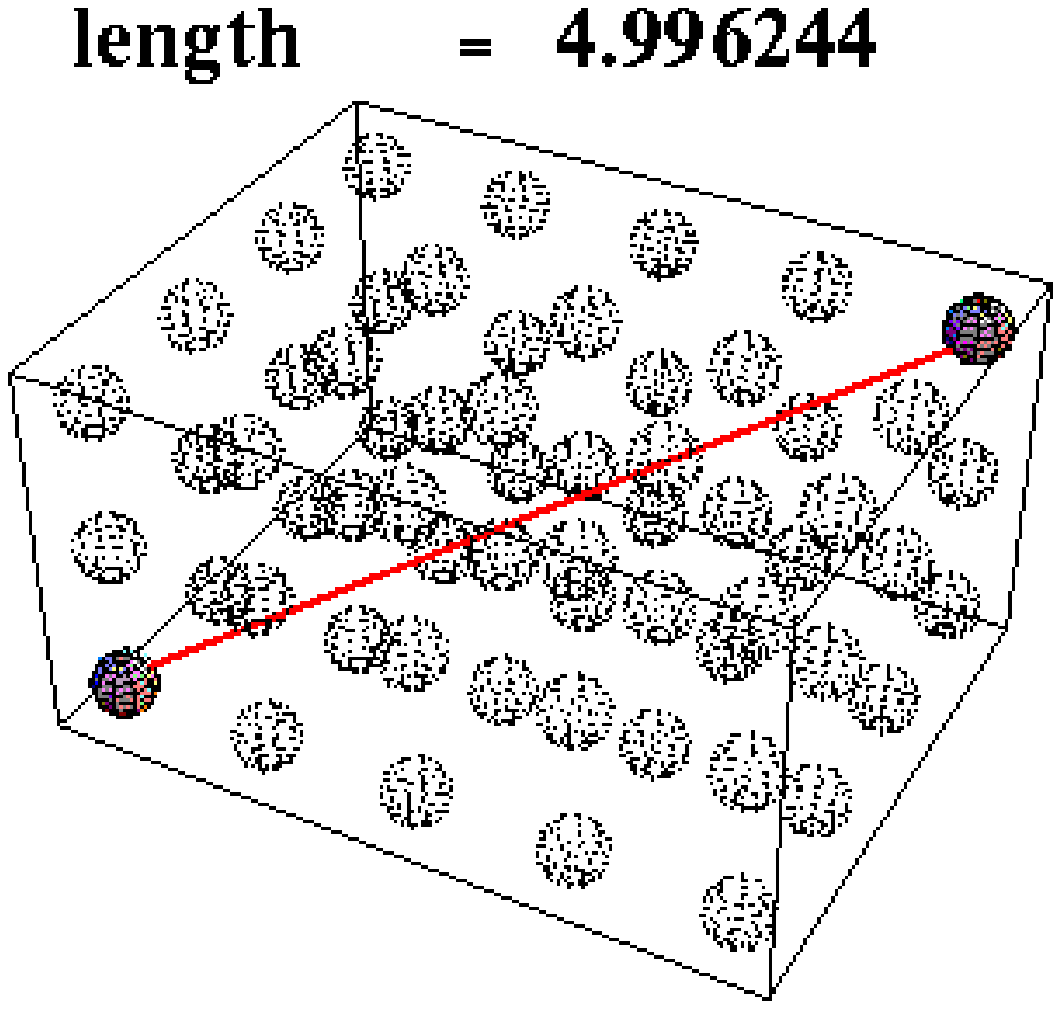,height=5cm}
    \end{tabular}

    \caption{A sample of periodic orbits of the desymmetrized 3D
      Sinai billiard with $S=1$, $R=0.2$ with a single reflection.
      The periodic orbits are shown in the unfolded representation.
      The ``full'' spheres are those from which the periodic orbit
      reflects. The ``faint'' dotted spheres are those from which
      there is no reflection.}

    \label{fig:pos-gallery1}

  \end{center}
\end{figure}

\begin{figure}[p]
  \begin{center}
    \leavevmode
    \begin{tabular}{ll}
      \psfig{figure=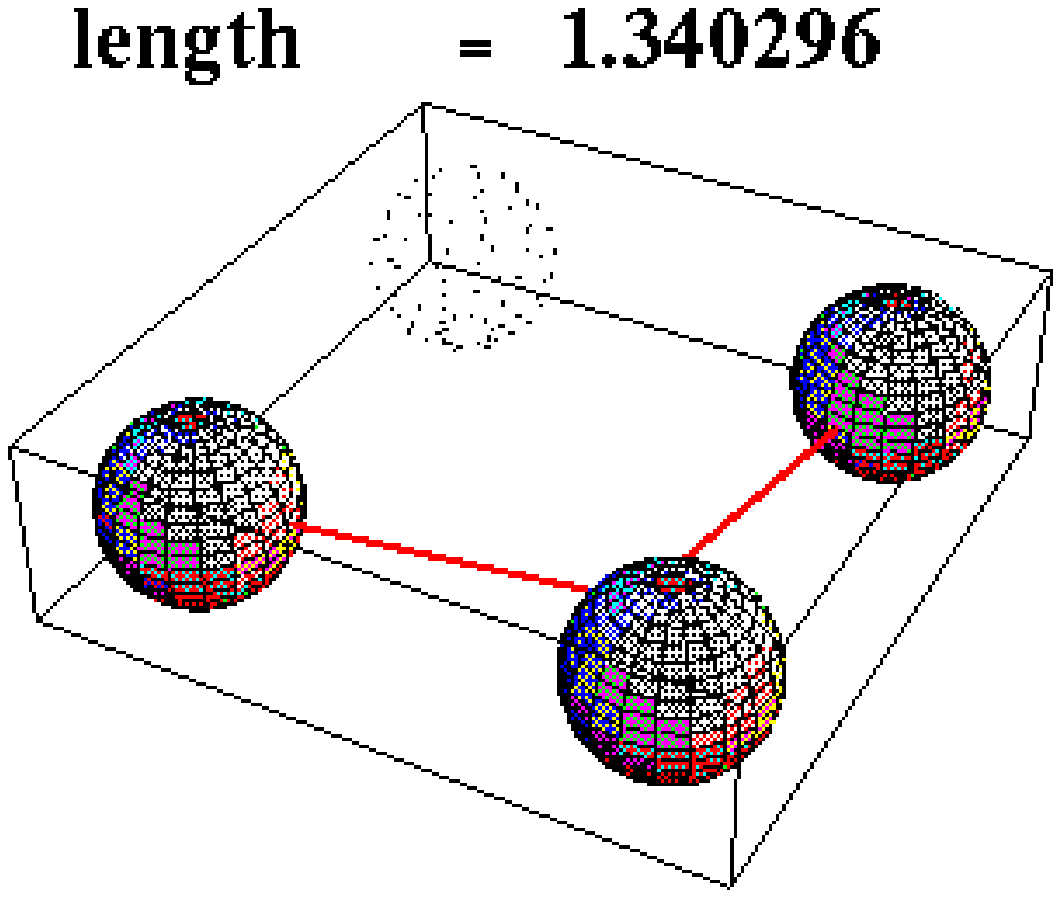,height=5cm} &
      \psfig{figure=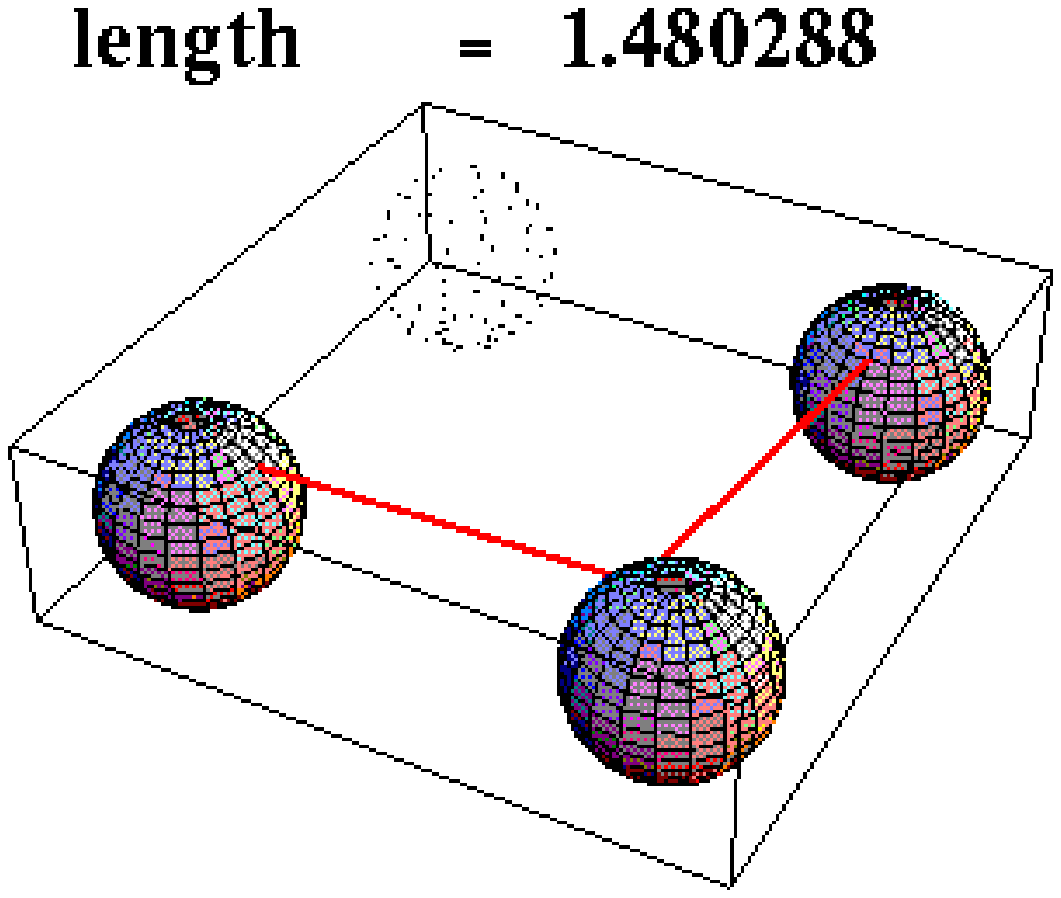,height=5cm} \\
      \psfig{figure=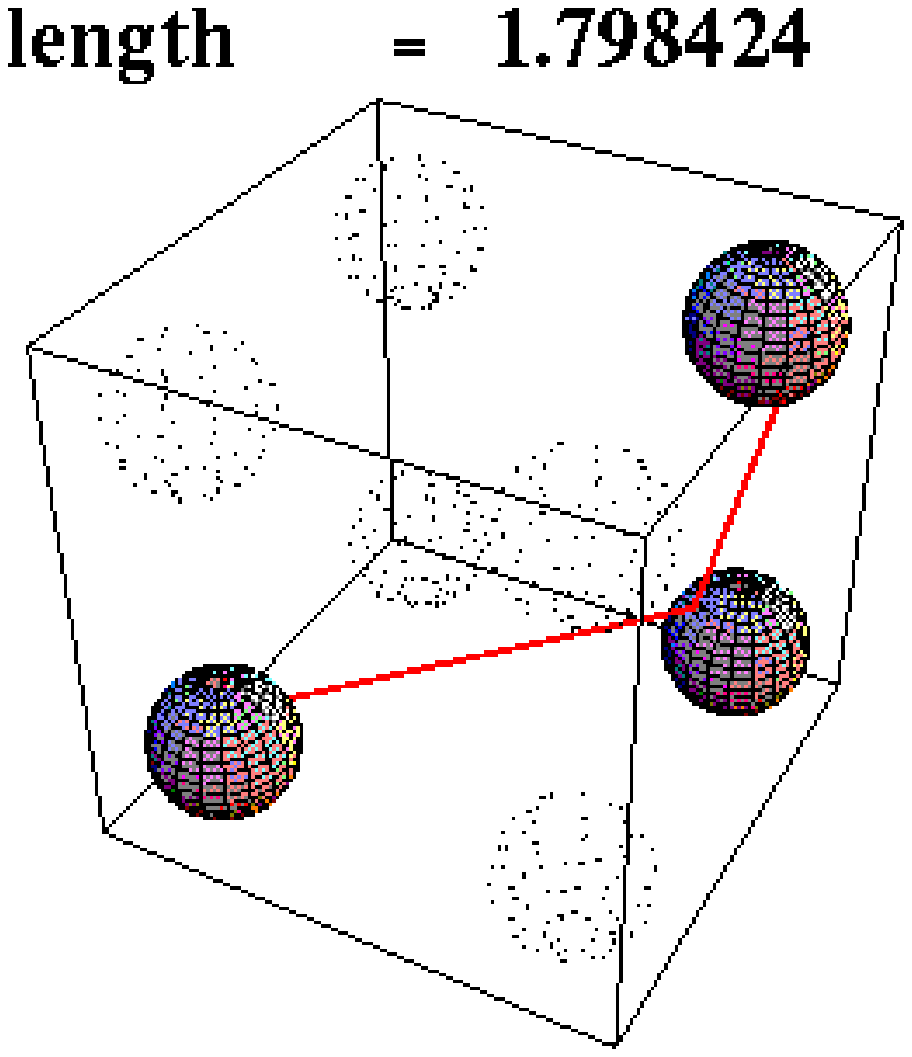,height=5cm} &
      \psfig{figure=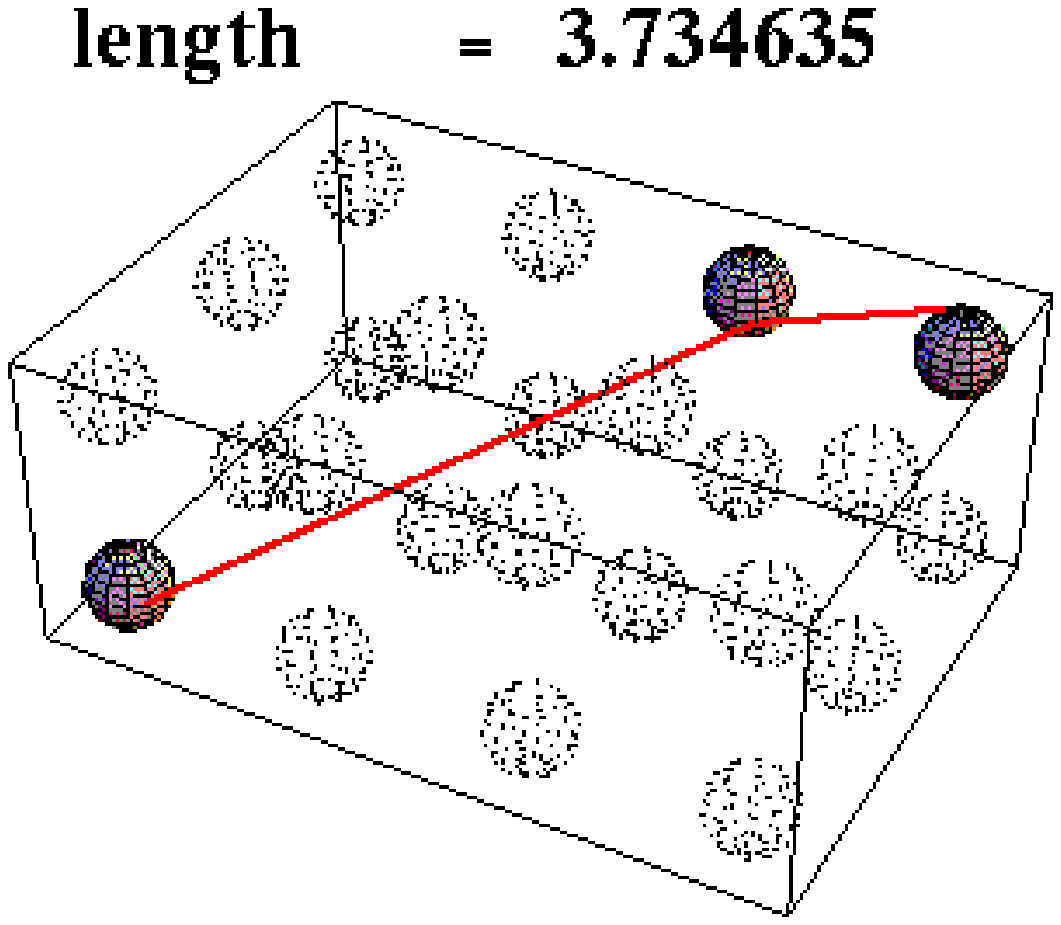,height=5cm}
    \end{tabular}
    \psfig{figure=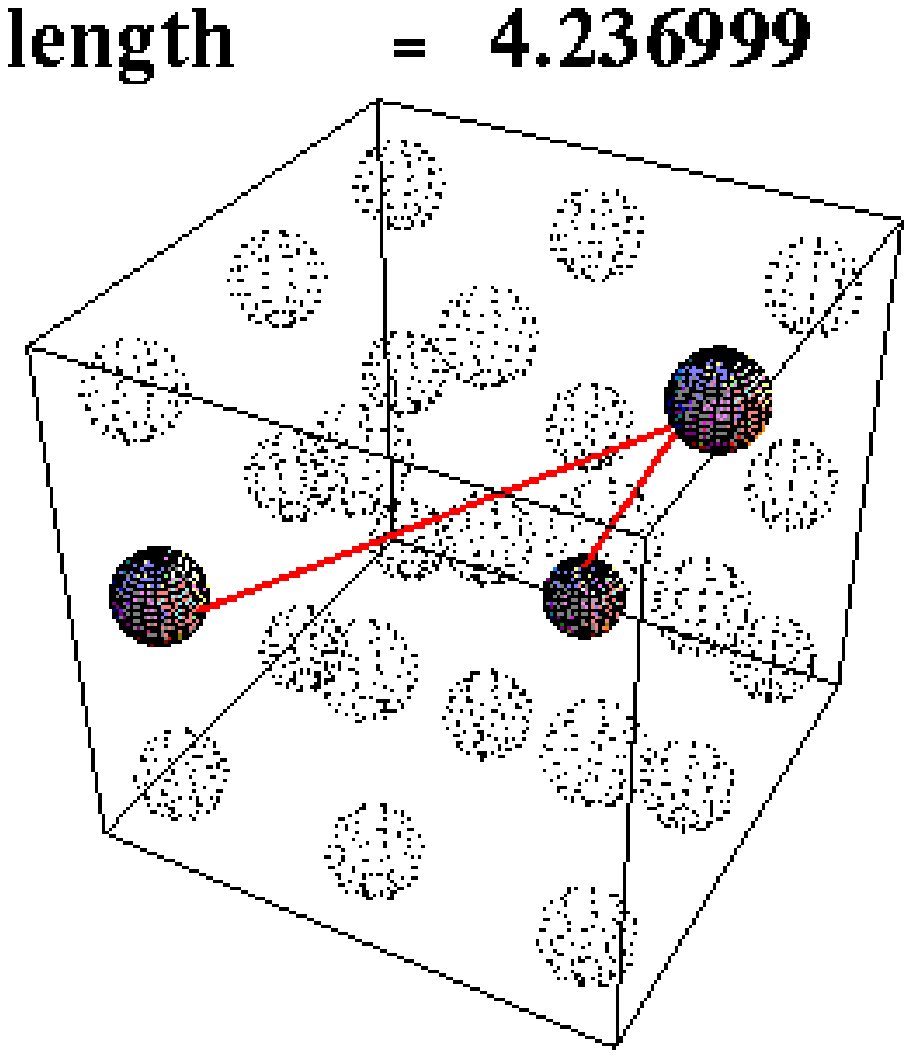,height=5cm}

    \caption{A sample of periodic orbits of the 3D SB
      with 2 reflections.}

    \label{fig:pos-gallery2}

  \end{center}
\end{figure}

\begin{figure}[p]
  \begin{center}
    \leavevmode
    \begin{tabular}{ll}
      \psfig{figure=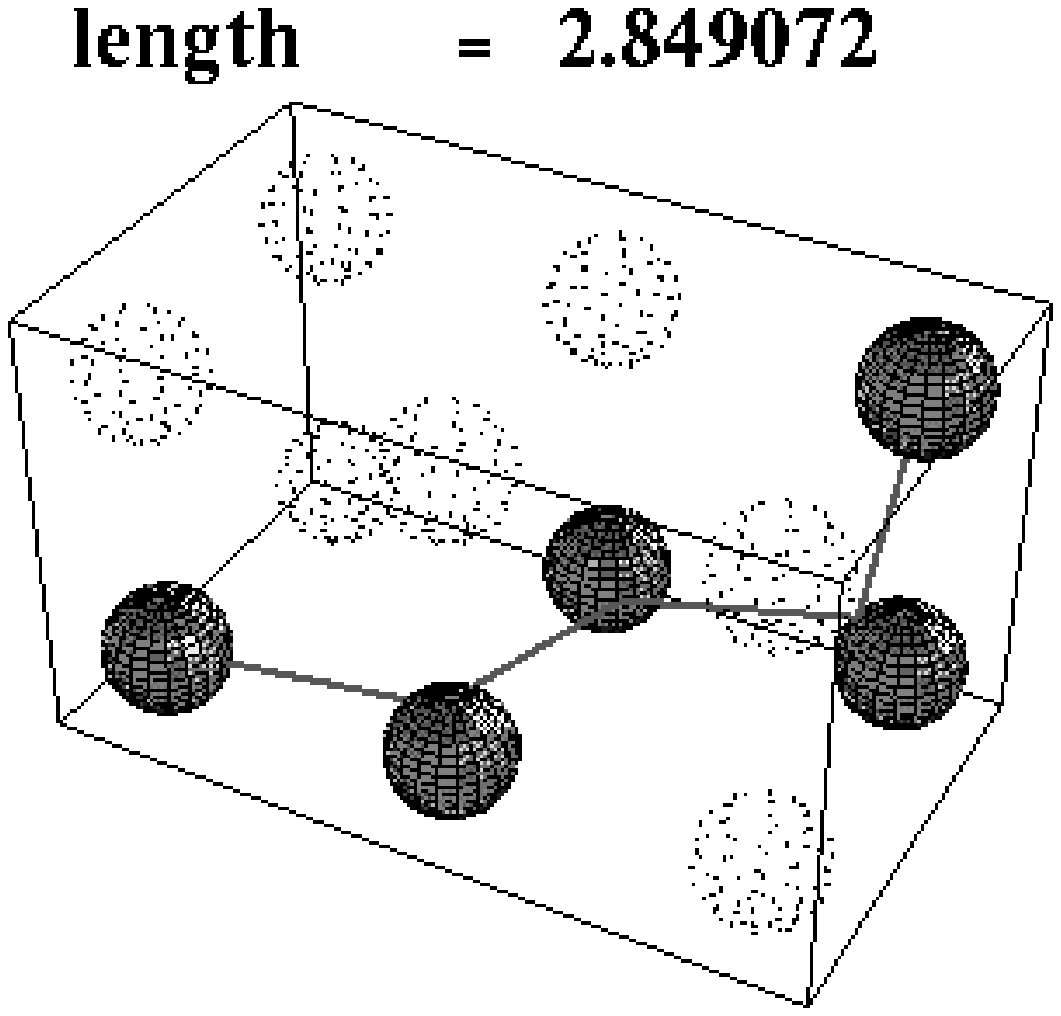,height=5cm} &
      \psfig{figure=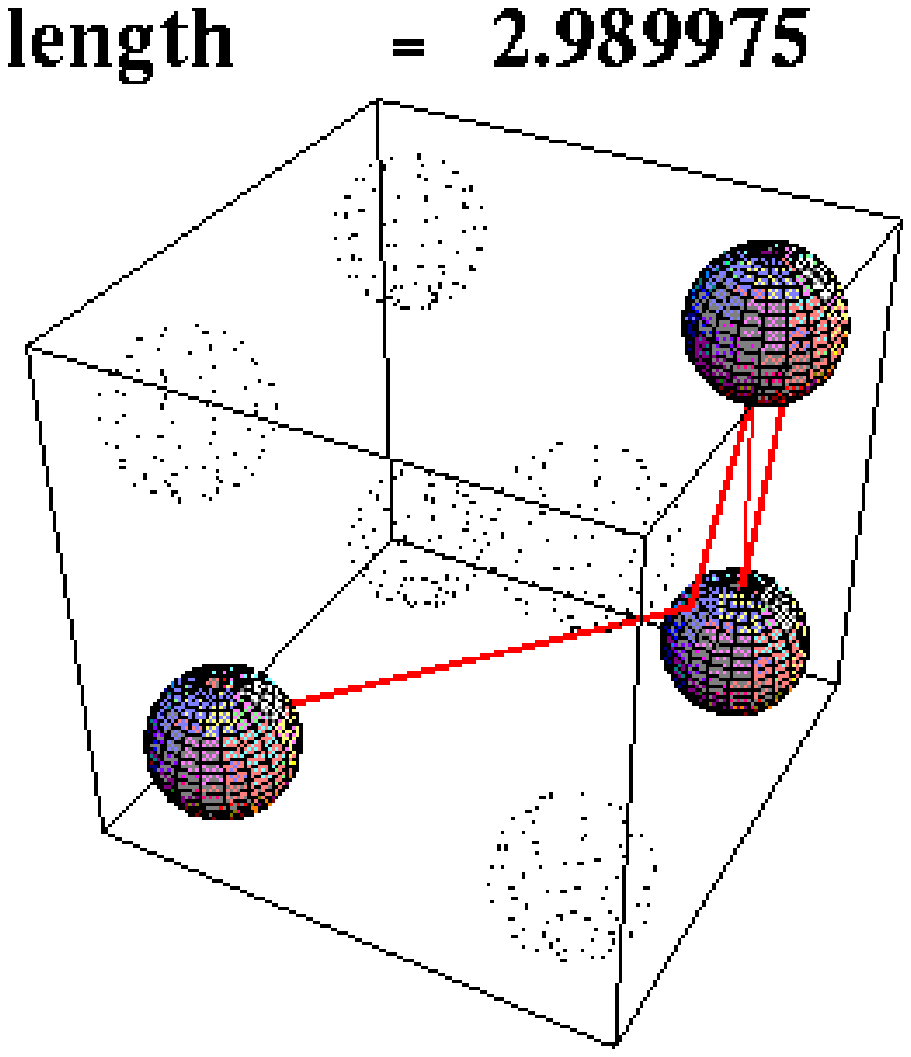,height=5cm} \\
      \psfig{figure=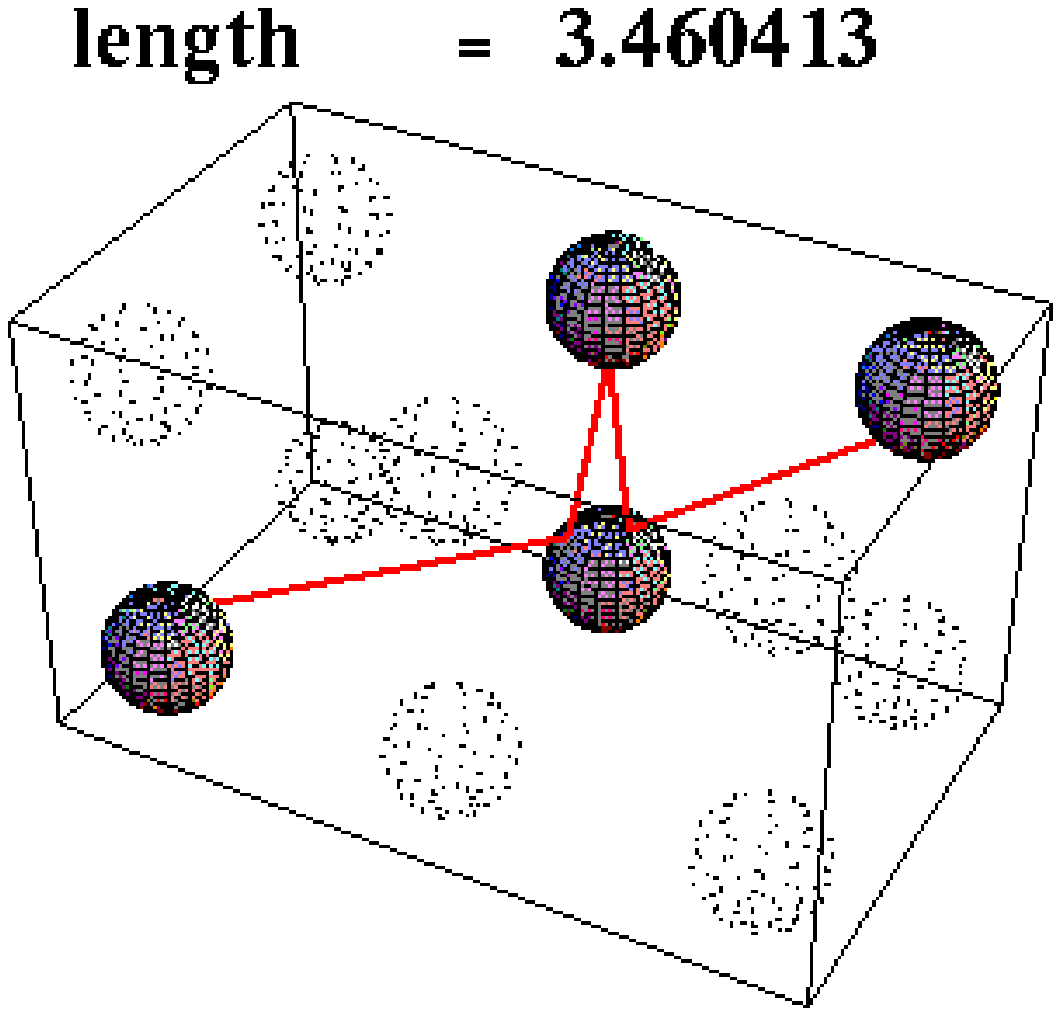,height=5cm} &
      \psfig{figure=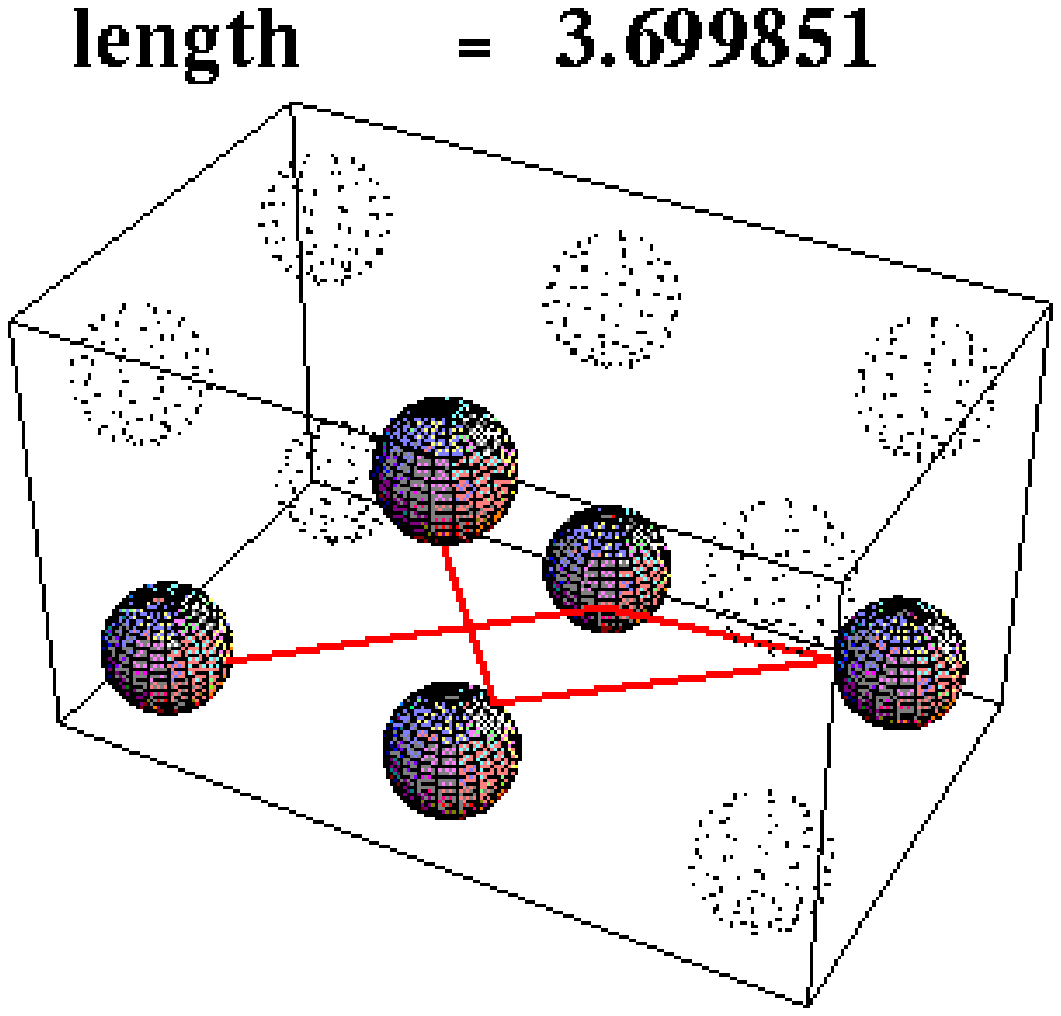,height=5cm} \\
      \psfig{figure=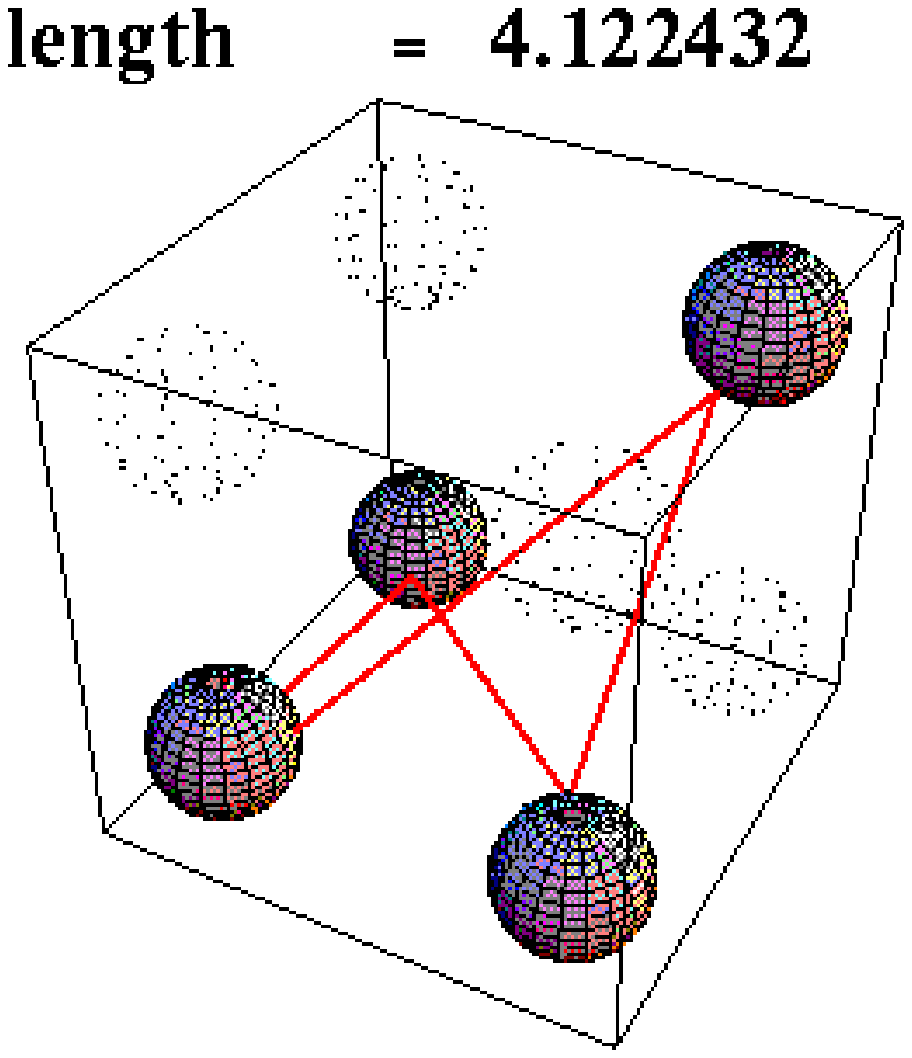,height=5cm} &
      \psfig{figure=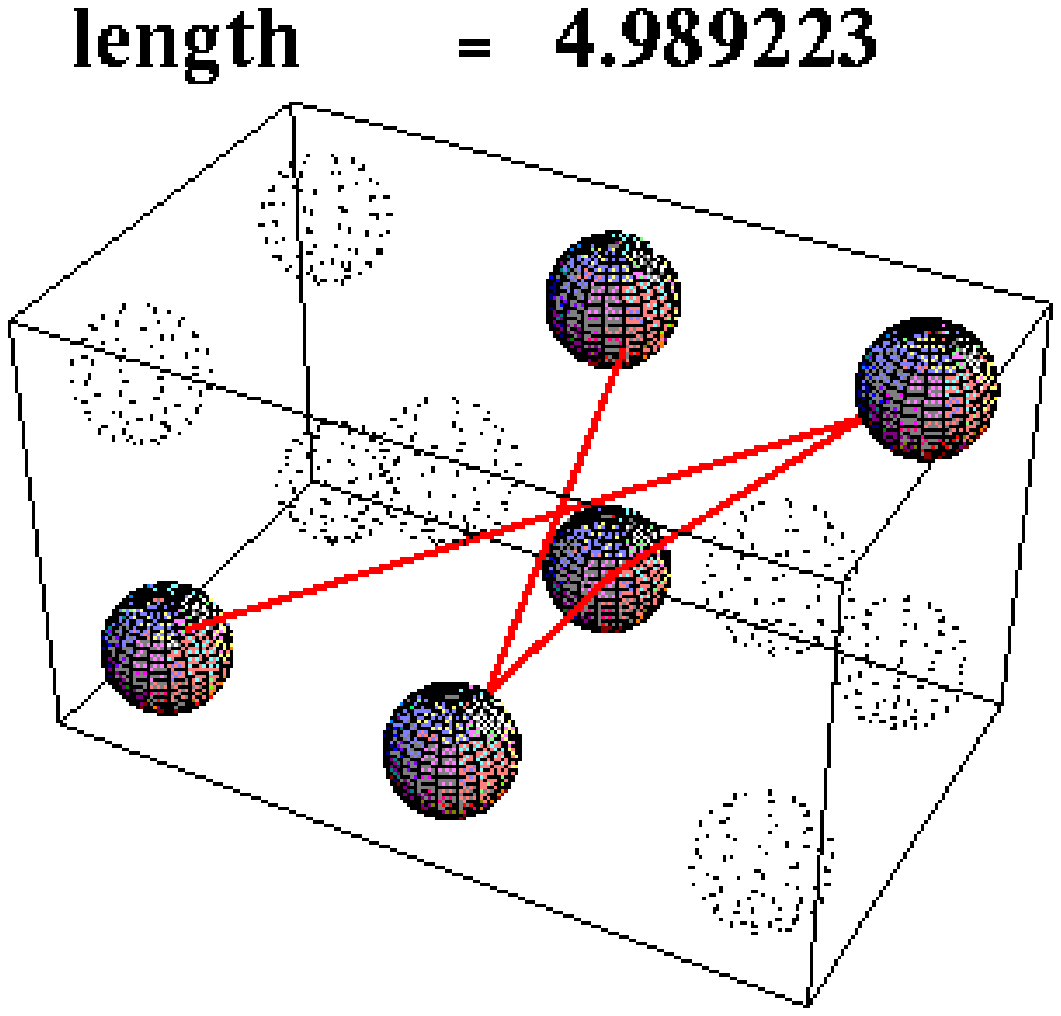,height=5cm}
    \end{tabular}

    \caption{A sample of periodic orbits of the 3D SB
      with 4 reflections.}

    \label{fig:pos-gallery3}

  \end{center}
\end{figure}

\begin{figure}[p]
  \begin{center}
    \leavevmode
    \begin{tabular}{ll}
      \psfig{figure=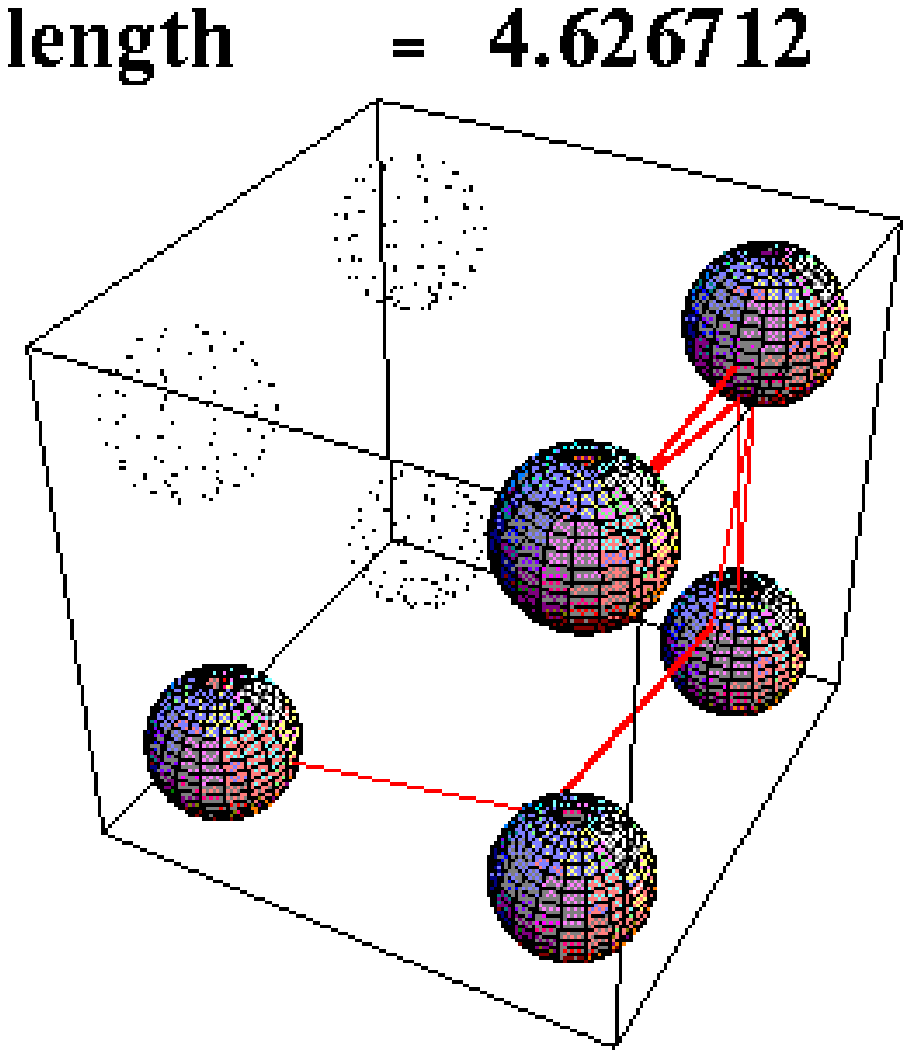,height=5cm} &
      \psfig{figure=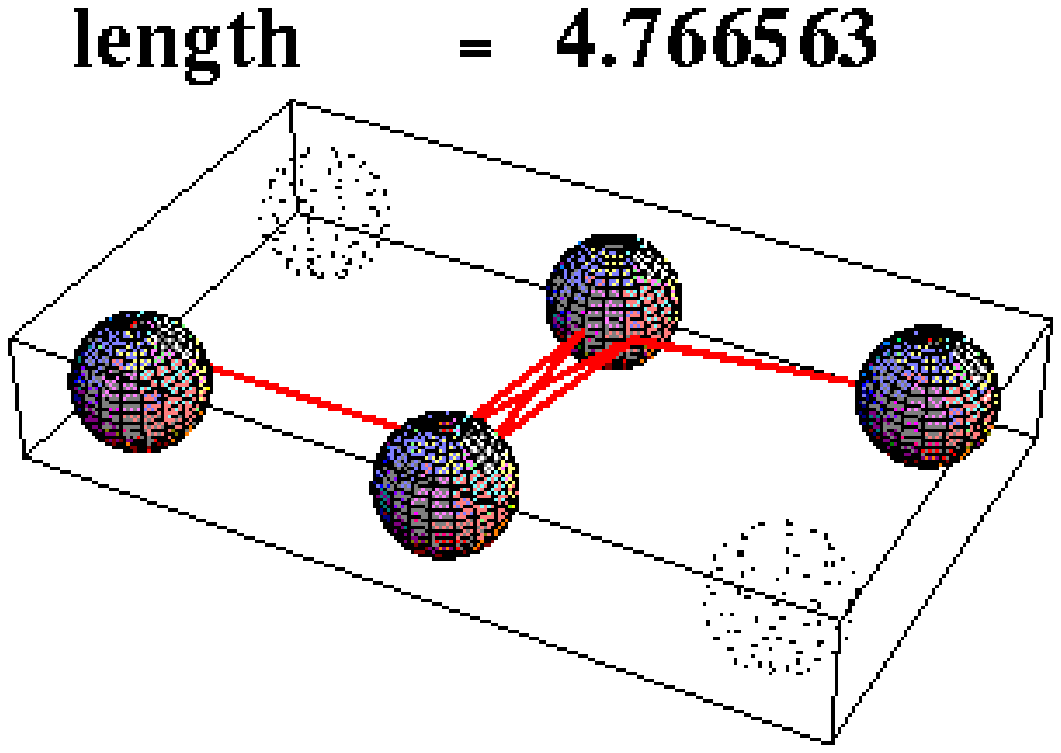,height=5cm} \\
      \psfig{figure=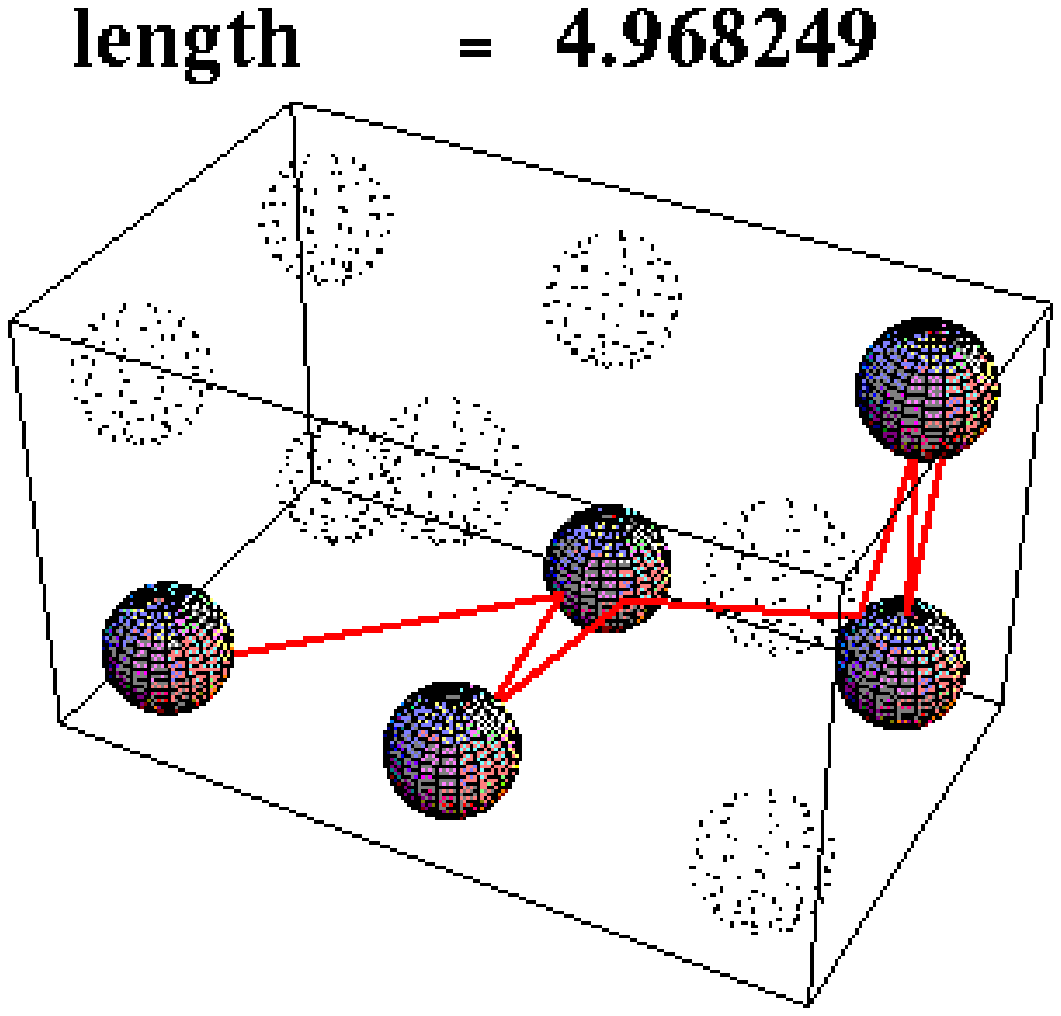,height=5cm} &
      \psfig{figure=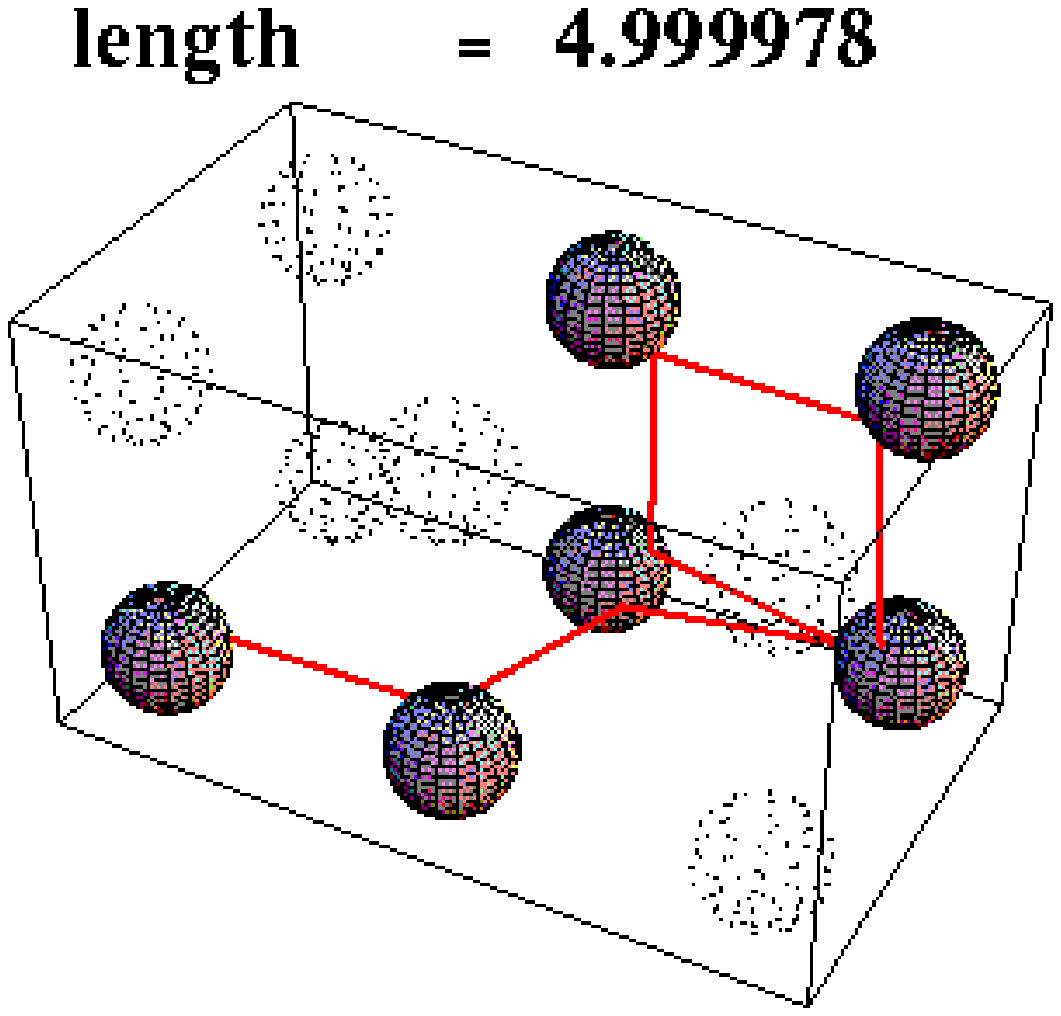,height=5cm}
    \end{tabular}
    \psfig{figure=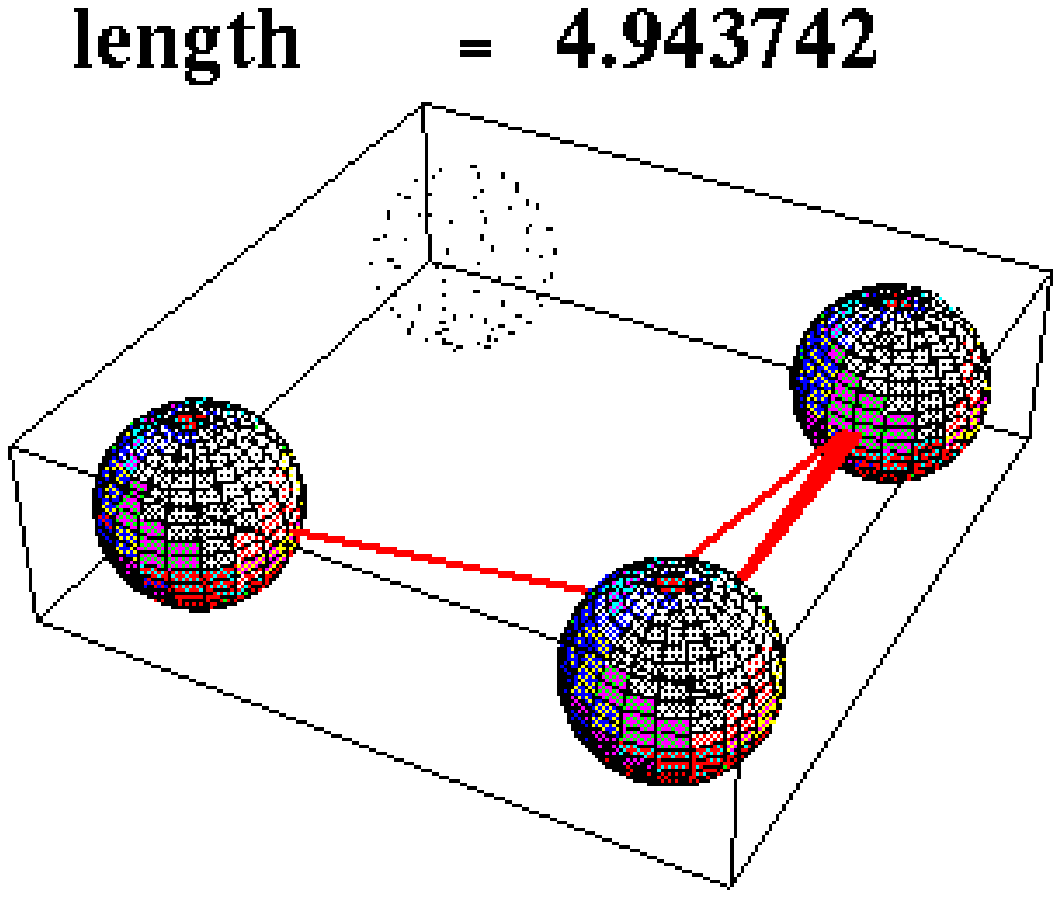,height=5cm}

    \caption{A sample of periodic orbits of the 3D SB
      with 7 reflections. The bottom periodic orbit undergoes 8
      reflections.}

    \label{fig:pos-gallery4}

  \end{center}
\end{figure}

The algorithm described above is capable of finding all of the
periodic orbits up to any desired length. Before discussing the
properties of this set, we find it appropriate to display a few
typical periodic orbits, which were computed for the desymmetrized
billiard with $R=0.2$ (and $S=1.$). The orbits are represented in an
unfolded way in figures \ref{fig:pos-gallery1}--\ref{fig:pos-gallery4}.

\begin{figure}[p]
  \begin{center}
    \leavevmode
    \psfig{figure=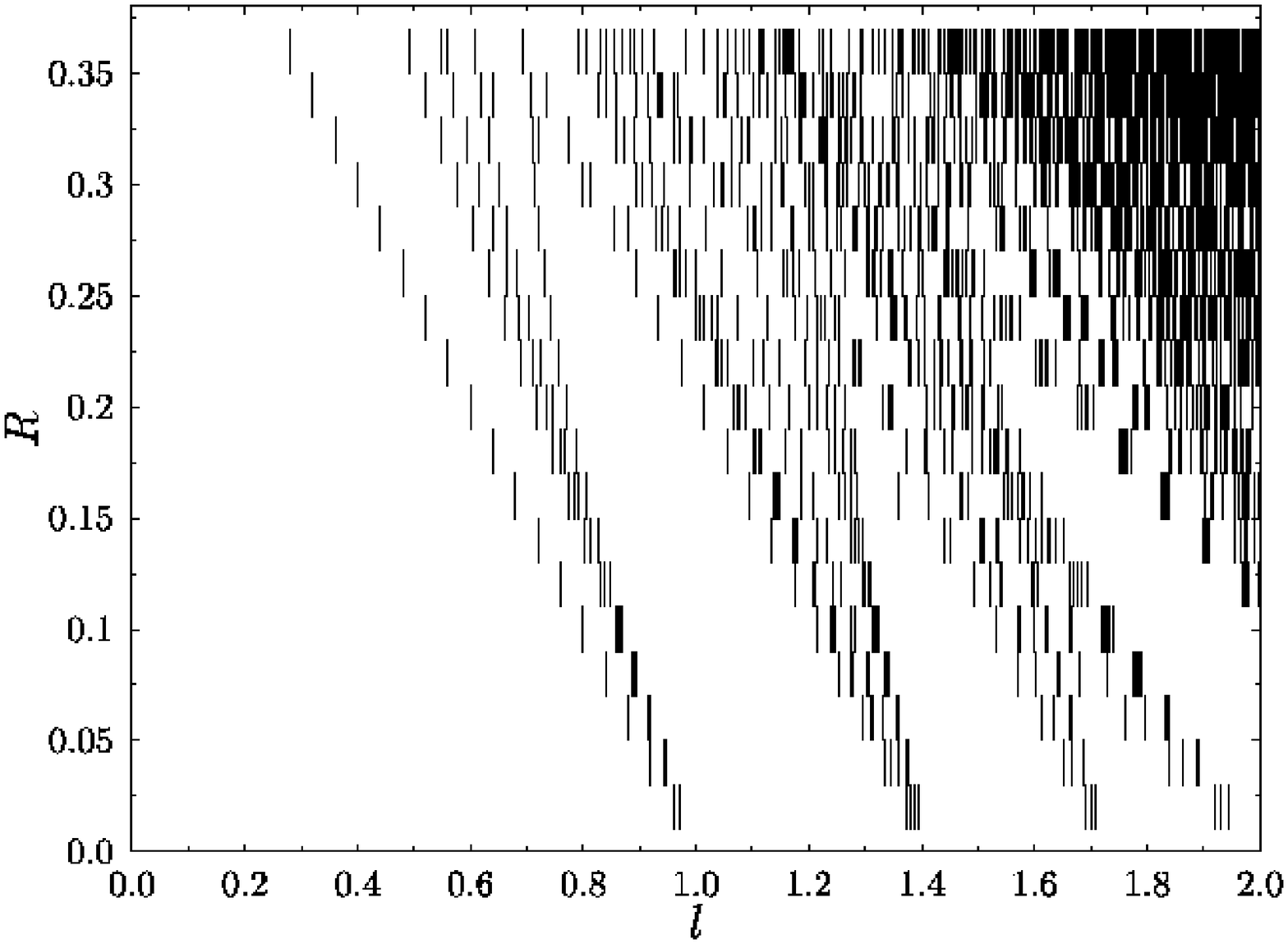,width=16cm}

    \caption{Length spectra of periodic orbits for Sinai billiards
      with $R$ values between 0.02 and 0.36 in steps of $\Delta R =0.02.$
      The vertical bars indicate the lengths of periodic orbits. }

    \label{fig:pos-radius}

  \end{center}
\end{figure}

In this subsection we shall study in detail the spectrum of lengths of
periodic orbits, a small interval thereof is shown in figure
\ref{fig:pos-radius}. Each horizontal strip provides the lower end of
the length spectrum of  Sinai billiards with $0.02 \leq R \leq 0.36$.
The spectrum corresponding to the lowest value of $R$ shows clustering
of the lengths near the typical distances of points of the $\bbbz^3$
lattice $(1, \sqrt{2}, \sqrt{3}, 2, \ldots)$. Once $R$ is increased,
some of the periodic orbits which were allowed for the smaller $R$ are
decimated because of the increased effect of shadowing. However, their
lengths become shorter, resulting in the proliferation of the periodic
orbits with their length. This is best seen in the spectrum which
corresponds to the largest value of $R$ --- the graphics is already
not sufficiently fine to resolve the individual lengths.

After these introductory comments, we now study the length spectrum in
detail, and compare the theoretical expectations with the numerical
results. The exponential proliferation of the periodic orbits puts a
severe limit on the length range which we could access with our finite
computer resources. However, we were able to compute the periodic
orbits for a few values of the radius $R$, and concentrated on the
$R=0.2$ case in order to be able to perform a semiclassical analysis
of the longest quantal spectrum (see next section). For this radius we
found all the 586,965 periodic orbits up to length 5. This number of
periodic orbits includes repetitions and time-reversed conjugates. We
also computed for this radius all the 12,928,628 periodic orbits up to
length 10 which have no more than 3 reflections. This comprises the
database on which we based our further numerical studies and
illustrations. The systematic algorithm which was used to produce this
data set, together with a few tests which will be described here and in
the next section, lead us to believe that the data set is both accurate
and complete.

\begin{figure}[p]
  \begin{center}
    \leavevmode

    \psfig{figure=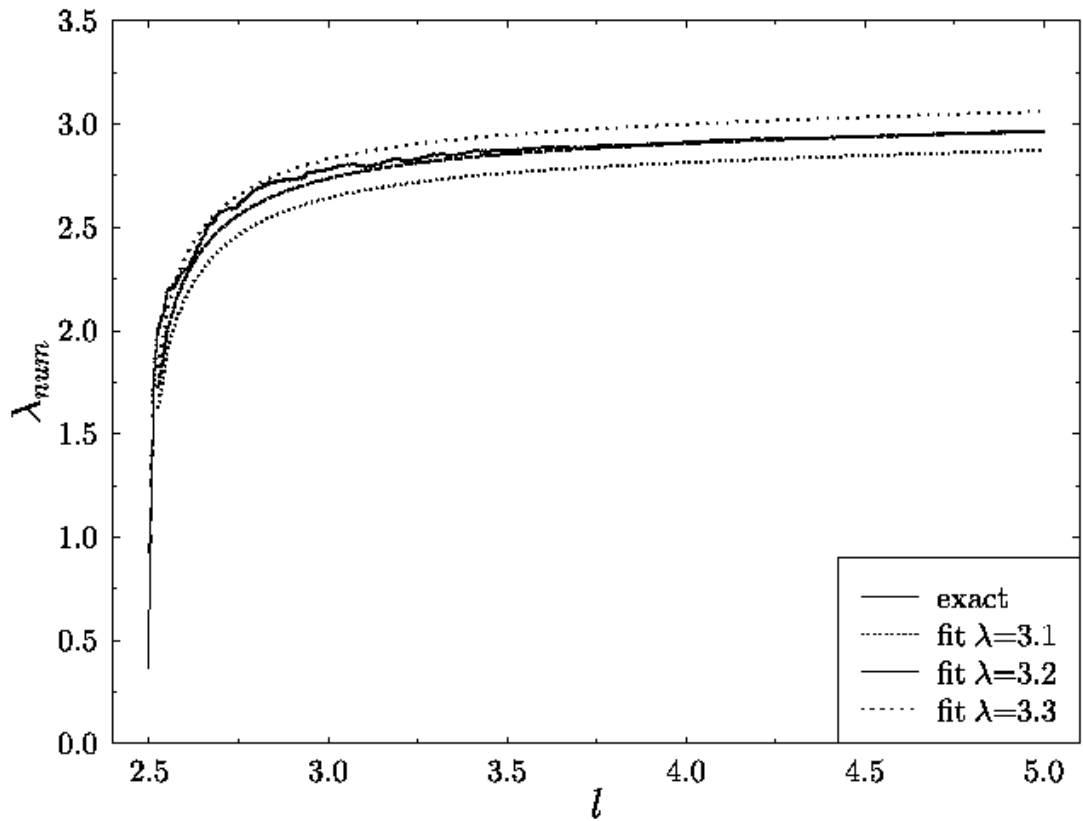,width=16cm}

    \caption{The quantity $\lambda_{\rm num}$ (c.f.\ RHS of equation
      (\protect\ref{eq:lambda-num-def})) computed from the periodic
      orbit database of $R=0.2$. We used $L_{\rm erg}=2.5$. The
      theoretical fit is according to equation
      (\protect\ref{eq:lambda-num-theory}).}

    \label{fig:lyap}

  \end{center}
\end{figure}

Periodic orbits are expected to proliferate exponentially (e.g.,
\cite{Gut90}). That is, the number $N_{\rm len}(l)$ of periodic orbits
of length less than $l$ should approach asymptotically \cite{Gut90}:
\begin{equation}
  N_{\rm len}(l) \approx \frac{\exp(\lambda l)}{\lambda l} \; ,
  \; \; \; l \rightarrow \infty \: ,
  \label{eq:dos-pos}
\end{equation}
where $\lambda$ is the topological entropy (per unit length). To
examine the validity of the above formula in our case we use the
numerical data to compute:
\begin{equation}
  \lambda_{\rm num}(l)
  \equiv
  \frac{1}{l} \ln
    \left( \sum_{L_{\rm erg} \leq  L_j \leq l} L_j \right) \: ,
  \label{eq:lambda-num-def}
\end{equation}
where $L_{\rm erg}$ is a length below which we do not expect
universality (i.e, the law (\ref {eq:dos-pos})) to hold. The exponential
proliferation implies:
\begin{equation}
  \lambda_{\rm num}(l)
  \approx
  \frac{1}{l} \ln \left|  e^{\lambda l} - e^{\lambda L_{\rm erg}}
    \right| - \frac{\ln \lambda}{l} \rightarrow \lambda \; ,
    \; \; \;
    l \rightarrow \infty \, .
    \label{eq:lambda-num-theory}
\end{equation}
Therefore, we expect $\lambda_{\rm num}(l)$ to approach a constant
value $\lambda$ when $l$ is sufficiently larger than $L_{\rm erg}$. In
figure \ref{fig:lyap} we show the results of the numerical computation
of $\lambda_{\rm num}$ for the $R=0.2$ database and for $L_{\rm
  erg}=2.5$. The figure clearly indicates a good agreement between the
data and the theory (\ref{eq:lambda-num-theory}) for $\lambda = 3.2$.

\begin{figure}[p]
  \begin{center}
    \leavevmode
    \psfig{figure=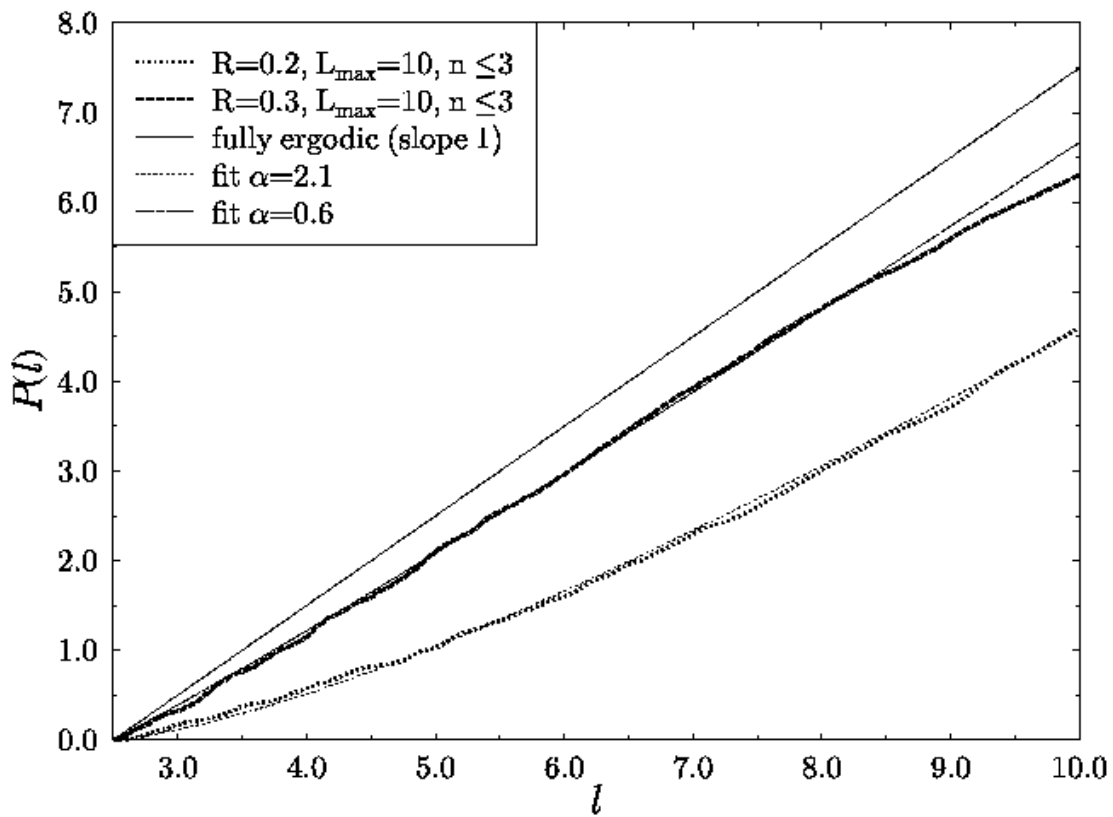,width=16cm}

    \caption{The function $P(l)$ (c.f.\ RHS of equation
      (\protect\ref{eq:ihoz-def})) computed for $R=0.2$ and $0.3$ and
      fitted according to equation (\protect\ref{eq:ihoz-theory}). We
      also show the asymptotic prediction
      (\protect\ref{eq:hannay-ozorio}).}

    \label{fig:hoz}

  \end{center}
\end{figure}

One of the hallmarks of classically ergodic systems is the balance
between the proliferation of periodic orbits and their stability
weights due to ergodic coverage of phase space. This is a
manifestation of the uniform coverage of phase space and is frequently
referred to as the ``Hannay -- Ozorio de Almeida sum rule''
\cite{HO84}. It states that:
\begin{equation}
  p(l)
  \equiv
  \sum_{\rm PO} \frac{L_p}{\left| \det (I - M_j) \right|}
    \, \delta ( l - L_j ) \rightarrow 1 \; ,
    \; \; \; l \rightarrow \infty \: ,
  \label{eq:hannay-ozorio}
\end{equation}
where $L_p$ is the primitive length and $M_j$ is the stability
(monodromy) matrix \cite{Gut90} (see appendix \ref{app:3d-monodromy}
for explicit expressions). The above relation is meaningful only after
appropriate smoothing. For generic billiards the only classical length
scale is the typical length traversed  between reflections, and we expect
(\ref{eq:hannay-ozorio}) to approximately hold after a few
reflections. In the Sinai billiard we are faced with the problem of an
``infinite horizon'', that is, that the length of free flight between
consecutive reflections is unbounded. This is just another manifestation
of the existence of the bouncing--ball families.  According to
\cite{DA96,FS95} this effect is responsible for a non-generic
power--law tail in $p(l)$:
\begin{equation}
  p(l) \approx 1 - \frac{\alpha(R)}{l} \: ,
  \label{eq:hoz-dahl}
\end{equation}
where $\alpha(R)$ is a parameter that depends on the radius $R$. When
$R$ increases the influence (measure in configuration space) of the
bouncing--balls is reduced, and we expect $\alpha(R)$ to decrease. To
check (\ref{eq:hoz-dahl}) we computed numerically the cumulant:
\begin{equation}
  P(l)
  =
  \int_{L_{\rm erg}}^{l} {\rm d}l' \ p(l')
  \approx
  \sum_{L_{\rm erg} \leq L_j \leq l}
    \frac{L_p}{\left| I - M_j \right|} \: ,
  \label{eq:ihoz-def}
\end{equation}
which should be compared to the theoretical expectation:
\begin{equation}
  P(l)
  =
  ( l - L_{\rm erg} ) -
    \alpha(R) \ln \left( \frac{l}{L_{\rm erg}} \right) \: .
  \label{eq:ihoz-theory}
\end{equation}
The results are shown in figure \ref{fig:hoz}. We considered $R =
0.2$ and $0.3$ and included periodic orbits up to $L_{\rm max} = 10$ with
number of reflections $n \leq 3$. The restriction on $n$ facilitates
the computation and is justified for moderate values of $l$ since the
contributions from higher $n$'s are small. The observed deviation
between the theoretical and numerical curves for $R = 0.3$ at $l
\gtrsim 8$ is due to the fact that periodic orbits with $n = 4$ become
significant in this region. The above numerical tests confirm the
validity of (\ref{eq:hoz-dahl}), with $\alpha(R)$ which is a
decreasing function of $R$. In particular, for the length interval
considered here, there is a significant deviation from the fully
ergodic behavior (\ref{eq:hannay-ozorio}).

The sum--rule (\ref{eq:hannay-ozorio}) which formed the basis of the
previous analysis is an expression of the ergodic nature of the
billiards dynamics. In the next subsection we shall make use of similar
sum--rules which manifest the ergodicity of the Poincar\'e map obtained
from the billiard flow by, e.g., taking the surface of the sphere and the
tangent velocity vector as the Poincar\'e section. The resulting
return--map excludes the bouncing--ball manifolds since they do not
intersect the section. However, their effect is noticed because between
successive collisions with the sphere the trajectory may reflect off the
planar faces of the billiard an arbitrary number of times. Thus, the
number of periodic orbits which bounce $n$ times off the sphere
($n$-periodic orbits of the map) is unlimited, and the topological
entropy is not well defined. Moreover, the length spectrum of
$n$-periodic orbits is not bounded. These peculiarities, together with
the fact that the symbolic code of the map consists of an infinite number
of symbols, are the manifestations of the infinite horizon of the
unfolded Sinai billiard. The return map itself is discontinuous but it
remains area preserving, so the formulas which we use below, and which
apply to generic maps, can be used here as well.

\begin{figure}[p]

  \centerline{\psfig{figure=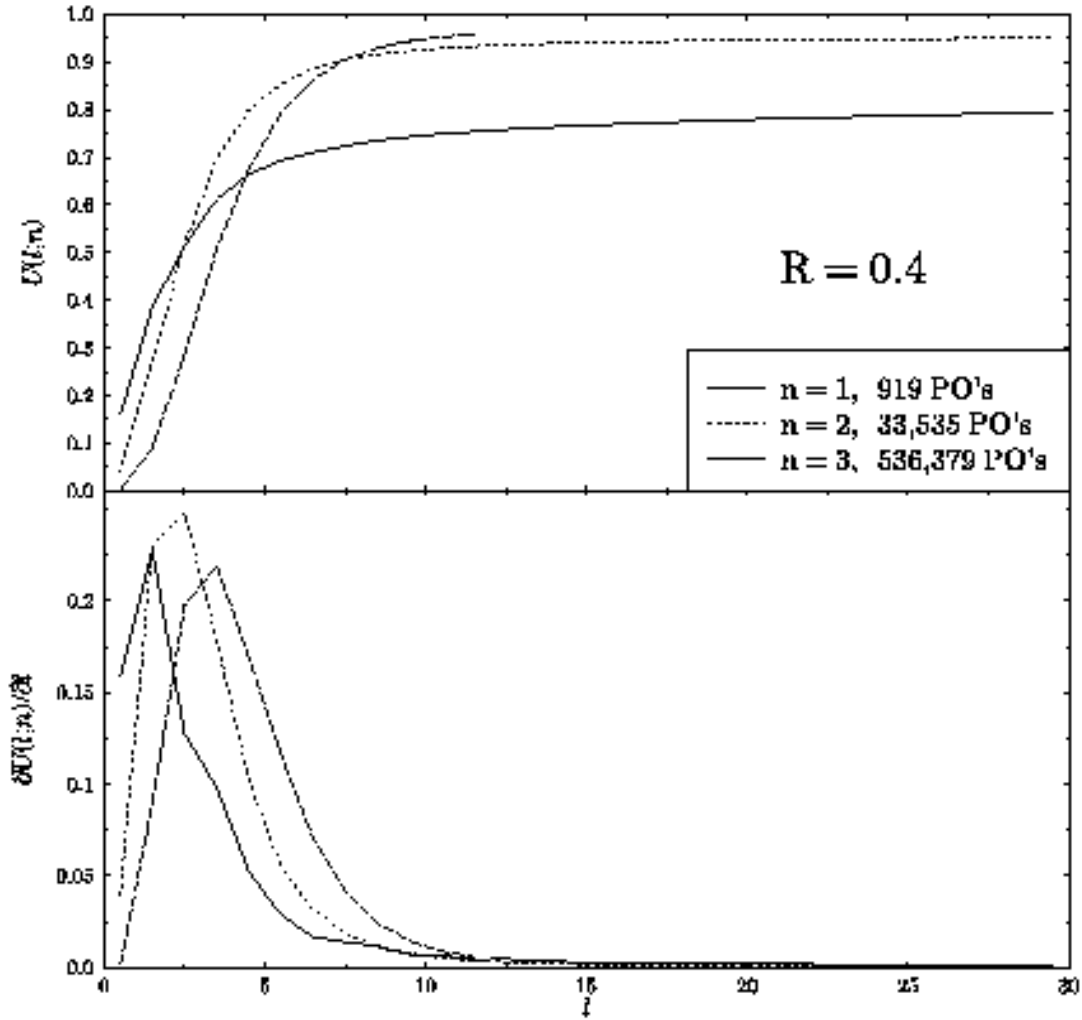,width=16cm}}

  \caption{Upper plot: The function $U(l; n)$
    (c.f.\ equation (\protect\ref{eq:uln})) for the cases $R = 0.4$,
    $n = 1, 2, 3$. Lower plot: The function $\partial U(l; n) /
    \partial l$ for the same cases. Both plots indicate the saturation
    of the classical return probability in spite of the infinitely
    many periodic orbits in ${\cal P}_n$.}

  \label{fig:uln}

\end{figure}

The classical return probability is defined as the trace of the
$n$-step classical evolution operator (see, e.g., \cite{Smi94} and
references therein). It is given by:
\begin{equation}
  U(n)
  \equiv
  \sum_{j \in {\cal P}_n} \frac{n_{p,j}}{|\det (I-M_j)|} \: ,
  \label{eq:class-prob}
\end{equation}
where $n$ is the number of times the periodic orbit reflects from the
sphere, ${\cal P}_n $ is the set of all $n$-periodic orbits, $n_{p,j}$
is the period of the primitive periodic orbit of which $j$ is a
repeated traversal. As a consequence of the ergodic nature of the map
$U(n) \rightarrow 1$ in the limit $n \rightarrow \infty$. However, due
to the effect of the infinite horizon, the number of periodic orbits
in ${\cal P}_n$ is infinite, and in any numerical simulation it is
important to check to what degree the available data set satisfies the
sum rule. For this purpose we define the function:
\begin{equation}
  U(l; n)
  \equiv
  \sum_{j \ in {\cal P}_n} \frac{n_{p,j}}{|\det (I-M_j)|}
    \, \Theta(l - L_j) \: ,
  \label{eq:uln}
\end{equation}
which takes into account only $n$-periodic orbits with $L_j \leq l$.
In figure \ref{fig:uln} we plot $U(l; n)$ for $R = 0.4$ and $n = 1, 2,
3$. The results clearly indicate that for the present data saturation
is reached, and once $n \geq 2$ the asymptotic value is very close to
$1$. Even at $n=1$ one gets $U(n=1) \approx 0.8$ which is surprisingly
close to $1$, bearing in mind that we are dealing with the fixed
points of the map! It should be noted that to reach saturation in the
case $R=0.4$, $n=3$ one needs 536,379 periodic orbits up to $l = 12$,
whose computation consumes already an appreciable amount of time.
Thus, we are practically restricted to the few lowest $n$'s in our
computations. As can be seen in figure \ref{fig:uln} the function
$\partial U(l; n) / \partial l$  is mostly supported on a finite interval
of $L$ values. Its width will be  denoted by  $\Delta L(n)$.        

\subsection{Periodic orbit correlations}
\label{subsec:pos-correlations}

In the previous subsection we discussed various aspects of the
one--point statistics of the classical periodic orbits, and
demonstrated their consistency with the standard results of ergodic
theory. Here, we shall probe the length spectrum further, and show
that this spectrum is not Poissonian. Rather, there exist correlations
between periodic orbits which have far--reaching effects on the
semiclassical theory of spectral statistics of the quantum billiard.
The semiclassical theory will be dealt with in section
\ref{sec:sc-specstat}, and here we restrict ourselves to purely
classical investigations.

Above we introduced the Poincar\'e return map of the sphere, and have
shown that the ergodicity of this map implies a sum rule for the set
of $n$-periodic orbits of the map. We define the weighted density of
lengths of $n$-periodic orbits as follows:
\begin{equation}
  d_{\rm cl} (l; n)
  \equiv
  \sum_{j \in {\cal P}_n } \tilde{A}_j \delta(l - L_j) \: ,
  \label{eq:classdens}
\end{equation}
where $\tilde{A}_j$ are given by:
\begin{equation}
  \tilde A_j
  =
  \frac{n_{p,j} (-1)^{b_{j}}}
       {|\det (I-M_j)|^{1/2}}  \: ,
  \label{eq:weightstilde}
\end{equation}
where $b_{j}$ is the number of times the trajectory reflects from the
planar boundaries. The amplitudes $\tilde{A}_j$ are related to the
standard semiclassical amplitudes $A_j$ defined in (\ref{eq:aw}) by
$\tilde{A}_j = \pi n \sigma_j A_j / L_j$.

The density (\ref{eq:classdens}) is different from the density $p(l)$
defined previously (\ref{eq:hannay-ozorio}) since: (a) it relates to
the subset of the $n$-periodic orbits of the return map of the sphere,
(b) it assigns a signed weight to each of the $\delta$-functions
located at a particular length, and (c) the absolute value of the
weights in (\ref{eq:classdens}) are the square roots of the weights in
(\ref{eq:hannay-ozorio}). Densities with signed weights are not
encountered frequently in spectral theory, but they emerge naturally
in the present context. At this point the definition of $d_{\rm cl}
(l; n)$ might look unfamiliar and strange, but the reason for this
particular choice will become clear in the sequel.

To examine the possible existence of correlations in the length
spectrum, we study the corresponding autocorrelation function:
\begin{equation}
  R_{\rm cl}(\delta l; n)
  \equiv
  \int_{0}^{\infty}
    {\rm d}l \, d_{\rm cl}(l+\delta l/2; n)
             \, d_{\rm cl}(l-\delta l/2; n) \: .
  \label{eq:Rfunction}
\end{equation}
The two--point form factor is the Fourier transform of $R_{\rm
  cl}(\delta l; n)$, and it reads explicitly as:
\begin{equation}
  K_{\rm cl}(k; n) 
  = 
   \int_{-\infty}^{+\infty}
    {\rm e}^{i k x} R_{\rm cl}(x; n) {\rm d}x
  =
  \left| \sum_{j \in {\cal P}_n}
    \tilde{A}_j \exp (i k L_j) \right|^2 \: .
  \label{eq:classform}
\end{equation}
The form factor has the following properties:
\begin{itemize}
  
\item $K_{\rm cl}(k; n)$ is a Fourier transform of a distribution and
  therefore it displays fluctuations, which become stronger as the
  number of contributing orbits increases. Therefore, any discussion
  of this function requires some smoothing or averaging. We shall
  specify the smoothing we apply in the sequel.

\item At $k=0$,
\begin{equation}
  K_{\rm cl} (0; n)
  =
  \left |\sum _{j \in {\cal P}_n} \tilde{A}_j  \right |^2 \: .
  \label{eq:classformk=0}
\end{equation}
Because of the large number of periodic orbits, the sum of the signed
amplitudes is effectively reduced due to mutual cancellations. Its value
can be estimated by assuming that the signs are random. Hence,
\begin{equation}
  K_{\rm cl}(0; n)
  \approx
  \sum_{j \in {\cal P}_n} \left| \tilde{A}_j \right|^2 \: ,
\end{equation}
which will be shown below to be bounded.

\item At large values of $k$,
\begin{equation}
  K_{\rm cl}(k; n)
  \approx
  \sum_{j \in {\cal P}_n}
    g_j \left| \tilde{A}_j  \right|^2 \: ,
    \ \ \ {\rm for } \ \ \ k \rightarrow \infty \: ,
\end{equation}
where $g_j$ is the number of isometric periodic orbits of length
$L_j$. Since large fluctuations are endemic to the form factor, this
relation is meaningful when $k$-averaging is applied. Comparing the
last sum with (\ref{eq:class-prob}) we can write:
\begin{equation}
  K_{\rm cl}(k; n)
  \approx
  \langle n_{p, j} g_j \rangle \, U(n) \: ,
  \ \ \ {\rm for } \ \ k \rightarrow \infty \: .
  \label{eq:kcl-asy}
\end{equation}
In our case of the 3D SB, $n_{p, j} = n$ for the large majority of the
periodic orbits in ${\cal P}_n$, which is the generic situation for
chaotic systems. Also, $g_j = 2$ for almost all the periodic orbits
with $n \geq 3$. Thus, one can safely replace $\langle n_{p, j} g_j
\rangle$ with $2n$ for large $n$. Moreover, as we saw above, $U(n)
\rightarrow 1$ for large $n$, hence $K_{\rm cl} \rightarrow 2 n$ for
large $k$ and $n$.

\item If the length spectrum as defined above were constructed by a
  random sequence of lengths with the same smooth counting  function
$U(l;n)$, or if the phases were picked at random, one
  would obtain the Poisson behavior of the form factor, namely, a
  constant:
  \begin{equation}
    K_{\rm cl}(k; n)
    \approx 
    \langle n_p g_p \rangle \, U(n) \ ,
      \ \ \ {\rm for} \ \ k > {2\pi\over \Delta L(n)}  \: .
    \label{eq:poissondist}
  \end{equation}
 Here, $\Delta L(n)$ is the effective width of the length distribution
defined above. 

\end{itemize}
Thus, we could identify two--point correlations in the classical
length spectrum by computing $K_{\rm cl}(k; n)$ and observing
deviations from the $k$-independent expression (\ref{eq:poissondist}).

\subsubsection{Numerical tests}

We used the periodic orbit database at our disposal to compute the
form factors for several values of $n$ and $R$. In each case presented
we made sure that the function $U(l; n)$ is numerically saturated.
This guarantees that the (infinitely many) neglected periodic orbits
have very small weight, and are thus insignificant.

\begin{figure}[p]

  \centerline{\psfig{figure=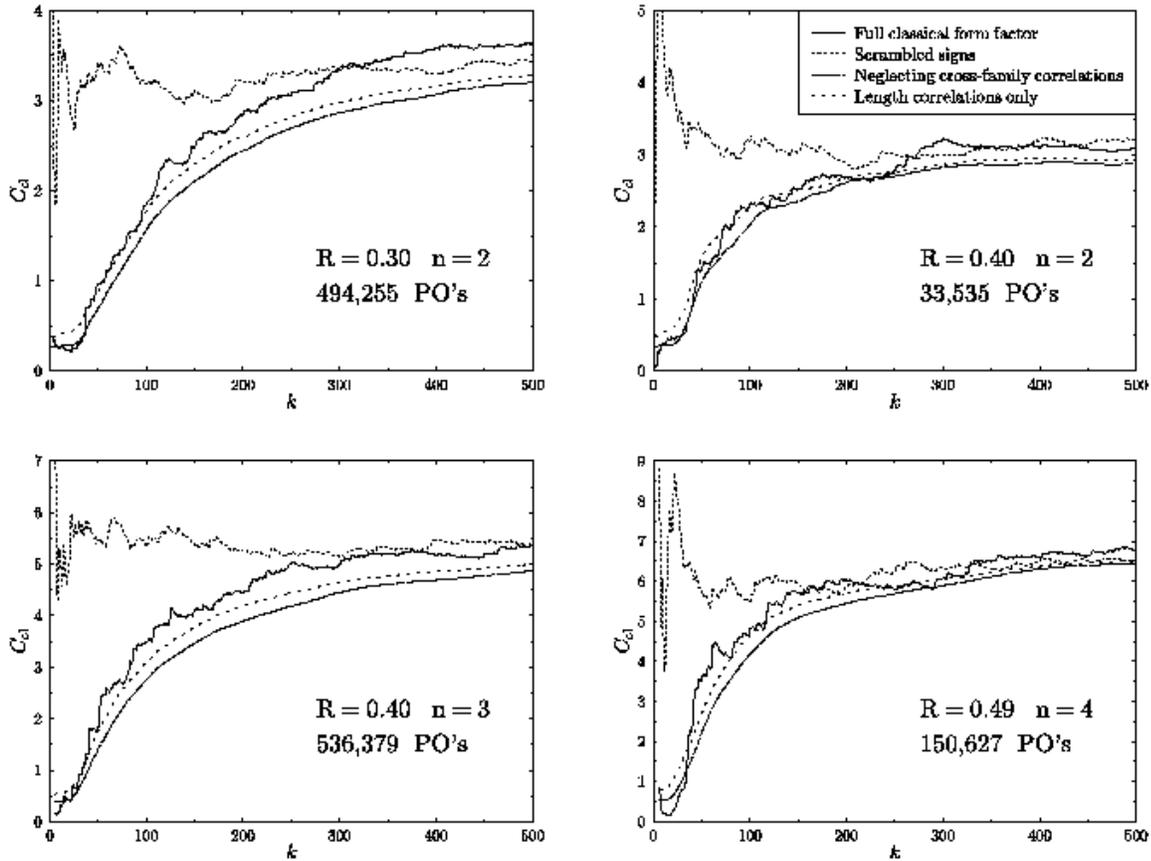,width=16cm}}

  \caption{The averaged classical form factor $C_{\rm cl}(k; n)$
    (c.f.\ (\ref{eq:cdef})) of the 3D SB for several values of $n$ and
    $R$. We also plot the averaged form factors with signs of the
    amplitudes scrambled, without cross--family terms, and with
    amplitudes averaged over family (length--correlation only). See
    text for details.}

  \label{fig:kcl}

\end{figure}

In figure \ref{fig:kcl} we present the numerical results, where we
plotted the function:
\begin{equation}
  C_{\rm cl}(k;n)
  \equiv
  \frac{1}{k-k_{\rm min}} \int_{k_{\rm min}}^{k}
    {\rm d}k' \, K_{\rm cl}(k'; n) \: ,
  \label{eq:cdef}
\end{equation}
designed to smooth the fluctuations in $K_{\rm cl}(k; n)$
\cite{CPS98}. We started the integration at $k_{\rm min} > 0$ to avoid
the large peak near $k=0$, which otherwise overwhelms the results. In
any case, the neglected small-$k$ region is irrelevant for the
semiclassical theory of quantal spectral correlations.
Analyzing the results, we note that the asymptotic form factors
(denoted as ``Full classical form factor'') approach constant values,
which are indeed close to $2n$, as predicted. More importantly, the
deviations from the constant (Poissonian) result at low $k$
demonstrate unambiguously the {\em existence of correlations}\/ in the
classical spectra. The structure of the form factor indicates that the
classical spectrum is rigid on the scale of a correlation length
$\lambda(n; R)$, which can be defined as the inverse of the $k$ value
at which the form factor makes its approach to the asymptotic value
\cite{CPS98}. In the following we shall describe a few tests which
prove that the observed correlations are real, and not a numerical
artifact or a trivial consequence of the way in which the length
spectral density is defined.

The spectral density $d_{\rm cl}(l; n)$ has an effective finite width
$\Delta L(n)$ which was defined above.   
The fact that the lengths are constrained to this
interval induces trivial correlations which appear on the  scale 
$\Delta L(n)$, and we should check that this scale is sufficiently
remote from the correlation scale $\lambda(n; R)$. To this
end, and to show that the observed classical correlations are
numerically significant, we scrambled the signs of the weights
$\tilde{A}_j$ by multiplying each of them with a randomly chosen sign. We
maintained, however, the time--reversal symmetry by multiplying
conjugates by the same sign. The resulting form factors, shown in figure
\ref{fig:kcl} (denoted as ``Scrambled signs''), are consistent with the
Poissonian value $2n$ for essentially all $k$ values, and the difference
between the scrambled and unscrambled data is large enough to add
confidence to the existence of the classical correlations. This indicates
also that the correlations are not due to the effective width of $d_{\rm
  cl}(l; n)$, since both the scrambled and unscrambled data have the
same effective width.

On the other extreme, one might suspect that the classical
correlations are due to rigidity on the scale of one mean spacing
between lengths of periodic orbits. This is certainly not the case,
since the typical mean length spacing for the cases shown in figure
\ref{fig:kcl} is $10^{-3}$--$10^{-4}$, which implies a transition to
the asymptotic value for much larger $k$-values than observed. We
therefore conclude, that the correlation length $\lambda(n; R)$ is
{\em much larger}\/ than the mean spacing between neighboring lengths.
This is the reason why various studies of the length--spectrum
statistics \cite{Sie91,HS92} claimed that it is Poissonian. Indeed it
is Poissonian on the scale of the mean spacing where these studies
were conducted. The correlations become apparent on a very different
(and much larger) scale, and there is no contradiction. The
coexistence of a Poissonian behavior on the short length scales, and
apparent rigidity on a larger scale was discussed and explained in
\cite{CPS98}. It was suggested there that a possible way to construct
such a spectrum is to form it as a union of $N \gg 1$ statistically
independent spectra, all having the same mean spacing $\bar{\Delta}$,
and which show spectral rigidity on the scale of a single spacing. The
combined spectrum with a mean spacing $\bar{\Delta}/N$ will be
Poissonian when tested on this scale, since the spectra are
independent. However, the correlations on the scale $\bar{\Delta}$
will persist in the combined spectrum. A simple example will
illustrate this construction. Take a random (Poissonian) spectrum with
a mean spacing $1$. Generate a shifted spectrum by adding
$\bar{\lambda} \gg 1$ to each spectral point and combine the original
and the shifted spectra to a single spectrum. On the scale $1$ the
combined spectrum is Poissonian. However, the fact that each spectral
point is (rigidly) accompanied by another one, a distance
$\bar{\lambda}$ apart, is a correlation which will be apparent at the
scale $\bar{\lambda}$ only. We use this picture in our attempt to
propose a dynamical origin of the length correlations.

\subsubsection{The dynamical origin of the correlations}

As was already mentioned, the idea that periodic orbit correlations
exist originates from the quantum theory of spectral statistics which
is based on trace formulas. The classical correlations are shown to be
a manifestation of a fundamental duality between the quantum and the
classical descriptions \cite{ADDKK93,CPS98}. However, the effect is
purely classical, and hence should be explained in classical terms,
without any reference to the quantum mechanical analogue. The
essential point is to find the classical origin of the partition of
the periodic orbits to independent and uncorrelated families, as was
explained in the previous section. So far, all the attempts to find
the classical roots of these correlations failed, and till now there
is no universal theory which provides the classical foundations for
the effect. For the Sinai billiard in 3D there seems to exist a
physical--geometrical explanation, which is consistent with our data,
and which is supported by further numerical tests.

\begin{figure}[p]
  \begin{center}
    \leavevmode

    \centerline{\psfig{figure=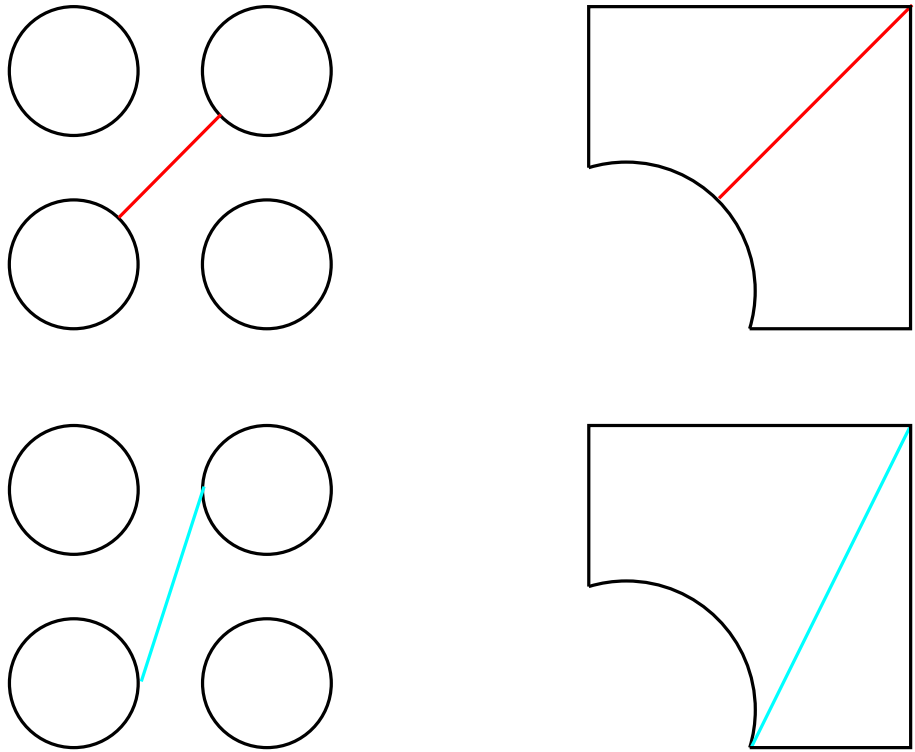,height=10cm}}

    \caption{Two periodic orbits which are members of the same
      family of the quarter 2D SB. The two periodic orbits have the
      same $W$, hence they reflect off the same discs. But they
      correspond to two {\em different}\/ symmetry elements, and hence
      are different. For simplicity the illustration is made for the
      2D SB, but the same principle applies also to the 3D Sinai
      billiard.  Left: Unfolded representation, right: Standard
      representation.}

    \label{fig:sb2d-family}

  \end{center}
\end{figure}

\begin{figure}[p]

  \centerline{\psfig{figure=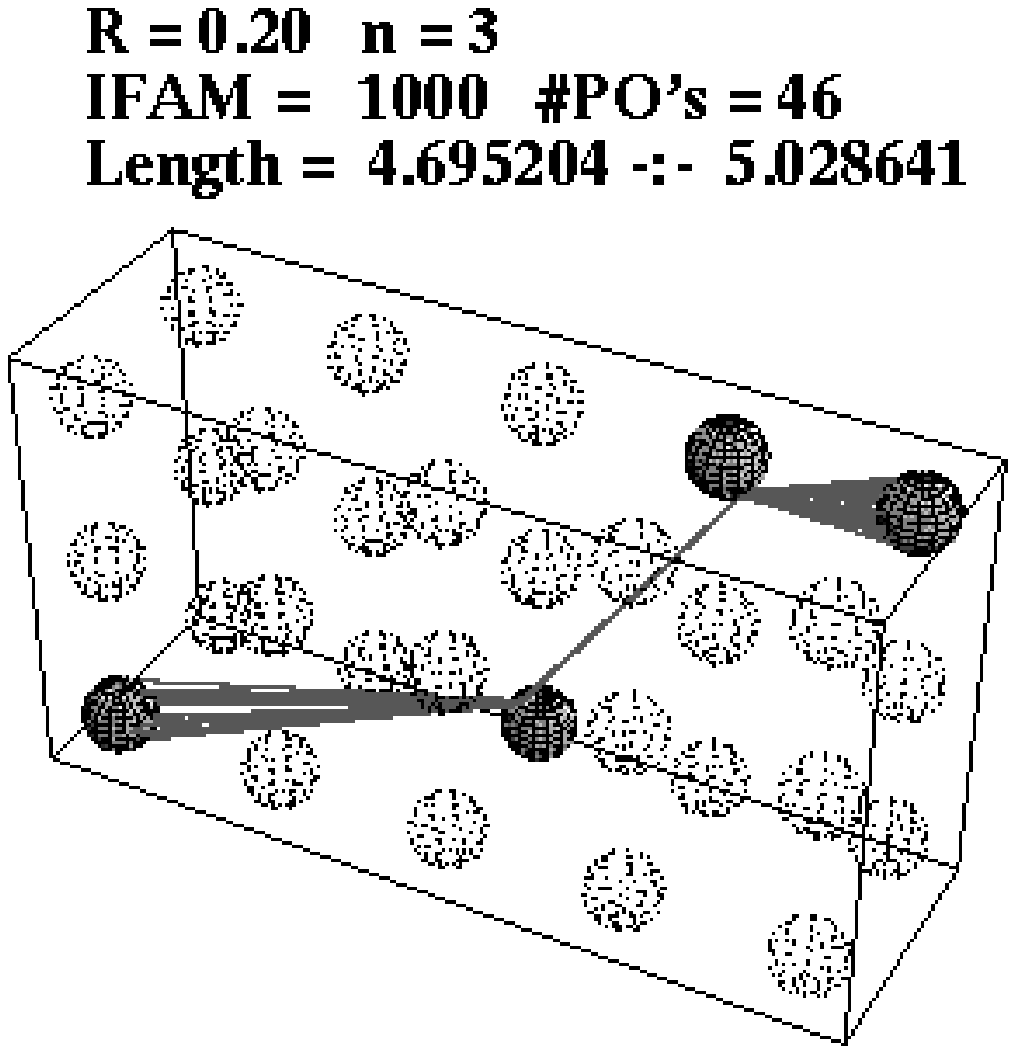,height=10cm}}

  \centerline{\psfig{figure=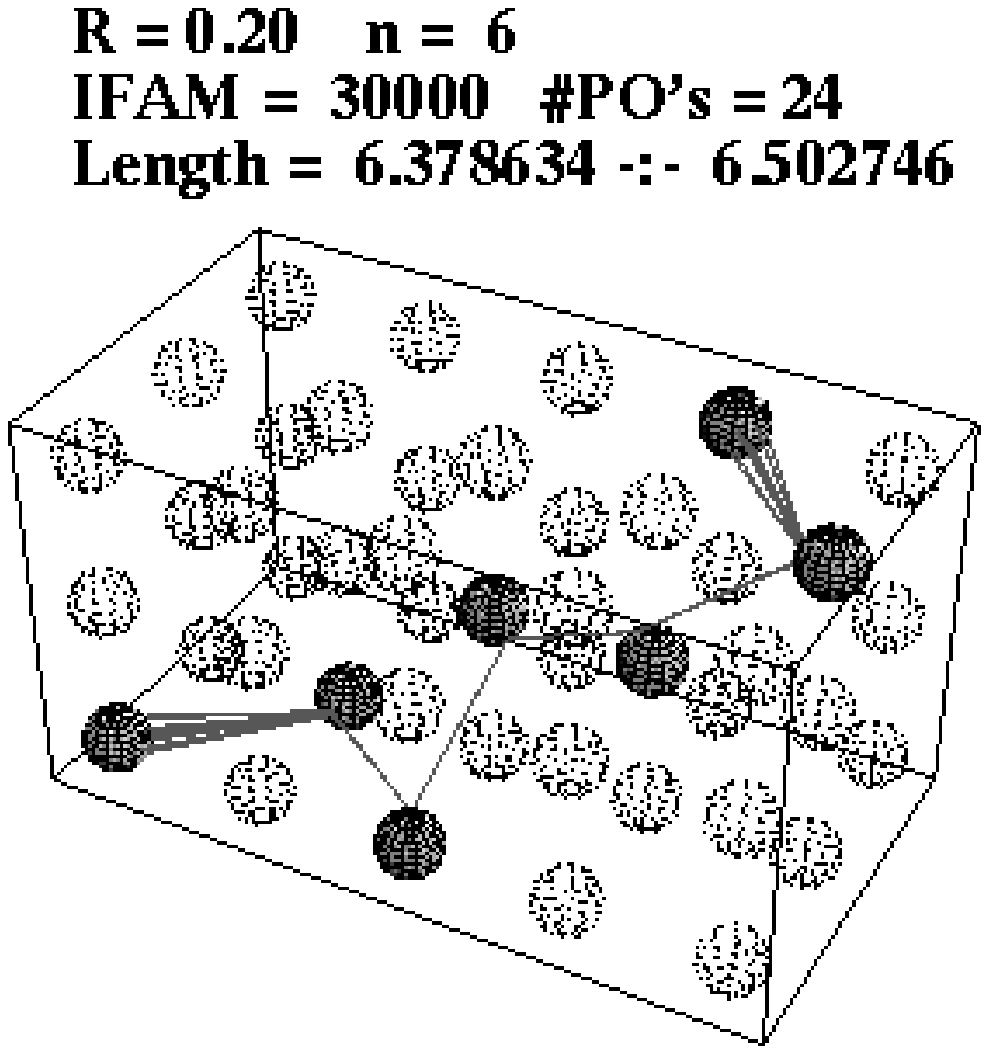,height=10cm}}

  \caption{Two families of periodic orbits of the 3D SB,
    represented in the unfolded representation of the SB. The faint
    spheres do not participate in the code.}

  \label{fig:3dsb-families}

\end{figure}

Consider the Sinai billiard with a sphere with a vanishingly small
radius. In this case, all the periodic orbits which are encoded by
words $W$ built of the same letters $w_i$ are isometric, independently
of the ordering of the letters or the attached symmetry element $\hat
g$. This phenomenon can be clearly seen in the spectrum of lengths
corresponding to $R=0.02$ in figure \ref{fig:pos-radius}. In this
case, it is clear that the spectrum of lengths is a union of
``families'' of periodic orbits, each family is characterized by a
unique set of building blocks $w_i$, which are common to the family
members. When the radius $R$ increases and becomes comparable to the
linear dimension of the billiard, the approximate isometry and the
resulting correlations breaks down, and one should use a more refined
and restrictive definition of a family. The aim is to find a partition
to families which will restrict the membership in a family to the
smallest set, without losing any of the correlation features. The most
restrictive definition of a family in the present context will be to
include all the periodic orbits which share the same $W = (w_1, w_2,
\ldots, w_n)$ part of the code and have different admissible $\hat{g}$
symmetry elements. Words which are built of the same letters but in a
different order define different families. Since there are 48 possible
$\hat{g}$'s, each family consists of at most 48 members and will be
denoted by $\Omega(W)$. It should also be noted that the signs of the
weights $\tilde A_j$ within a family do not change with $R$ since they
reflect the parity of $\hat{g}$. The partition of the set of periodic
orbits in families is not particular to just a few orbits, but rather,
is valid for the entire set. This partition is the proposed source of
the correlations that were observed in the form factor. This concept
is illustrated in figure \ref{fig:sb2d-family}, and graphic
representations of two families are displayed in figure
\ref{fig:3dsb-families}.
The most outstanding feature which emerges from figure
(\ref{fig:3dsb-families}) is that the orbits occupy a very narrow
volume of phase--space throughout most of their length, and they fan
out appreciably only at a single sphere.

The above arguments suggest that the main source of correlations are
the similarities of orbits within each family $\Omega(W)$. To test
this argument we performed a numerical experiment, in which we
excluded the inter--family terms of the form factor, leaving only the
intra--family terms. This excludes family--family correlations and
maintains only correlations within the families. The results are shown
in figure \ref{fig:kcl} (denoted by ``Neglecting cross--family
contributions''). The obvious observation is that the form factors
were only slightly affected, proving that periodic--orbit correlations
do not cross family lines! Thus, the main source of correlations is
within the families $\Omega(W)$. We mention that very similar results
are obtained if inter--family sign randomization is applied instead of
the exclusion of cross--terms. We note, that in most cases a periodic
orbit and its time--reversal conjugate do not belong to the same
family. Thus, neglecting the cross--family terms leads to partial
breaking of time--reversal, which we compensated for by rectifying the
intra--family form factor such that it will have the same asymptotic
value as the full one.

It is interesting to check whether the correlations are due to the
lengths or due to the size of the amplitudes. To examine that, we not
only neglected the cross--family terms, but also replaced the
amplitudes $\tilde{A}_j$ within each family by constants multiplied by
the original signs, such that the overall asymptotic contribution of
the family does not change. The results are also plotted in figure
\ref{fig:kcl} (denoted as ``Length correlations only''). The resulting
(rectified) form factors display slightly diminished correlations.
However there is no doubt that almost all correlations still persist.
This proves, that the correlations between the magnitudes of the
weights play here a relatively minor role, and the correlations are
primarily due to the lengths.

There are a few points in order. First, the numerical results
presented here concerning the classical correlations are similar to
those of reference \cite{CPS98}. However, here we considered the
classical mapping rather than the flow, and this reduces the numerical
fluctuations significantly. Using the mapping also enables the
quantitative comparison to the semiclassical theory, which will be
discussed in section \ref{sec:sc-specstat}. Second, it is interesting
to enquire whether the average number of family members $N_{\rm
  fam}(n; R)$ increases or decreases with $n$. Since, if it decreases,
our explanation of the origin of correlations becomes invalid for
large $n$. The numerical results clearly indicate that $N_{\rm fam}(n;
R)$, computed as a weighted average with the classical weights,
increases with $n$, which is encouraging. For example, for the case $R
= 0.4$ we obtained $N_{\rm fam} = 9.64$, $18.31$, $21.09$, $28.31$ for
$n = 1$, $2$, $3$, $4$, respectively.

Thus, we were able to identify the grouping of orbits into
``families'' with the same code word $W$ but with different symmetry
$\hat{g}$ as the prominent source of the classical correlations in the
3D Sinai billiard. In each family the common geometric part of the
code $W$ sets the mean length and the different group elements
$\hat{g}$ introduce the modulations. This pattern repeats for all the
families, but the lengths of different families are not correlated.
This finding conforms very well with the general scheme which was
proposed to explain the typical correlations in the classical spectrum
\cite{ADDKK93,HKS99}. However, a classical derivation of a
quantitative expression for the correlations length $\lambda$ is yet
to be done.

\subsubsection{Length correlations in the 3--Torus}

The ideas developed above about the correlations between periodic
orbits in the fully chaotic billiard, have an analogue in the spectrum
of lengths of periodic tori in the integrable case of the 3--torus.
In section \ref{sec:quantal-spectral-statistics} we studied the
quantum 3--torus of size $S$ and showed in section
\ref{subsubsec:two-point-int} that due to number theoretical
degeneracies, the quantum form factor is not Poissonian. The form
factor displays a negative (repulsive) correlation which levels off at
$\tau^{*} = 1 / \gamma = 1 / (2 S k)$. This can be transcribed into an
expression for the correlation length of the classical spectrum in the
following way.

Expressing  $\tau^{*}$ in  units of length we obtain:
\begin{equation}
  L^{*} = 2 \pi \bar{d}(k) \tau^{*} = \frac{S^2 k}{2 \pi} \: .
\end{equation}
Consequently:
\begin{equation}
  k^{*}(L) = \frac{2 \pi L}{S^2} \: ,
\end{equation}
from which we read off
\begin{equation}
  \lambda(L) = \frac{S^2}{L} \: .
\end{equation}
Since the lengths of the periodic orbits are of the form $L_j = S
\times \sqrt{\mbox{integer}}$, the minimal spacing between
periodic orbits near length $L$ is:
\begin{equation}
  \Delta_{\rm min}(L) = \frac{S^2}{2 L} \: ,
\end{equation}
and therefore
\begin{equation}
  \lambda(L) = 2 \Delta_{\rm min}(L) \: .
\end{equation}
In other words, the classical correlation length of the 3--torus
coincides (up to a factor 2) with the minimal spacing between the
periodic orbits. Therefore, $\lambda(L)$ indeed signifies the
correlation length scale between periodic orbits, which is imposed by
their number--theoretical structure.


\section{Semiclassical analysis}
\label{sec:sc-analysis}

In the previous sections we accumulated information about the quantum
spectrum and about the periodic orbits of the 3D Sinai billiard. The
stage is now set for a semiclassical analysis of the billiard. We
shall focus on the analysis of the semiclassical Gutzwiller trace
formula \cite{Gut90} that reads in the case of the Sinai billiard:
\begin{equation}
  d(k) 
  \equiv 
  \sum_{n=1}^{\infty} \delta(k-k_n) 
  \approx
  \bar{d}(k) + d_{\rm bb}(k) + \sum_{\rm PO} A_j \cos(k L_j) \: .
  \label{eq:gtf}
\end{equation}
The quantum spectral density on the LHS is expressed as the sum of
three terms. The term $\bar{d}$ is the smooth density of states (see
section \ref{app:weyl}). The term $d_{\rm bb}$ consists of the
contributions of the non-generic bouncing--ball manifolds. It contains
terms of the form (\ref{eq:dkint}) with different prefactors (which
are possibly 0) due to partial or complete shadowing of the
bouncing--ball family by the sphere. The last term is the contribution
of the set of generic and unstable periodic orbits, where $L_j$ denote
their lengths and $A_j$ are semiclassical amplitudes. One of the main
objective of the present work was to study the accuracy of
(\ref{eq:gtf}) by a direct numerical computation of the difference
between its two sides. This cannot be done by a straightforward
substitution, since three obstacles must be removed:

\begin{itemize}
  
\item The spectrum of wavenumbers $k_n$ was computed for the fully
  desymmetrized Sinai billiard. To write the corresponding trace
  formula, we must remember that the folding of the Sinai {\em
    torus}\/ into the Sinai {\em billiard}\/ introduces new types of
  periodic orbits due to the presence of symmetry planes, edges and
  corners. Strictly speaking, the classical dynamics of these orbits
  is singular, and becomes meaningful only if proper limits are taken.
  As examples we mention periodic orbits that bounce off a corner, or
  that are wholly confined to the symmetry planes. These periodic
  orbits are isolated and unstable, and should not be confused with
  the bouncing--ball families which are present both in the ST and in
  the SB. For periodic orbits that reflect from a corner but are not
  confined to symmetry planes, the difficulty is resolved by unfolding
  the dynamics from the SB to the ST as was described in the previous
  section. Periodic orbits which are confined to symmetry planes are
  more troublesome since there is more than one code word $\hat{W}$
  which correspond to the same periodic orbit. We denote the latter as
  ``improper''.  The 3D Sinai billiard is abundant with improper
  periodic orbits, and we cannot afford treating them individually as
  was done e.g.\ by Sieber \cite{Sie91} for the 2D hyperbola billiard.
  Rather, we have to find a general and systematic method to identify
  them and to calculate their semiclassical contributions. This will
  be done in the next subsection. (The semiclassical contributions of
  the improper periodic manifolds for the integrable case $R = 0$ were
  discussed in section \ref{subsec:integrable}).

\item As it stands, equation (\ref{eq:gtf}) is a relation between {\em
    distributions}\/ rather than between {\em functions}\/, and hence
  must be regulated when dealing with actual computations. Moreover,
  even though our quantum and classical databases are rather
  extensive, the sums on the two sides of the equation can never be
  exhausted. We overcome these problems by studying the weighted
  ``length spectrum'' obtained from the trace formula by a proper
  smoothing and Fourier-transformation. It is defined in subsection
  \ref{subsec:length-spectrum}.
  
\item Finally, we must find ways to rid ourselves from the large, yet
  non-generic contributions of the bouncing--ball families. This was
  achieved using rather elegant tricks which are described in
  subsections \ref{subsec:d-n}, \ref {subsec:mbc} below.

\end{itemize}

\subsection{Semiclassical desymmetrization}
\label{subsec:sc-desym}

To derive the spectral density of the desymmetrized Sinai billiard we
make use of its expression in terms of the (imaginary part of the)
trace of the SB Green function. This Green function satisfies the
prescribed boundary conditions on all the boundaries of the
fundamental domain, and the trace is taken over its volume. In the
following we shall show how to transform this object into a trace over
the volume of the entire ST, for which all periodic orbits are proper
(no symmetry planes). This will eliminate the difficulty of treating
the improper orbits. To achieve this goal we shall use
group--theoretical arguments \cite{Tin64,Sie91,Gut89,CE93,Lau91}. The
final result is essentially contained in \cite{Rob89}.

When desymmetrizing the ST into SB, we have to choose one of the
irreps of $O_h$ to which the eigenfunctions of the SB belong (see
section \ref{subsec:kkr-desym}). We denote this irrep by $\gamma$. We
are interested in the trace of the Green function of the SB over the
volume of the SB which is essentially the density of states:
\begin{equation}
  T
  \equiv
  \mbox{Tr}_{\rm SB} G_{\rm SB}^{(\gamma)} (\vec{r}, \vec{r}\,') \: .
\end{equation}
One can apply the projection operation \cite{Tin64} and express
$G_{\rm SB}^{(\gamma)}$ using the Green function of the ST:
\begin{equation}
  G_{\rm SB}^{(\gamma)} (\vec{r}, \vec{r}\,') 
  =
  \frac{1}{l_{\gamma}} \sum_{\hat{g} \in O_h} \chi^{(\gamma)*}(\hat{g})
    G_{\rm ST} (\vec{r}, \hat{g} \vec{r}\,') \: ,
\end{equation}
where $\chi^{(\gamma)}(\hat{g})$ is the character of $\hat{g}$ in the
irrep $\gamma$ and $l_{\gamma}$ is the dimension of $\gamma$. It can
be verified that the above $G_{\rm SB}$ satisfies the inhomogeneous
Helmholtz equation with the correct normalization, and it is composed
only of eigenfunction that transform according to $\gamma$. Thus:
\begin{equation}
  T 
  = 
  \frac{1}{l_{\gamma}} \sum_{\hat{g} \in O_h}
    \chi^{(\gamma)*}(\hat{g})
    \mbox{Tr}_{\rm SB} G_{\rm ST} (\vec{r}, \hat{g} \vec{r}\,') \: .
  \label{eq:t-sb}
\end{equation}
To relate $\mbox{Tr}_{\rm SB}$ with $\mbox{Tr}_{\rm ST}$ we use the
relation:
\begin{equation}
  G_{\rm ST}(\vec{r}, \vec{r}\,') 
  =
  G_{\rm ST}(\hat{h} \vec{r}, \hat{h} \vec{r}\,') \; \; \; \;
    \forall \hat{h} \in O_h
\end{equation}
which can be proven by e.g.\ using the spectral representation of
$G_{\rm ST}$. In particular, we can write:
\begin{equation}
  G_{\rm ST}(\vec{r}, \vec{r}\,') 
  =
  \frac{1}{48} \sum_{\hat{h} \in O_h}
    G_{\rm ST}(\hat{h} \vec{r}, \hat{h} \vec{r}\,') \, .
  \label{eq:gst}
\end{equation}
Combining (\ref{eq:gst}) with (\ref{eq:t-sb}) we get:
\begin{eqnarray}
  T
  & = &
  \frac{1}{48 l_{\gamma}} \sum_{\hat{g}, \hat{h} \in O_h}
  \chi^{(\gamma)*}(\hat{g})
  \mbox{Tr}_{\rm SB} G_{\rm ST} (\hat{h} \vec{r}, 
                                 \hat{h} \hat{g} \vec{r}\,')
  \nonumber \\
  & = &
  \frac{1}{48 l_{\gamma}} \sum_{\hat{g}, \hat{h} \in O_h}
    \chi^{(\gamma)*}(\hat{h} \hat{g} \hat{h}^{-1})
    \mbox{Tr}_{\rm SB} G_{\rm ST} (\hat{h} \vec{r},
    (\hat{h} \hat{g} \hat{h}^{-1}) \hat{h} \vec{r}\,')
  \nonumber \\
  & = &
  \frac{1}{48 l_{\gamma}} \sum_{\hat{h}, \hat{k} \in O_h}
    \chi^{(\gamma)*}(\hat{k})
    \mbox{Tr}_{\rm SB} G_{\rm ST} (\hat{h} \vec{r},
    \hat{k} \hat{h} \vec{r}\,') \, .
  \label{eq:t-hk}
\end{eqnarray}
To obtain the second line from the first one, we recall that the
character is the trace of the irrep matrix, and we have in general
$\mbox{Tr}(ABC) = \mbox{Tr}(CAB)$, therefore $\chi(\hat{g}) =
\chi(\hat{h} \hat{g} \hat{h}^{-1})$. The third line is obtained from
the second one by fixing $\hat{h}$ and summing over $\hat{g}$. Since
$\hat{h} \hat{g}_1 \hat{h}^{-1} = \hat{h} \hat{g}_2 \hat{h}^{-1}
\Longleftrightarrow \hat{g}_1 = \hat{g}_2$ the summation over
$\hat{g}$ is a rearrangement of the group. We now apply the
geometrical identity:
\begin{equation}
  \sum_{\hat{h} \in O_h} \int_{\rm SB} {\rm d}^3 
    r f(\hat{h} \vec{r}) 
  =
  \int_{\rm ST} {\rm d}^3 r f(\vec{r})
\end{equation}
to cast (\ref{eq:t-hk}) into the desired form:
\begin{equation}
  T 
  =
  \frac{1}{48 l_{\gamma}} \sum_{\hat{g} \in O_h}
    \chi^{(\gamma)*}(\hat{g})
    \mbox{Tr}_{\rm ST} G_{\rm ST} (\vec{r}, \hat{g} \vec{r}\,') \, ,
  \label{eq:g-sbst}
\end{equation}
where we relabelled $\hat{k}$ as $\hat{g}$ for convenience. The result
(\ref{eq:g-sbst}) is the desired one, since $T$ is now expressed using
traces over ST which involve no symmetry planes. Semiclassically, the
formula (\ref{eq:g-sbst}) means that we should consider all the
periodic orbits of the ST {\em modulo}\/ a symmetry element $\hat{g}$
to get the density of states of the SB. Therefore, the difficulty of
handling improper orbits is eliminated, since in the ST all of the
isolated periodic orbits are proper.

Let us elaborate further on (\ref{eq:g-sbst}) and consider the various
contributions to it. A proper periodic orbit of the SB with code $(W;
\hat{g})$ has 48 realizations in the ST which are geometrically
distinct. They are obtained from each other by applying the operations
of $O_h$. These conjugate periodic orbits are all related to the same
$\hat{g}$ and thus have the same lengths and monodromies. Consequently
they all have the same semiclassical contributions. Hence, their
semiclassical contribution to $T$ is the same as we would get from
naively applying the Gutzwiller trace formula to the SB, considering
only proper periodic orbits. This result is consistent with our
findings about classical desymmetrization (section
\ref{subsec:cl-desym} above). For the improper periodic orbits there
is a difference, however. There are genuine semiclassical effects due
to desymmetrization for unstable periodic orbits that are confined to
planes or to edges, notably large reduction in the contributions for
Dirichlet conditions on the symmetry planes.

To demonstrate this point, let us consider in some detail an example
of the periodic orbit that traverses along the $8$-fold edge $AE$ in
figure \ref{fig:sb}. For the ST (no desymmetrization) its
semiclassical contribution is:
\begin{equation}
  A_1 = \frac{R}{2 \pi} \: .
\end{equation}
For the SB there are 8 code words that correspond to the periodic
orbit(s) which traverses along this $8$-fold edge. A calculation
yields for the semiclassical contribution:
\begin{equation}
  A_{8}
  =
  \frac{R}{8 \pi} \left[ 2 \pm 2 \sqrt{1 - 2 \beta} \pm
    \beta \left( \frac{2 - \beta}{1 - \beta} \right) \right] \: ,
\end{equation}
where $\beta \equiv R / S$. The upper sign is for the case of the
totally symmetric irrep, and the lower one for the totally
antisymmetric irrep. In the antisymmetric case we get for $\beta \ll
1$:
\begin{equation}
  \frac{A_8}{A_1} \approx \left( \frac{\beta}{2} \right)^4 \: ,
\end{equation}
which means that the desymmetrization {\em greatly reduces}\/ the
contribution of this periodic orbit in case of Dirichlet boundary
conditions on the planes. For the case of our longest spectrum
($R=0.2$, $S=1$) this reduction factor is approximately $2 \times
10^{-4}$ which makes the detection of this periodic orbit practically
impossible. For Neumann boundary conditions the contribution is
comparable to the ST case and is appreciable.

The formula (\ref{eq:g-sbst}) together with the algorithm described
above are the basis for our computations of the semiclassical
contributions of the periodic orbits of the SB. Specifically, the
contribution of a code $\hat{W}$ is given by:
\begin{equation}
  A_{\hat{W}} 
  =
  \frac{L^{\rm po}_{\hat{W}}
    K_{\hat{W}} \chi^{(\gamma)*}(\hat{g}) \sigma_{\hat{W}}}
    {\pi l_{\gamma} r
    \left| \det \left( I - M_{\hat{W}} \right) \right|^{1/2}} \: ,
  \label{eq:aw}
\end{equation}
where $L^{\rm po}_{\hat{W}}$ is the length of the periodic orbit,
$K_{\hat{W}} = $ (\# of distinct realizations of $\hat{W}$ under
$O_h$)/48 and $r$ is the repetition index. The term $\sigma_{\hat{W}}$
is due to the reflections from the spheres and is determined by the
boundary conditions on them. For Neumann boundary conditions
$\sigma_{\hat{W}} = 1$, for Dirichlet boundary conditions
$\sigma_{\hat{W}} = (-1)^n$, where $n$ is the number of bounces.

\subsection{Length spectrum}
\label{subsec:length-spectrum}

Having derived the explicit expression for the semiclassical
amplitudes for the SB (\ref{eq:aw}), we are in position to transform
the trace formula (\ref{eq:gtf}) to a form which can be used for
numerical computations which test its validity. We define the {\em
  length spectrum}\/ as the Fourier transform of the density of
states:
\begin{equation}
  D(l)
  \equiv
  \frac{1}{\sqrt{2 \pi}}
    \int_{-\infty}^{+\infty} d(k) \, e^{i k l} \, {\rm d}k =
    \frac{1}{\sqrt{2 \pi}} \sum_{n} e^{i k_n l} \, .
\end{equation}
For convenience we define $d(-k) \equiv d(k) \Longrightarrow k_{-n} =
-k_n$ and the sum is carried out for all $n \in \bbbz \backslash
\{0\}$. Using the trace formula (\ref{eq:gtf}) we obtain
semiclassically:
\begin{equation}
  D_{\rm sc}(l) 
  = 
  \bar{D}(l) + D_{\rm bb}(l) +
    \sum_{\rm PO} \sqrt{\frac{\pi}{2}} A_j
    \left[ \delta(l - L_j) + \delta(l + L_j) \right] \, .
  \label{eq:length-spectrum}
\end{equation}
In the above $\bar{D}(l)$ is a singularity at $l=0$ which is due to
the smooth density of states. The length spectrum is sharply peaked
near lengths of periodic orbits hence its name. To regularize
(\ref{eq:length-spectrum}) such that it can be applied to finite
samples of the quantum spectrum, we use a weight function and
construct the weighted length spectrum \cite{SS90}:
\begin{equation}
  D^{(w)}(l; k)
  \equiv
  \frac{1}{\sqrt{2 \pi}}
    \int_{-\infty}^{+\infty} w(k{-}k') \, d(k') \, e^{i k' l} \, {\rm d}k'
  =
  \frac{1}{\sqrt{2 \pi}} \sum_{n} w(k{-}k_n) \, e^{i k_n l}
\end{equation}
where $w$ is a weight function (with an effective finite support) that
is concentrated at the origin. The corresponding semiclassical
expression is:
\begin{equation}
  D^{(w)}_{\rm sc}(l; k)
  =
  \bar{D}^{(w)}(l) + D_{\rm bb}^{(w)}(l) +
  \sum_{PO} \frac{A_j}{2}
  \left[ \hat{w}(l - L_j) e^{i k (l-L_j)} +
    \hat{w}(l + L_j) e^{i k (l+L_j)} \right] \, ,
\end{equation}
where $\hat{w}(l) \equiv (1/\sqrt{2 \pi}) \int_{-\infty}^{+\infty}
w(k) \, e^{i k l} \, {\rm d}k$ is the Fourier transform of $w(k)$.

In principle, $d(k)$ and $D(l)$ contain the same information and are
therefore equivalent. However, for our purposes, it is advantageous to use
the length spectrum $D(l)$ (and in practice
$D^{(w)}(l; k))$ rather than the spectral density (\ref{eq:gtf}) for the
following reasons:
\begin{itemize}
  
\item The regularized semiclassical length spectrum, $D^{(w)}_{\rm
    sc}$, is absolutely convergent for suitably chosen windows
  \cite{SS90} (e.g.\ Gaussians). This is in contrast with the original
  trace formula (\ref{eq:gtf}).
  
\item There is an exact mathematical result \cite{AM77} that states
  that for billiards the singular supports of $D(l)$ and of $D_{\rm
    sc}(l)$ are the same, if the infinite spectra are considered. This
  exact quantum--classical result specifically relates to the length
  spectra. It is therefore useful to identify and treat transient
  effects (e.g.\ diffraction contributions) for finite spectra using
  the length coordinate.
  
\item The trace formula (\ref{eq:gtf}) can be considered as a means to
  quantize a chaotic system, since it expresses the quantal density of
  states in terms of the classical length spectrum. However, in
  practice this is not convenient because the semiclassical amplitudes
  are only leading terms in asymptotic series in $k$ (equivalently in
  $\hbar$). For finite values of $k$ there can be large deviations due
  to sub-leading corrections \cite{GA93,AG93} and also due to
  significant diffraction corrections \cite{VWR94,PSSU96,PSSU97}.
  Treating the trace formula the other way (``inverse quantum
  chaology'') is advantageous because the quantal amplitudes have all
  equal weights 1.
  
\item The appearance of peaks in both $d(k)$ and $D(l)$ comes as a
  result of the constructive interference of many oscillatory
  contributions. Any missing or spurious contribution can blur the
  peaks (see figure \ref{fig:ls-shortest-bb} for an example with a
  single energy level missing). For the energy levels we have a good
  control on the completeness of the spectrum due to Weyl's law (see
  section \ref{app:weyl}). As discussed above, this is not the case
  for periodic orbits where we do not have an independent verification
  of their completeness. Hence it is advantageous to use the energy
  levels which are known to be complete in order to reproduce peaks
  that correspond to the periodic orbits.
  
\item For the Sinai billiard the low--lying domain of the spectrum is
  peculiar due to effects of desymmetrization (see section
  \ref{subsec:low-lying}). For Dirichlet boundary conditions on the
  planes, the levels $k_n R < 9$ are very similar to those of the
  integrable case $(R=0)$. The ``chaotic'' levels for which the
  semiclassical approximation is valid ($k_n R > 9$) thus start higher
  up, which makes the semiclassical reproduction of them very
  difficult in practice even with the use of Berry--Keating
  resummation techniques \cite{Kea93}. On the other hand, using the
  quantum levels we can reproduce a few isolated length peaks, as will
  be seen in the sequel.

\end{itemize}

In the following we shall demonstrate a stringent test of the
completeness and of the accuracy of the quantal spectrum using the
length spectrum. Then we shall investigate the agreement between the
quantal and the semiclassical length spectra. We shall employ a
technique to filter the effects of the bouncing balls, such that only
generic contributions remain.

\subsection{A semiclassical test of the quantal spectrum}
\label{subsec:sc-test}

\begin{figure}[p]
  \begin{center}
    \leavevmode

    \psfig{figure=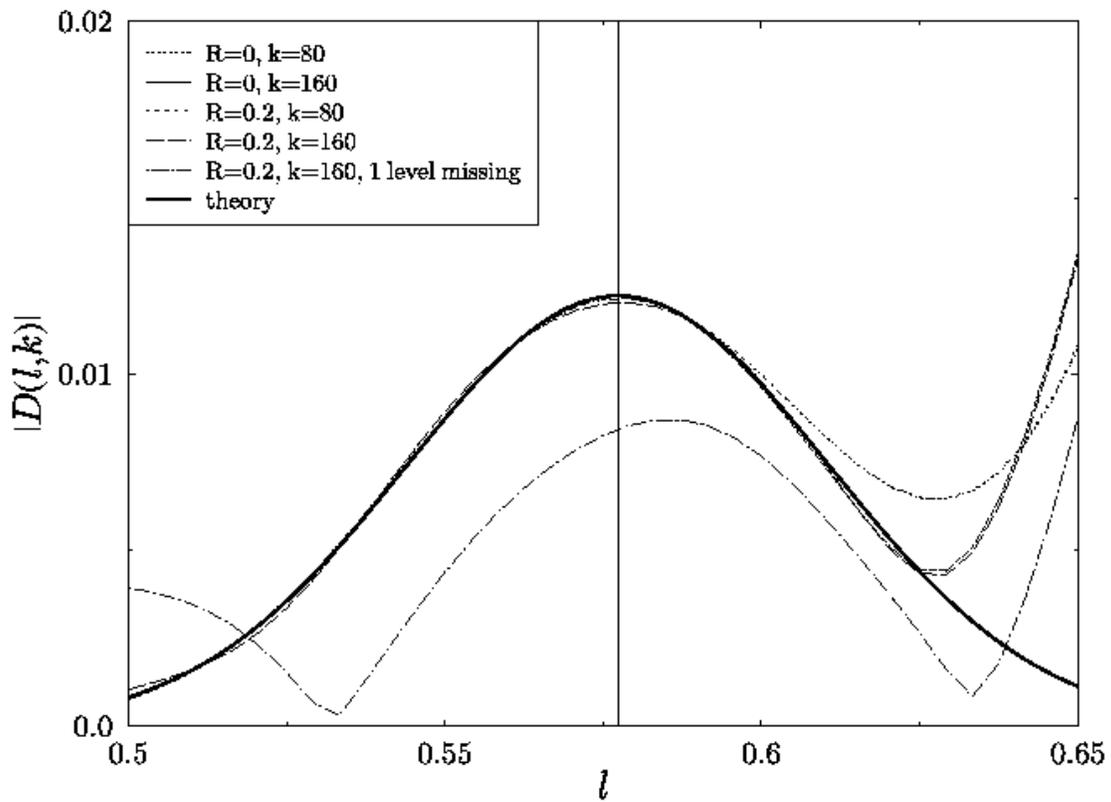,width=16cm}

    \caption{Absolute value of the quantal length spectra
      $|D^{(w)}|$ with a Gaussian window, $\sigma=30$, compared to the
      theoretical prediction (\protect\ref{eq:lssc-shortest-bb}). The
      location of the shortest bouncing ball is indicated by the
      vertical line.}

    \label{fig:ls-shortest-bb}

  \end{center}
\end{figure}

In the following we use the length spectrum in order to develop a
stringent test of the completeness and integrity of the quantal
spectrum. This supplements the integrity and completeness analysis of
the quantal spectrum done in subsections
\ref{subsec:low-lying}--\ref{subsec:verify-weyl}. The idea is to focus
on an isolated contribution to the length spectrum that can be
compared to an analytical result. In section \ref{subsec:integrable}
we discussed the integrable billiard ($R=0$) and observed that there
are contributions to the density of states due to isolated but {\em
  neutral}\/ periodic orbits. The shortest periodic orbit of this kind
has length $S/\sqrt{3} \approx 0.577 S$ and was shown in figure
\ref{fig:pyramid}. Its contribution must prevail for $R>0$ until it is
shadowed by the inscribed sphere, which occurs at $R = S / \sqrt{6}
\approx 0.41 S$. Being the shortest bouncing ball, its length is
distant from the other bouncing balls. As for the isolation from other
generic periodic orbits, for $R=0.2$ there is a nearby contribution of
the shortest unstable periodic orbit of length $0.6 S$. However, for
Dirichlet boundary conditions on the planes the latter is practically
eliminated due to symmetry effects as was discussed in section
\ref{subsec:sc-desym}. Since other periodic orbits are fairly distant,
this shortest bouncing ball is an ideal test--ground of the length
spectrum. Using (\ref{eq:dkint}) and a Gaussian window:
\begin{equation}
  w(k-k') 
  = 
  \frac{1}{\sqrt{2 \pi \sigma^2}}
    \exp \left[ -\frac{(k-k')^2}{2 \sigma^2} \right] \: ,
\end{equation}
one obtains the contribution of the shortest bouncing ball to the
length spectrum:
\begin{equation}
  D^{(w)}_{\rm sc, shortest-bb}(l; k)
  =
  \frac{e^{i k (l- S/\sqrt{3})}}{(6 \pi)^{3/2}}
    \exp \left[ -(l-S/\sqrt{3})^2 \sigma^2 / 2 \right] \, .
  \label{eq:lssc-shortest-bb}
\end{equation}
Due to its isolation, one expects that the shortest bouncing ball
gives the dominant contribution to the length spectrum near its
length. Thus, for $l \approx S/\sqrt{3}$, one has $|D^{\rm (w)}_{\rm
  sc}| \approx |D^{(w)}_{\rm sc, shortest-bb}|$. The latter is
independent of $k$. To test the above relation, we computed the
quantal length spectrum $D^{(w)}$ for $R=0$ and $R=0.2$ for two
different values of $k$, and compared with
(\ref{eq:lssc-shortest-bb}). The results are shown in figure
\ref{fig:ls-shortest-bb}, and the agreement is very satisfactory.

To show how sensitive and stringent this test is, we removed from the
$R=0.2$ quantal spectrum a {\em single}\/ level, $k_{1500} =
175.1182$, and studied the effect on the length spectrum. As is seen
in the figure, this is enough to severely damage the agreement between
the quantum data and the theoretical expectation. Therefore we
conclude that our spectrum is complete and also accurate to a high
degree.

\subsection{Filtering the bouncing-balls I:
         Dirichlet--Neumann difference}
\label{subsec:d-n}

\begin{figure}[p]

    \centerline{\psfig{figure=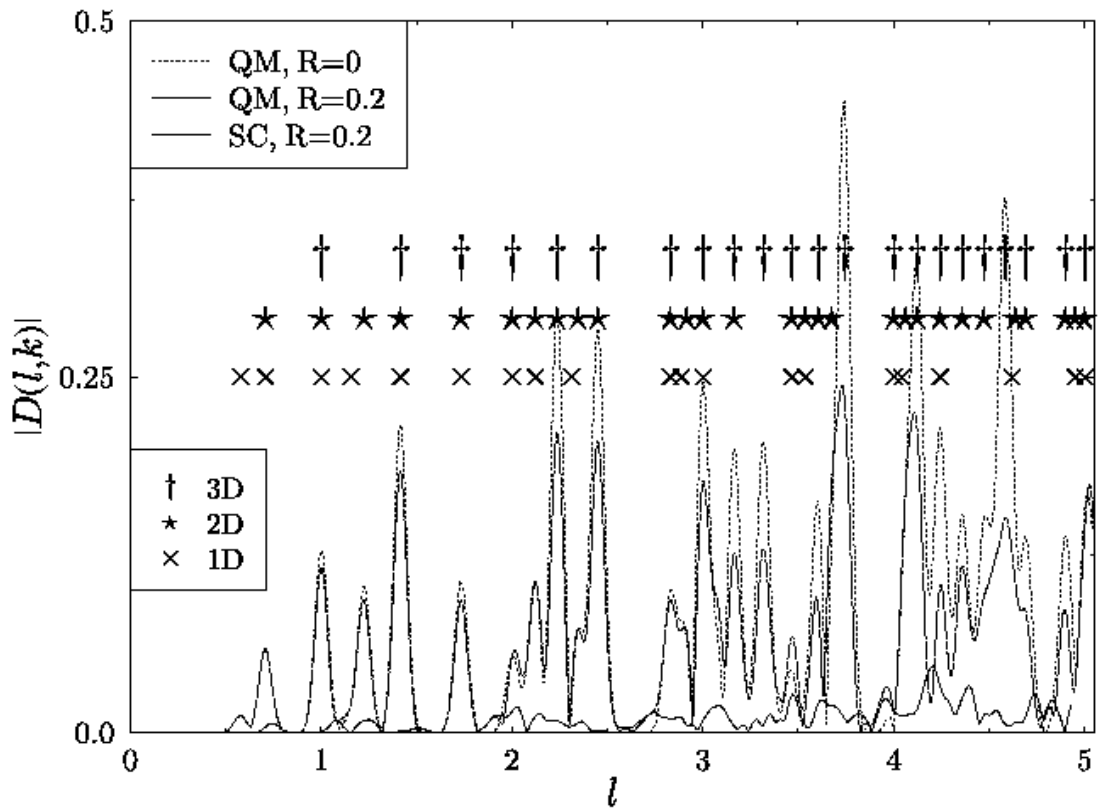,width=16cm}}

    \caption{Quantal length spectra for $R=0$ and $R=0.2$
      compared to semiclassical length spectrum for $R=0.2$ that
      contains only generic, unstable periodic orbits. In all cases
      $k=160$, $\sigma=30$. The locations of the bouncing balls are
      indicated: Daggers for $2$-parameters bouncing balls that occupy
      3D volume in configuration space, stars for 2D bouncing balls
      and crosses for 1D bouncing balls.}

  \label{fig:bare-ls}

\end{figure}

The final goal of our semiclassical analysis is to test the
predictions due to Gutzwiller's trace formula. Since the 3D Sinai is
meant to be a paradigm for 3D systems, we must remove the influence of
the non-generic bouncing-ball families and find a way to focus on the
contributions of the generic and unstable periodic orbit. This is
imperative, because in the 3D Sinai billiard the bouncing balls have
contributions which are much larger than those of the generic periodic
orbits. Inspecting equations (\ref{eq:dkint}) and (\ref{eq:aw}), we
find that the contributions of the leading--order bouncing balls are
stronger by a factor of $k$ than those of the generic periodic orbits.
This is worse than in the 2D case, where the factor is $\sqrt{k}$. To
show how overwhelming is the effect of the bouncing balls, we plot in
figure \ref{fig:bare-ls} the quantal lengths spectra $|D^{(w)}|$ for
$R=0$ and $R=0.2$ (Dirichlet everywhere) together with $|D_{\rm
  sc}^{(w)}|$ which contains contributions {\em only}\/ from generic
and unstable periodic orbits.
One observes that all the peaks in the quantal length spectra are near
lengths of the bouncing balls. Contributions of generic periodic
orbits are completely overwhelmed by those of the bouncing balls and
cannot be traced in the quantal length spectrum of $R=0.2$. Also, we
see that for $R=0.2$ the peaks are in general lower than for $R=0$.
This is because of the (partial or complete) shadowing effect of the
inscribed sphere that reduces the prefactors of the bouncing balls as
$R$ increases.

In the case of the 2D Sinai billiard it was possible to analytically
filter the effect of the bouncing balls from the semiclassical density
of states \cite{SSCL93,SS95}. In three dimensions this is much more
difficult. The functional forms of the contributions to the density of
states of the bouncing balls are given in (\ref{eq:dkint}), but it is
a difficult geometric problem to calculate the prefactors which are
proportional to the cross sections of the bouncing--ball manifolds in
configuration space. The desymmetrization makes this difficulty even
greater and the calculations become very intricate. In addition, there
is always an {\em infinite}\/ number of bouncing--ball manifolds in
the 3D Sinai. This is in contrast with the 2D Sinai, in which a finite
(and usually quite small for moderate radii) number of bouncing--ball
families exist. All this means, that an analytical subtraction of the
bouncing--ball contributions is very intricate and vulnerable to
errors which are difficult to detect and can have a devastating effect
on the quantal--semiclassical agreement.

\begin{figure}[p]
  \begin{center}
    \leavevmode

    \psfig{figure=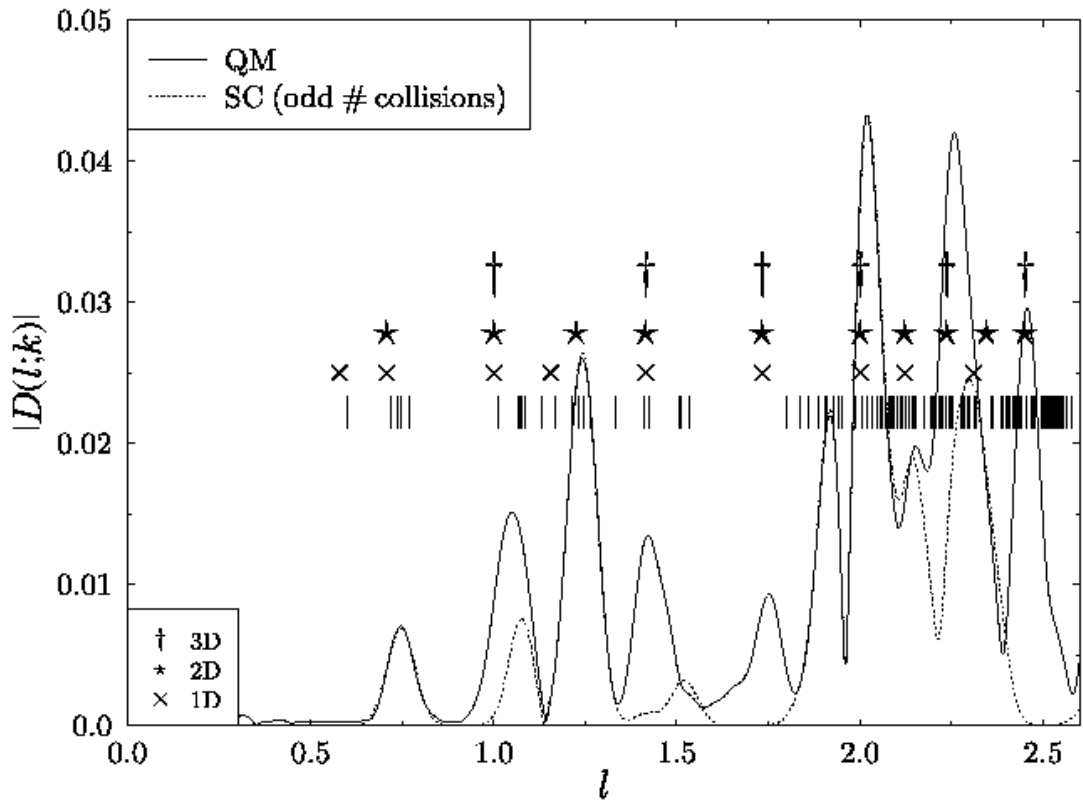,width=16cm}

    \caption{Dirichlet--Neumann difference length spectra for $R=0.2$,
      with $k=100$, $\sigma=30$. The semiclassical length spectrum is
      computed according to (\protect\ref{eq:aj-d-n}). The daggers,
      stars and crosses indicate the positions of the bouncing balls
      (refer to figure \protect\ref{fig:bare-ls}) and the vertical
      bars indicate the positions of the generic, unstable periodic
      orbits.}

    \label{fig:d-n}

  \end{center}
\end{figure}

In order to circumvent these difficulties, we present in the following
an efficient and simple method to get rid of the bouncing balls. The
idea is simple: The bouncing balls are exactly those periodic orbits
that {\em do not}\/ reflect from the sphere. Therefore, changing the
boundary conditions {\em on the sphere}\/ does not affect the
bouncing--ball contributions. Thus, the semiclassical density of
states for Dirichlet / Neumann boundary conditions on the sphere is:
\begin{equation}
  d_{\rm D/N}
  =
  \bar{d}_{\rm D/N} + d_{\rm bb} + d^{\rm (osc)}_{\rm D/N} \ .
\end{equation}
The difference $d_{\rm D}-d_{\rm N}$ is hence independent (in leading
approximation in $k$) of $d_{\rm bb}$ and has the standard form of a
trace formula:
\begin{equation}
  d_{\rm D-N}
  \equiv
  d_{\rm D}(k) - d_{\rm N}(k)
  =
  \left[ \bar{d}_{\rm D}(k) - \bar{d}_{\rm N}(k) \right] +
    \sum_{\rm PO} \left( A_j^{\rm (D)} -
    A_j^{\rm (N)} \right) \cos(k L_j) \ .
\end{equation}
Here $A_j^{\rm (D)}$, $A_j^{\rm (N)}$ are the coefficients that
correspond to Dirichlet and Neumann cases, respectively. In fact, for
Dirichlet, each reflection with the sphere causes a sign change, while
for Neumann there are no sign changes. Therefore:
\begin{equation}
  A_j^{\rm (D-N)}
  \equiv
  A_j^{\rm (D)} - A_j^{\rm (N)}
  =
  \left\{
    \begin{array}{ll}
      2 A_j^{\rm (D)} & \mbox{odd number of reflections} \\
      0               & \mbox{even number of reflections}
    \end{array}
    \right. \: ,
    \label{eq:aj-d-n}
\end{equation}
and we expect to observe in the length spectrum of $d_{\rm D-N}$
contributions only due to generic periodic orbits with an odd number
of reflections. The results of the numerical computations are
presented in figure \ref{fig:d-n} where we compare the quantal (exact)
vs.\ semiclassical (theoretical) length spectra.
We observe on the outset that in contrast to figure \ref{fig:bare-ls}
the quantal and semiclassical length spectra are of similar magnitudes
and the bouncing balls no longer dominate. The peaks near lengths that
correspond to the bouncing balls are greatly diminished, and in fact
we see that the peak corresponding to the shortest bouncing ball ($l
\approx 0.577$) is completely absent, as predicted by the theory. Even
more important is the remarkable agreement between the quantal and the
semiclassical length spectra which one observes near various peaks
(e.g.\ near $l = 0.75$, $1.25$, $2$). Since the semiclassical length
spectrum contains only generic contributions from unstable periodic
orbits, this means that we demonstrated the existence and the
correctness of these Gutzwiller contributions in the quantal levels.
Therefore, one can say that at least as far as length spectra are
concerned, the semiclassical trace formula is partially successful.
There are, however, a few locations for which there is no agreement
between the quantal and the semiclassical length spectra. The places
where this discrepancy takes place are notably located near 3D
bouncing--ball lengths. This suggest that there are ``remnants'' of
the bouncing--ball contributions that are not filtered by the
Dirichlet -- Neumann difference procedure. It is natural to expect
that these remnants are most prominent for the strongest (3D) bouncing
balls. The origin of these remnants are the periodic orbits that are
exactly tangent to the sphere. As an example, consider the 3D
bouncing--ball families that are shown in figure \ref{fig:bb} (upper
part). The tangent orbits that are related to them constitute a
1-parameter family that surrounds the sphere like a ``corona''. For a
single tangent traversal their contributions acquire opposite signs
for Dirichlet and Neumann boundary conditions on the sphere. Hence the
Dirichlet -- Neumann difference procedure still include these
contributions which is apparent in the large discrepancy near $l=1$.
For two tangent traversals the Dirichlet and Neumann contributions
have the same sign and hence cancel each other. This is indeed
confirmed in figure \ref{fig:d-n} where we observe that near $l=2$
there is no discrepancy between the quantal and the semiclassical
length spectra.

The above mentioned tangent orbits belong to the set of points in
phase space in which the classical mapping is discontinuous.
Semiclassically they give rise to diffraction effects. Tangent orbits
were treated for the 2D case in our work \cite{PSSU96,PSSU97}. To
eliminate their effects we hence need to sharpen our tools and to find
a better filtering method than the present Dirichlet -- Neumann
difference procedure. This is performed in the following using mixed
boundary conditions.

\subsection{Filtering the bouncing-balls II:
         Mixed boundary conditions}
\label{subsec:mbc}

The idea behind the Dirichlet--Neumann difference method was to
subtract two spectra which differ only by their boundary conditions on
the sphere. This can be generalized, if one replaces the discrete
``parameter'' of Dirichlet or Neumann conditions by a continuous
parameter $\alpha$, and studies the differences of the corresponding
densities of states $d(k; \alpha_1) - d(k; \alpha_2)$. In section
\ref{subsec:kkr-det} we discussed the mixed boundary conditions
regarding the exact quantization of the 3D SB and gave the
$\alpha$-dependent expressions for the quantal phase shifts. Mixed
boundary conditions were extensively discussed in \cite{BB70,SPSUS95}.

To include the mixed boundary conditions in the semiclassical trace
formula we generalize the results of Berry \cite{Ber81}. There, he
derived the trace formula for the 2D Sinai billiard from an expansion
of the KKR determinant in terms of traces. If one uses the 3D KKR
matrix with (\ref{eq:pl}) and perform a similar expansion, the result
is a modification of the Gutzwiller terms as follows:
\begin{eqnarray}
  A_j \cos (k L_j)
  & \longrightarrow &
  A_j \cos (k L_j + n_j \pi + \phi_j) \; ,
  \label{eq:mbc-aj} \\
  \phi_j
  & = &
  (-2) \sum_{i=1}^{n_j} \arctan \left(
    \frac{ \kappa \cot \alpha}{k \cos \theta_i^{(j)}} \right) \: .
  \label{eq:phi-mbc}
\end{eqnarray}
Here $A_j$ are the semiclassical coefficients for the Dirichlet
conditions on the sphere (c.f.\ equation (\ref{eq:aw})) and $n_j$
counts the number of reflections from the sphere. The angles
$\theta_{i}^{(j)}$ are the reflection angles from the sphere measured
from the normal of the $j$'th periodic orbit. It is instructive to
note that the phases (\ref{eq:phi-mbc}) above are exactly the same as
those obtained by a plane wave that reflects from an infinite plane
with mixed boundary conditions (\ref{eq:mbc-alpha}). This is
consistent with the local nature of the semiclassical approximation. A
prominent feature of the mixed boundary conditions which is manifest
in (\ref{eq:mbc-aj}) is that they do not affect the geometrical
properties (length, stability) of the periodic orbits. Rather, they
only cause a change of a phase which depends on the geometry of the
periodic orbit. This is due to the fact that the mixing parameter
$\alpha$ has no classical analogue. The invariance of periodic orbits
with respect to $\alpha$ renders the mixed boundary conditions an
attractive parameter for e.g.\ investigations of parametric
statistics. This was discussed and demonstrated in detail in
\cite{SPSUS95}.

We are now in a position to apply the mixed boundary conditions to get
an efficient filtering of the bouncing--ball contributions. We first
note that if we fix $\kappa$, then the levels are functions of
$\alpha$: $k_n = k_n(\alpha)$. Let us consider the derivative of the
quantal counting function at $\alpha = 0$:
\begin{eqnarray}
    \tilde{d}(k)
    & \equiv &
    \left. \frac{\partial N(k; \alpha)}
                {\partial \alpha} \right|_{\alpha=0} =
    \sum_n \left. \frac{\partial}{\partial \alpha}
      \Theta \left[ k - k_n(\alpha) \right] \right|_{\alpha=0}
    \nonumber \\
    & = &
    \sum_{n} \left(
      -\frac{{\rm d}k_n(\alpha)}{{\rm d} \alpha} \right)_{\alpha=0}
    \delta ( k - k_n ) \: ,
  \label{eq:mbc-qm}
\end{eqnarray}
where $k_n = k_n(0)$ are the Dirichlet eigenvalues. Hence, the
quantity $\tilde{d}$ is a weighted density of states with delta--peaks
located on the Dirichlet eigenvalues. 

The semiclassical expression for $\tilde{d}$ does not contain the
leading contribution of the bouncing balls, since this contribution is
independent of $\alpha$. The semiclassical contributions of the
isolated periodic orbits to $\tilde{d}$ are of the form $A_j B_j \cos
(k L_j)$, where
\begin{equation}
  B_j
  =
  \frac{2 k}{\kappa}
    \sum_{i=1}^{n_j} \cos \theta_i^{(j)} \: .
  \label{eq:mbc-bj}
\end{equation}
This is easily derived from (\ref{eq:mbc-aj}) and (\ref{eq:phi-mbc}).
Since the reflection angles $\theta_i^{(j)}$ are in the range $[0,
\pi/2]$, the coefficient $B_j$ vanish {\em if and only if}\/
$\theta_i^{(j)} = \pi/2$ for all $i = 1, \ldots, n_j$, which is an
exact tangency. Therefore, exactly tangent periodic orbits are also
eliminated by the derivative method. This is the desired effect of the
mixed boundary conditions method that serves to further clean the
spectrum from sub-leading contributions of the bouncing balls. We
summarize equations (\ref{eq:mbc-qm}) and (\ref{eq:mbc-bj}):
\begin{equation}
  \tilde{d}(k)
  =
  \sum_{n} v_n \delta (k - k_n)
  \approx
  \left(\hspace{-0.2cm}
    \mbox{\small \begin{tabular}{c} smooth \\ term \end{tabular}}
  \hspace{-0.2cm}\right)
  + \sum_{PO} A_j B_j \cos (k L_j) \: ,
  \; \; v_n \equiv \left( - \frac{\partial k_n}{\partial \alpha}
  \right)_{\alpha=0} \, .
  \label{eq:gutz-mbc}
\end{equation}

To check the utility of $\tilde{d}$ and to verify (\ref{eq:gutz-mbc})
we computed both sides of (\ref{eq:gutz-mbc}) for $R=0.2$ and
$\kappa=100$. The quantal spectrum was computed for $\alpha=0.003$ and
the derivatives $v_n$ were obtained by the finite differences from the
$\alpha=0$ (Dirichlet) spectrum. The coefficients $B_j$ were extracted
from the geometry of the periodic orbits. In figures \ref{fig:ls-mbc1}
and \ref{fig:ls-mbc2} the length spectra are compared. The agreement
between the quantal and the semiclassical data is impressive,
especially for the lower $l$-values. There are no significant remnants
of peaks near the bouncing--ball locations, and the peaks correspond
to the generic and unstable periodic orbits. This demonstrates the
utility of using $\tilde{d}$ as an efficient means for filtering the
spectrum from the non-generic effects.
\begin{figure}[p]
  \begin{center}
    \leavevmode

    \psfig{figure=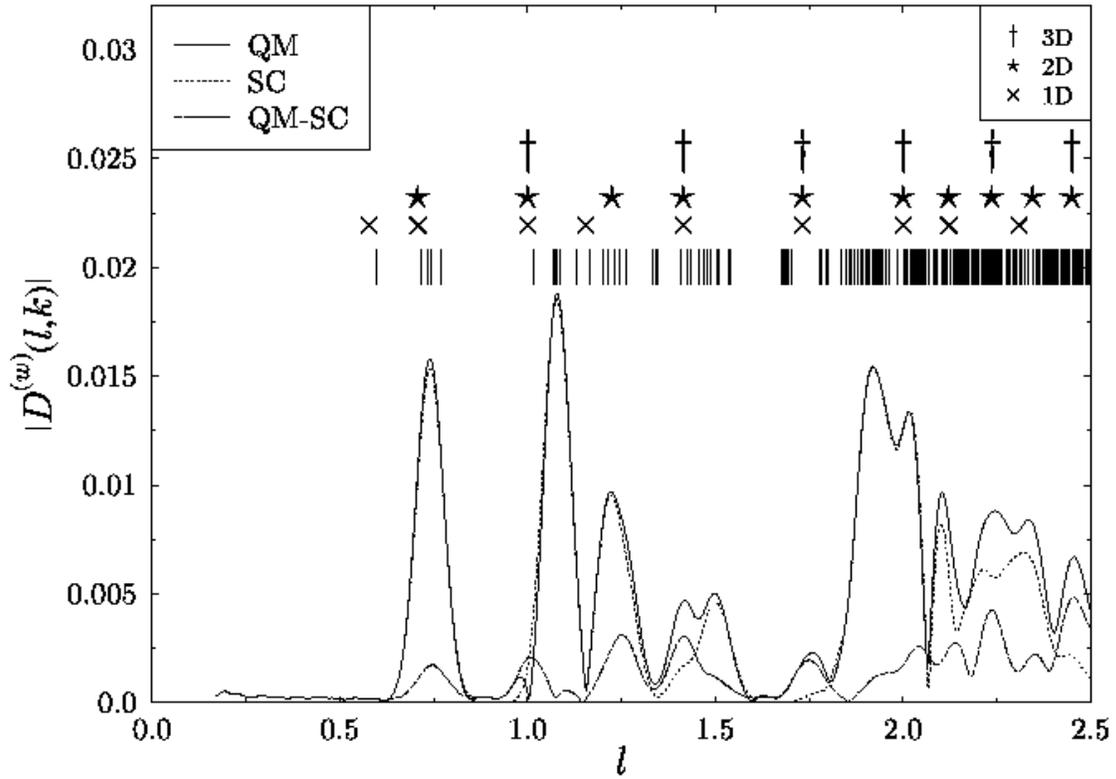,width=16cm}

    \caption{Length spectra for the mixed boundary conditions derivative
      method (\protect\ref{eq:gutz-mbc}). Data are for $R=0.2$,
      $k=150$, $\sigma=30$, $\kappa=100$. The dashed line represents
      $|D^{(w)}(l)-D_{\rm sc}^{(w)}(l)|$. The daggers, stars and
      crosses indicate the positions of the bouncing balls (refer to
      figure \protect\ref{fig:bare-ls}) and the vertical bars indicate
      the positions of the generic, unstable periodic orbits.}

    \label{fig:ls-mbc1}

  \end{center}
\end{figure}
\begin{figure}[p]
  \begin{center}
    \leavevmode

    \psfig{figure=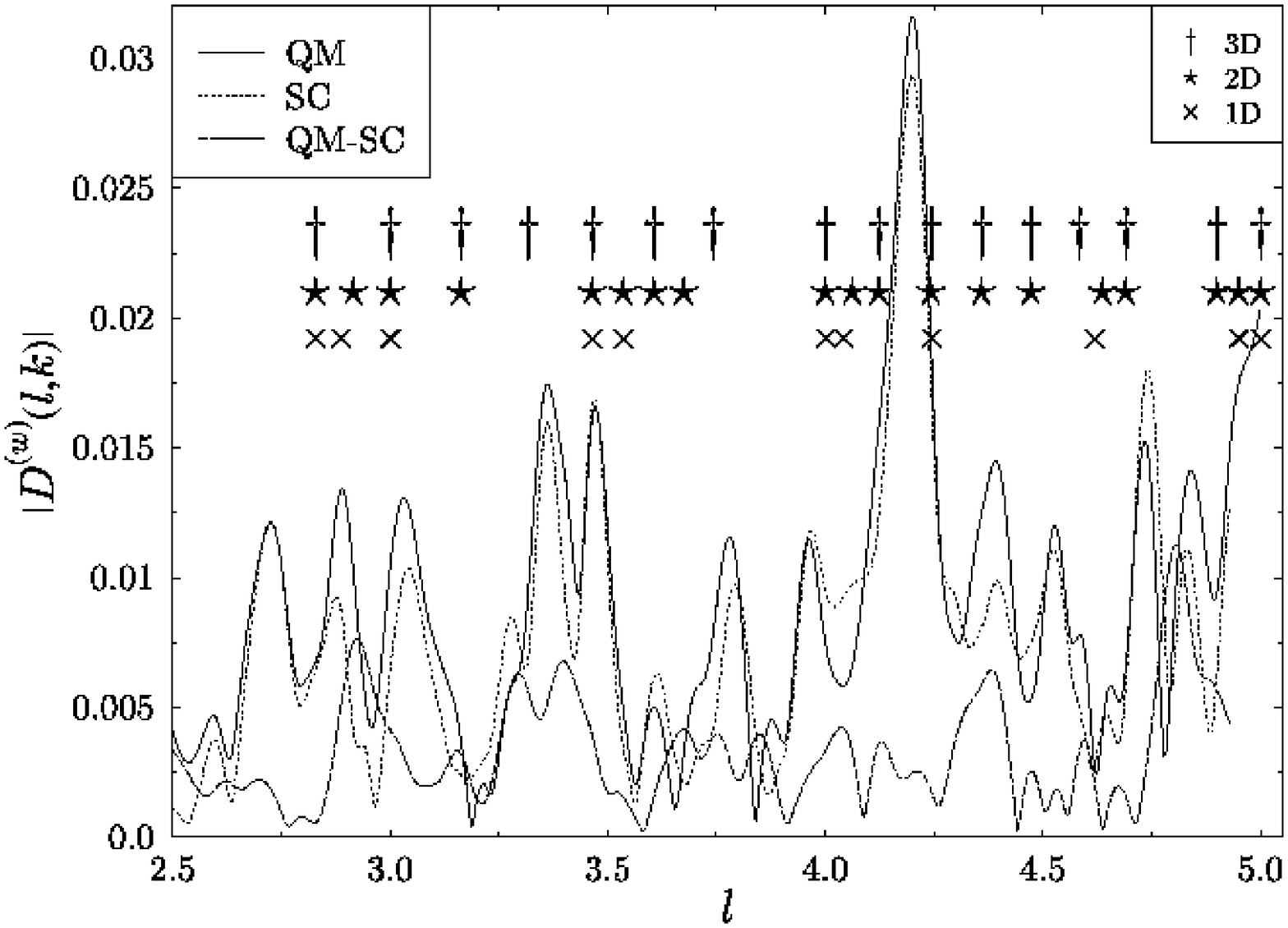,width=16cm}

    \caption{Continuation of figure \protect\ref{fig:ls-mbc1} to
      $2.5 \leq l \leq 5$. We did not indicate the locations of
      unstable periodic orbits due to their enormous density.}

    \label{fig:ls-mbc2}

    \end{center}
\end{figure}

The quantal--semiclassical agreement of the length spectra is not
perfect, however, and it is instructive to list possible causes of
this disagreement. We first recall that the semiclassical amplitudes
$A_j$ are the leading terms in an asymptotic series, hence we expect
corrections of order $1/k$ to the weights of periodic orbits. They are
denoted as {\em $\hbar$ corrections}\/ and were treated in detail by
Gaspard and Alonso \cite{GA93} and by Alonso and Gaspard \cite{AG93}.
In our case, however, $1/k \approx 1/100$ and these corrections are
not expected to be dominant. More important are diffraction
corrections which are also finite $k$ effects that stem from the
existence of a concave component (the sphere) in the billiard. Several
kinds of diffraction corrections to the trace formula were analyzed
for 2D billiards. Vattay, Wirzba and Rosenqvist \cite{VWR94}
considered {\em creeping orbits}\/, and we considered in
\cite{PSSU96,PSSU97} {\em penumbra corrections}\/. (The penumbra is
the region in phase space which is close to tangency: $|\ell - kR|
\approx (kR)^{1/3}$, where $\hbar \ell$ is the angular momentum.) We
list the various diffraction corrections in the following:
\begin{description}
  
\item[Creeping orbits:] These are orbits which are classically
  forbidden. They ``creep'' over concave parts of the billiard, and
  their semiclassical contribution is {\em exponentially small}\/ in
  $k^{1/3}$, which should be negligible for the $k$ values that we
  consider.
  
\item[Exactly tangent orbits:] These were already mentioned above, and
  we showed that their contributions are eliminated to a large extent
  by the mixed boundary conditions procedure. For 2D systems we found,
  however, that this is true in leading order only, and there are
  small remnants of the tangent orbits in the weighted density
  $\tilde{d}$ \cite{PSSU97}. The magnitude of the remnants in 2D is
  ${\cal O}(1/\sqrt{k})$, which is smaller than ${\cal O}(k^0)$ of a
  generic unstable periodic orbit. In 3D, a similar analysis shows
  that the remnants of each family of tangent orbits is ${\cal
    O}(k^0)$ which is the same magnitude as for unstable periodic
  orbits. Reviewing figures \ref{fig:ls-mbc1} and \ref{fig:ls-mbc2},
  we can observe some of the peaks of the quantal--semiclassical
  difference near lengths that correspond to exactly tangent orbits.
  
\item[Unstable and isolated periodic orbits that traverse the
  penumbra:] We have shown in \cite{PSSU96,PSSU97} that for periodic
  orbits which just miss tangency with a concave component of the
  billiard boundary, there is a correction to the semiclassical
  amplitude $A_j$ which is of the same magnitude as $A_j$ itself.
  These ${\cal O}(1)$ diffraction corrections are the most important
  corrections to the trace formula for generic billiards. For periodic
  orbits which reflect at an extreme forward direction from a concave
  component, the amplitude $A_j$ is very small due to the extreme
  classical instability. If we include diffraction corrections, the
  semiclassical contributions of these orbits get much larger.
  Therefore, the semiclassical contributions of periodic orbit that
  traverse the penumbra must be radically corrected. Moreover, we
  found that if one considers all the periodic orbits up to the
  Heisenberg length $L_{\rm H} \equiv 2 \pi \bar{d}(k)$ (necessary to
  obtain a resolution of one mean level spacing), then {\em almost
    all}\/ of the periodic orbits are vulnerable to penumbra
  diffraction corrections.
  
\item[Classically forbidden periodic orbits] {\bf that traverse 
  the shaded part of the penumbra:} 
  Penumbra diffraction effects lead to
  semiclassical contributions from periodic orbits that slightly
  traverse through a concave component. Since they do not relate to
  classically allowed orbits, they represent new contributions to the
  trace formula rather than corrections of existing ones. Their
  magnitudes are comparable to those of generic unstable periodic
  orbits.

\end{description}
The above list of corrections, which was compiled according to studies
of 2D billiards, suggests that there is a wealth of effects that must
be considered if one wishes to go beyond the Gutzwiller trace formula.
It is very difficult to implement these corrections systematically
even for 2D billiards, and it goes beyond the scope of the present
work to study them further for the 3D Sinai billiard. We mention in
passing that except exact tangency, the penumbra effects are transient
and depend on $k$.

According to the mathematical theorem \cite{AM77} mentioned above, the
quantal and the semiclassical length spectra are asymptotically the
same. The significance of our findings in this section is that we have
shown that the quantal--semiclassical agreement is achieved already
for finite and moderate values of $k$ and that the corrections are not
very large (for the $l$-range we looked at). This is very encouraging,
and justifies an optimistic attitude to the validity of the
semiclassical approximation in 3D systems. However, obtaining accurate
energy levels from the trace formula involves many contributions from
a large number of periodic orbits. Therefore, one cannot directly
infer at this stage from the accuracy of the peaks of the length
spectrum to the accuracy of energy levels. There is a need to quantify
the semiclassical error and to express it in a way which makes use of
the above semiclassical analysis. This is done in 
section~\ref{sec:sc-accuracy}.  


\section{The accuracy of the semiclassical energy spectrum}
\label{sec:sc-accuracy}

%
One of the most important applications of the trace formula is to
explain the spectral statistics and their relation to the universal
predictions of Random Matrix Theory (RMT) \cite{Ber85,BK96}. However,
a prerequisite for the use of the semiclassical approximation to
compute short--range statistics is that it is able to reproduce the
exact spectrum within an error comparable to or less than the mean
level spacing! This is a demanding requirement, and quite often it is
doubted that the semiclassical approximation is able to reproduce
precise levels for high--dimensional systems on the following grounds.
The mean level spacing depends on the dimensionality (number of
freedoms) of the system, and it is ${\cal O}(\hbar^d)$ \cite{LLqm}.
Gutzwiller \cite{Gut90} quotes an argument by Pauli \cite{Pau51} to
show that in general the error margin for the semiclassical
approximation scales as ${\cal O}(\hbar^2)$ {\em independently of the
  dimensionality}. Applied to the trace formula, the expected error in
units of the mean spacing, which is the figure of merit in the present
context, is therefore expected to be ${\cal O}(\hbar^{2-d})$. We shall
refer to this as the ``traditional estimate''. It sets $d=2$ as a
critical dimension for the applicability of the semiclassical trace
formula and hence for the validity of the conclusions which are drawn
from it. The few systems in $d>2$ dimensions which were numerically
investigated display spectral statistics which adhere to the
predictions of RMT as accurately as their counterparts in $d=2$
\cite{PS95,Pri97,Pro97}. Thus, the traditional estimate cannot be
correct in the present context, and we shall explain the reasons why
it is inadequate.

%
In this section we shall develop measures for the accuracy of the
semiclassical energy levels. We shall then derive formulas to evaluate
these measures. Using our quantal and classical (periodic orbits)
databases for the 2D and 3D Sinai billiards, we shall apply the
formulas and get numerical bounds for the semiclassical errors.

%
The problem of the accuracy of the energy spectrum derived from the
semiclassical trace formula was hardly discussed in the literature.
Gutzwiller quotes the traditional estimate of ${\cal O}(\hbar^{2-d})$
\cite{Gut90,Gut89}. Gaspard and Alonso \cite{GA93}, Alonso and Gaspard
\cite{AG93} and Vattay, Wirzba and Rosenqvist \cite{VWR94} derived
explicit and generic $\hbar$ corrections for the trace formula, but do
not address directly the issue of semiclassical accuracy of energy
levels. Boasman \cite{Boa94} estimates the accuracy of the Boundary
Integral Method (BIM) \cite{BW84} for 2D billiards in the case that
the exact kernel is replaced by its asymptotic approximation. He finds
that the resulting error is of the same magnitude as the mean spacing,
in agreement with the traditional estimate. However, the dependence of
the semiclassical error on the dimensionality is not established. We
also mention a recent work by Dahlqvist \cite{Dah98} in which the
semiclassical error due to penumbra (diffraction) effects is
analytically estimated for the 2D Sinai billiard. The results are
compatible with the ones reported here.

\subsection{Measures of the semiclassical error}
\label{subsec:accuracy-measures}
%
In order to define a proper error measure for the semiclassical
approximation of the energy spectrum one has to clarify a few issues.
In contrast with the EBK quantization which gives an {\em explicit}
formula for the spectrum, the semiclassical spectrum for chaotic
systems is {\em implicit} in the trace formula, or in the
semiclassical expression for the spectral determinant.  To extract the
semiclassical spectrum we recall that the exact spectrum, $\{E_n \}$,
can be obtained from the exact counting function:
\begin{equation}
  N(E) \equiv \sum_{n=1}^{\infty} \Theta ( E - E_n ) \ ,
  \label{eq:qmstair}
\end{equation}
by solving the equation
\begin{equation}
  N(E_n) = n - \frac{1}{2} \; ,
  \; \; n = 1, 2, \ldots \; \; \; .
  \label{eq:qc}
\end{equation}
In the last equation, an arbitrarily small amount of smoothing must be
applied to the Heavyside function. In analogy, one obtains the
semiclassical spectrum $\{E_n^{\rm sc}\}$ as \cite{AM96}:
\begin{equation}
  N_{\rm sc}(E_n^{\rm sc}) = n - \frac{1}{2} \; ,
  \; \; n = 1, 2, \ldots \; \; \; ,
  \label{eq:scqc}
\end{equation}
where $N_{\rm sc}$ is a semiclassical approximation of $N$. Note that
$N_{\rm sc}$ with which we start is not necessarily a sharp counting
function. However, once $\{E_n^{\rm sc}\}$ is known, we can
``rectify'' the smooth $N_{\rm sc}$ into the sharp counting function
$N^{\#}_{\rm sc}$ \cite{BK96}:
\begin{equation}
  N^{\#}_{\rm sc}(E) \equiv 
  \sum_{n=1}^{\infty} \Theta ( E - E_n^{\rm sc} ) \, .
  \label{eq:nsc-sharp}
\end{equation}
The most obvious choice for $N_{\rm sc}$ is the Gutzwiller trace
formula \cite{Gut90} truncated at the Heisenberg time, which is what
we shall use. Alternatively, one can start from the regularized
Berry--Keating Zeta function $\zeta_{\rm sc}(E)$ \cite{Kea93}, and
define $N_{\rm sc} = \bar{N} - (1/\pi) \, \mbox{Im} \, \log \,
\zeta_{\rm sc} (E+i0)$, which yields $N_{\rm sc} = N^{\#}_{\rm sc}$.

Next, in order to define a quantitative measure of the semiclassical
error, one should establish a one-to-one {\em correspondence} between
the quantal and the semiclassical levels, namely, one should identify
the semiclassical counterparts of the exact quantum levels. In
classically chaotic systems, for which the Gutzwiller trace formula is
applicable, the only constant of the motion is the energy. This is
translated into a single ``good'' quantum number in the quantum
spectrum, which is the ordinal number of the levels when ordered by
their magnitude. Thus, the only correspondence which can be
established between the exact spectrum $\{E_n\}$ and its semiclassical
approximation, $\{E^{\rm sc}_n\}$, is
\begin{equation}
  E_n \longleftrightarrow E^{\rm sc}_n \; .
\end{equation}
This is to be contrasted with integrable systems, where it is
appropriate to compare the exact and approximate levels which have the
same quantum numbers.

The natural scale on which the accuracy of semiclassical energy levels
should be measured is the mean level spacing $(\bar d(E))^{-1}$. We
shall be interested here in the mean semiclassical error, and proper
measures are the mean absolute difference:
\begin{equation}
  \epsilon^{\rm (1)}(E)   \equiv
  \langle \  \bar{d}(E_n) \left| E_{n} - E^{\rm sc}_{n} \right| \
    \rangle_{E}
\end {equation}
or the variance:
\begin{equation}
  \epsilon^{\rm (2)}(E) \equiv
  \langle \ \left ( \bar{d}(E_n) 
    \left( E_{n} - E^{\rm sc}_{n} \right ) \right)^2 \
  \rangle_{E} \ ,
\end{equation}
where $\langle \cdot \rangle$ denotes averaging over a spectral interval 
$\Delta E$ centered at $E$. The interval $\Delta E$ is large enough so 
that the mean number of levels $\Delta E \cdot \bar{d}(E) \gg 1$. Yet,
$\Delta E$ is small enough on the classical scale, such that $\bar{d}(E) 
\approx \mbox{constant}$ over the interval considered.

We shall now compare two different estimates for the semiclassical error. 
The first one is the traditional estimate:
\begin{equation}
  \epsilon^{\rm traditional} = {\cal O}(\hbar^{2-d})
  \longrightarrow
  \left\{
  \begin{array}{lcc}
     {\rm const} & , & d = 2 \\
     \infty      & , & d \ge 3
  \end{array}
  \right. \; \; \;
  \mbox{as } \ \  \hbar \rightarrow 0
  \label{eq:epstrad}
\end{equation}
(c.f.\ section \ref{sec:intro}). It claims that  the semiclassical 
approximation is (marginally) accurate in two dimensions, 
but it fails to predict accurate energy levels for three
dimensions or more. We emphasize that the traditional estimate is a 
qualitative error measure, emerging from global error estimate of the
time propagator. Hence, it cannot be directly connected to either 
$\epsilon^{(1)}$ or $\epsilon^{(2)}$. We mention it here since it is
the one usually quoted in the literature.

One may get a different estimate of the semiclassical error, if the
Gutzwiller Trace Formula (GTF) is used as a starting point. Suppose
that we have calculated $N_{\rm sc}$ to a certain degree of precision,
and we compute from it the semiclassical energies $E_n^{\rm sc}$ using
(\ref{eq:scqc}). Denote by $\Delta N_{\rm sc}$ the higher order terms 
which were neglected in the calculation of $N_{\rm sc}$. The expected error 
in  $E_n^{\rm sc}$ can be estimated by including $\Delta N_{\rm sc}$ and 
calculating the energy differences $\delta_n$. That is, we consider:
\begin{equation}
  N_{\rm sc}(E_n^{\rm sc}+\delta_n) +
  \Delta N_{\rm sc}(E_n^{\rm sc}+\delta_n) =
  n - \frac{1}{2} \, .
  \label{eq:impscqc}
\end{equation}
Combining (\ref{eq:scqc}) and (\ref{eq:impscqc}) we get (to first order
in $\delta_n$):
\begin{equation}
  \delta_n 
  \approx 
  \frac{- \Delta N_{\rm sc}(E_n^{\rm sc})}
       {\partial N_{\rm sc}(E_n^{\rm sc}) / \partial E} 
  \approx
  \frac{- \Delta N_{\rm sc}(E_n^{\rm sc})}
       {\bar{d}(E_n^{\rm sc})} \: .
\end{equation}
In the above we assumed that the fluctuations of $N_{\rm sc}$ around
its average are not very large. Thus,
\begin{equation}
  \epsilon^{(1), {\rm GTF}} 
  \approx
  \bar{d}(E_n^{\rm sc}) | \delta_n |
  \approx
  \Delta N_{\rm sc}(E_n^{\rm sc}) \: .
\end{equation}

Let us apply the above formula and consider the case in which we take
for $N_{\rm sc}$ its mean part $\bf\bar{\it N}$, and that we include
in $\bf\bar{\it N}$ terms of order up to (and including) $\hbar^{-m},
m \leq d$. For $\Delta N_{\rm sc}$ we use both the leading correction
to $\bf\bar{\it N}$ and the leading order periodic orbit sum which is
formally (termwise) of order $\hbar^0$. Hence,
\begin{equation}
  \epsilon^{(1), {\rm GTF}}_{\bar{N}} 
  =
  {\cal O}(\hbar^{-m+1}) + {\cal O}(\hbar^0) 
  =
  {\cal O}\left( \hbar^{\min(-m+1,\,0)} \right) \: .
  \label{eq:tf-estimate}
\end{equation}
We conclude, that approximating the energies only by the mean counting
function $\bf\bar{\it N}$ up to (and not including) the constant term,
is already sufficient to obtain semiclassical energies which are
accurate to ${\cal O}(\hbar^0) = {\cal O}(1)$ with respect to the mean
density of states. Note again, that no periodic orbit contributions 
 were included in $N_{\rm sc}$. Including less terms in 
$\bf\bar{\it N}$ will lead to a diverging semiclassical error, 
while more terms will be masked by
the periodic orbit (oscillatory) term. One can do even better if one
includes in $N_{\rm sc}$ the smooth terms up to and including the
constant term (${\cal O}(\hbar^0)$) together with the leading order
periodic orbit sum which is formally also ${\cal O}(\hbar^0)$. The
semiclassical error is then:
\begin{equation}
  \epsilon^{(1), {\rm GTF}}_{\rm po} 
  = 
  {\cal O}(\hbar^1) \: .
  \label{eq:epsgtf}
\end{equation}
That is, the semiclassical energies measured in units of the mean
level spacing are asymptotically accurate independently of the
dimension! This estimate grossly contradicts the traditional
estimate (\ref{eq:epstrad}) and calls for an explanation.

The first point that should be noted is that the order of magnitude
(power of $\hbar$) of the periodic orbit sum, which we considered
above to be ${\cal O}(\hbar^0)$, is only a formal one. Indeed, each
term which is due to a {\em single}\/ periodic orbit is of order ${\cal
O}(\hbar^0)$. However the periodic orbit sum {\em absolutely
diverges}, and at best it is only {\em conditionally convergent}. To
give it a numerical meaning, the periodic orbit sum must therefore be
regularized. This is effectively achieved by truncating the trace
formula or the corresponding spectral $\zeta$ function
\cite{DS92,Bog92b,Kea93,PG95}. However, the cutoff itself
depends on $\hbar$. One can conclude, that the simple-minded estimate
(\ref{eq:epsgtf}) given above is at best a lower bound, and the error
introduced by the periodic orbit sum must be re-evaluated with more
care. This point will be dealt with in great detail in the sequel, and
we shall eventually develop a meaningful framework for evaluating the
magnitude of the periodic orbit sum.

The disparity between the traditional estimate
of the semiclassical error and the one based on the trace formula can
be further illustrated by the following argument. The periodic orbit
formula is derived from the semiclassical propagator $K_{\rm sc}$
using further approximations \cite{Gut90}. Therefore one wonders, how can
it be that {\em further} approximations of $K_{\rm sc}$ actually {\em
reduce} the semiclassical error from (\ref{eq:epstrad}) to
(\ref{eq:epsgtf})?  The puzzle is resolved if we recall, that in order
to obtain $\epsilon^{(1), {\rm GTF}}_{\rm po}$ above we separated the density
of states into a smooth part and an oscillating part, and we required that
the smooth part is accurate enough. To achieve this, we have to go
beyond the leading Weyl's term and to use specialized methods to
calculate the smooth density of states beyond the leading order. These
methods are mostly developed for billiards \cite{BB70,BH76,BH94}. In
any case, to obtain $\epsilon^{(1), {\rm GTF}}_{\rm po}$ we have added
{\em additional} information which goes beyond the leading
semiclassical approximation.

A direct check of the accuracy of the semiclassical spectrum
using the error measures $\epsilon^{\rm (1)}$, $\epsilon^{\rm (2)}$ 
is exceedingly difficult due to the exponentially large number of
periodic orbits needed. The few cases where such tests were carried out
involve 2D systems and it was possible to check only the lowest (less
than a hundred) levels (e.g.\ \cite{Sie91,HS92}). The good agreement
between the exact and the semiclassical values confirmed the
expectation that in 2D the semiclassical error is small. In 3D, the
topological entropy is typically much larger \cite{AM96,Pri97}, and
the direct test of the semiclassical spectrum becomes prohibitive.

Facing with this grim reality, we have to introduce alternative error
measures which yield the desired information, but which are more
appropriate for a practical calculation. We construct the measure:
\begin{equation}
  \delta^{\rm (2)}(E) \equiv \left\langle
    \left| N(E) - N_{\rm sc}^{\#}(E) \right|^2
    \right\rangle_{E} \ .
  \label{eq:delta2def}
\end{equation}
As before, the triangular brackets indicate averaging over an energy
interval $\Delta E$ about $E$. We shall now show that $\delta^{\rm
(2)}$ faithfully reflects the deviations between the spectra, and is
closely related to $\epsilon^{\rm (1)}$ and $\epsilon^{\rm
(2)}$. Note, that the following arguments are purely statistical and
apply to every pair of staircase functions.

\begin{figure}[p]
  \begin{center}
    \leavevmode

    \psfig{figure=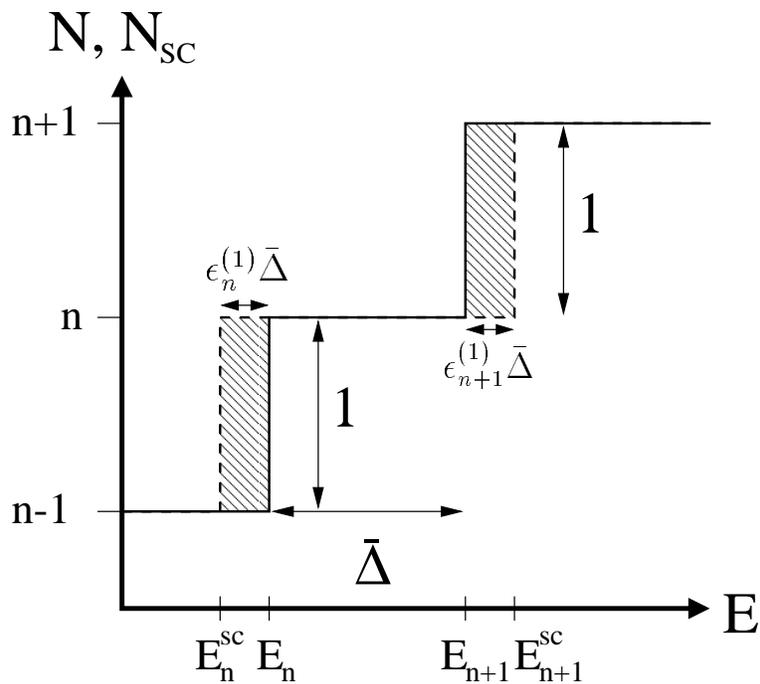,width=10cm}

    \caption{Illustration of $|N(E) - N_{\rm sc}^{\#}(E)|$ for small
      deviations between quantum and semiclassical energies:
      $\epsilon^{\rm (1)} \ll \bar{d}^{-1} \equiv \bar{\Delta}$. The
      quantum staircase $N(E)$ is denoted by the full line and the
      semiclassical staircase $N_{\rm sc}^{\#}(E)$ is denoted by the
      dashed line. The difference is shaded.}

    \label{fig:delta2small}

  \end{center}
\end{figure}

Suppose first, that all the differences $E_n^{\rm sc} - E_n$ are smaller
than the mean spacing. Then, $|N - N_{\rm sc}^{\#}|$ is either 0 or 1
in most of the cases (see figure \ref{fig:delta2small}). Hence,
$|N - N_{\rm sc}^{\#}| = |N - N_{\rm sc}^{\#}|^2$ along most of the $E$ axis. 
Consequently,
\begin{equation}
  \delta^{\rm (2)}(E) \approx \left\langle
    \left| N(E) - N_{\rm sc}^{\#}(E) \right|
   \right\rangle_{E} \; , \; \; \;
   \mbox{for small deviations} \: .
\end{equation}
The right hand side of the above equation (the fraction of
non--zero contributions) equals $\epsilon^{\rm (1)}$. Thus,
\begin{equation}
  \delta^{\rm (2)} \approx \epsilon^{\rm (1)} \: , 
  \; \; \;
  \mbox{for small deviations} \: .
  \label{eq:delta2small}
\end{equation}
If, on the other hand, deviations are much larger than one mean
spacing, the typical horizontal distance $\bar{d}|E-E_n|$ should be
comparable to the vertical distance $|N - N_{\rm sc}^{\#}|$, and hence,
in this limit
\begin{equation}
  \delta^{\rm (2)} \approx \epsilon^{\rm (2)} \: , 
  \; \; \;
  \mbox{for large deviations} \: .
  \label{eq:delta2large}
\end{equation}
Therefore, we expect $\delta^{\rm (2)}$ to interpolate between
$\epsilon^{\rm (1)}$ and $\epsilon^{\rm (2)}$ throughout the entire
range of deviations. This behavior was indeed observed in a numerical
tests which were performed to check the above expectations \cite{PS98}. 
Moreover, it was shown in \cite{PS98} that $\delta^{\rm (2)}$ is 
completely equivalent to $\epsilon^{\rm (2)}$ when the
spectral counting functions are replaced with their smooth counterparts,
provided that the smoothing width is of the order of 1 mean level
spacing and the same smoothing is applied to both counting
functions. That is,
\begin{equation}
  \delta^{\rm (2)}_{\rm smooth} 
  \approx 
  \epsilon^{\rm (2)} \: , 
  \; \; \;
  \mbox{for all deviations} \: .
  \label{eq:delta2epsilon2}
\end{equation}
In testing the semiclassical accuracy, this kind
of smoothing is essential and will be introduced by truncating the
trace formula at the Heisenberg time $t_{\rm H} \equiv h \bar{d}$. 
These properties of the measure $\delta^{\rm (2)}$, and its complete
equivalence to $\epsilon^{\rm (2)}$ for smooth counting functions,
renders it a most appropriate measure of the semiclassical error.

We now turn to the practical evaluation of $\delta^{\rm (2)}$. To
perform the energy averaging, we choose a positive window function
$w(E'-E)$ which has a width $\Delta E$ near $E$ and is normalized by
$\int_{-\infty}^{+\infty} {\rm d}E' w^{2}(E') = 1$. It falls off
sufficiently rapidly so that all the expressions which follow are well
behaved. We consider the following counting functions that have an
effective support on an interval of size $\Delta E$ about $E$:
\begin{eqnarray}
  \hat{N}(E'; E)
   & \equiv &
   w(E'{-}E) N(E') \\
   \hat{N}_{\rm sc}^{\#}(E'; E)
   & \equiv &
   w(E'{-}E) N_{\rm sc}^{\#}(E') \, .
\end{eqnarray}
The functions $\hat{N}$ and $\hat{N}_{\rm sc}^{\#}$ are sharp
staircases, since the multiplication with~$w$ preserves the
sharpness of the stairs (it is not a convolution!). We now explicitly
construct $\delta^{\rm (2)}(E)$ as:
\begin{eqnarray}
  \delta^{\rm (2)}(E)
  & = &
  \int_{-\infty}^{+\infty} {\rm d}E' \,
  \left| \hat{N}(E'; E) - \hat{N}_{\rm sc}^{\#}(E'; E) \right|^2
  \nonumber \\
  & = &
  \int_{-\infty}^{+\infty} {\rm d}E' \,
  \left| N(E') - N_{\rm sc}^{\#}(E') \right|^2
   w^2 (E'-E).
   \label{eq:delta2int}
\end{eqnarray}
To obtain $\delta^{\rm (2)}_{\rm smooth}$ we need to smooth $N$ and
$N_{\rm sc}^{\#}$ over a scale of order of one mean spacing. 
One can, e.g., replace the sharp stairs by error functions. As for
$N_{\rm sc}^{\#}$, we prefer to simply replace it with the original
$N_{\rm sc}$, which we assume to be smooth over one mean spacing. That
is, we suppose that $N_{\rm sc}$ contains periodic orbits up to
Heisenberg time. Hence,
\begin{equation}
  \delta^{\rm (2)}_{\rm smooth}(E) =
  \int_{-\infty}^{+\infty} {\rm d}E' \,
  \left| N^{\rm smooth}(E') - N_{\rm sc}(E') \right|^2
   w^2 (E'-E)  .
  \label{eq:delta2smoothint}
\end{equation}
A comment is in order here. Strictly speaking, to satisfy
(\ref{eq:delta2epsilon2}) we need to apply the same smoothing to $N$
and to $N_{\rm sc}^{\#}$, and in general $N_{\rm sc}^{\#, {\rm smooth}} \ne
N_{\rm sc}$, but there are differences of order 1 between the two
functions. However, since our goal is to determine whether the
semiclassical error remains finite or diverges in the semiclassical
limit $\hbar \rightarrow 0$, we disregard such inaccuracies of order
1. If a more accurate error measure is needed, then more care should be
practised in this and in the following steps.

Applying Parseval's theorem to (\ref{eq:delta2smoothint}) we get:
\begin{equation}
  \delta^{\rm (2)}_{\rm smooth}(E) =
  \frac{1}{\hbar} \int_{-\infty}^{+\infty} {\rm d}t \,
  \left| \hat{D}(t; E) - \hat{D}_{\rm sc}(t; E) \right|^2
  \label{eq:delta2ls}
\end{equation}
where
\begin{eqnarray}
  \hat{D}(t; E)
  & \equiv &
  \frac{1}{\sqrt{2 \pi}} \int_{-\infty}^{+\infty} {\rm d}E' \,
  \hat{N}^{\rm smooth}(E'; E) \exp(i E' t / \hbar) \\
  \hat{D}_{\rm sc}(t; E)
  & \equiv &
  \frac{1}{\sqrt{2 \pi}} \int_{-\infty}^{+\infty} {\rm d}E' \,
  \hat{N}_{\rm sc}(E'; E) \exp(i E' t / \hbar).
\end{eqnarray}
We shall refer to $\hat{D}$, $\hat{D}_{\rm sc}$ as the (regularized)
quantal and semiclassical time spectra, respectively. These functions are 
the analogs of the length spectra $D(l; k)$ used in section 
\ref{sec:sc-analysis} for the billiard problem. The analogue becomes clear
by invoking the Gutzwiller trace formula and expressing the
semiclassical counting function as a mean part plus a sum over
periodic orbits. We have:
\begin{equation}
  N_{\rm sc}(E) =
  \bar{N}(E) +
  \sum_{\rm po} \frac{\hbar A_j(E)}{T_j(E)}
  \sin [ S_j(E) / \hbar - \nu_j \pi / 2 ] \ ,
  \label{eq:gtf-n}
\end{equation}
where $A_j = T_j / (\pi \hbar r_j \sqrt{|\det(I - M_j)|})$ is the
semiclassical amplitude of the j'th periodic orbit, and $T_j, S_j,
\nu_j, M_j, r_j$ are its period, action, Maslov index, monodromy matrix and
repetition index, respectively. Then, the corresponding time spectrum
reads:
\begin{eqnarray}
  \hat{D}_{\rm sc}(t; E)
  & \approx &
  \bar{D}(t; E) 
  \label{eq:scltspt} \\
  & + &
  \frac {1}{2i}\sum_{\rm po}
  \frac{\hbar A_j(E)}{T_j(E)}
  \left\{{\rm e}^{(i/\hbar)[E t + S_j(E)]} 
    \hat{w}([t + T_j(E)]/\hbar) - 
  \right. \nonumber \\
  & &
  \left. \; \; \; \; \; \; \; \; \; \; \; \; \; \; \; \; \; \;
         \; \; \; \; \;
  {\rm e}^{(i/\hbar)[E t - S_j(E)]} 
    \hat{w}([t - T_j(E)]/\hbar)
  \right\} .
  \nonumber
\end{eqnarray}
In the above, the Fourier transform of $w$ is denoted by $\hat{w}$. It
is a localized function of $t$ whose width is $\Delta t \approx \hbar
/ \Delta E$. The sum over the periodic orbits in $D_{\rm sc}$
therefore produces sharp peaks centered at times that correspond to
the periods $T_j$. The term $\bar{D}$ corresponds to the smooth part 
and is sharply peaked
near $t=0$. To obtain (\ref{eq:scltspt}) we expanded the actions near
$E$ to first order: $S_j(E') \approx S_j(E) + (E'-E)T_j(E)$. We note
in passing, that this approximate expansion of $S_j$ can be avoided
altogether if one performs the Fourier transform over $\hbar^{-1}$
rather than over the energy. This way, an action spectrum will emerge,
but also here the action resolution will be finite, because the range
of $\hbar^{-1}$ should be limited to the range where $\bar d(E;\hbar)$
is approximately constant. It turns out therefore, that the two
approaches are essentially equivalent, and for billiards they are
identical.

The manipulations done thus far were purely formal, and did not
manifestly circumvent the difficult task of evaluating $\delta^{\rm
(2)}_{\rm smooth}$. However, the introduction of the time spectra and
the formula (\ref{eq:delta2ls}) put us in a better position than
the original expression (\ref{eq:delta2int}). The advantages of using
the time spectra in the present context are the following:
\begin{itemize}

\item The semiclassical time spectrum $\hat{D}_{\rm sc}(t;E)$ is absolutely
  convergent for all times (as long as the window function~$w$ is well
  behaved, e.g.\ it is a Gaussian). This statement is correct even if
  the sum (\ref{eq:scltspt}) extends over the entire set of periodic
  orbits! This is in contrast with the trace formula expression for
  $N_{\rm sc}$ (and therefore $\hat{N}_{\rm sc}$) which is absolutely
  divergent if all of the periodic orbits are included.

\item Time scale separation: As we noted above, the time spectrum
  is  peaked at times that correspond to periods of the classical
  periodic orbits. This allows us to distinguish between various
  qualitatively different types of contributions to
  $\delta^{\rm (2)}_{\rm smooth}$.

\end{itemize}

We shall now pursue the separation of the time scales in detail. We
first note, that due to $\hat{N}, \hat{N}_{\rm sc}$ being real, there
is a $t \leftrightarrow (-t)$ symmetry in (\ref{eq:delta2ls}), and
therefore the time integration can be restricted to the limits 
$0$ to $+\infty$: $\delta^{\rm (2)} = 
(1/\hbar) \int_{-\infty}^{+\infty} \cdots =
(2/\hbar) \int_0^\infty \cdots$. We now divide the time axis 
into four intervals :
\begin{description}

\item{$\bf 0 \leq t \leq \Delta t$}: The shortest time scale in our
  problem is $\Delta t = \hbar / \Delta E$. The contributions to this
  time interval are due to the differences between the exact and the
  semiclassical {\em mean} densities of states. This is an important
  observation, since it allows us to distinguish between the two
  sources of semiclassical error --- the error that emerges from the
  mean densities and the error that originates from the fluctuating
  part (periodic orbits). Since we are interested only in the
  semiclassical error that results from the fluctuating part of the
  spectral density, we shall ignore this regime in the following.
  
\item[$\bf \Delta t \leq t \leq t_{\rm erg}$]: This is the
  non--universal regime \cite{Ber89}, in which periodic orbits are
  still sparse, and cannot be characterized statistically. The
  ``ergodic'' time scale $t_{\rm erg}$ is purely classical and is
  independent of $\hbar$.
  
\item{$\bf t_{\rm erg} \leq t \leq t_{\rm H} $}: In this time regime
  periodic orbits are already in the universal regime and are dense
  enough to justify a statistical approach to their proliferation and
  stability. The upper limit of this interval is the Heisenberg time
  $t_{\rm H} = h\bar{d}(E)$, which is the time that is needed to
  resolve the quantum (discrete) nature of a wavepacket with energy
  concentrated near $E$. The Heisenberg time is ``quantal'' in the
  sense that it is dependent of $\hbar$: $t_{\rm H} = {\cal
    O}(\hbar^{1-d})$.
  
\item{$\bf t_{\rm H} \leq t < \infty$}: This is the regime of ``long''
  orbits which is effectively truncated from the integration as a
  result of the introduction of a smoothing of the quantal and
  semiclassical counting functions, with a smoothing scale of the
  order of a mean level spacing.

\end{description}
Dividing the integral (\ref{eq:delta2ls}) according to the above time
intervals, we can rewrite $\delta^{\rm (2)}_{\rm smooth}$:
\begin{eqnarray}
  \delta^{\rm (2)}_{\rm smooth}(E)
  & = &
  \left( \int_{\Delta t}^{t_{\rm erg}} +
  \int_{t_{\rm erg}}^{t_{\rm H}} + \int_{t_{\rm H}}^{\infty} \right)
  \frac{2 {\rm d}t}{\hbar} \,
  \left| \hat{D}(t; E) - \hat{D}_{\rm sc}(t; E) \right|^2
  \nonumber \\
  & \equiv &
  \delta^{\rm (2)}_{\rm short} +
  \delta^{\rm (2)}_{\rm m} +
  \delta^{\rm (2)}_{\rm long} \; .
  \label{eq:delta2split}
\end{eqnarray}
As explained above, $\delta^{\rm (2)}_{\rm long}$ can be ignored due
to smoothing on the scale of a mean level spacing. The integral
$\delta^{\rm (2)}_{\rm short}$ is to be neglected for the following
reason.  The integral extends over a time interval which is finite and
independent of $\hbar$, and therefore it contains a fixed number of
periodic orbits contributions. The semiclassical approximation
provides, for each individual contribution, the leading order in
$\hbar$, and therefore \cite{AM77} we should expect:
\begin{equation}
  \delta^{\rm (2)}_{\rm short} \longrightarrow 0 \; \; \;
  \mbox{as } \; \hbar \longrightarrow 0.
  \label{eq:delta2-short}
\end{equation}
Our purpose is to check whether the semiclassical error
is finite or divergent as $\hbar \longrightarrow 0$, and to study whether
the rate of divergence depends on dimensionality. Equation
(\ref{eq:delta2-short}) implies that $\delta^{\rm (2)}_{\rm short}$
cannot affect $\delta^{\rm (2)}$ in the semiclassical limit and we
shall neglect it in the following.

We remain with:
\begin{equation}
  \delta^{\rm (2)}_{\rm smooth} 
  \approx 
  \delta^{\rm (2)}_{\rm m} \: ,
\end{equation}
which will be our object of interest from now on.

The fact that $t_{\rm H}$ is extremely large on the classical scale
renders the calculation of all the periodic orbits with periods less
than $t_{\rm H}$ an impossible task. However, sums over periodic
orbits when the period is longer than $t_{\rm erg}$ tend to meaningful
limits, and hence, we would like to recast the expression for
$\delta^{\rm (2)}_{\rm m}$ in the following way.  Write $\delta^{\rm
(2)}_{\rm m}$ as:
\begin{eqnarray}
  \delta^{\rm (2)}_{\rm m}
  & = &
  \frac{2}{\hbar} \int_{t_{\rm erg}}^{t_{\rm H}} {\rm d}t \,
  \left\langle \left| \hat{D}(t) - \hat{D}_{\rm sc}(t) \right|^2
  \right\rangle_{t} \\
  & = &
  \frac{2}{\hbar} \int_{t_{\rm erg}}^{t_{\rm H}} {\rm d}t \,
  \left\langle \left| \hat{D}(t) \right|^2 \right\rangle_{t}
  \times \left[
    \frac{\left\langle \left| \hat{D}(t) - \hat{D}_{\rm sc}(t)
        \right|^2 \right\rangle_{t}}
    {\left\langle \left| \hat{D}(t) \right|^2 \right\rangle_{t}}
  \right]
  \label{eq:ct} \\
  & \equiv &
  \frac{2}{\hbar} \int_{t_{\rm erg}}^{t_{\rm H}} {\rm d}t \,
  \left\langle \left| \hat{D}(t) \right|^2
  \right\rangle_{t} \times C(t) \\
  & = &
  \int_{t_{\rm erg}}^{t_{\rm H}} \mbox{\large envelope} \times
  \mbox{\large correlation} \nonumber \,
\end{eqnarray}
where the parametric dependence on $E$ was omitted for brevity. The
smoothing over $t$ is explicitly indicated to emphasize that one may
use a statistical interpretation for the terms of the integrand. This
is so because in this domain, the density of periodic orbits is so
large, that within a time interval of width $\hbar/\Delta E$ there are
exponentially many orbits whose contributions are averaged due to the
finite resolution.

We note now that we can use the following relation between the time
spectrum and the spectral form factor $K(\tau)$:
\begin{equation}
  \frac{\left\langle \left| \hat{D}(t) \right|^2
    \right\rangle_{t}}{\hbar}
  \, {\rm d}t =
  \frac{K(\tau)}{4 \pi^2 \tau^2} \, {\rm d}\tau
  \label{eq:dtktau}
\end{equation}
where $\tau \equiv t/t_{\rm H}$ is the scaled time. The above form
factor is smoothed according to the window function ~$w$. Hence:
\begin{equation}
  \delta^{\rm (2)}_{\rm smooth} \approx \frac{1}{2 \pi^2}
  \int_{\tau_{\rm erg}}^{1} {\rm d}\tau \,
  \frac{K(\tau) C(\tau)}{\tau^2} \, .
\end{equation}
For generic chaotic systems we expect that $K(\tau)$ agrees with the
results of RMT in the universal regime $\tau > \tau_{\rm erg}$
\cite{BGS84,Ber85,Ber89}. Therefore
\begin{equation}
  \tau \leq K(\tau) \leq g \tau \;\;\;
  \mbox{for } \tau_{\rm erg} < \tau \leq 1 \: ,
\end{equation}
where $g=1$ for systems which violate time reversal symmetry, and
$g=2$ if time reversal symmetry is respected. This implies that the
evaluation of $\delta^{\rm (2)}_{\rm smooth}$ reduces to
\begin{equation}
  \delta^{\rm (2)}_{\rm smooth} \approx \frac{g}{2 \pi^2}
  \int_{\tau_{\rm erg}}^{1} {\rm d}\tau \, \frac{C(\tau)}{\tau} \: ,
  \label{eq:delta2final}
\end{equation}
where we took the upper bound $g \tau$ for $K(\tau)$. 
The dependence on $\hbar$ in this expression comes from
the lower integration limit which is proportional to $\hbar^{d-1}$
as well as from the implicit dependence of the function $C$ on $\hbar$.

Formula (\ref{eq:delta2final}) is our main theoretical
result. However, we do not know how to evaluate the correlation
function $C(\tau)$ from first principles. The knowledge of the $\hbar
$ corrections to each of the terms in the semiclassical time spectrum
is not sufficient since the resulting series which ought to be summed
is not absolutely convergent. Therefore we have to recourse to a numerical
analysis, which will be described in the next section. The numerical
approach requires one further approximation, which is imposed by the
fact that the number of periodic orbits with $t<t_{\rm H}$ is
prohibitively large. We had to limit the data base of periodic orbits
to the domain $t< t_{\rm cpu}$ with $t_{\rm erg} \ll t_{\rm cpu}
\ll t_{\rm H}$. The time $t_{\rm cpu}$ has no physical origin, 
it represents only the limits of our computational
resources. Using the available numerical data we were able to compute
$C(t)$ numerically for all $ t_{\rm erg}<t<t_{\rm cpu}$ and we then
{\em extrapolated} it to the entire domain of interest. We consider
this extrapolation procedure to be the main source of
uncertainty. However, since the extrapolation is carried out in the
{\em universal} regime, it should be valid if there are no other time
scales between $t_{\rm erg}$ and $t_{\rm H}$.

\subsection{Numerical results}
\label{subsec:sinai-numerics}
%
We used the formalism and definitions presented above to check the
accuracy of the semiclassical spectra of the 2D and 3D Sinai
billiards. The most important ingredient in this numerical study is
that we could apply the {\em same} analysis to the two systems, and by
comparing them to give a reliable answer to the main question posed in
this work, namely, how does the semiclassical accuracy depend on
dimensionality.

The classical dynamics in billiards depends on the energy
(velocity) trivially, and therefore the relevant parameter 
is the length rather
than the period of the periodic orbits. Likewise,
the quantum wavenumbers $k_n$ are the relevant variables in the 
quantum description. From now on we shall
use the variables $(l, k)$ instead of $(t, E)$, and use ``length
spectra'' rather than ``time spectra''. The semiclassical limit is
obtained for $k \rightarrow \infty$ and ${\cal O}(\hbar)$ is
equivalent to ${\cal O}(k^{-1})$. Note also that for a billiard
$\bar{N}(k) \approx A k^{d}$ where $A$ is proportional to the
billiard's volume.

We start with the 2D Sinai billiard, which is the free space between a
square of edge $S$ and an inscribed disc of radius $R$, with $2R <
S$. Specifically, we use $S=1$ and $R=0.25$ and consider the quarter
desymmetrized billiard with Dirichlet boundary conditions for the 
quantum calculations. The quantal database consists of the lowest 
27645 eigenvalues in the range $0 < k < 1320$, with eigenstates which 
are either symmetric or antisymmetric with respect to reflection 
on the main diagonal. The classical database consists of the 
shortest 20273 periodic orbits (including time
reversal, reflection symmetries and repetitions) in the length range
$0 < l < 5$. For each orbit, the length, the stability determinant and
the reflection phase were recorded. The numerical work is based on 
the quantum spectra and on the classical periodic orbits which were 
computed by Schanz and Smilansky \cite{SS95,Sch96a} for the 2D billiard.

\begin{figure}[p]
  \begin{center}
    \leavevmode
    
    \psfig{figure=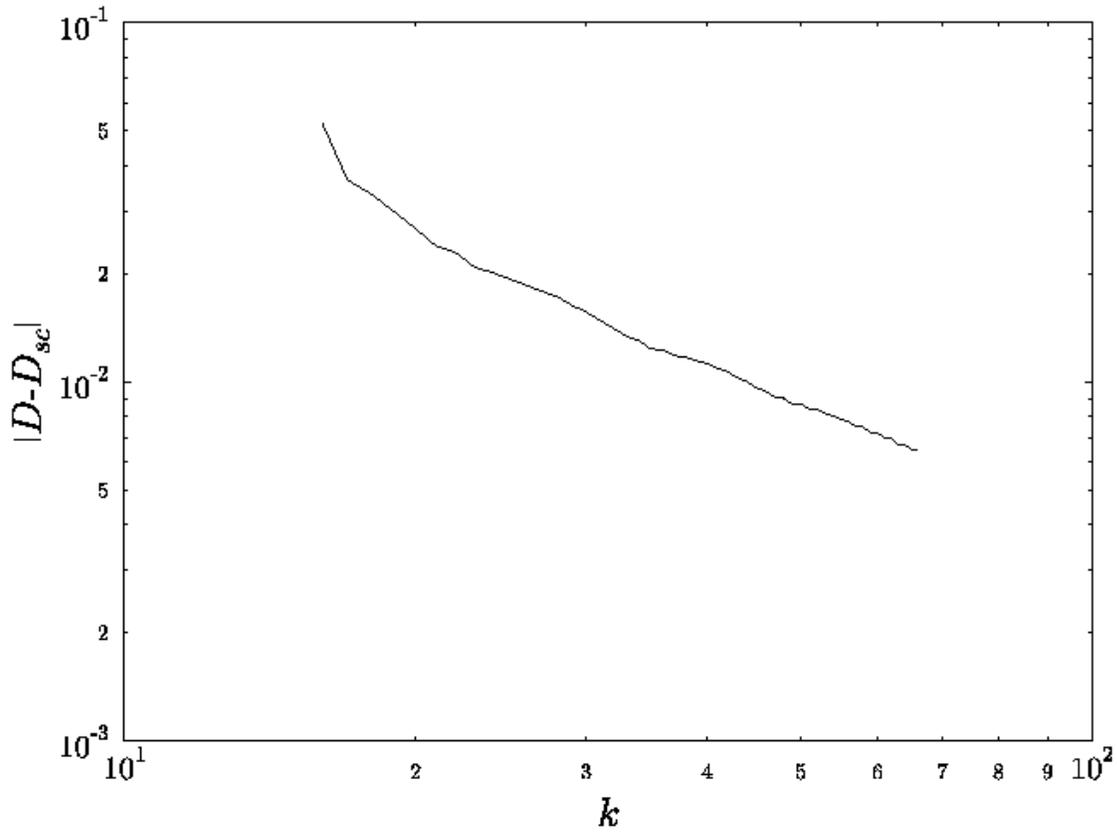,width=16cm}

    \caption{The absolute difference between the quantal and the
      semiclassical (Gutzwiller) length spectra for the 2D Sinai
      billiard at $l=0.5$. This length corresponds to the shortest
      unstable periodic orbit. The average log--log slope is about
      $-1.1$, indicating approximately $k^{-1}$ decay. The data were
      averaged with a Gaussian window.}

    \label{fig:sb2d-qm-sc}

  \end{center}
\end{figure}

We begin the numerical analysis by
demonstrating numerically the correctness of equation
(\ref{eq:delta2-short}). That is, that for each {\em individual}\/ 
contribution of a periodic orbit, 
the semiclassical error indeed vanishes in the
semiclassical limit. In figure \ref{fig:sb2d-qm-sc} we plot $|D -
D_{\rm sc}|$ for $l=0.5$ as a function of $k$. This length corresponds
to the shortest periodic orbit, that is, the one that runs along the
edge that connects the circle with the outer square. For $D_{\rm sc}$
we used the Gutzwiller trace formula. As is clearly seen from the
figure, the quantal--semiclassical difference indeed vanishes
(approximately as $k^{-1}$), in accordance with
(\ref{eq:delta2-short}). We emphasize again, that this behavior does
not imply that $\delta^{\rm (2)}$ vanishes in the semiclassical limit,
since the number of periodic orbits included depends on $k$. 
It implies {\em only} that
$\delta^{\rm (2)}_{\rm short}$ vanishes in the limit, since it
consists of a fixed and finite number of periodic orbit
contributions. We should also comment that penumbra
corrections to individual grazing orbits introduce errors which are of
order $k^{-\gamma}$ with $0< \gamma < 1$ \cite
{PSSU96,PSSU97}. However, since the definition of ``grazing'' is in
itself $k$ dependent, one can safely neglect penumbra corrections in
estimating the large $k$ behavior of $\delta^{\rm (2)}_{\rm short}$.

\begin{figure}[p]
  \begin{center}
    \leavevmode

    \psfig{figure=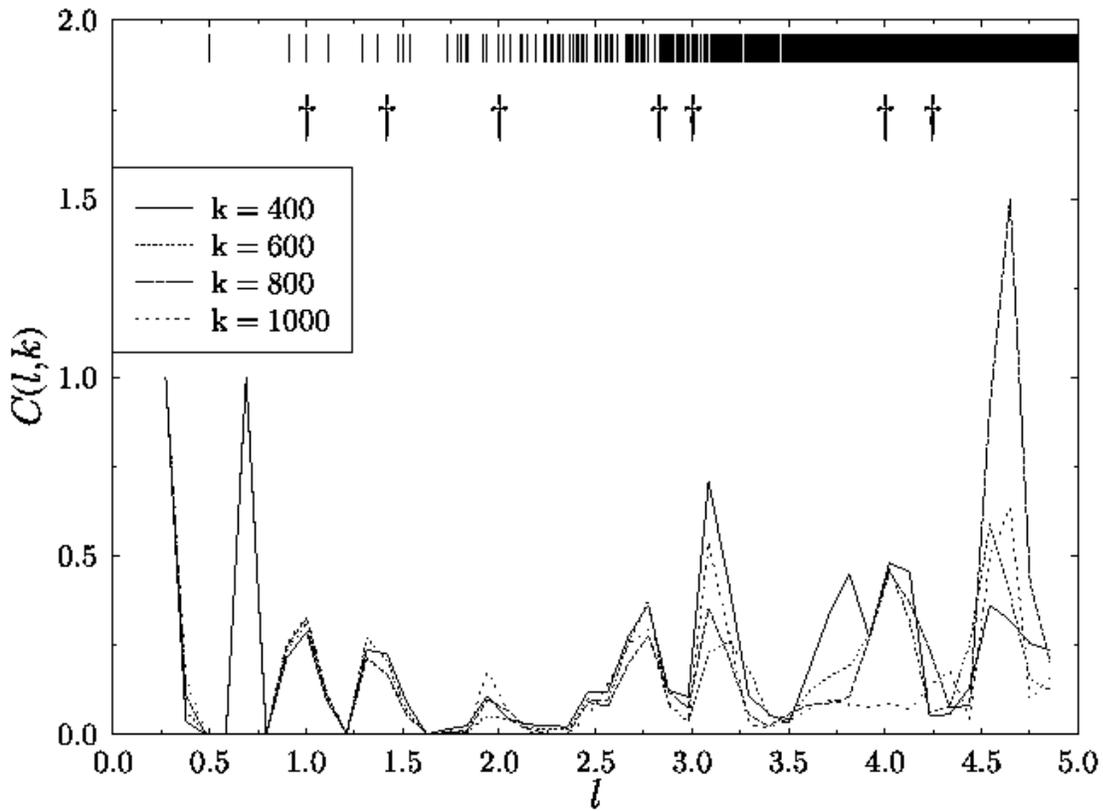,width=16cm}

    \caption{The functions $C(l; k)$ for quarter 2D Sinai billiard
      $S=1, R=0.25$ with Dirichlet boundary conditions. The window
      $w(k'-k)$ was taken to be a Gaussian with standard deviation
      $\sigma = 60$. We averaged $C(l; k)$ over $l$-intervals of
      $\approx 0.2$ in accordance with (\protect\ref{eq:ct}) to avoid
      sharp peaks due to small denominators. The averaging, however,
      is fine enough not to wash out all of the features of $C(l; k)$.
      The vertical bars indicate the locations of primitive periodic
      orbits, and the daggers indicate the locations of the
      bouncing--ball families.}

    \label{fig:dcxavg.2}

  \end{center}
\end{figure}

\begin{figure}[p]
  \begin{center}
    \leavevmode

    \psfig{figure=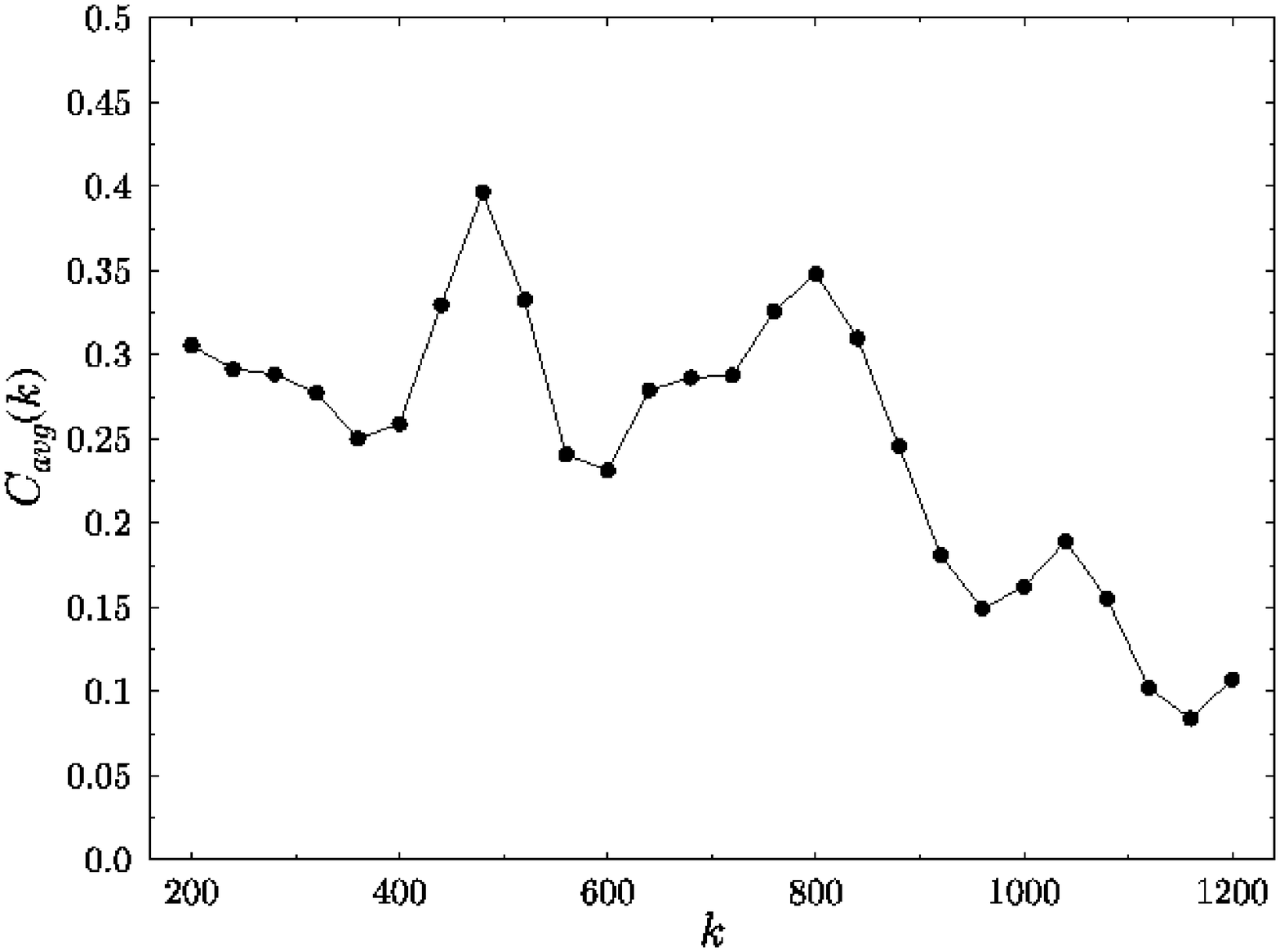,width=16cm}

    \caption{Averaging in $l$ of $C(l; k)$ for 2D Sinai billiard
      as a function of $k$.}

    \label{fig:dcavg.2}

  \end{center}
\end{figure}

We now turn to the main body of the analysis, which is the evaluation
of $\delta^{\rm (2)}_{\rm m}$ for the 2D Sinai billiard. Based on the
available data sets, we plot in figure \ref{fig:dcxavg.2} the function
$C(l; k)$ in the interval $2.5 < l < 5$ for various values of $k$. One
can observe, that as a function of $l$ the functions $C(l; k)$
fluctuate in the interval for which numerical data were available,
without exhibiting any systematic mean trend to increase or to
decrease. We therefore approximate $C(l; k)$ by
\begin{equation}
  C(l; k) \approx {\rm const} \cdot f(k) \equiv C_{\rm avg}(k).
\end{equation}
As mentioned above, we
extrapolate this formula in $l$ up to the Heisenberg length $L_{\rm H}
= 2 \pi \bar{d}(k)$ and using (\ref{eq:delta2final}) we obtain:
\begin{equation}
  \delta^{\rm (2), 2D}_{\rm smooth} = 
  \frac{C_{\rm avg}(k)}{2 \pi^2} \ln(L_{\rm H} /
  L_{\rm erg}) = C_{\rm avg}(k) \ {\cal O}(\ln k).
\end{equation}
The last equality is due to $L_{\rm H} = {\cal O}(k^{d-1})$. To
evaluate $C_{\rm avg}(k)$ we averaged $C(l; k)$ over the interval
$L_{\rm erg} = 3.5 < l < 5 = L_{\rm cpu}$ and the results are shown in
figure \ref{fig:dcavg.2}. We choose $L_{\rm erg} = 3.5$ because the
density of periodic orbits is large enough for this length (see
figure \ref{fig:dcxavg.2}) to expect universal behavior of the periodic
orbits. (For the Sinai billiard described by flow the approach to the
invariant measure is algebraic rather than exponential
\cite{FS95,DA96}, and thus one cannot have a well-defined $L_{\rm erg}$.
An any rate, the specific choice of $L_{\rm erg}$ did not affect the
results in any appreciable way.) Inspecting $C_{\rm avg}(k)$, it is
difficult to arrive at firm conclusions, since it seems to fluctuate
around a constant value up to $k \approx 900$ and then to decline. If
we approximate $C_{\rm avg}(k)$ by a constant, we get a
``pessimistic'' value of $\delta^{\rm (2)}$:
\begin{equation}
  \delta^{\rm (2), 2D}_{\rm smooth}(k) =
  {\cal O}(\ln k) = {\cal O}(\ln \hbar) \; \; \;
  \mbox{``pessimistic''}
  \label{eq:delta2-pessimistic}
\end{equation}
while if we assume that $C_{\rm avg}(k)$ decays as a power-law,
$C_{\rm avg}(k) = k^{-\beta}, \beta > 0$, then
\begin{equation}
  \delta^{\rm (2), 2D}_{\rm smooth}(k) = 
  {\cal O}(k^{-\beta} \ln k)
  \longrightarrow 0 \; \; \;
  \mbox{``optimistic''} \ .
\end{equation}
Collecting the two bounds we get:
\begin{equation}
  {\cal O}(k^{-\beta} \ln k) \leq
  \delta^{\rm (2), 2D}_{\rm smooth}(k) \leq
  {\cal O}(\ln k) \, .
  \label{eq:delta2-2d}
\end{equation}
Our estimates for the 2D Sinai billiard can be summarized by stating
that the semiclassical error diverges no worse than logarithmically
(meaning, very mildly). It may well be true that the semiclassical
error is constant or even vanishes in the semiclassical limit.  To
reach a conclusive answer one should invest exponentially larger
amount of numerical work.

\begin{figure}[p]
  \begin{center}
    \leavevmode
    
    \psfig{figure=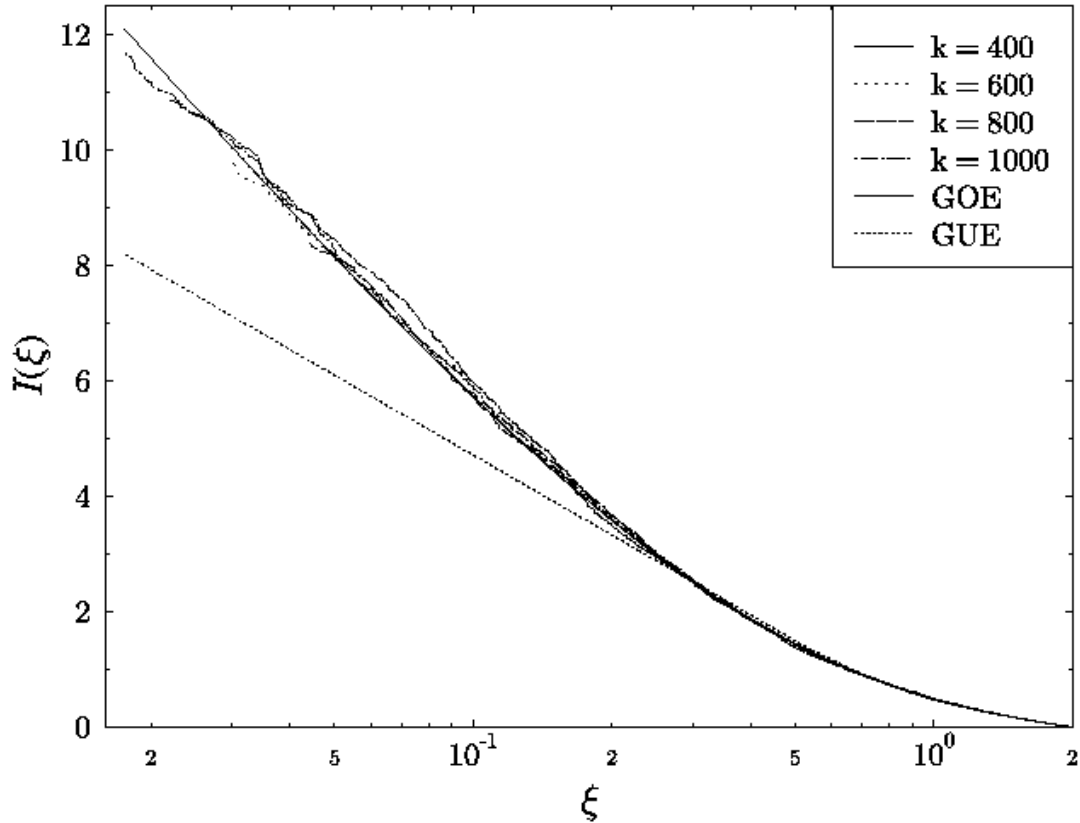,width=16cm}

    \caption{Verification of equation (\protect\ref{eq:dxkxi}) for the
      quarter 2D Sinai billiard. We plot $I(\xi) \equiv \int_{\xi}^{2}
      {\rm d}\xi' \ K(\xi')/\xi'\,^{2}$ and compare the quantum data
      with RMT. The minimal $\xi$ corresponds to $L_{\rm erg}=3.5$.
      The integration is done for smoothing, and we fix the {\em
        upper} limit to avoid biases due to non--universal regime.
      Note the logarithmic scale.}

    \label{fig:dikt.2}

  \end{center}
\end{figure}

\begin{figure}[p]
  \begin{center}
    \leavevmode
    
    \psfig{figure=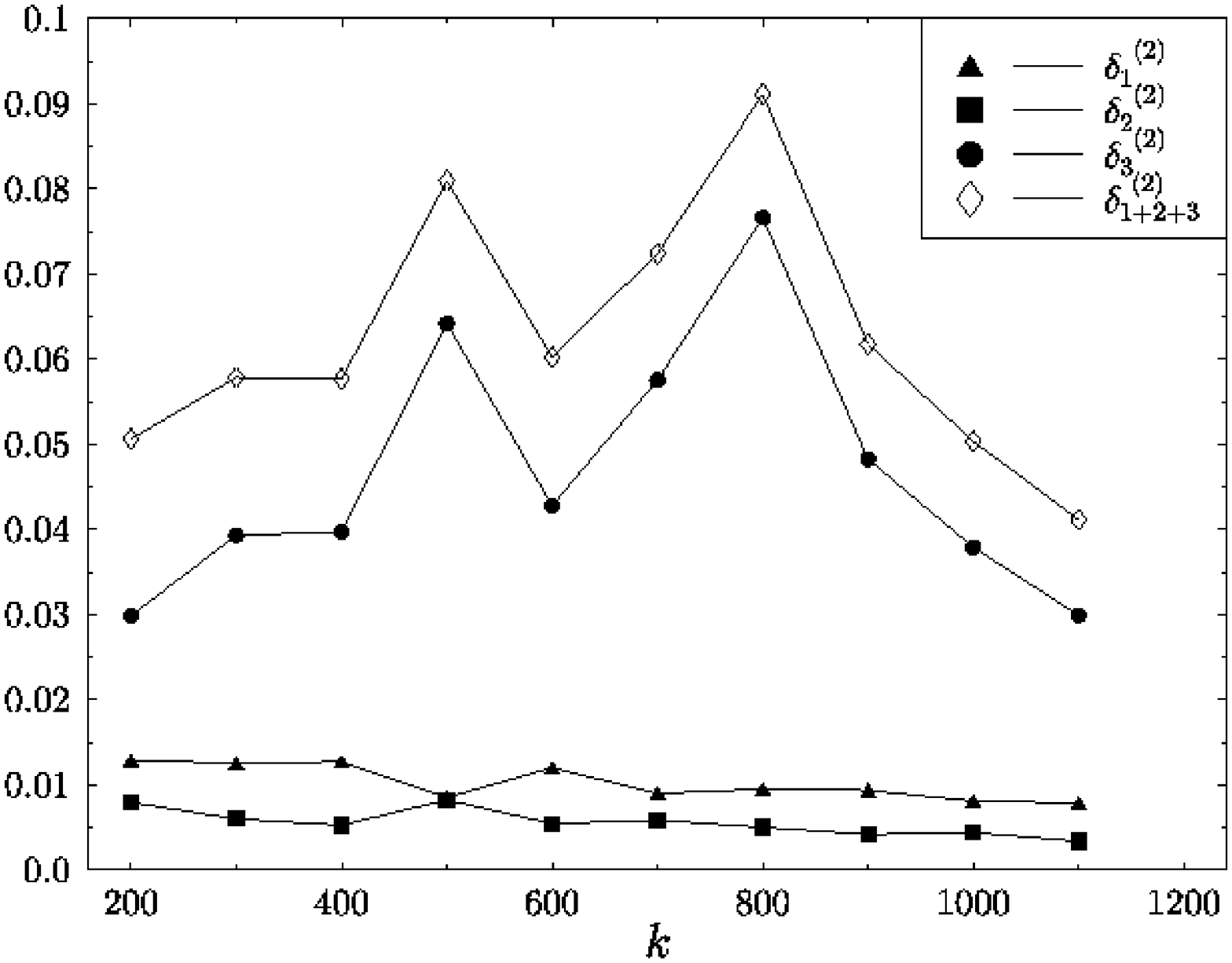,width=16cm}

    \caption{The numerical values of $\delta^{\rm (2)}_{\rm smooth}$
      for the quarter 2D Sinai billiard. We included also the
      contribution $\delta^{\rm (2)}_{\rm short}$ of the
      non--universal regime. The contributions from the time interval
      $t_{\rm erg} \leq t \leq t_{\rm cpu}$ are contained in
      $\delta^{\rm (2)}_{\rm m, cpu}$, and $\delta^{\rm (2)}_{\rm m,
        ext}$ is the extrapolated value for $t_{\rm cpu} \leq t \leq
      t_{\rm H}$ (refer to equation (\protect\ref{eq:delta2split}) and
      to the end of subsection
      \protect\ref{subsec:accuracy-measures}).}

    \label{fig:ddelta2.2}

  \end{center}
\end{figure}

There are a few comments in order here. Firstly, the quarter
desymmetrization of the 2D Sinai billiard does not exhaust its
symmetry group, and in fact, a reflection symmetry around the diagonal
of the square remains. This means, that the spectrum of the quarter 2D
Sinai billiard is composed of two independent spectra, which differ by
their parity with respect to the diagonal. If we assume that the
semiclassical deviations of the two spectra are not correlated, the
above measure is the sum of the two independent measures. It is
plausible to assume also that both spectra have roughly the same
semiclassical deviation, and thus $\delta^{\rm (2), 2D}_{\rm smooth}$
is twice the semiclassical deviation of each of the spectra. Secondly,
we recall that the 2D Sinai billiard contains ``bouncing--ball''
families of neutrally stable periodic orbits
\cite{Ber81,SSCL93,SS95}. We have subtracted their leading-order
contribution from $\hat{D}$ such that it includes (to leading order)
only contributions from generic, isolated and unstable periodic
orbits. This is done since we would like to deduce from the 2D Sinai
billiard on the 2D generic case in which the bouncing--balls are not
present. (In the Sinai billiard, which is concave, there are also
diffraction effects
\cite{PSSU96,PSSU97}, but we did not treat them here.) Thirdly, the
analogue of (\ref{eq:dtktau}) for billiards reads:
\begin{equation}
  \left\langle \left| \hat{D}(l) \right|^2 \right\rangle_{l} {\rm d}l =
  \frac{K(\xi)}{4 \pi^2 \xi^2} \ {\rm d}\xi
  \label{eq:dxkxi}
\end{equation}
when $\xi \equiv l/L_{\rm H}$. In figure \ref{fig:dikt.2} we
demonstrate the compliance of the form factor with RMT GOE using the
integrated version of the above relation, and taking into account the
presence of two independent spectra. Finally, it is interesting to
know the actual numerical values of $\delta^{\rm (2), 2D}_{\rm
smooth}(k)$ for the $k$ values that we considered. We carried out the
computation, and the results are presented in figure
\ref{fig:ddelta2.2}. One observes that for the entire
range we have $\delta^{\rm (2), 2D}_{\rm smooth}(k) \approx 0.1 \ll
1$, which is very encouraging from an ``engineering'' point of view.

We now turn to the analysis of the 3D Sinai billiard. We use the 
longest quantal spectrum ($R=0.2$, Dirichlet) 
and the classical periodic orbits with length $0 < l < 5$.

To treat the 3D Sinai billiard we have to somewhat modify the
formalism which was presented above. This is due to
the fact that in the 3D case the contributions of the various non--generic
bouncing--ball manifolds overwhelm the spectrum
\cite{PS95,Pri97}, and unlike the 2D case, it is difficult to 
explicitly eliminate their (leading--order) contributions 
(c.f.\ the discussion in section \ref{subsec:d-n}).
Since our goal is to give an indication of the
semiclassical error in generic systems, it is imperative to avoid 
this dominant and non-generic effect.

We shall use the mixed boundary conditions, which were discussed in section
\ref{subsec:mbc} and were shown to largely filter the bouncing--ball effects.
Specifically, we consider $\tilde{d}$ (c.f.\ (\ref{eq:gutz-mbc})) 
for our purposes. Let us construct the weighted counting function:
\begin{equation}
  \tilde{N}(k) 
  \equiv
  \int_{0}^{k} {\rm d}k' \  \tilde{d}(k') 
  =
  \sum_{n} v_n \Theta(k-k_n) \: .
\end{equation}
The function $\tilde{N}$ is a staircase with stairs of variable height
$v_n$. As explained above, its advantage over $N$ is that it is
semiclassically free of the bouncing balls (to leading order) and
corresponds only to the generic periodic orbits
\cite{SPSUS95}. Similarly, we construct from $\tilde{d}_{\rm sc}$
the function $\tilde{N}_{\rm sc}$. Having defined $\tilde{N},
\tilde{N}_{\rm sc}$, we proceed in analogy to the Dirichlet case. We form
from $\tilde{N}, \tilde{N}_{\rm sc}$ the functions $\hat{N},
\hat{N}_{\rm sc}$, respectively, by multiplication with a window
function $w(k'-k)$ and then construct the measure $\delta^{\rm (2)}$
as in (\ref{eq:delta2int}). The only difference is that the
normalization of $w$ must be modified to account for the ``velocities"
$v_n$ such as:
\begin{equation}
   \bar{d}^{-1}(k) \sum_n v_n^2 |w(k_n-k)|^2 = 1.
\end{equation}
The above considerations are meaningful provided the
``velocities'' $v_n$ are narrowly distributed around a well-defined
mean $v(k)$, and we consider a small enough $k$-interval, such that
$v(k)$ does not change appreciably within this interval. Otherwise,
$\delta^{\rm (2)}$ is greatly affected by the fluctuations of $v_n$
(which is undesired) and the meaning of the normalization is
questionable. We shall check this point numerically.

\begin{figure}[p]
  \begin{center}
    \vspace*{-0.8cm}
    \leavevmode
    \vspace*{-0.5cm}

    \centerline{\psfig{figure=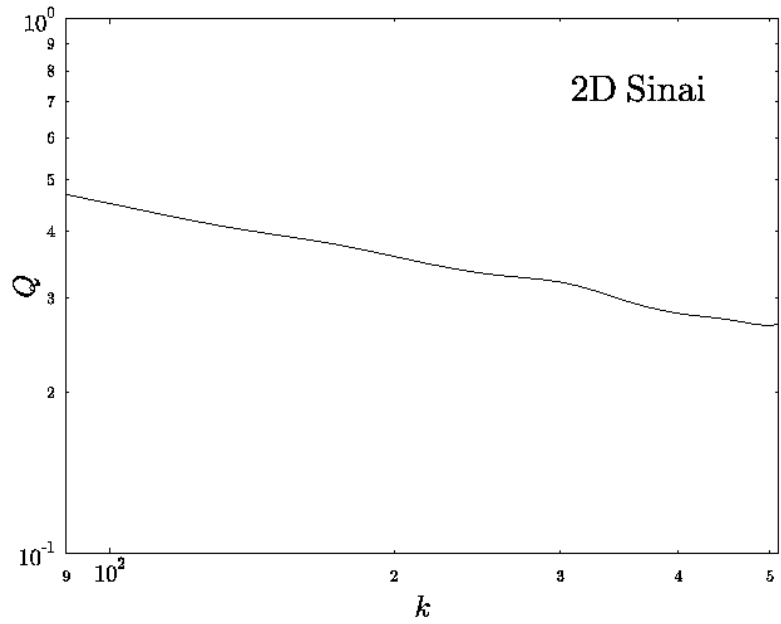,height=9cm}}
    \centerline{\psfig{figure=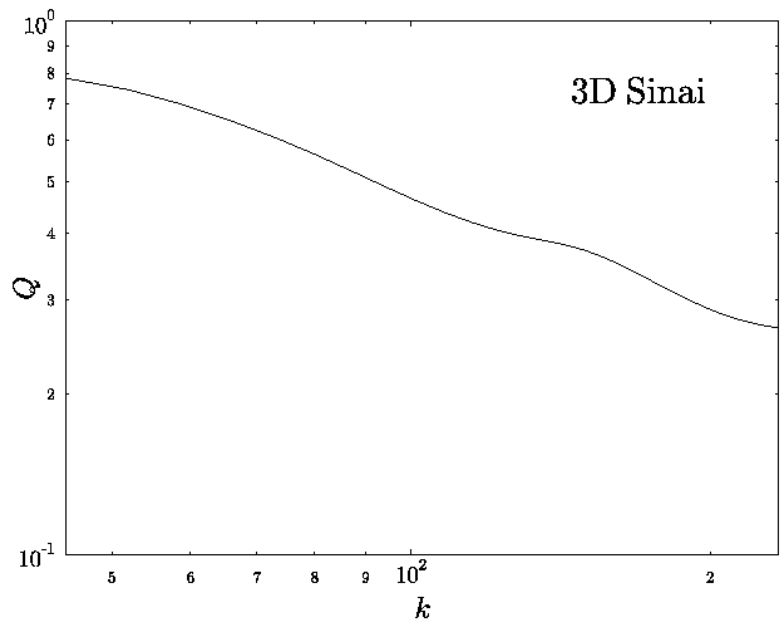,height=9cm}}

    \caption{Calculation of $Q \equiv \protect\sqrt{\langle
        v_n^2 \rangle - \langle v_n \rangle^2} / |\langle v_n
      \rangle|$ for quarter 2D Sinai billiard (up) and for the
      desymmetrized 3D Sinai billiard (down).}

    \label{fig:q}

  \end{center}
\end{figure}

\begin{figure}[p]
  \begin{center}
    \leavevmode 

    \psfig{figure=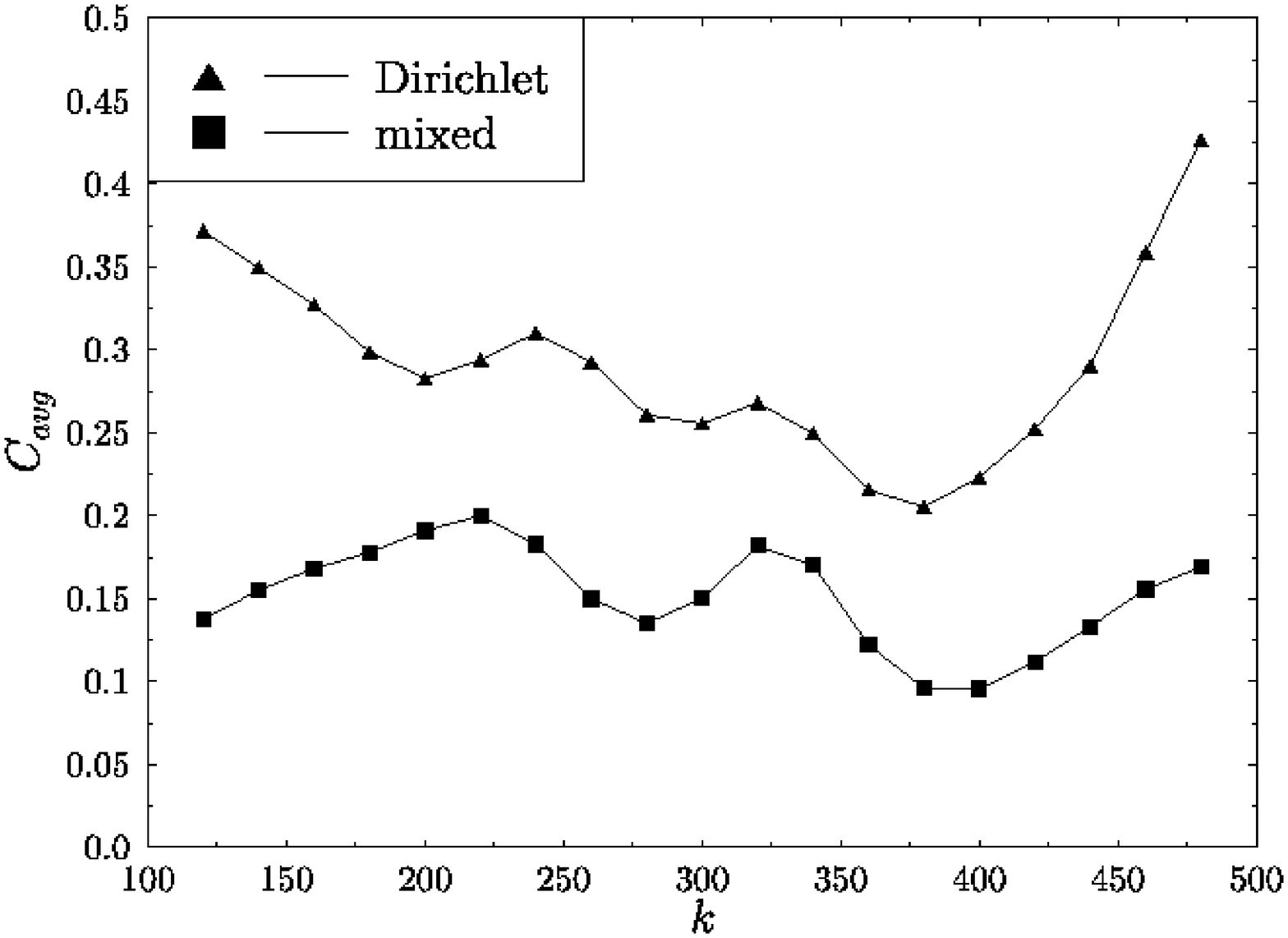,width=16cm}

    \caption{Comparison of $C_{\rm avg}(k)$ for Dirichlet
      and mixed boundary conditions for the quarter 2D Sinai billiard.
      We used a Gaussian window with $\sigma = 40$.}

    \label{fig:dmcavg.2}

  \end{center}
\end{figure}

To demonstrate the utility of the above construction using the mixed
boundary conditions, we return to the 2D case. We set $\kappa = 100
\pi$, and note that the spectrum at our disposal for the mixed case 
was confined to the interval $0 < k < 600$. First, we want
to examine the width of the distribution of the $v_n$'s. In figure
\ref{fig:q} we plot the ratio of the standard deviation of $v_n$ to
the mean, averaged over the $k$-axis using a Gaussian window. We use
the same window also in the calculations below. The observation is
that the distribution of $v_n$ is moderately narrow and the width
decreases algebraically as $k$ increases. This justifies the use of
the mixed boundary conditions as was discussed above. One also needs
to check the validity of (\ref{eq:dxkxi}), and indeed we found
compliance with GOE also for the mixed case (results not shown). We
next compare the functions $C(l;k)$ for both the Dirichlet and the
mixed boundary conditions. It turns out, that also in the mixed case
the functions $C(l; k)$ fluctuate in $l$ with no special
tendency (not shown). The averages $C_{\rm avg}(k)$ for the 
Dirichlet and mixed
cases are compared in figure \ref{fig:dmcavg.2}. The values in the
mixed case are systematically smaller than in the Dirichlet case which
is explained by the efficient filtering of tangent and close to
tangent orbits that are vulnerable to large diffraction corrections
\cite{PSSU96,PSSU97}. However, from $k = 250$ on the two graphs show
the same trends, and the values of $C_{\rm avg}$ in both cases are of
the same magnitude. Thus, the qualitative behavior of $\delta^{\rm
(2)}_{\rm smooth}$ is shown to be equivalent in the Dirichlet and
mixed cases, which gives us confidence in using $\delta^{\rm (2)}_{\rm
smooth}$ together with the mixed boundary conditions procedure.

\begin{figure}[p]
  \begin{center}
    \leavevmode

    \psfig{figure=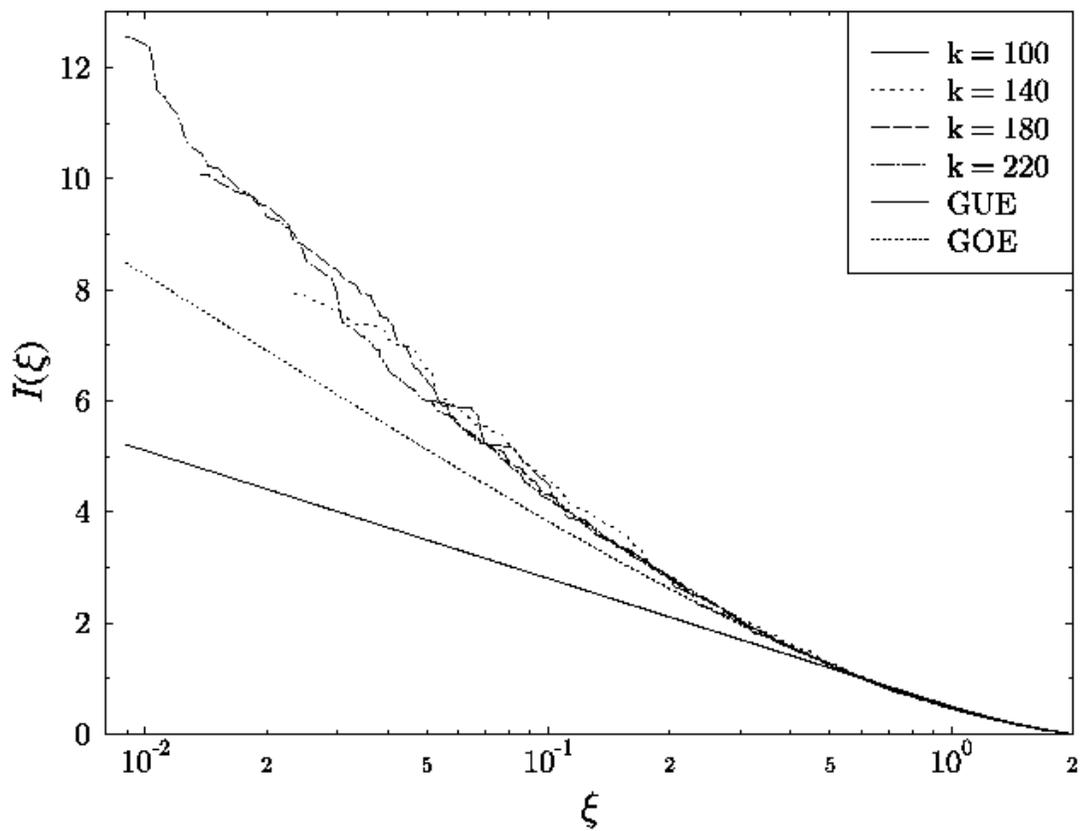,width=16cm}

    \caption{Check of equation (\protect\ref{eq:dxkxi}) for the
      desymmetrized 3D Sinai billiard. The minimal $\xi$ corresponds
      to $L_{\rm erg}=2.5$. The function $I(\xi)$ is defined as in
      figure \protect\ref{fig:dikt.2}. Note the logarithmic scale.}

    \label{fig:mikt.3}

  \end{center}
\end{figure}

\begin{figure}[p]
  \begin{center}
    \leavevmode

    \psfig{figure=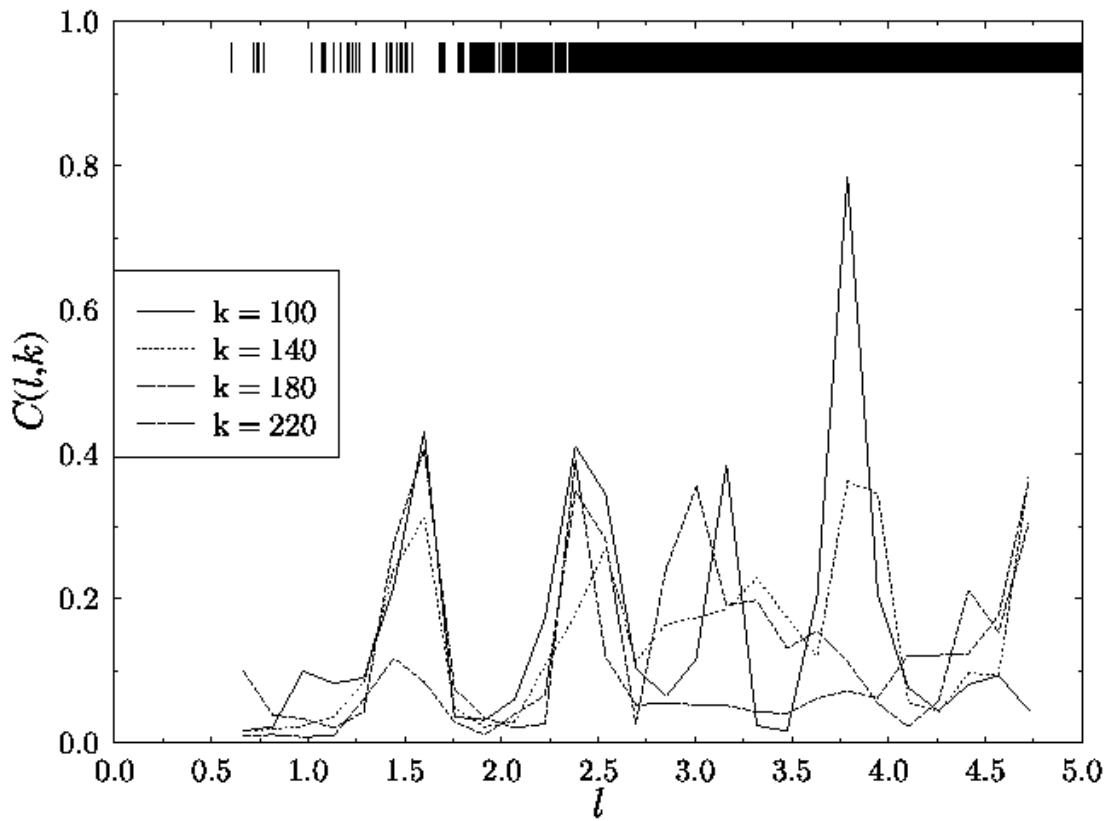,width=16cm}

    \caption{The functions $C(l; k)$ for desymmetrized 3D Sinai
      billiard $S=1, R=0.2$ with mixed boundary conditions. We took a
      Gaussian window with $\sigma=20$, and smoothed over
      $l$-intervals of $\approx 0.3$. The upper vertical bars indicate
      the locations of primitive periodic orbits.}

    \label{fig:mcxavg.3}

  \end{center}
\end{figure}

\begin{figure}[p]
  \begin{center}
    \leavevmode

    \psfig{figure=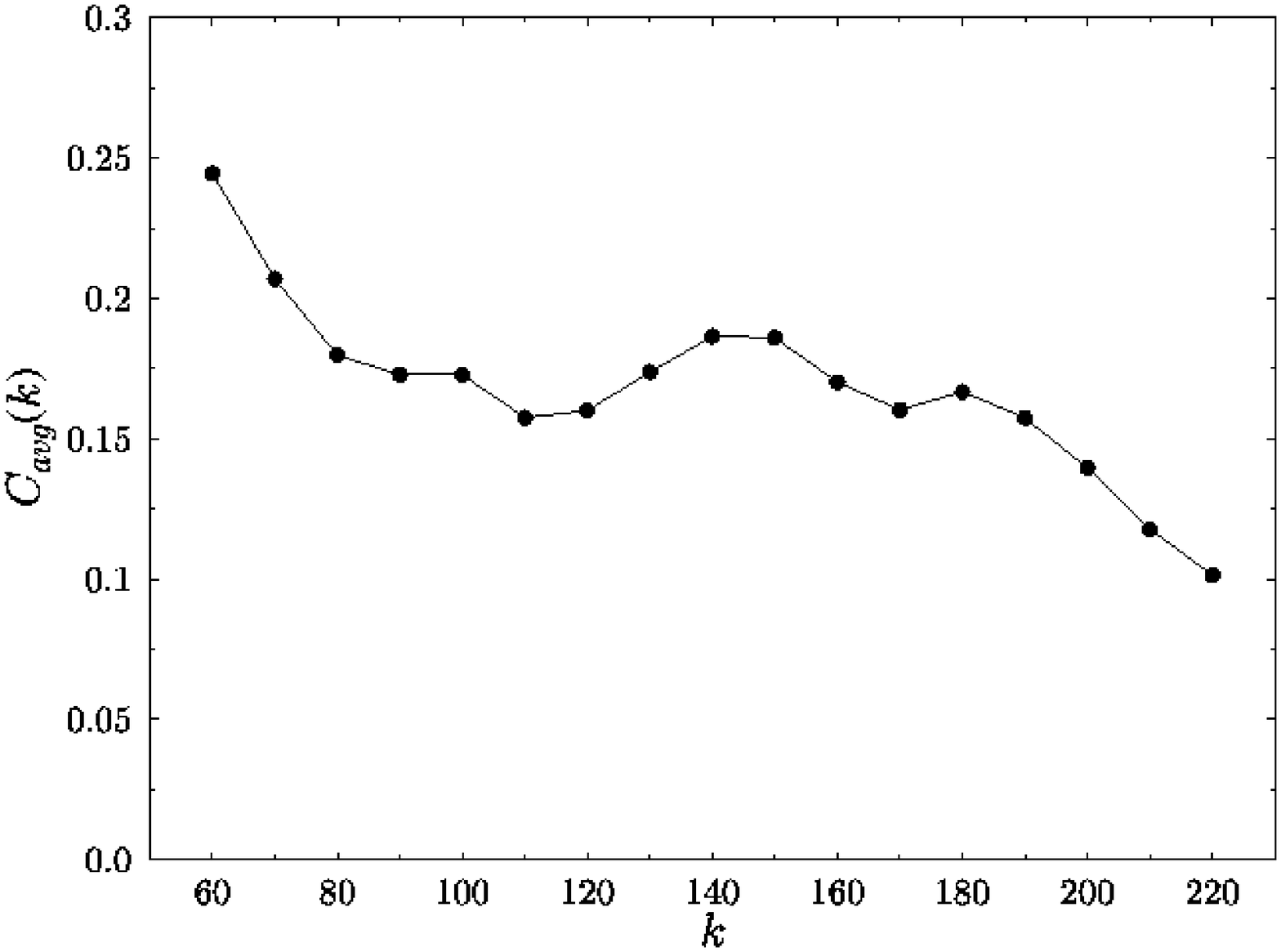,width=16cm}

    \caption{Averaging in $l$ of $C(l; k)$ for 3D Sinai billiard
      as a function of $k$. The averaging was performed in the
      interval $L_{\rm erg}=2.5 < l < 5 = L_{\rm cpu}$.}

    \label{fig:mcavg.3}

  \end{center}
\end{figure}

We finally applied the mixed boundary conditions procedure to compute
$\delta^{\rm (2)}_{\rm smooth}$ for the desymmetrized 3D Sinai with
$S=1, R=0.2$ and set $\kappa = 100$. We first verified that also in
the 3D case the velocities $v_n$ have a narrow distribution --- see
figure \ref{fig:q}. Next, we examined equation (\ref{eq:dxkxi}) using
quantal data, and discovered that there are deviations form GOE
(figure \ref{fig:mikt.3}). We have yet no satisfactory explanation of
these deviations, but we suspect that they are caused because the
ergodic limit is not yet reached for the length regime under
consideration due to the effects of the infinite horizon which are
more acute in 3D. Nevertheless, from observing the figure as well as
suggested by semiclassical arguments, it is plausible to assume that
$K(\xi) \propto \xi$ for small $\xi$. Hence, this deviation should not
have any qualitative effect on $\delta^{\rm (2)}$ according to
(\ref{eq:delta2final}). Similarly to the 2D case, the behavior of the
function $C(l; k)$ is fluctuative in $l$, with no special tendency
(figure \ref{fig:mcxavg.3}). If we average $C(l; k)$ over the
universal interval $L_{\rm erg} = 2.5 \leq l \leq L_{\rm cpu} = 5$, we
obtain $C_{\rm avg}(k)$ which is shown in figure
\ref{fig:mcavg.3}. The averages $C_{\rm avg}(k)$ are fluctuating with a
mild decrease in $k$, and therefore we can again conclude that
\begin{equation}
  {\cal O}(k^{-\beta} \ln k) \leq
  \delta^{\rm (2), 3D}_{\rm smooth}
  \leq {\cal O}(\ln k)
  \label{eq:delta2-3d}
\end{equation}
where the ``optimistic'' measure (leftmost term) corresponds to
$C_{\rm avg}(k) = {\cal O}(k^{-\beta}), \beta > 0$, and the
``pessimistic'' one (rightmost term) is due to $C_{\rm avg}(k) = {\rm
const}$. In other words, the error estimates (\ref{eq:delta2-2d},
\ref{eq:delta2-3d}) for the 2D and the 3D cases, respectively, are the
same, and in sharp contrast to the traditional error estimate
which predicts that the errors should be different by a factor ${\cal
O}(\hbar^{-1})$. On the basis of our numerical data, and in spite of
the uncertainties which were clearly delineated, we can safely rule out
the traditional error estimate.

Our main finding is that the upper bound on the semiclassical error
is a logarithmic divergence, both for a generic 2D and 3D systems
(equations (\ref{eq:delta2-2d}), (\ref{eq:delta2-3d})). In this
respect, there are a few points which deserve discussion.

To begin, we shall try to evaluate $\delta^{\rm (2)}_{\rm smooth}$
using the explicit expressions for the leading corrections to the
semiclassical counting function of a 2D generic billiard system, as
derived by Alonso and Gaspard \cite{AG93}:
\begin{equation}
  N(k) = \bar{N}(k) + \sum_{j} \frac{A_j}{L_j} \sin \left[ k L_j +
    \frac{Q_j}{k} + {\cal O}(1/k^2) \right]
\end{equation}
where $A_j$ are the standard semiclassical amplitudes,
$L_j$ are the lengths of periodic orbits and $Q_j$
are the $k$-independent amplitudes of the $1/k$ corrections. The
$Q_j$'s are explicitly given in \cite{AG93}. 
We ignored in the above equation the
case of odd Maslov indices. If we calculate from $N(k)$ the
corresponding length spectrum $\hat{D}(l; k)$ using a (normalized)
Gaussian window $w(k'-k) = (1 / \sqrt[4]{\pi \sigma^2}) \exp [
-(k'-k)^2/(2 \sigma^2) ]$, we obtain:
\begin{equation}
  \hat{D}(l; k)
  \approx
  \frac{i \sqrt{\sigma}}{2 \sqrt[4]{\pi}}
  \sum_j \frac{A_j}{L_j}
  \left[
    {\rm e}^{i k (l - L_j) - i \frac{Q_j}{k}}
    {\rm e}^{-(l - L_j)^2 \frac{\sigma^2}{2}} -
    {\rm e}^{i k (l + L_j) + i \frac{Q_j}{k}}
    {\rm e}^{-(l + L_j)^2 \frac{\sigma^2}{2}}
  \right] .
\end{equation}
In the above we regarded the phase ${\rm e}^{i Q_j / k}$ as slowly
varying.  The results of Alonso and Gaspard \cite{AG93} suggest that
the $Q_j$ are approximately proportional to the length of the
corresponding periodic orbits:
\begin{equation}
  Q_j \approx Q L_j .
  \label{eq:mj-lj}
\end{equation}
 We can therefore well--approximate $\hat{D}$ as:
\begin{equation}
  \hat{D}(l; k) \approx 
  \frac{i \sqrt{\sigma}}{2 \sqrt[4]{\pi}} {\rm e}^{-i Q l / k}
  \sum_j \frac{A_j}{L_j} \left[ \cdots \right] =
  {\rm e}^{- i Q l / k} \hat{D}_{\rm sc-GTF} \, ,
\end{equation}
where $\hat{D}_{\rm sc-GTF}$ is the length spectrum which corresponds
to the semiclassical Gutzwiller trace formula for the counting
function (without $1/k$ corrections). We are now in a position to
evaluate the semiclassical error, indeed:
\begin{eqnarray}
  \delta^{\rm (2)}_{\rm smooth}(k)
  & = &
  2 \int_{L_{\rm min}}^{L_{\rm H}} {\rm d}l
  \left| \hat{D}(l; k) - \hat{D}_{\rm sc-GTF}(l; k) \right|^2 =
  \nonumber \\
  & = &
  8 \int_{L_{\rm min}}^{L_{\rm H}} {\rm d}l \,
  \sin^2 \left( \frac{Q l}{2k} \right)
  \left| \hat{D}(l; k) \right|^2 \, .
\end{eqnarray}
If we use equation (\ref{eq:dxkxi}) and $K(l) \approx g l / L_{\rm
H}$ (which is valid for $l < L_{\rm H}$ for chaotic systems), we get:
\begin{equation}
  \delta^{\rm (2)}_{\rm smooth}(k) 
  \approx 
  \frac{2 g}{\pi^2} \int_{L_{\rm min}}^{L_{\rm H}}
  \frac{{\rm d}l}{l} \sin^2 \left( \frac{Q l}{2 k} \right) =
  \frac{2 g}{\pi^2} \int_{Q L_{\rm min} / (2k)}^{Q L_{\rm H} / (2 k)}
  {\rm d}t \frac{\sin^2 (t)}{t} \, .
  \label{eq:delta2-analytical1}
\end{equation}
For $k \rightarrow \infty$ we have that
\begin{equation}
  \int_{0}^{Q L_{\rm min} / (2k)} {\rm d}t
  \frac{\sin^2 (t)}{t} \approx
  \int_{0}^{Q L_{\rm min} / (2k)}
  {\rm d}t \cdot t = {\cal O}(1/k^2)
\end{equation}
which is negligible, hence we can replace the lower limit in
(\ref{eq:delta2-analytical1}) with 0:
\begin{equation}
  \delta^{\rm (2)}_{\rm smooth}(k) \approx
  \frac{2 g}{\pi^2} \int_{0}^{\frac{Q L_{\rm H}}{2 k}}
  {\rm d}t \frac{\sin^2 (t)}{t} \, .
  \label{eq:delta2-analytical2}
\end{equation}
This is the desired expression. The dimensionality enter in
$\delta^{\rm (2)}_{\rm smooth}(k)$ only through the power of $k$ in
$L_{\rm H}$.

Let us apply equation (\ref{eq:delta2-analytical2}) to the 2D and the 3D
cases. For 2D we have to leading order that $L_{\rm H} = A k$, where
$A$ is the billiard's area, thus,
\begin{equation}
  \delta^{\rm (2), 2D}_{\rm analytical}(k) \approx
  \frac{2 g}{\pi^2} \int_{0}^{Q A / 2}
  {\rm d}t \frac{\sin^2 (t)}{t} = {\rm const} = {\cal O}(k^0)
  \label{eq:delta2-analytical-2d}
\end{equation}
which means, that the semiclassical error in 2D billiards is of the
order of the mean spacing, and therefore the semiclassical trace
formula is (marginally) accurate and meaningful. This is compatible
with our numerical findings.

For 3D, the coefficients $Q_j$ were not obtained explicitly, but we
shall assume that they are still proportional to $L_j$
(equation~(\ref{eq:mj-lj})) and therefore that
(\ref{eq:delta2-analytical2}) holds.  For 3D billiards $L_{\rm H} = (V
/ \pi) k^2$ to leading order, where $V$ is the billiard's volume. Thus
the upper limit in (\ref{eq:delta2-analytical2}) is $Q V k / (2 \pi)$
which is large in the semiclassical limit. In this case, we can
replace $\sin^2 (t)$ with its mean value $1/2$ and the integrand
becomes essentially $1/t$ which results in:
\begin{equation}
  \delta^{\rm (2), 3D}_{\rm analytical}(k) = {\cal O}(\ln k) \, .
  \label{eq:delta2-analytical-3d}
\end{equation}
That is, in contrast to the 2D case, the semiclassical error diverges
logarithmically and the semiclassical trace formula becomes
meaningless as far as the prediction of individual levels is
concerned. This statement is compatible with our numerical results 
within the numerical dispersion. However, it relies heavily on the 
assumption that $Q_j \approx Q L_j$, for which we can 
offer no justification. We note in passing, 
that the logarithmic divergence persists also for $d > 3$.

Another interesting point relates to integrable systems. It can
happen, that for an integrable system it is either difficult or
impossible to express the Hamiltonian as an explicit function of the
action variables. In that case, we cannot assign to the levels other
quantum numbers than their ordinal number, and the semiclassical error
can be estimated using $\delta^{\rm (2)}$. However, since for
integrable systems $K(\tau) = 1$, we get that:
\begin{equation}
  \delta^{\rm (2), int}_{\rm smooth} 
  \approx \frac{1}{2 \pi^2}
  \int_{\tau_{\rm erg}}^{1} {\rm d}\tau \, 
  \frac{C(\tau)}{\tau^2}.
  \label{eq:delta2-integrable}
\end{equation}
Therefore, for deviations which are comparable to the chaotic cases,
$C(\tau) = {\cal O}(1)$, we get $\delta^{\rm (2), int}_{\rm smooth} =
{\cal O}(\hbar^{1-d})$ which is much larger than for the chaotic case
and diverges for $d \geq 1$.

The formula (\ref{eq:delta2final}) for the semiclassical error
contains semiclassical information in two respects. Obviously,
$C(\tau)$, which describes the difference between the quantal and the
semiclassical length spectra, contains semiclassical information. But
also the fact that the lower limit of the integral in
(\ref{eq:delta2final}) is finite is a consequence of semiclassical
analysis. If this lower limit is replaced by $0$, the integral
diverges for finite values of $\hbar$.
Therefore, the fact that the integral has a lower cutoff, or rather,
that $D$ is exactly $0$ below the shortest period, is a crucial
semiclassical ingredient in our analysis. 

Finally, we consider the case in which the semiclassical error is
estimated with no periodic orbits taken into account. That is, we want
to calculate $\langle | N(E) - \bar{N}(E) |^2 \rangle_{E}$ which is
the number variance $\Sigma^{2}(x)$ for the large argument $x = \Delta
E \, \bar{d}(E) \gg 1$. This implies $C(\tau) = 1$, and using
(\ref{eq:delta2final}) we get that $\delta^{\rm (2)}_{\rm smooth} =
g/(2 \pi^2) \ln (t_{\rm H}/t_{\rm erg})$, which in the semiclassical
limit becomes $g/(2 \pi^2) \ln (t_{\rm H}) = {\cal O}(\ln
\hbar)$. This result is fully consistent and compatible with previous
results for the asymptotic (saturation) value of the number variance
$\Sigma^{2}$ (see for instance \cite{Ber89,BS93,ABS94}). It implies
also that the pessimistic error bound (\ref{eq:delta2-pessimistic})
is of the same magnitude as if periodic orbits were not taken into
account at all. (Periodic orbits improve, however, quantitatively,
since in all cases we obtained $C_{\rm avg} < 1$.) Thus, if we assume
that periodic orbit contributions do not make $N_{sc}$ worse than
$\bar{N}$, then the pessimistic error bound ${\cal O} (\ln \hbar)$ is
the {\em maximal}\/ one in any dimension $d$. This excludes, in
particular, algebraic semiclassical errors, and thus refutes the
traditional estimate ${\cal O} (\hbar^{2-d})$.


\section{Semiclassical theory of spectral statistics}
\label{sec:sc-specstat}

In section \ref{sec:quantal-spectral-statistics} we studied several
quantal spectral statistics of the Sinai billiard and have shown that
they can be reproduced to a rather high accuracy by the predictions of
Random Matrix Theory (RMT). In the present section we would like to
study the spectral two-point correlation function in the semiclassical
approximation, and to show how the classical sum rules and
correlations of periodic orbits, which were defined in section
\ref{sec:classical-pos}, can be used to reconstruct, within the
semiclassical approximation, the predictions of RMT.

The starting point of the present discussion is the observation that
the semiclassical spectrum can be derived from a secular equation of
the form \cite{DS92,Bog92b}:
\begin{equation}
  Z_{\rm sc}(k)
  \equiv
  \det (I - S(k))
  = 0 \: ,
  \label {eq:sclz}
\end{equation}
where $S(k)$ is a (semiclassically) unitary matrix which depends
parametrically on the wavenumber $k$. In the semiclassical
approximation, the unitary operator $S(k)$ can be considered as the
quantum analogue of a classical Poincar\'e mapping, which for billiard
systems in $d$ dimensions, is the classical billiard bounce map. The
dimension $N(k)$ of the Hilbert space on which $S(k)$ acts, can be
expressed within the semiclassical approximation, in terms of the
phase-space volume of the Poincar\'e section ${\cal M}$ as follows:
\begin{equation}
  N(k)   
  =   
  \left[ {\cal N}(k) \right] \: , 
  \; \;
  {\cal N}(k) 
  =  
  \frac{\cal M}{(2 \pi \hbar)^{d-1}} \: ,
\end{equation}
where $[ \cdot ]$ stands for the integer value. For a billiard in two
dimensions ${\cal N}(k) = {\cal L} k / \pi $, where ${\cal L}$ is the
circumference of the billiard. In the case of the fully desymmetrized
3D Sinai billiard, for which we consider the sphere return map, ${\cal
  N}(k) = k^2 R^2 / 48$. The reason why we defined the smooth function
${\cal N}(k)$ will become clear in the sequel.

The eigenvalues of $S(k)$ are on the unit circle: $\{ \exp (i
\theta_l(k) ) \}_{l=1}^{N(k)} $. If for a certain $k$, one of the
eigenphases is an integer multiple of $2 \pi$, then equation (\ref
{eq:sclz}) is satisfied, and this value of $k$ belongs to the
spectrum. 
Because of this connection between the billiard spectrum on
the $k$ axis and the eigenphase spectrum on the unit circle, the
statistics of $k$-intervals can be read off the corresponding
statistics of the eigenphase intervals averaged over an appropriate
$k$-interval where $N(k)$ is constant \cite{DS92,Bog92b}. For this
reason, it is enough to study the eigenphase statistics, and if they
can be reproduced by the predictions of RMT for the relevant {\em
  circular}\/ ensemble, the wavenumber spectral statistics will
conform with the prediction of RMT for the corresponding {\em
  Gaussian}\/ ensemble.

The spectral density of the matrix $S(k)$ can be written as:
\begin{equation}
  d_{\rm qm}(\theta; k)
  \equiv
  \sum_{l=1}^{N(k)} \delta(\theta - \theta_l(k))
  =
  {N(k) \over 2 \pi} + {1 \over 2 \pi} \sum_{n=1}^{\infty}
    \left( {\rm e}^{-i n \theta}{\rm tr}S^n +
           {\rm e}^{ i n \theta}{\rm tr}(S^{\dagger})^n
  \right) \: .
  \label{eq:dtheta}
\end{equation}
The corresponding two-point correlation function is derived by
computing:
\begin{equation}
  C_2 (\eta)
  =
  {2\pi \over N} \left \langle \int_{0}^{2\pi}
    {{\rm d} \theta \over 2\pi} \
    d_{\rm qm}(\theta + {\eta \over 2}; k) \
    d_{\rm qm}(\theta - {\eta \over 2}; k) \right \rangle \: ,
  \label{eq:R_2}
\end{equation}
where $\left \langle \cdot \right \rangle$ denotes an average over a
$k$-interval where $N(k)$ takes the constant value $N$. The two-point
spectral form factor is defined as the Fourier coefficients of
$C_2(\eta)$, and by substituting (\ref{eq:dtheta}) in (\ref{eq:R_2}),
one finds that they are equal to ${1\over N} \left \langle | {\rm
    tr}S^n(k) |^2 \right \rangle$. RMT provides an explicit
expression:
\begin{equation}
  {1 \over N} \left \langle | {\rm tr}S^n(k) |^2
    \right \rangle_{\rm RMT}
  =
  K_{\beta} \left( {n \over N(k)} \right) \: ,
  \label{eq:Kdef}
\end{equation}
where $\beta$ is the standard ensemble label \cite{Boh89}. The most
important fact to be noticed is that $n$, the ``topological time", is
scaled by $N$, which plays here the r\^ole of the Heisenberg time.
For a Poisson ensemble:
\begin{equation}
  {1\over N} \left \langle | {\rm tr}S^n(k) |^2
    \right \rangle_{\rm Poisson}
  =
  1 \: .
  \label{eq:Kpoiss}
\end{equation}

From now on we shall be concerned with the Circular Orthogonal
Ensemble (COE: $\beta=1$). We define $\tau equiv n/N$. The function
$K_{\rm COE}(\tau)$ is a monotonically increasing function which
starts as $2 \tau$ near the origin, and bends towards its asymptotic
value $1$ in the vicinity of $\tau = 1$. For an explicit expression
consult, e.g.\ \cite{Smi94}.  Our aim is to show that the
semiclassical expression for ${1\over N} \left \langle | {\rm
    tr}S^n(k) |^2 \right \rangle$ reproduces this behaviour when the
correlations of periodic orbits are properly taken into account.

Recalling that the unitary matrix $S(k)$ is the quantum analogue of
the Poincar\'e map, one can express ${\rm tr}S^n(k) $ in terms of the
$n$-periodic orbits of the mapping. If the semiclassical mapping is
hyperbolic, and the billiard bounce map is considered, one gets
\cite{Smi94}:
\begin{equation}
  {\rm tr}S^n(k)
  \approx
  \sum_{j \in {\cal P}_n} {n_{p,j} \over
    | \det (I-M_j) |^{1 \over 2}}
    {\rm e}^{i k L_j} (-1)^{b_j} \: .
  \label{eq:scltrace}
\end{equation}
Here ${\cal P}_n$ is the set of all $n$-periodic orbits of the bounce
map, $n_{p,j}$ is the period of the primitive orbit of which the
$n$-periodic orbit is a multiple. The monodromy matrix is denoted
$M_j$, $L_j$ is the length, and $b_j$ is the number of bounces from
the boundaries (for a Dirichlet boundary condition). Note that when
the Poincar\'e section consists of a part of the boundary (as is the
case for the sphere return map in the 3D Sinai billiard), $b_j$ can be
different from $n$. Recalling the definition of the classical density
$d_{\rm cl}(l; n)$ (\ref{eq:classdens}) in subsection
\ref{subsec:pos-correlations}, and realizing that the pre-exponential
factors are just the $\tilde{A}_j$ coefficients
(\ref{eq:weightstilde}), we deduce that within the semiclassical
approximation,
\begin {equation}
  \int_{0}^{2\pi} {\rm e}^{i n \theta} \, d_{\rm qm}(\theta; k)
  \, {\rm d}\theta
  =
  {\rm tr}S^n(k)
  \approx
  \int_{0}^{\infty} {\rm e}^{i k l}\, d_{\rm cl}(l; n) \, {\rm d}l \: .
  \label{eq:duality}
\end{equation}
This equation is of fundamental importance, because it expresses the
duality between the quantum mechanical spectral density and the
classical length density via their Fourier transforms \cite{CPS98}.
Hence, the spectral form factors of the classical and the quantum
spectral distributions are also related by:
\begin{equation}
  {1 \over N} \left \langle | {\rm tr}S^n(k) |^2 \right \rangle
  =
  {1 \over N} \left \langle K_{\rm cl} (k; n) \right \rangle
  =
  {1 \over N} \left \langle \left| \sum_{j \in {\cal P}_n }
    \tilde{A}_j {\rm e}^{i k L_j} \right|^2 \right \rangle \: .
  \label{eq:ffactordual}
\end{equation}
We have shown already in section \ref{subsec:pos-correlations} that
the length spectrum as defined by the classical density
(\ref{eq:classdens}) contains non-trivial correlations. They appear on
a scale $\lambda(n;R)$ which is inversely proportional to the value of
$k$ where the classical correlation function approaches it asymptotic
value $g n$. What remains to be seen now is the extent by which the
semiclassical expression (\ref{eq:ffactordual}) reproduces the
expected universal scaling and the detailed functional dependence on
the scaled topological time $\tau = n / N$ as predicted by RMT.

The large $k$ limit of $K_{\rm cl}(k; n)$ was written explicitly in
(\ref{eq:kcl-asy}) and verified numerically:
\begin{equation}
  K_{\rm cl} (k \rightarrow \infty; n)
  \approx \langle n_p g_p \rangle \cdot U(n)
  \approx 2n \: .
\end{equation}
This limit corresponds to the limit ${n \over N(k)} \rightarrow 0$ so
that
\begin{equation}
  {1 \over N} \left \langle | {\rm tr}S^n(k) |^2  \right \rangle
  \approx {2n \over N} \: ,
  \label{eq:smalltau}
\end{equation}
which is identical to the behavior of $K_{\rm COE}(\tau)$ in the small
$\tau$ limit \cite{Boh89}. Therefore, the classical uniform coverage
of phase space guarantees the adherence to RMT in the limit $\tau
\rightarrow 0$. This result was derived originally by Berry in his
seminal paper \cite{Ber85}. It is the ``diagonal approximation'' which
can be used as long as the range of $k$ values is larger than
$\lambda(n;R)^{-1}$.  In other words, this approximation is valid on
the scale on which the classical length spectrum looks uncorrelated.
This observation shows that the domain of validity of the diagonal
approximation has nothing to do with the ``Ehrenfest time'', sometimes
also called the ``$\log \hbar$ time''. Rather, it depends on the
correlation length in the classical spectrum $\lambda(n; R)$, as
displayed by the classical form factor.

Given the classical correlation function, $K_{\rm cl}(k; n)$, it
cannot be meaningfully compared to the COE result at all values of the
parameters. This is because once $N(k) = 1$, one cannot talk about
quantum two-point correlations, since the spectrum consists of a
single point on the unit circle. In other words, this is the extreme
quantum limit, where the Hilbert space consists of a single state.
Therefore, the $k$-values to be used must exceed in the case of the 3D
Sinai billiard $k_{\rm min} = \sqrt{96}/R$, which corresponds to
${\cal N}(k) = 2$. Hence, the values of $\tau = n/{\cal N}$ which are
accessible are restricted to the range $0 \le \tau \le n/2$.

\begin{figure}[p]

  \centerline{\psfig{figure=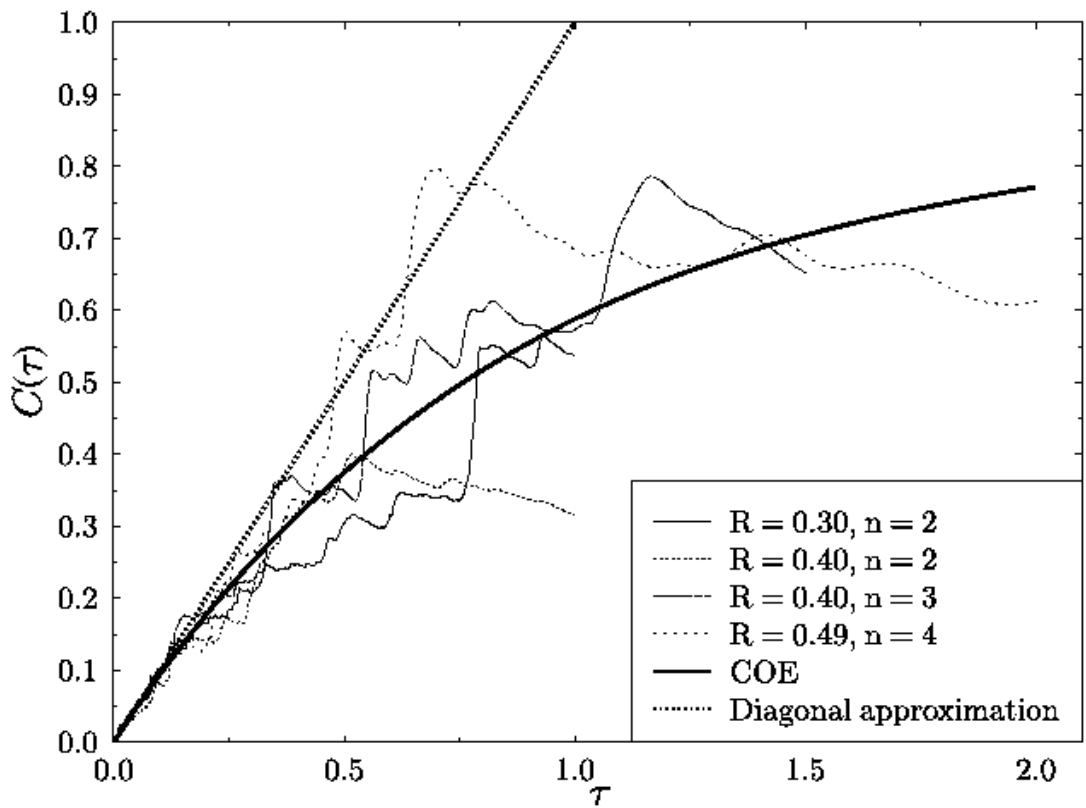,width=16cm}}

  \caption{Comparing the classical form factor with the universal RMT
    predictions for various cases of the 3D SB.}

  \label{fig:uiff}

\end{figure}

\begin{figure}[p]

  \centerline{\psfig{figure=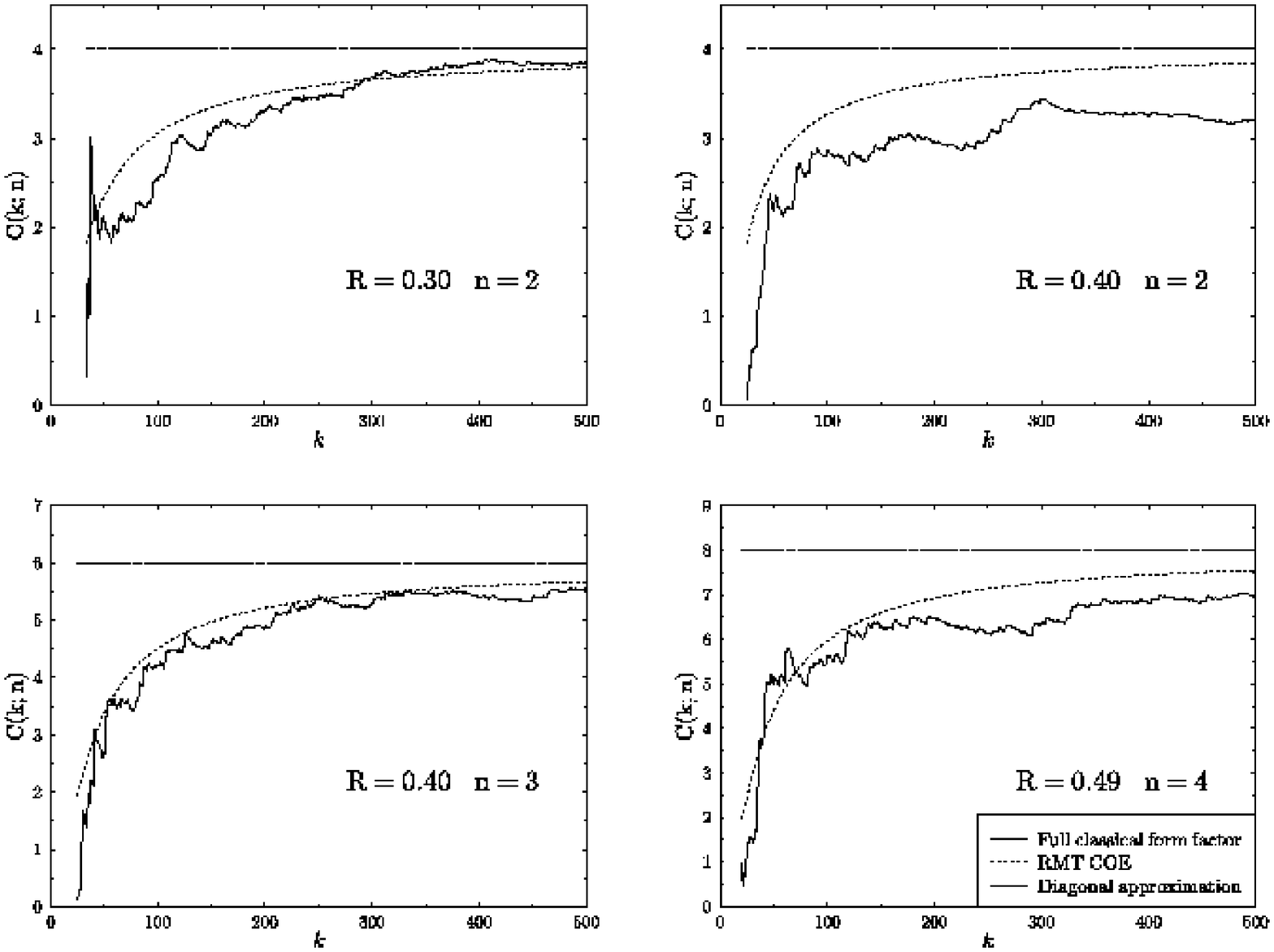,width=16cm}}

  \caption{The classical form factor compared with the universal RMT
    predictions for various cases of the 3D SB in the variable $k$.}

  \label{fig:ciff-tau}

\end{figure}

In figure \ref{fig:uiff} we summarize our numerical results by
comparing the the form factor obtained from periodic orbit theory
$K_{\rm cl}$ with the theoretical RMT prediction $K_{\rm COE}$. What
we actually show is the running average,
\begin{equation}
  C(\tau)
  \equiv
  \frac{1}{\tau} \int_{0}^{\tau} {\rm d}{\tau}' \,
    \frac{\tau'}{n} K_{\rm cl}({\tau}') \: ,
\end{equation}
where $K_{\rm cl}(\tau) \equiv K_{\rm cl}(k(\tau; n); n)$. The
corresponding COE curve (c.f.\ equations (\ref{eq:Kdef}) and
(\ref{eq:ffactordual})) is given by:
\begin{equation}
  C(\tau)
  \equiv
  \frac{1}{\tau} \int_{0}^{\tau} {\rm d}{\tau}' \,
    K_{\rm COE}({\tau}') \: .
\end{equation}
The ``diagonal approximation'' curve is obtained by replacing $K_{\rm
  COE}(\tau)$ by $2 \tau$, namely, classical correlations are ignored.
The data sets which were chosen are those for which sufficiently many
periodic orbits were computed so that the sum rule $U(n;l) \approx 1$
was satisfied. We did not include the $n=1$ data because they are
non-generic. As clearly seen from the figure, the data are consistent
with the RMT expression and they deviate appreciably from the diagonal
approximation. This is entirely due to the presence of classical
correlations, and it shows that the classical correlations are indeed
responsible for the quantitative agreement. Note also that the data
represents four different combinations of $n$ and $R$, which shows
that the classical scaling is indeed consistent with the universal
scaling implied by RMT. In figure \ref{fig:ciff-tau} we present
essentially the same data, but integrated and plotted using the
variable $k$, similarly to section \ref{sec:classical-pos}. The
integration started at $k_{\rm min}$ for a meaningful comparison with
RMT. Again, we observe the quantitative agreement, which is especially
good for the higher $n$ values ($n = 3$, $4$).

\begin{figure}[p]

  \centerline{\psfig{figure=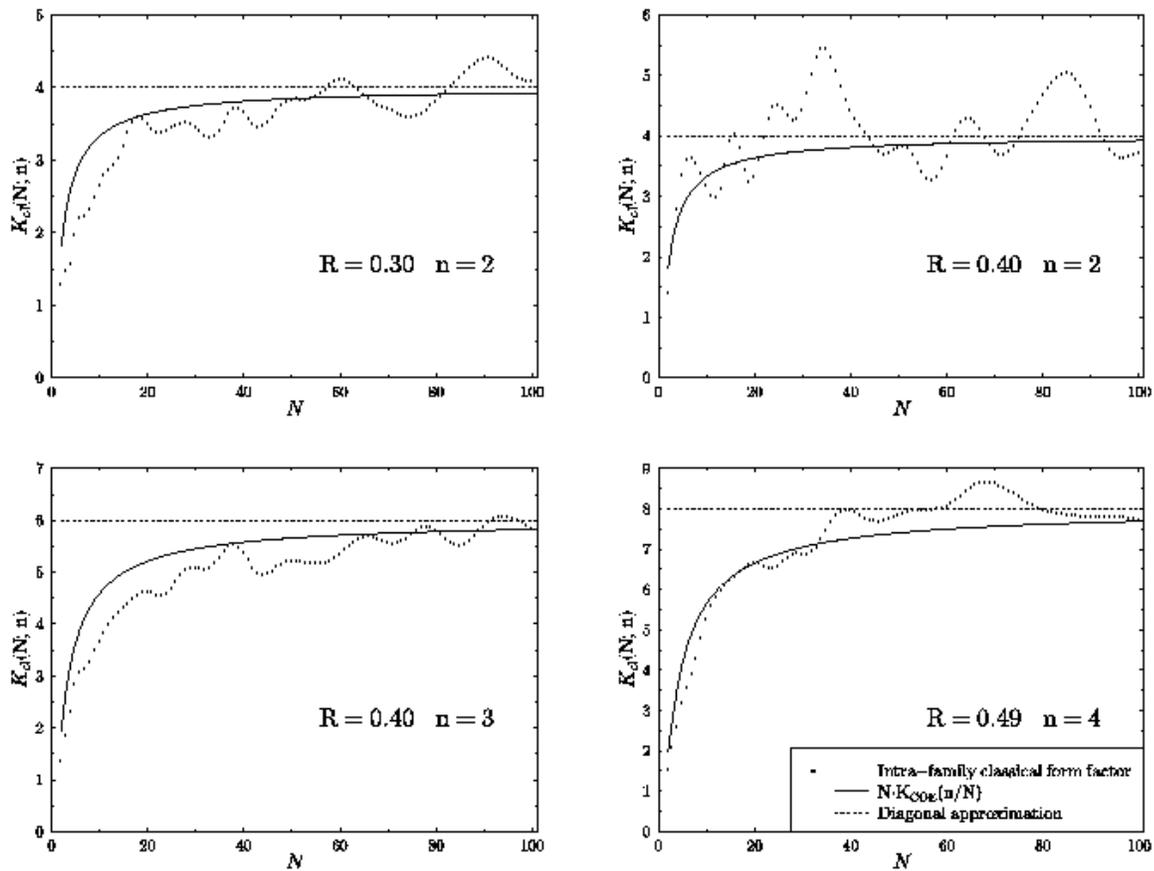,width=16cm}}

  \caption{The intra-family classical form factor $K_{\rm cl}(N; n)$ 
    compared to RMT COE. The variable is $ N$.}

  \label{fig:cff-fam-nk}

\end{figure}

\begin{figure}[p]

  \centerline{\psfig{figure=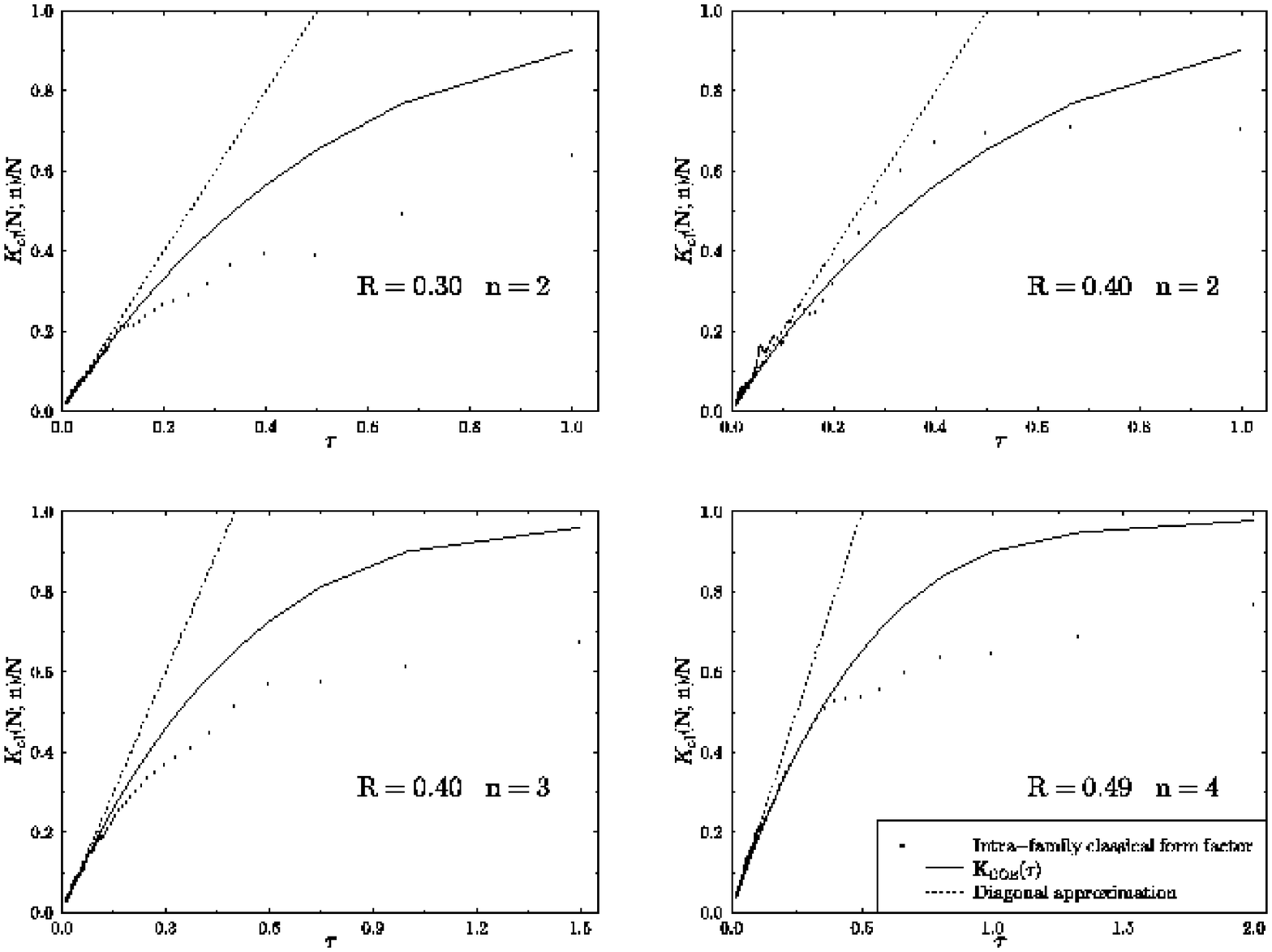,width=16cm}}

  \caption{The intra-family classical form factor $K_{\rm cl}(N; n)$ 
    compared to RMT COE. The control variable is $\tau = n/N$.}

  \label{fig:uff-fam-nk}

\end{figure}

In section \ref{subsec:pos-correlations} we showed that the classical
correlations originate to a large extent from the $\Omega(W)$ families
of periodic orbits. Moreover, the form factor which was calculated by
neglecting cross-family contributions was much smoother than the
original one. It is therefore appealing to take advantage of this
smoothness and compare the numerical and theoretical form factors
themselves instead of their running averages. We define:
\begin{equation}
  K_{\rm cl}(N; n)
  \equiv
  \langle K_{\rm cl}(k; n) \rangle_{N}
  =
  \frac{1}{k(N+1)-k(N)} \int_{k(N)}^{k(N+1)} {\rm d}k' \,
    K_{\rm cl}(k'; n)
\end{equation}
which is the semiclassical ensemble average of the form factor. In
figure \ref{fig:cff-fam-nk} we compare $K_{\rm cl}(N; n)$ with $N
\cdot K_{\rm COE}(n/N)$. The classical form factor included
intra-family contribution only, and we multiplied it by a factor such
that asymptotically it will match the theoretical value $2 n$. This
factor compensates for the partial breaking of time-reversal symmetry
and for the fact that the classical saturation is to values slightly
below $2 n$ for the $n$'s under consideration. One observes that the
agreement is quite good, and in any case the classical form factor is
sharply different from the diagonal approximation, meaning that
classical correlations are important. In figure \ref{fig:uff-fam-nk}
we present the same results with $\tau = n/N$ as the variable. It
again shows that the classical from factor agrees with the COE
expression beyond the validity range of the diagonal approximation.
The range of $\tau$ where a good agreement is observed increases with
$n$ as expected, but the estimated domain of valid comparison $\tau <
2n$ seems to be too optimistic.

In summary we can say that the present results show that the
semiclassical theory based on the Gutzwiller trace formula is capable
to reproduce the COE form factor beyond the ``diagonal
approximation''. To do this, one has to include the classical
correlations in the way which was done here, and once this is done,
there is no need to augment the theory by uncontrolled ``higher
order'' or ``diffractive'' corrections as was done in
\cite{AL96,WLS99} and by others. The results obtained in the present
section are corroborated by a recent analysis of periodic orbit
lengths correlations in billiards constructed from octagonal modular
domains in the hyperbolic plain \cite{HKS99}. The same quality of
agreement was obtained between the classical form factor and the
corresponding RMT result. These billiards are in two dimensions, and
therefore the scaling laws depend differently on $k$, and the fact
that the resulting scaled quantities agree with the expressions
derived from RMT gives further support to the line of thinking
developed here. We have grounds to believe that the classical
correlations are universal in hyperbolic systems, and have to do with
the self-similar organization of the set of periodic orbits. The
previous numerical studies which were conducted also on different
systems support this conjecture \cite{ADDKK93,CPS98}.


\section{Summary}
\label{sec:summary}


In the present paper we tried to provide a complete description of a
paradigmatic three dimensional quantum system which is chaotic in the
classical limit --- the three dimensional Sinai billiard. This study
is called for especially because most of the detailed investigations
in the field were carried out for systems in two dimensions.

Our main purpose in this study was to emphasize and clarify issues
which are genuinely related to the three dimensional character of the
system. The question which concerned us most was whether the
semiclassical approximation --- the main theoretical tool in the field
--- is sufficiently accurate for the spectral analysis of systems in
three dimensions.

We were able to obtain accurate and extensive data bases for the
quantum energy levels and for the classical periodic orbits. These
allowed us to check various properties of the quantum spectrum, and in
particular to study the applicability of the semiclassical
approximation. The main conclusion from our work is that contrary to
various expectations, the semiclassical accuracy, measured in units of
the mean spacing, does not diverge as a $\hbar^{2-d}$. Our numerical
tests and analytical arguments indicate an error margin which at worst
diverges weakly (logarithmically) with $\hbar$.

One of the main problems which we had to overcome was how to separate
the generic features which are common to all chaotic systems, from the
system specific attributes, which in the present case are the
``bouncing ball'' manifolds of periodic orbits. We should emphasize
that in $d$ dimensions the bouncing--ball manifolds contribute terms
of order $k^{(d-1)/2}$, which are much larger than the order $1$
contributions due to generic periodic orbits. Hence it is clear that
as the dimension increases, the extracting of generic features becomes
more difficult, and one has also to control higher $\hbar$
corrections, such as, e.g., diffraction corrections to the bouncing
ball contributions. We developed a method to circumvent some of these
difficulties which was sufficient for the 3D Sinai billiard case,
namely, we focused on the derivative of the spectrum with respect to
the boundary condition. This method is a powerful means which can also
be used in other instances, where non-generic effects should be
excluded.

One of the issues which are essential to the understanding of trace
formulae and their application, was first mentioned by Gutzwiller in
his book, under the title of the ``third entropy'' \cite{Gut90}.
Gutzwiller noticed that in order that the series over periodic orbits
can be summed up (in some sense) to a spectral density composed of
$\delta$ functions, the phases of the contributing terms should have
very special relations. The more quantitative study of this problem
started when Argaman {\it et al}.\ \cite{ADDKK93} defined the
concept of periodic orbit correlations. The dual nature of the quantum
spectrum of energies and the classical spectrum of periodic orbit was
further developed in \cite{CPS98}. It follows that the universality of
the quantum spectral fluctuations implies that the correlation length
in the spectrum of the classical actions depends on the dimensionality
in a specific way. This was tested here for the first time, and the
mechanism which induces classical correlations was discussed.

Our work on the Sinai billiard in three dimensions proved beyond
reasonable doubt that the methods developed for two dimensional
chaotic systems can be extended to higher dimensions. Of utmost
importance and interest is the study of classical chaos and its
quantum implications in many body systems. This is probably the
direction to which the research in ``quantum chaos'' will be
advancing.


\section*{Acknowledgements}

The research performed here was supported by the Minerva Center for
Non-Linear Physics, and by the Israel Science Foundation. Many
colleagues helped us during various stages of the work. We are
indebted in particular to
Michael Berry,
Eugene Bogomolny,
Leonid Bunimovich,
Doron Cohen,
Barbara Dietz,
Eyal Doron,
Shmuel Fishman,
Klaus Hornberger,
Jon Keating,
Dieter Klakow,
Daniel Miller,
Zeev Rudnick,
Holger Schanz,
Martin Sieber and
Iddo Ussishkin
for numerous discussions, for suggestions and for allowing us to use
some results prior to their publication.
In particular we thank Klaus Hornberger and Martin Sieber for their
critical reading of the manuscript and for their remarks and
suggestions.
HP is grateful for a MINERVA postdoctoral fellowship, and wishes to
thank Reinhold Bl\"umel and John Briggs for their hospitality in
Freiburg.
The Humboldt foundation is acknowledged for supporting US stay in
Marburg, Germany during the summer of 1998 when much of the work on
this manuscript was carried out.

\appendix
\section{Efficient quantization of billiards: 
  BIM vs.\ full diagonalization}
\label{app:bimvsfull}
%
In this Appendix we wish to compare two possible quantization schemes
for billiards: Direct Diagonalization (DD) of the Hamiltonian matrix
vs.\ the Boundary Integral Method (BIM) (see e.g.\ \cite{BW84,Boa94}).
The diagonalization is a generic method to solve the time independent
Schr\"odinger equation, while the BIM is specialized for billiards. To
compare the two methods, we estimate the complexity of computing all
of the eigenvalues up to a given wavenumber $k$.

To find the matrix elements of the Hamiltonian we treat the billiard
boundaries as very high potential walls. The linear dimension $M(k)$
of the Hamiltonian matrix that is needed for finding eigenvalues
around $k$ is:
\begin{equation}
  M_{\mbox{\tiny DD}}(k) = 
  {\cal O} \left( \left(\frac{S}{\lambda}\right)^{d} \right) = 
  {\cal O} \left( (kS)^d \right)
  \label{eq:mdd}
\end{equation}
where $S$ is the typical linear dimension of the billiard, $\lambda =
2\pi/k$ is the wavelength and $d$ is the dimensionality of the
billiard. The above estimate is obtained by enclosing the billiard in
a hypercube with edge $S$ and counting the modes up to wavenumber $k$.
The numerical effort to find eigenvalues of a matrix is of order of
its linear dimension to the power 3. Thus, the numerical effort to
find all the eigenvalues of the billiard up to $k$ using DD is
estimated as:
\begin{equation}
  C_{\mbox{\tiny DD}}(k) = 
  {\cal O}(M_{\mbox{\tiny DD}}^3(k)) = 
  {\cal O}((kS)^{3d}).
\end{equation}
The expected number of eigenvalues up to $k$ is given to a good
approximation by Weyl's law, which for billiards reads:
\begin{equation}
  N(k) = {\cal O}((kS)^{d}).
  \label{eq:weyld}
\end{equation}
Thus, the numerical effort to calculate the first (lowest) $N$
eigenvalues of a billiard in $d$ dimension in the direct Hamiltonian
diagonalization is
\begin{equation}
  C_{\mbox{\tiny DD}}(N) = {\cal O}(N^3)
  \label{eq:cdd}
\end{equation}
which is independent of the dimension.

As for the BIM, one traces the $k$-axis and searches for eigenvalues
rather than obtaining them by one diagonalization. This is done by
discretizing a kernel function on the boundary of the billiard and
looking for zeroes of the resulting determinant. The linear dimension
of the BIM matrix is
\begin{equation}
  M_{\mbox{\tiny BIM}} = 
  {\cal O} \left( \left(\frac{S}{\lambda}\right)^{d-1} \right) = 
  {\cal O} \left( (kS)^{d-1} \right) =
  {\cal O} \left( N^{1-1/d} \right) \, .
  \label{eq:mbim}
\end{equation}
This estimate is obtained from discretizing the boundary of the
billiards which is of dimension $d-1$ by hypercubes of edge $\lambda$.
The numerical effort of calculating the determinant once is:
\begin{equation}
  c_{\mbox{\tiny BIM}}(k) = 
  {\cal O}(M_{\mbox{\tiny BIM}}^3(k)) = 
  {\cal O}((kS)^{3(d-1)})\ .
\end{equation}
(In practice, one often uses the SVD algorithm \cite{NAG90}, which is
much more stable than a direct computation of the determinant and has
the same complexity.) Using the relation (\ref{eq:weyld}) we find that
the numerical effort to find an eigenvalue near the $N$th one is
estimated by
\begin{equation}
  c_{\mbox{\tiny BIM}}(N) = {\cal O}(N^{3-3/d}).
\end{equation}
To get the above result we assumed that a fixed number of iterations
(evaluations of the determinant) is needed to detect each eigenvalue,
which is justified at least for the case where level repulsion is
expected. Thus, the complexity to calculate all the eigenvalues up to
the $N$th is
\begin{equation}
  C_{\mbox{\tiny BIM}}(N) = 
  {\cal O}(N) c_{\mbox{\tiny BIM}}(N) = 
  {\cal O}(N^{4-3/d}).
  \label{eq:cbim}
\end{equation}
In particular:
\\
\begin{tabular}{ll}
  \hspace{2cm} $C_{\mbox{\tiny BIM}}(N) = {\cal O}(N^{5/2})$, 
  & for $d = 2$ 
  \\
  \hspace{2cm} $C_{\mbox{\tiny BIM}}(N) = {\cal O}(N^3)$, 
  & for $d = 3$.
\end{tabular}
\\

We conclude that the BIM is more efficient than DD for $2$ dimensions,
and for $3$ dimensions they are of the same level of complexity. In
practice, however, it seems that the BIM is better also in $3$
dimensions, since the DD matrices can be prohibitly large, and
manipulating them (if possible) can be very expensive due to memory
limitations (paging). Also one has to take into account, that due to
evanescent modes, the numerical proportionality factor in
(\ref{eq:mbim}) is actually close to $1$, while for (\ref{eq:mdd}) the
factor can be large if high accuracy is desired. This is due to the
fact that the off-diagonal matrix elements of the Hamiltonian decay
only like a power-law due to the sharp potential and hence very large
matrices are needed in order to obtain accurate eigenvalues.

\section{Symmetry reduction of the numerical effort 
         in the quantization of billiards}
\label{app:symm-reduction}

Consider a $d$-dimensional billiard which is invariant under a group
$\cal G$ of geometrical symmetry operations. We want to compare the
numerical effort that is needed to compute the lowest $N$ eigenvalues
of the fully symmetric billiard with that of computing the lowest $N$
eigenvalues of the desymmetrized billiard. In ``desymmetrized'' we
mean the following: if $\Omega$ is the full billiard domain, then the
desymmetrized billiard $\omega$ is such that $\bigcup_{\hat{g} \in
  \cal{G}} \hat{g} \omega = \Omega$. If one uses the direct
diagonalization (DD) of the Hamiltonian matrix, then there is no
advantage to desymmetrization, because the prefactor in (\ref{eq:cdd})
should not depend on the shape of the billiard if its aspect ratio is
close to 1. Therefore, the numerical effort of computing the lowest
$N$ levels of either the fully symmetric or the desymmetrized billiard
is more or less the same using DD. On the other hand, as we show in
the sequel, desymmetrization is very advantageous within the framework
of the BIM,.

We first note, that considering a particular irreducible
representation $\gamma$ of $\cal{G}$ is equivalent to desymmetrization
of the billiard together with imposing boundary conditions that are
prescribed by $\gamma$. The dimension of $\gamma$ is denoted as
$d_{\gamma}$ and the order of $\cal{G}$ is denoted as $N_{\cal{G}}$.
Given a complete basis of functions in which the functions are
classified according to the irreps of $\cal{G}$, then the fraction of
the basis functions that belong to the irrep $\gamma$ is $d_{\gamma}^2
/ N_{\cal G} \equiv F_{\gamma}$. This is also the fraction of
eigenvalues that belong to $\gamma$ out of the total number of levels,
when we consider a large number of levels. Using the notations of
appendix \ref{app:bimvsfull}, we thus have:
\begin{eqnarray}
  M_{\mbox{\tiny BIM}}^{(\gamma)}(k) 
  & = &
  F_{\gamma} M_{\mbox{\tiny BIM}}(k)
  \nonumber \\
  N^{(\gamma)}(k) 
  & = & 
  F_\gamma N(k)
\end{eqnarray}
where the quantities with superscript $\gamma$ correspond to the
desymmetrized billiard, and the others to the fully symmetric one.
Using (\ref{eq:mbim}) and repeating arguments from appendix
\ref{app:bimvsfull} results in:
\begin{equation}
  C_{\mbox{\tiny BIM}}^{(\gamma)}(N) = 
  F_{\gamma}^{\frac{3}{d}} C_{\mbox{\tiny BIM}}(N) \: .
\end{equation}
In the equation above we replaced $N^{(\gamma)} \rightarrow N$. Thus,
the decrease in the density of states is more than compensated by the
reduction in the size of the secular matrix and the overall numerical
effort is diminished by a factor of $F_{\gamma}^{\frac{3}{d}}$. For
example, in the case of the 3D Sinai billiard and for a
one--dimensional irrep, the saving factor is:
\begin{equation}
  F_{\gamma}^{\frac{3}{d}} = 
  \left( \frac{1^2}{48} \right)^{\frac{3}{3}} = 
  \frac{1}{48}
\end{equation}
which is a very significant one.

\section{Resummation of $D_{LM}$ using
         the {E}wald summation technique}
\label{app:ewald}

In general, the Ewald summation technique is used to calculate
(conditionally convergent) summations over lattices $\{\vec{\rho}\,\}$:
\begin{equation}
  S = \sum_{\vec{\rho}} f(\vec{\rho}\,) \: .
  \label{eq:latsum}
\end{equation}
One splits the sum $S$ into two sums $S_1$, $S_2$ which depend on a
parameter $\eta$:
\begin{equation}
S = S_1 + S_2 = \sum_{\vec{\rho}} f_1(\vec{\rho}\,; \eta) + 
                \sum_{\vec{\rho}} f_2(\vec{\rho}\,; \eta) \: .
\end{equation}
This splitting is usually performed by representing $f(x)$ as an integral, and
splitting the integral at $\eta$. The idea is to resum $S_1$ on the
reciprocal lattice $\{\vec{g}\,\}$ using the Poisson summation formula:
\begin{equation}
  S_1 
  = 
  \sum_{\vec{g}} \int {\rm d}^d \rho \ 
    \exp (2 \pi i \vec{\rho} \, \vec{g})
    f_1(\vec{\rho}; \eta) \equiv
    \sum_{\vec{g}} \hat{f}_1(\vec{g}\,; \eta) \: ,
\end{equation}
and to choose $\eta$ such that both $S_1$ and $S_2$ will rapidly
converge.

We need to apply the Ewald summation technique to $D_{LM}(k)$, given
explicitly in equation (\ref{eq:dlm}), which constitute the main
computational load. This is because they include summations over the
$\bbbz^3$ lattice which need to be computed afresh for each new value
of $k$. It is possible to apply the Ewald technique directly to each
$D_{LM}$, but it is much simpler to take an indirect route: We shall
Ewald resum the free Green function on the 3-torus, and then read off
the $D_{LM}$'s as expansion coefficients.

\begin{figure}[p]
  \begin{center}
    \leavevmode 

    \centerline{\psfig{figure=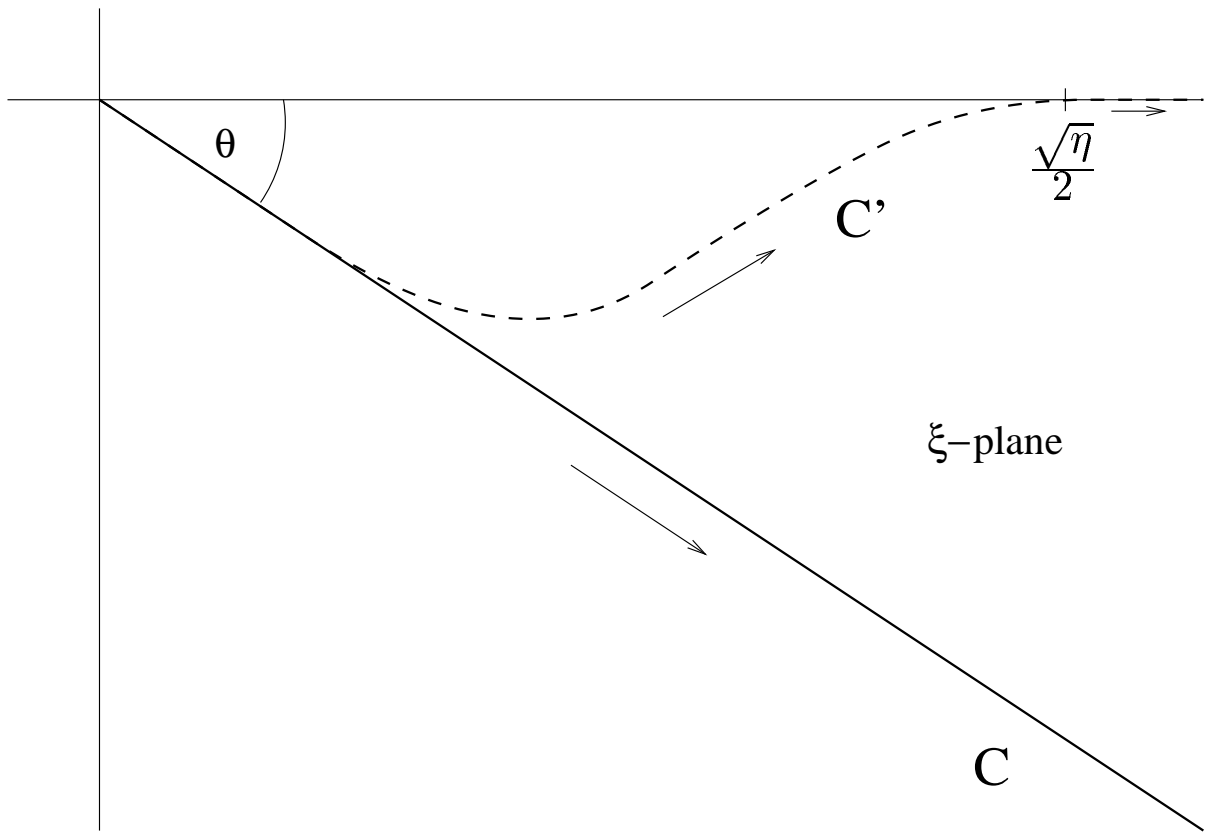,width=16cm}}

    \caption{Contours for the integral evaluation of $G_0^T$. 
      In the above $\theta \equiv \arctan [ {\rm Im}(k) / {\rm Re}(k)
      ] / 2 - \pi /4$.}

    \label{fig:ewald-contour}
  \end{center}
\end{figure}

We start with the free outgoing Green function on the
three--dimensional torus:
\begin{equation}
  G_0^T(\vec{q}\,) 
  = 
  -\frac{1}{4 \pi} \sum_{\vec{\rho} \in \bbbz^3} 
    \frac{\exp (ik|\vec{q}-\vec{\rho}\,|)}
         {|\vec{q}-\vec{\rho}\,|} \: ,
\end{equation}
where we took the side of the torus to be $1$ for simplicity and
defined $\vec{q} \equiv \vec{r} - \vec{r}\,'$. To split the sum we use
an integral representation of the summands \cite{Ewa21,HS61}:
\begin{equation}
  \frac{\exp (ik|\vec{q}-\vec{\rho}\,|)}
       {|\vec{q}-\vec{\rho}\,|} 
  = 
  \frac{2}{\sqrt{\pi}} \int_{0(C)}^{\infty}
    \exp \left[ -(\vec{q}-\vec{\rho}\,)^2 \xi^2 +
    \frac{k^2}{4 \xi^2} \right] {\rm d}\xi \: ,
  \label{eq:intrep}
\end{equation}
where the integration contour $C$ is shown in figure
\ref{fig:ewald-contour}. It is assumed that $k$ has an infinitesimal
positive imaginary part, which is taken to $0$ at the end of the
calculation. We now deform the contour into $C'$ (see figure
\ref{fig:ewald-contour}), such that it runs along the real axis for
$\xi > \sqrt{\eta}/2$, and split the integral at this point as
follows:
\begin{eqnarray}
  G_0^T(\vec{q}\,) 
  & = & 
  G_1^T(\vec{q}\,) + G_2^T(\vec{q}\,) \: ,
  \\
  G_1^T(\vec{q}\,)
  & = &
  -\frac{1}{2 \pi \sqrt{\pi}} \sum_{\vec{\rho}} 
    \int_{0(C')}^{\sqrt{\eta}/2} 
    \exp \left[ -(\vec{q}-\vec{\rho})^2 \xi^2 +
                \frac{k^2}{4 \xi^2} \right] {\rm d}\xi \: ,
  \\
  G_2^T(\vec{q}\,)
  & = &
  -\frac{1}{2 \pi \sqrt{\pi}} \sum_{\vec{\rho}} 
    \int_{\sqrt{\eta}/2}^{\infty}
    \exp \left[ -(\vec{q}-\vec{\rho})^2 \xi^2 +
                 \frac{k^2}{4 \xi^2} \right] {\rm d}\xi \: .
  \label{eq:gt2} \\
\end{eqnarray}
The summation in $G_2^T$ is rapidly convergent, due to the fact that
we integrate over the tails of a rapidly decaying function in $\xi$
(faster than a Gaussian), and we start further on the tail when $\rho$
grows. In order to make $G_1^T$ also rapidly convergent, we need to
Poisson resum it. We use the identity:
\begin{equation}
  \sum_{\vec{\rho}} \exp \left[ -(\vec{q}-\vec{\rho})^2 \xi^2 \right] 
  =
  \frac{\pi \sqrt{\pi}}{\xi^3} \sum_{\vec{g}} \exp \left[
    \frac{(2 \pi g)^2}{4 \xi^2} + i (2 \pi \vec{g}) \vec{q} \right] \: ,
\end{equation}
which is obtained by explicitly performing the integrals of the
Poisson summation. Thus,
\begin{eqnarray}
  G_1^T(\vec{q}\,) 
  & = &
  -\frac{1}{2} \sum_{\vec{g}} \exp (2 \pi i \vec{g} \vec{q}\,)
    \int_{0(C')}^{\sqrt{\eta}/2} \frac{{\rm d}\xi}{\xi^3} 
    \exp \left[ \frac{k^2 - (2 \pi g)^2}{4 \xi^2} \right]
  \nonumber \\
  & = &
  \sum_{\vec{g}} \frac{\exp (2 \pi i \vec{g} \vec{q}\,)
    \exp \left[ \frac{k^2 - (2 \pi g)^2}{\eta} \right]}
  {k^2 - (2 \pi g)^2} \: .
  \label{eq:gt1}
\end{eqnarray}
The second line was obtained from the first one by performing the
integrals explicitly. The expression obtained for $G_1^T$ is also
rapidly convergent, and is suitable for computations. We thus
succeeded in rewriting $G_0^T$ as two rapidly converging sums
(\ref{eq:gt1}), (\ref{eq:gt2}). We note that the results
(\ref{eq:gt1}), (\ref{eq:gt2}) are valid for general lattices, the
cubic lattice being a special case \cite{HS61}.

The heart of the above resummation of $G_0^T$ was the integral
representation (\ref{eq:intrep}) which is non-trivial. In appendix
\ref{app:ewald-physical} we present an alternative derivation of the
above results using more intuitive, physical arguments.

It remains to extract the $D_{LM}$'s from the resummed $G_0^T$. The
basic relation is \cite{HS61}:
\begin{equation}
  G_0^T(\vec{q}\,) 
  =
  \sum_{LM} j_L(kq) Y_{LM}(\Omega_{\vec{q}}) 
    \left[ D_{LM}(k) + 
      \frac{k}{\sqrt{4 \pi}} \frac{n_0(kq)}{j_0(kq)} \delta_{L0} 
    \right] \: .
  \label{eq:gt-dlm}
\end{equation}
Using expansion theorems \cite{VMK88} applied to (\ref{eq:gt2}) and
(\ref{eq:gt1}) one can rewrite $G_0^T$ as:
\begin{eqnarray}
  G_0^T(\vec{q}\,) 
  & = &
  \sum_{LM} Y_{LM} (\Omega_{\vec{q}}) \left\{
    \sum_{\vec{g}} 4 \pi i^L e^{k^2 / \eta}  Y^*_{LM}(\Omega_{\vec{g}})
    j_L (2 \pi g q) \frac{e^{-(2 \pi g)^2 / \eta}}{k^2 - (2 \pi g)^2} 
    - \right. \nonumber \\
  & &
  \left. \sum_{\vec{\rho}} \frac{2 i^L}{\sqrt{\pi}}  
    Y^*_{LM}(\Omega_{\vec{\rho}})
    \int_{\frac{\sqrt{\eta}}{2}}^{\infty} 
    {\rm d}\xi \, j_L(-2 i \rho q \xi^2)
    \exp \left[ -(\rho^2 {+} q^2) \xi^2 {+} 
      \frac{k^2}{4 \xi^2} \right] \right\}. 
  \nonumber \\
  & &
  \label{eq:gt-expanded}
\end{eqnarray}
Comparing equations (\ref{eq:gt-dlm}) and (\ref{eq:gt-expanded}), and
using the orthogonality of the spherical harmonics
$Y_{LM}(\Omega_{\vec{q}})$, one obtains:
\begin{equation}
  D_{LM}(k) 
  = 
  \frac{1}{j_L(kq)} \left[
    \sum_{\vec{g}} \cdots + \sum_{\vec{\rho}} \cdots -
    \frac{k}{\sqrt{4 \pi}} n_0(k q) \delta_{L0} \right] \: .
  \label{eq:dlmewa}
\end{equation}
This is the Ewald--resumed expression of $D_{LM}(k)$. It has the
interesting feature, that even though each of the terms explicitly
depends on $q$, the total expression is independent of $q$. The same
applies also to $\eta$. This freedom can be used to simplify the
expression (\ref{eq:dlmewa}), since for $q \rightarrow 0$ the
spherical Bessel functions simplify to powers \cite{HS61}:
\begin{equation}
  j_{L}(aq) \longrightarrow \frac{(aq)^L}{(2L+1)!!} \: ,
\end{equation}
which are computationally less demanding. Taking the limit is
straightforward for $L \neq 0$, while for $D_{00}$ there is a
complication due to the singularity of $n_0(kq)$. As shown in appendix
\ref{app:dlm3} this singularity is exactly cancelled by the
$\vec{\rho} = \vec{0}$ term, resulting in a finite expression also for
$D_{00}$. The final result is:
\begin{eqnarray}
  D_{LM}
  & = &
  D_{LM}^{(1)} + D_{LM}^{(2)} + D_{00}^{(3)} \delta_{L0} \: ,
  \label{eq:dlmewa1} 
  \\
  D_{LM}^{(1)} 
  & = &
  4 \pi i^L k^{-L} e^{k^2 / \eta} \sum_{\vec{g}}
  (2 \pi g)^L Y^*_{LM}(\Omega_{\vec{g}}) \frac{e^{-(2 \pi g)^2 / \eta}}
  {k^2 - (2 \pi g)^2} \: ,
  \label{eq:dlm1} 
  \\
  D_{LM}^{(2)} 
  & = &
  \frac{2^{L+1}k^{-L}}{\sqrt{\pi}} \sum_{\vec{\rho} \neq \vec{0}}
  \rho^L Y^*_{LM}(\Omega_{\vec{\rho}}) 
  \int_{\frac{\sqrt{\eta}}{2}}^{\infty} {\rm d}\xi \ \xi^{2L} 
  \exp \left[ -\rho^2 \xi^2 + \frac{k^2}{4 \xi^2} \right]  \: ,
  \label{eq:dlm2} 
  \\
  D_{00}^{(3)} & = & - \frac{\sqrt{\eta}}{2 \pi} \sum_{n=0}^{\infty}
  \frac{(k^2 / \eta)^n}{n! (2n-1)} \: ,
  \label{eq:dlm3ewa1}
\end{eqnarray}
with the convention $g^L|_{g=0, L=0} = 1$. This completes the task of
Ewald--resuming the building blocks $D_{LM}(k)$ into rapidly
convergent series.


\section{``Physical'' Ewald summation of $G_0^T(\vec{q}\,)$}
\label{app:ewald-physical}

In this appendix we present a derivation of the results
(\ref{eq:gt1}), (\ref{eq:gt2}) by a method that is different than the
one used in appendix \ref{app:ewald}. The present method is physically
appealing and does not require the use of complicated integral
representations.  It is inspired by Appendix B of Kittel's book
\cite{Kit53} which deals with the problem of calculating Madelung
constants (electrostatic potentials) of ion crystals.

In the sequel we use $\vec{q} \equiv \vec{r} - \vec{r}\,'$ and adopt
the following notational convention: For any quantity $X(\vec{q}\,)$
we add a superscript $T$ to denote its lattice sum:
\begin{equation}
  X^T(\vec{q}\,) 
  \equiv 
  \sum_{\vec{\rho}} X(\vec{q} - \vec{\rho}\,) \: .
\end{equation}

We start from the Helmholtz equation for $G_0$:
\begin{equation}
  (\nabla_{\vec{r}}^2 + k^2) G_0(\vec{q}\,) 
  = 
  \delta(\vec{q}\,) \: .
\end{equation}
Due to linearity, the function $G_0^T$ satisfies:
\begin{equation}
  (\nabla_{\vec{r}}^2 + k^2) G_0^T(\vec{q}\,) 
  = 
  \delta^T(\vec{q}\,) \: .
  \label{eq:helmg0t}
\end{equation}
The RHS of (\ref{eq:helmg0t}) can be interpreted as a ``charge
distribution'' which is composed of point charges on a lattice. Each
such point charge $\delta(\vec{q}-\vec{\rho}\,)$ induces a
``potential'' $G_0(\vec{q}-\vec{\rho}\,) = - \exp(i k
|\vec{q}-\vec{\rho}\,|)/(4 \pi |\vec{q}-\vec{\rho}\,|)$ which is
long--ranged due to the sharpness of the charge. (This is in analogy
to the electrostatic case.) Hence, the lattice sum of potentials
$G_0^T$ is conditionally convergent. To overcome this difficulty we
introduce an arbitrary charge distribution $\lambda(\vec{q}\,)$ and
rewrite (\ref{eq:helmg0t}) as:
\begin{eqnarray}
  G_0^T(\vec{q}\,) 
  & = &
  G_1^T(\vec{q}\,) + G_2^T (\vec{q}\,) \: ,
  \label{eq:split1} 
  \\
  (\nabla^2 + k^2) G_1^T (\vec{q}\,) 
  & = &
  \lambda^T (\vec{q}\,) \: ,
  \label{eq:split2} 
  \\
  (\nabla^2 + k^2) G_2^T (\vec{q}\,) 
  & = &
  \delta^T (\vec{q}\,) - \lambda^T (\vec{q}\,) \: .
  \label{eq:split3}
\end{eqnarray}
We want $\lambda(\vec{q}-\vec{\rho}\,)$ to effectively screen the
$\delta(\vec{q}-\vec{\rho}\,)$ charges, making $G_2$ short--ranged.
This will result in rapid convergence of $G^T_2$. (Note, that the
equations (\ref{eq:split1})-(\ref{eq:split3}) hold also for the
quantities without the T superscript due to linearity.) On the other
hand, $\lambda(\vec{q}\,)$ must be smooth enough, such that $G_1^T$
will rapidly converge when Poisson resumed. It is hence plausible to
choose a (spherically symmetric) Gaussian charge distribution for
$\lambda (\vec{q}\,)$:
\begin{equation}
  \lambda(\vec{q}\,) 
  = 
  A \exp (-\alpha q^2) \: ,
\end{equation}
where $A$ and $\alpha$ are yet arbitrary parameters.

We calculate first $G_2 (\vec{q}\,)$ by rewriting the inducing charge
as an integral over $\delta$ charges, and using the fact that each
$\delta$ charge contributes $G_0$ to the potential:
\begin{equation}
  (\nabla^2 + k^2) G_2(\vec{q}\,) 
  = 
  \delta (\vec{q}\,) - \lambda(\vec{q}\,) 
  =
  \delta (\vec{q}\,) - 
  \int {\rm d}^3 Q \lambda(\vec{Q}\,) \delta(\vec{q}-\vec{Q}\,) \: .
\end{equation}
Hence,
\begin{eqnarray}
  G_2 (\vec{q}\,) 
  & = &
  G_0(\vec{q}\,) -
  \int {\rm d}^3 Q \lambda(\vec{Q}\,) G_0(\vec{q}-\vec{Q}\,) 
  \nonumber \\
  & = &
  G_0(\vec{q}\,) \left[ 1 - A \left( \frac{\pi}{\alpha} \right)^3 
    e^{-\frac{k^2}{4 \alpha}} \right] +
  \frac{A}{2 \alpha q} \int_{0}^{\infty} {\rm d}t \ 
  e^{-\alpha (t+q)^2} \cos(k t).
\end{eqnarray}
The first term is long-ranged due to $G_0$, and the second term is
short-ranged due to the integral that is rapidly decreasing as a
function of $q$. To make $G_2$ short ranged, we thus have to set the
coefficient of $G_0$ to $0$, which is satisfied if we choose
\begin{equation}
  A = A(k, \alpha) = \left( \frac{\alpha}{\pi} \right)^3 
  \exp \left( \frac{k^2}{4 \alpha} \right).
\end{equation}
Therefore, we get for $G_2^T$ a rapidly convergent sum:
\begin{equation}
  G_2^T(\vec{q}\,) 
  = 
  - \frac{\sqrt{\alpha} e^{\frac{k^2}{4 \alpha}}}
  {2 \pi \sqrt{\pi}} \sum_{\vec{\rho}} \frac{1}{|\vec{q} - \vec{\rho}\,|}
  \int_0^{\infty} {\rm d}t \ 
  \exp \left[ - \alpha \left( t + |\vec{q} - \vec{\rho}\,| \right)^2 \right]
  \cos (k t).
  \label{eq:g2tp}
\end{equation}
This can be re-expressed in a more compact form using complement error
functions with complex arguments:
\begin{equation}
  G_2^T(\vec{q}\,) 
  =
  - \frac{1}{2 \pi} \sum_{\vec{\rho}} \frac{1}{|\vec{q} - \vec{\rho}\,|}
  {\rm Re} \left[ \exp (-i k |\vec{q} - \vec{\rho}\,|) \cdot
    {\rm erfc} \left( \sqrt{\alpha} |\vec{q} - \vec{\rho}\,| - 
      \frac{i k}{2 \sqrt{\alpha}} \right) \right] \: ,
  \label{eq:g2tpp}
\end{equation}
where
\begin{equation}
  {\rm erfc}(z) \equiv
  \frac{1}{\sqrt{\pi}} \int_{z}^{\infty} e^{-u^2} {\rm d}u \: .
\end{equation}

To calculate $G_1^T$ we can directly Poisson resum (\ref{eq:g2tp}).
Alternatively, we can use again the Helmholtz equation for $G_1^T$
(\ref{eq:split2}) to simplify the calculations. We expand $G_1^T$ in
the reciprocal lattice:
\begin{eqnarray}
  G_1^T(\vec{q}\,) 
  & = & 
  \sum_{\vec{g}} \int {\rm d}^3 \rho \ 
  \exp ( 2 \pi i \vec{\rho} \vec{g} ) G_1(\vec{q}-\vec{\rho}\,) 
  \nonumber \\
  & = &
  \sum_{\vec{g}} \exp ( 2 \pi i \vec{q} \vec{g}\,)
  \int {\rm d}^3 \rho \ 
  \exp ( - 2 \pi i \vec{\rho} \vec{g}\,) G_1(\vec{\rho}\,) 
  \nonumber \\
  & \equiv & 
  \sum_{\vec{g}} \exp ( 2 \pi i \vec{q} \vec{g}\,) G_{1 \vec{g}}
  \label{eq:g1tg}
\end{eqnarray}
where the second line was obtained from the first one by shifting the
origin of the integration. Similarly for $\lambda^T(\vec{q}\,)$:
\begin{equation}
  \lambda(\vec{q}\,) 
  = 
  \sum_{\vec{g}} \exp ( 2 \pi i \vec{q} \vec{g}\,) 
    \lambda_{\vec{g}} \: .
  \label{eq:lamg}
\end{equation}
Inserting (\ref{eq:g1tg}, \ref{eq:lamg}) into (\ref{eq:split2}) and
using the orthogonality of the Fourier components, we get the simple
relation between $G_{1 \vec{g}}$ and $\lambda_{\vec{g}}$:
\begin{equation}
  G_{1 \vec{g}} 
  = 
  \frac{\lambda_{\vec{g}}}{k^2 - (2 \pi g)^2}.
\end{equation}
When inserted back into (\ref{eq:g1tg}) we finally get for $G_1^T$:
\begin{equation}
  G_1^T(\vec{q}\,) 
  = 
  \sum_{\vec{g}} \frac{\exp (2 \pi i \vec{g} \vec{q}\,)
    \exp \left[ \frac{k^2 - (2 \pi g)^2}{4 \alpha} \right]}
  {k^2 - (2 \pi g)^2}.
\end{equation}
This expression is identical with (\ref{eq:gt1}) if we set $4 \alpha =
\eta$. It can be shown \cite{Ewa21} that also the expressions for
$G_2^T$, (\ref{eq:gt2}) and (\ref{eq:g2tpp}) are identical. However,
equation (\ref{eq:g2tpp}) is more convenient if one needs to compute
$G_0^T(\vec{q}\,)$, since it involves well-tabulated computer-library
functions \cite{NAG90} and saves the burden of numerical integrations.
On the other hand, the expression (\ref{eq:gt2}) is more convenient as
a starting point for calculating $D_{LM}(k)$.

To summarize, we re-derived the Ewald--resummed form of
$G_0^T(\vec{q}\,)$ using the underlying Helmholtz equation. We used a
physically intuitive argument of screening potentials, that was shown
to be equivalent to the more abstract integral representation of
$G_0(\vec{q}\,)$, equation (\ref{eq:intrep}).


\section{Calculating $D_{00}^{(3)}$}
\label{app:dlm3}

We need to calculate (refer to equation (\ref{eq:dlmewa}) and its
subsequent paragraph):
\begin{equation}
  D_{00}^{(3)} 
  \equiv
  \lim_{q \rightarrow 0} \left\{ \frac{1}{j_0 (kq)}
    \left[ \frac{1}{\sqrt{4 \pi}} \frac{\cos (kq)}{q} -
      \frac{1}{\pi} \int_{\sqrt{\eta} / 2}^{\infty} {\rm d}\xi \ 
      \exp \left( - q^2 \xi^2 + \frac{k^2}{4 \xi^2} \right)
    \right] \right\} \: ,
  \label{eq:dlm3}
\end{equation}
where we used the explicit expression $n_0(x) = -\cos(x)/x$. Taking
the limit of the denominator is trivial, since $j_0(kq) \rightarrow
1$. For $q \rightarrow 0$ we can write,
\begin{equation}
  \frac{1}{\sqrt{4 \pi}} \frac{\cos (kq)}{q} 
  = 
  \frac{1}{\sqrt{4 \pi} q} + {\cal O}(q) \: ,
\end{equation}
which contains $1/q$ singularity. As for the term with the integral,
we expand $\exp (k^2 / 4 \xi^2)$ in a Taylor series, and transforming
to the variable $t = q \xi$ one gets:
\begin{equation}
  -\frac{1}{\pi} \int_{\sqrt{\eta} / 2}^{\infty} {\rm d}\xi \,
  \exp \left( - q^2 \xi^2 + \frac{k^2}{4 \xi^2} \right) =
  - \frac{1}{\pi q} \sum_{n=0}^{\infty} 
  \frac{(k q)^{2n}}{4^n n!} \int_{q \sqrt{\eta} / 2}^{\infty}
  {\rm d}t \ t^{-2n} e^{-t^2} \, .
\end{equation}
For $n=0$:
\begin{equation}
  \int_{q \sqrt{\eta} / 2}^{\infty} {\rm d}t \ e^{-t^2} =
  \left( \int_{0}^{\infty}  - \int_{0}^{q \sqrt{\eta} / 2} \right)
  {\rm d}t \ e^{-t^2} = 
  \frac{\sqrt{\pi}}{2} - \frac{1}{2} \sqrt{\eta}q + {\cal O}(q^2).
\end{equation}
For $n > 0$ we integrate by parts:
\begin{equation}
  \int_{q \sqrt{\eta} / 2}^{\infty} {\rm d}t \ t^{-2n} e^{-t^2} =
  \frac{1}{2n-1} \left( \frac{1}{2} \sqrt{\eta} q \right)^{-2n + 1} 
  e^{- \eta q^2 / 4} + {\cal O}(q^{-2n + 3}) \, .
\end{equation}
Collecting everything together back to (\ref{eq:dlm3}), the $1/q$
singularities cancel, and we remain with the finite expression:
\begin{equation}
  D_{00}^{(3)} = - \frac{\sqrt{\eta}}{2 \pi}
  \sum_{n=0}^{\infty} \frac{(k^2 / \eta)^n}{n! (2n-1)} \: .
\end{equation}


\section{The ``cubic harmonics'' $Y_{LJK}^{(\gamma)}$}
\label{app:ylj}

\subsection{Calculation of the transformation 
  coefficients $a^{(L)}_{\gamma J K, M}$}

We want to calculate the linear combinations of spherical harmonics
that transform according to the irreducible representations of the
cubic group $O_h$. This problem was addressed by von der Lage and
Bethe \cite{LB47} which coined the term ``cubic harmonics'' for these
combinations. They gave an intuitive scheme that was used to calculate
the first few cubic harmonics, but their arguments are difficult to
extend for large $L$'s. Moreover, their method is recursive, because
one has to orthogonalize with respect to all lower lying combinations.
This is cumbersome to implement numerically and might result in
instabilities for large $L$'s. The only other work on the subject that
we were are aware of \cite{GT90} specializes in the symmetric
representation and gives only part of the combinations. It also
expresses the results not in terms of spherical harmonics, but rather
as polynomials that are difficult to translate to $Y_{LM}$'s.

We describe in the following a simple and general numerical method to
calculate the cubic harmonics in a non-recursive way. This is based on
a general theorem, that states that a function $f^{(\gamma)}$
transforms according to the irrep $\gamma$ iff it satisfies
\cite{Tin64}:
\begin{equation}
  \hat{P}^{(\gamma)} f^{(\gamma)} =  f^{(\gamma)}
  \label{eq:projeqn}
\end{equation}
where $\hat{P}^{(\gamma)}$ is the projection operator onto the
subspace that belongs to $\gamma$:
\begin{equation}
  \hat{P}^{(\gamma)} = \frac{l_\gamma}{N_G} \sum_{\hat{g} \in G} 
  \chi^{(\gamma)*}(\hat{g}) \hat{g}.
  \label{eq:projector}
\end{equation}
We denoted by $l_\gamma$ the dimension of $\gamma$, $N_G$ is the
number of elements in the group $G$, and $\chi^{(\gamma)}(\hat{g})$
are the characters. The realization of $\hat{P}^{(\gamma)}$ as a
matrix in an arbitrary basis results in general in an infinite matrix.
However, in the case of the cubic harmonics, we know that $O_h \subset
O(3)$, thus the operations of $\hat{g} \in O_h$ do not mix different
$L$'s. Hence, working in the $Y_{LM}$ basis, we can write the cubic
harmonics as the finite combinations:
\begin{equation}
  Y_{LJ}^{(\gamma)}(\Omega) = 
  \sum_{M = -L}^{+L} a^{(L)*}_{\gamma J, M} Y_{LM}(\Omega)
  \label{eq:ch}
\end{equation}
where $J$ enumerates the irreps $\gamma$ in $L$. For simplicity we
consider 1-dimensional irreps. Applying (\ref{eq:projeqn}),
(\ref{eq:projector}) to (\ref{eq:ch}) and using the Wigner matrices
${\cal D}^{(L)}(\hat{g})$ to express the operations of $\hat{g}$ on
$Y_{LM}$ \cite{VMK88}, we get the following $(2L+1) \times (2L+1)$
linear system:
\begin{equation}
  \sum_{M'} \left [ P^{(\gamma, L)}_{MM'} - \delta_{MM'} \right] 
  a^{(L)*}_{\gamma J, M'} = 0
  \label{eq:cheqns}
\end{equation}
where:
\begin{equation}
  P^{(\gamma, L)}_{MM'} =
  \frac{1}{48} \sum_{\hat{g} \in G} \chi^{(\gamma)*}(\hat{g})
  {\cal D}^{(L)}_{MM'}(\hat{g}).
\end{equation}
The above equations are best solved using SVD algorithm \cite{NAG90},
and the (orthonormalized) eigenvectors that belong to the zero
singular values are the required coefficients $a^{(L)*}_{\gamma J,
  M'}$. For multi-dimensional irreps one needs to classify the cubic
harmonics also with respect to the row $K$ inside the irrep. This can
be done by simple modification of the above procedure, using the
appropriate projectors \cite{Tin64}.

The above general procedure can be simplified for specific irreps. In
the following we shall concentrate on the completely antisymmetric
irrep $\gamma = a$ and further reduce the linear system
(\ref{eq:cheqns}). We first note, that the antisymmetric cubic
harmonics must satisfy per definition:
\begin{equation}
  \hat{g} Y_{LJ}^{(a)}(\Omega) = 
  \chi^{(a)}(\hat{g}) Y_{LJ}^{(a)}(\Omega) =
  (-1)^{(\mbox{\tiny parity of }\hat{g})} Y_{LJ}^{(a)}(\Omega) \; \; \;
  \forall \hat{g} \in O_h.
  \label{eq:gasch}
\end{equation}
We then choose a few particular $\hat{g}$'s for which the operations
on $Y_{LM}(\Omega)$ are simple:
\begin{eqnarray}
  \begin{array}{ll}
    \hat{r_x} (xyz) \equiv (-xyz) : &
    \hat{r_x} Y_{LM}(\theta, \phi) = 
    Y_{LM}(\theta, -\phi) =
    (-1)^M Y_{L -M}(\theta, \phi) \\
    \hat{r_y} (xyz) \equiv (x-yz) : &
    \hat{r_y} Y_{LM}(\theta, \phi) = 
    Y_{LM}(\theta, \pi - \phi) =
    Y_{L -M}(\theta, \phi) \\
    \hat{r_z} (xyz) \equiv (xy-z) : &
    \hat{r_z} Y_{LM}(\theta, \phi) = 
    Y_{LM}(\pi - \theta, \phi) =
    (-1)^{L+M} Y_{L M}(\theta, \phi) \\
    \hat{p_{xy}} (xyz) \equiv (yxz) : &
    \hat{p_{xy}} Y_{LM}(\theta, \phi) = 
    Y_{LM}(\theta, \frac{\pi}{2} - \phi) =
    (-i)^{M} Y_{L -M}(\theta, \phi).
  \end{array}
  \label{eq:special-gs}
\end{eqnarray}
Applying (\ref{eq:gasch}) and (\ref{eq:special-gs}) to (\ref{eq:ch})
results in the following ``selection rules'':
\begin{eqnarray}
  a^{(L)*}_{a J, M} 
  & = &
  0, \; \; \; L \neq 2p+1, M \neq 4q, \; \; \; p, q \in  {\rm I\!N} 
  \nonumber \\
  a^{(L)*}_{a J, -M} 
  & = &
  - a^{(L)*}_{a J, M}
  \label{eq:sr}
\end{eqnarray}
which reduces the number of independent coefficients to be computed by
a factor of 16. The form of the projector matrix $P^{(a L)}$ can also
be greatly reduced, if we observe that the group $O_h$ can be written
as the following direct multiplication:
\begin{eqnarray}
  O_h 
  & = &
  G_3 \otimes G_{16} \\
  G_3
  & = &
  \{ \hat{e}, \hat{c}, \hat{c}^2 \} \; \; \;
  \hat{e} = \mbox{identity}, \; \; \; 
  \hat{c}(xyz) = (yzx) \\
  G_{16}
  & = &
  \{ \hat{e}, \hat{p}_{xy} \} \otimes
  \{ \hat{e}, \hat{r}_{x} \} \otimes
  \{ \hat{e}, \hat{r}_{y} \} \otimes
  \{ \hat{e}, \hat{r}_{z} \}
\end{eqnarray}
and consequently, the projector can be written as
\begin{eqnarray}
  \hat{P}^{(a)} 
  & = &
  \hat{P}_{3} \hat{P}_{16} \\
  \hat{P}_{3}
  & = &
  \hat{e} + \hat{c} + \hat{c}^2 \\
  \hat{P}_{16}
  & = &
  ( \hat{e} - \hat{p}_{xy} ) ( \hat{e} - \hat{r}_{x} )
  ( \hat{e} - \hat{r}_{y}  ) ( \hat{e} - \hat{r}_{z} ).
\end{eqnarray}
The operator $\hat{P}_{16}$ acts as the identity on the subspace
defined by (\ref{eq:sr}) and hence we need to consider only the
operation of $\hat{P}_{3}$. Simple manipulations give the following
set of equations:
\begin{equation}
  \sum_{q' > 0} \left[ 2 d^{(L)}_{4q, 4q'} \left( \frac{\pi}{2} \right)
    - \delta_{4q, 4q'} \right] a^{(L)*}_{a J, 4q'} = 0 , \; \; \;
  q > 0 \; , \; L = 2p+1 \ .
\end{equation}
The matrices $d^{(L)}_{4q, 4q'}$ are the ``reduced'' Wigner matrices,
which are real \cite{VMK88}, thus the resulting coefficients are also
real. The above is a square linear system, which is 8 times smaller
than the general one (\ref{eq:cheqns}).

\subsection{Counting the $Y_{LJ}^{(\gamma)}$'s}
%
The number of the irreps $\gamma$ of $O_h$ that are contained in the
irrep $L$ of $O(3)$ is given by the formula \cite{Tin64}:
\begin{equation}
  N_{\gamma L} = \frac{1}{48} \sum_{\hat{g} \in O_h} 
  \chi^{(\gamma)*}(\hat{g}) \chi_{L}(\hat{g})
\end{equation}
where $\chi_{L}(\hat{g})$ are the characters of the irrep $L$.  An
explicit calculation shows that the main contributions for large $L$'s
come from the identity and from the inversion operations, thus:
\begin{equation}
  N_{\gamma L} \approx 
  \left[ 1 \pm (-1)^L \right] \frac{l_{\gamma} (2L+1)}{48}.
\end{equation}
where the $\pm$ corresponds to the parity of $\gamma$. Since for
$l_{\gamma}$-dim irrep we have $l_{\gamma}$ basis functions, and there
are $2L+1$ basis functions in the irrep $L$, the fraction of cubic
harmonics that belong to $\gamma$ is:
\begin{equation}
  F_{\gamma} \approx \frac{l_{\gamma}^2}{48}
\end{equation}
in accordance with the general relation:
\begin{equation}
  \sum_{\gamma} l_{\gamma}^2 = 48.
\end{equation}
Consequently, the fraction of cubic harmonics that belong to the
$K$'th block of $\gamma$ is
\begin{equation}
  F_{\gamma K} \approx \frac{l_{\gamma}}{48} \, .
\end{equation}

\section{Evaluation of $l(\vec{\rho_p})$}
\label{app:lrho}

\subsection{Proof of equation (\protect{\ref{eq:dlm-sr2}})}

We need to prove the relation:
\begin{equation}
  \sum_{\hat{g} \in O_h} Y_{LM}(\Omega_{\hat{g} \vec{\rho}}) =
  l(\vec{\rho_p}) \sum_{{\vec{\rho}'} \in S(\vec{\rho_p})} 
                  Y_{LM}(\Omega_{\vec{\rho}'})
  \label{eq:lrho1}
\end{equation}
where $\vec{\rho_p} \equiv (i, j, k)$ is the unique vector in the set
$O_h \vec{\rho}$ which resides in the fundamental domain $i \geq j
\geq k \geq 0$, $S(\vec{\rho_p})$ is the collection of all distinct
vectors obtained by the operations $\hat{g} \vec{\rho_p}\, , \;
\hat{g} \in
O_h$, and $l(\vec{\rho_p})$ is an integer. \\

\noindent
{\bf Proof.}
Let ${\cal H}$ be the set of all $\hat{g} \in O_h$ under
which $\vec{\rho_p}$ is invariant: 
\begin{equation}
  \begin{array}{lcl}
    \hat{g} \vec{\rho_p} =  \vec{\rho_p} &
    \Longleftrightarrow &
    \hat{g} \in {\cal H}.
  \end{array}
\end{equation}
The set ${\cal H}$ is a subgroup since:
\begin{enumerate}
  \item The identity $\hat{e} \in {\cal H}$.
  \item The set ${\cal H}$ is closed under multiplication, since if
    $\hat{g}_1, \hat{g}_2 \in {\cal H}$ then $\hat{g}_1 ( \hat{g}_2
    \vec{\rho_p}) = \hat{g}_1 \vec{\rho_p} = \vec{\rho_p}$.
  \item The set ${\cal H}$ is closed under inversion: $\hat{g}^{-1}
    \vec{\rho_p} = \hat{g}^{-1} (\hat{g} \vec{\rho_p}) = \vec{\rho_p}$.
\end{enumerate}
The order of (number of terms in) ${\cal H}$ is denoted as $N_{\cal
  H}$. We assume that ${\cal H}$ is the maximal invariance subgroup,
and construct the right cosets $\hat{g} {\cal H} = \{ \hat{g}
\hat{h}_1, \ldots \}$. According to \cite{Tin64} there are $N_c = 48 /
N_{\cal H}$ mutually exclusive such cosets $C_1, \ldots, C_{N_c}$ (The
number $48$ is the order of $O_h$). Their union is $O_h$. For each
coset $C_i$ we can define
\begin{equation}
  \vec{\rho}_i \equiv C_i \vec{\rho_p}
\end{equation}
which is meaningful due to the invariance of $\vec{\rho_p}$ under
${\cal H}$. \\
We want to prove the following

\noindent
{\bf Lemma.} $\vec{\rho}_i \neq \vec{\rho}_j$ iff $i \neq j$. \\

\noindent
{\bf Proof.} Assume the opposite, then in particular 
\[
\begin{array}{l}
\hat{g}_i \vec{\rho_p} = \hat{g}_j \vec{\rho_p} \\
\Leftrightarrow (\hat{g}_j^{-1} \hat{g}_i) \vec{\rho_p} = \vec{\rho_p} \\
\Leftrightarrow \hat{g}_j^{-1} \hat{g}_i = h \in {\cal H} \\
\Leftrightarrow \hat{g}_i = \hat{g}_j h \\
\Leftrightarrow C_i = C_j
\end{array}
\]
in contradiction to the assumption. The last line was obtained using
the rearrangement theorem \cite{Tin64} applied to the group ${\cal
  H}$.
QED. \\
We now set
\begin{eqnarray}
  S(\vec{\rho_p}) 
  & = &
  \cup_{i=1}^{N_c} \vec{\rho}_i \\
  l(\vec{\rho_p}) 
  & = &
  N_{\cal H} = \mbox{integer}
\end{eqnarray}
and since $\sum_{\hat{g} \in O_h} = \sum_{i=1}^{N_c} 
\sum_{\hat{g} \in C_i}$ we proved (\ref{eq:lrho1}). QED.

\subsection{Calculating $l(\vec{\rho_p})$}
%
We give an explicit expression of $l(\vec{\rho_p})$. Consider
$\vec{\rho_p} = (i, j, k)$ such that $i \ge j \ge k \ge 0$ with no
loss of generality. Then
\begin{eqnarray}
  l(\vec{\rho_p}) 
  & = &
  l_p(\vec{\rho_p}) l_s(\vec{\rho_p}) \, ,
  \label{eq:lpls} \\
  l_p(\vec{\rho_p}) 
  & = &
  \left\{
    \begin{array}{ll}
      1, & i \neq j \neq k \neq i \\
      2, & i = j \neq k \mbox{ or } i \neq j = k \mbox{ or } i = k \neq j \\
      6, & i = j = k
    \end{array} 
  \right. 
  \label{eq:lp} \\
  l_s(\vec{\rho_p})
  & = &
  2^{\left( \# \ \mbox{zero indices} \right)}.
  \label{eq:ls}
\end{eqnarray}
We prove this formula in the following. First we observe, that $O_h$
can be decomposed as
\begin{eqnarray}
  O_h
  & = &
  {\cal P}_3 \otimes {\cal S}_3 
  \label{eq:oh-decomposition} \\
  {\cal P}_3 
  & = &
  \mbox{group of permutation of 3 numbers} \\
  {\cal S}_3
  & = &
  \{ \pm \pm \pm \} = \mbox{3 sign changes}.
\end{eqnarray}
Let ${\cal H}_P, {\cal H}_S$ be the subgroups of ${\cal P}_3, {\cal
  S}_3$, respectively, under which $\vec{\rho_p}$ is invariant.
\\
{\bf Lemma.} ${\cal H} = {\cal H}_P \otimes {\cal H}_S$.
\\
\\
{\bf Proof.} Let $\hat{g} = \hat{p} \hat{s}$, where $\hat{g} \in O_h$,
$\hat{p} \in {\cal P}_3$ and $\hat{s} \in {\cal S}_3$. This
representation of $\hat{g}$ is always possible according to
(\ref{eq:oh-decomposition}). If $\hat{s} \in\!\!\!\!\!\backslash {\cal
  H}_S$ then $\hat{s} \vec{\rho_p} \neq \vec{\rho_p}$, thus
necessarily there is at least one sign change in $\hat{s}
\vec{\rho_p}$ with respect to $\vec{\rho_p}$. Consequently, $\hat{g}
\vec{\rho_p} \neq \vec{\rho_p}$, because permutations only change the
order of indices and cannot restore the different sign(s). We conclude
that $\hat{g} \in\!\!\!\!\!\backslash {\cal H}$. Thus, $\hat{g} \in
{\cal H} \Rightarrow \hat{s} \in {\cal H}_S$. For every $\hat{g} \in
{\cal H}$ we must have therefore $\hat{g} \vec{\rho_p} = \hat{p}
\hat{s} \vec{\rho_p} = \hat{p} \vec{\rho_p} = \vec{\rho_p}$
which proves that also $\hat{p} \in {\cal H}_P$. QED. \\
We conclude that $N_{\cal H} = \mbox{order}({\cal H}_P) \cdot
\mbox{order}({\cal H}_S)$. This is manifest in equations
(\ref{eq:lpls}-\ref{eq:ls}).


\section{Number theoretical degeneracy of the cubic lattice}
\label{app:nt-deg}

\subsection{First moment}

The following arguments are due to J.\ Keating \cite{KeaPC}. We first
need to estimate the fraction of integers that can be expressed as a
sum of 3 squares. The key theorem is due to Gauss and Legendre and
states that:
\begin{equation}
  \begin{array}{ccc}
    q = i^2 + j^2 + k^2 , \; \; \; i, j, k \in \bbbn &
    \Longleftrightarrow &
    q \neq 4^m (8 l + 7) , \; \; \; m, l \in \bbbn.
  \end{array}
\end{equation}
From this we can estimate that the fraction of integers which {\em can
  not} be expressed as a sum of 3 squares of integers is:
\begin{equation}
  \frac{1}{8} \left( 1 + \frac{1}{4} + \frac{1}{4^2} + \cdots \right) =
  \frac{1}{6} \ .
\end{equation}
In the above we used the fact (which is easily proven) that if $q =
4^m (8 l + 7)$ then $m$, $l$ are uniquely determined. Therefore,
asymptotically only $5/6$ of the integers are expressible as a sum of
three squares.

Our object of interest is the degeneracy factor $g_{\rho}(\rho)$
defined as:
\begin{equation}
  g_{\rho}(\rho) \equiv 
  \# (\vec{\kappa} \in \bbbz^3 | \kappa =  \rho) \ .
\end{equation}
The number of $\bbbz^3$-lattice points whose distance from the origin
is between $\rho$ and $\rho + \Delta \rho$ is estimated by considering
the volume of the corresponding spherical shell:
\begin{equation}
  N_{\rho} \approx 4 \pi \rho^2 \Delta \rho \ .
\end{equation}
Since $\rho^2$ is an integer, the number of integers in the same
interval is:
\begin{equation}
  n_{\rho} \approx 2 \rho \Delta \rho \ .
\end{equation}
Taking into account that only $5/6$ of the integers are accessible, we
obtain:
\begin{equation}
  \langle g_{\rho}(\rho) \rangle = 
  \frac{N_{\rho}}{(5/6) n_{\rho}} = \frac{12 \pi}{5} \rho \ .
  \label{eq:g-rho}
\end{equation}

\subsection{Second moment}

Here we use a result due to Bleher and Dyson \cite{BD94}, brought to
our attention by Z. Rudnick:
\begin{equation}
  \sum_{k = 1}^{N} g^2_{\rho}(\sqrt{k}) = c N^2 + \mbox{\rm error} \; ,  
  \; \; \; c = \frac{16 \pi^2}{7} \frac{\zeta(2)}{\zeta(3)} 
  \approx 30.8706 \ .
\end{equation}
Differentiating by $N$ and considering only integers for which
$g_{\rho} \neq 0$ one obtains:
\begin{equation}
  \langle g^2_{\rho}(\rho) \rangle \approx \frac{12}{5} c \rho^2 
  \approx 74.0894 \rho^2 \ .
  \label{eq:g2-rho}
\end{equation}
Therefore,
\begin{equation}
  \frac{\langle g^2_{\rho}(\rho) \rangle}
       {\langle g_{\rho}(\rho) \rangle} 
  \approx
  \beta \rho \; , 
  \; \; \;
  \beta = \frac{c}{\pi} \approx 9.8264 \ .
\end{equation}


\section{Weyl's law}
\label{app:weyl}

A very important tool in the investigation of eigenvalues is the
smooth counting function, known as Weyl's law. For billiards it was
thoroughly discussed e.g.\ by Balian and Bloch \cite{BB70} and by
Baltes and Hilf \cite{BH76}. We construct in the following the
expression for the 3D Sinai billiard. In general, it takes on the
form:
\begin{equation}
  \bar{N}(k) = N_3 k^3 + N_2 k^2 + N_1 k + N_0 \ ,
\end{equation}
where we included terms up to and including the constant term. In
fact, for the nearest-neighbour and two-point spectral statistics the
constant term $N_0$ is unimportant, since it shifts the unfolded
spectrum $x_n \equiv \bar{N}(k_n)$ uniformly. Nevertheless, for
completeness we shall calculate this term. We enumerate the
contributions in the case of Dirichlet boundary conditions one by one
and then write down the full expression. Figure \ref{fig:pyramid}
should be consulted for the geometry of the billiard.
\begin{description}
  
\item[$\large \bf N_3$:] There is only one contribution due to the
  volume of the billiard:
  \begin{equation}
    N_3 = \frac{{\rm volume}}{6 \pi^2} = 
    \frac{1}{288 \pi^2} \left( S^3 - \frac{4}{3}\pi R^3 \right) \, .
  \end{equation}
  
\item[$\large \bf N_2$:] The contribution is due to the surface area
  of the planes + sphere:
  \begin{equation}
    N_2 = - \frac{\rm surface}{16 \pi} =
    - \frac{1}{384 \pi} \left[ 6 (1 + \sqrt{2}) S^2 - 
      7 \pi R^2 \right] \, .
  \end{equation}
  
\item[$\large {\bf N_1}$:] Here we have contributions due to the
  curvature of the sphere and due to 2-surface edges:
  
  \begin{description}
    
  \item[curvature:]
    \begin{equation}
      N_1^{\rm curvature} 
      = 
      \frac{1}{12 \pi^2} \int_{\rm surface} 
      {\rm d}s \, \left[ \frac{1}{R_1(s)} + \frac{1}{R_2(s)} \right] = 
      - \frac{R}{72 \pi} \, ,
    \end{equation}
    where $R_1$, $R_2$ are the principal local radii of curvature.
    
  \item[edges:] We have 6 plane-plane edges and 3 plane-sphere edges.
    Their contributions are given by:
    \begin{eqnarray}
      N_1^{\rm edges} 
      & = &
      \frac{1}{24 \pi} \sum_{\rm edges} \left( \frac{\pi}{\alpha_j} 
        - \frac{\alpha_j}{\pi} \right) L_j  \\
      & = &
      \frac{S}{144 \pi}(27+9\sqrt{2}+8\sqrt{3}) + 
      \frac{R}{24 \pi} \left( \frac{9 \pi}{8} - 
        \frac{95}{12} \right) \, ,
      \nonumber
    \end{eqnarray}
    where $L_j$ are the lengths of the edges, and $\alpha_j$ are the
    corresponding angles.
  
  \end{description}  
  
\item[$\large {\bf N_0}$:] There are three terms here due to square of
  the curvatures, 3-surface corners and curvature of the edges:
  \begin{description}
    
  \item[curvature$^2$:]
    \begin{equation}
      N_0^{\rm curvature^2} 
      = 
      \frac{1}{512 \pi} \int_{\rm surface} {\rm d}s
        \, \left[ \frac{1}{R_1(s)} - \frac{1}{R_2(s)} \right]^2 
      = 0 \, .
    \end{equation}
    
  \item[3-surface corners:] In the 3D Sinai billiard we have 6 corners
    due to intersection of 3 surfaces; 3 of them are due to
    intersection of 3 symmetry planes and the other 3 are due to
    intersection of 2 symmetry planes and the sphere. The corners are
    divided into 4 types as follows:
    \begin{eqnarray*}
      1 & \times & \alpha \equiv (45^\circ, 54.74^\circ, 36.26^\circ) \\
      3 & \times & \beta  \equiv (45^\circ, 90^\circ, 90^\circ) \\
      1 & \times & \gamma \equiv (60^\circ, 90^\circ, 90^\circ) \\
      1 & \times & \delta \equiv (90^\circ, 90^\circ, 90^\circ) \, .
    \end{eqnarray*}
    As for the corners $\beta, \gamma, \delta$ which are of the type
    $(\phi, 90^\circ, 90^\circ)$ there is a known expression for their
    contribution \cite{BH76}:
    \begin{equation}
      c_{\phi} = - \frac{1}{96} \left( \frac{\pi}{\phi} - 
        \frac{\phi}{\pi} \right) \, .
    \end{equation}
    Therefore:
    \begin{equation}
      c_{\beta}  = -\frac{5}{128} \; , \; \; \;
      c_{\gamma} = -\frac{1}{36} \; , \; \; \;
      c_{\delta} = -\frac{1}{64} \; .
    \end{equation}
    As for the corner $\alpha$, we calculate its contribution from the
    $R=0$ integrable case (``the pyramid''). The constant term in the
    case of the pyramid is $-5/16$ \cite{PS95} and originates only
    from 3-plane contributions (there are no curved surfaces or curved
    edges in the pyramid). The pyramid has 4 corners: 2 of type
    $\alpha$ and 2 of type $\beta$.  Using $c_{\beta}$ above we can
    therefore eliminate $c_{\alpha}$:
    \begin{equation}
      2 \cdot c_{\alpha} + 2 \cdot \left( -\frac{5}{128} \right)
      = -\frac{5}{16} \; \; 
      \Longrightarrow \; \; c_{\alpha} = -\frac{15}{128} \, .
    \end{equation}
    Hence, the overall contribution due corners in the 3D Sinai is:
    \begin{eqnarray}
      N_0^{\rm 3-surface} 
      & = &
      1 \cdot \left( - \frac{15}{128} \right) +
        3 \cdot \left( - \frac{5}{128}  \right) + 
      \nonumber
      \\
      & &
      1 \cdot \left( - \frac{1}{36}   \right) +
        1 \cdot \left( - \frac{1}{64}   \right) 
      \nonumber 
      \\
      & = &
      - \frac{5}{18} \ .
    \end{eqnarray}
      
  \item[curvature of edges:] We have 3 edges which are curved. They
    are $90^\circ$ edges that are due to plane-sphere intersections.
    Baltes and Hilf \cite{BH76} quote the constant term $(-1/12) +
    (1/256)(H/R)$ for the circular cylinder, where $H$ is the height
    and $R$ is the radius of the cylinder. We conclude from this that
    the $H$-independent term $-1/12$ is due to the curvature of the
    $90^\circ$ edges between the 2 bases and the tube. Assuming
    locality, it is then plausible to conjecture that the contribution
    due to the curvature of a $90^\circ$ edge is:
    \begin{equation} 
      - \frac{1}{48 \pi} \int_{\rm edge} \frac{{\rm d}l}{R(l)} \, ,
    \end{equation}
    where $R(l)$ is the local curvature radius of the edge. When
    applied to our case ($R(l) = -R$), we get:
    \begin{equation}
      N_0^{\rm curv.\ edge} = \frac{1}{64} \, .
    \end{equation}
    
  \end{description} 
  
\end{description}
Putting everything together we get:
\begin{eqnarray}
  \bar{N}(k) 
  & = &
  \frac{1}{288 \pi^2} \left( S^3 - \frac{4}{3}\pi R^3 \right) k^3 
  \nonumber \\
  & - & 
  \frac{1}{384 \pi} \left[ 6 (1 + \sqrt{2}) S^2 - 7 \pi R^2 \right] k^2 
  \nonumber \\
  & + & \left[ \frac{S}{144 \pi}(27+9\sqrt{2}+8\sqrt{3}) + 
    R \left( \frac{3}{64} - \frac{11}{32 \pi} \right) \right] k 
  \nonumber \\
  & - & \frac{151}{576} \: .
  \label{eq:weyl}
\end{eqnarray}


\section{Calculation of the monodromy matrix}
\label{app:3d-monodromy}

The monodromy matrix measures the linear response to infinitesimal
displacements of the initial conditions of a classical orbit. Its
eigenvalues determine the stability of the orbit. Due to the
symplectic form of the equations of motion, if $\lambda$ is an
eigenvalue of the monodromy matrix then also $\lambda^{*}$,
$1/\lambda$ and $1/\lambda^{*}$ \cite{Gut90}. Therefore, generically
the eigenvalues come in groups of four:
\begin{equation}
  \lambda 
  = 
  \exp ( \pm u \pm i v ) \: , 
  \; \; \; u, v \in \bbbr \: .
  \label{eq:generic-ev-monod}
\end{equation}
In $d$ dimensions the monodromy has $2(d-1)$ eigenvalues. Therefore,
only for $d \geq 3$ the generic situation (\ref{eq:generic-ev-monod})
can take place. In two dimensions there are only two eigenvalues and
consequently one obtains the following three possible situations
(which are special cases of (\ref{eq:generic-ev-monod}) with either
$u$ or $v$ set to 0):
\begin{description}
\item[Elliptic:] $\lambda_{1, 2} = \exp (\pm i v)$, stable orbit.
\item[Parabolic:] $\lambda_{1, 2} = 1$ or $\lambda_{1,2} = (-1)$,
  neutrally stable orbit.
\item[Hyperbolic:] $\lambda_{1, 2} = \exp ( \pm u )$ 
  or $\lambda_{1, 2} = -\exp ( \pm u )$, unstable orbit.
\end{description}
The parabolic case with the ``+'' sign is denoted as ``direct
parabolic'' and with ``-'' sign it is denoted as ``inverse
parabolic''. Similar terminology applies to the hyperbolic case. The
generic case (\ref{eq:generic-ev-monod}) is designated as ``loxodromic
stability'' \cite{Gut90}.

\subsection{The 3D Sinai torus case}

We wish to calculate explicitly the $4 \times 4$ monodromy matrices in
the case of the 3D Sinai torus. There are (at least) two possible ways
to tackle this problem. One possibility is to describe the classical
motion by a discrete (Hamiltonian, area--preserving) mapping between
consecutive reflections from the spheres. The mapping is generated by
the straight segment that connects the two reflection points, and the
monodromy can be explicitly calculated from the second derivatives of
the generating function. This straightforward calculation was
performed for the 2D case (for general billiards) e.g.\ in
\cite{Smi94} and it becomes very cumbersome for three dimensions.
Rather, we take the alternative view of describing the classical
motion as a continuous flow in time, as was done e.g.\ by Sieber
\cite{Sie91} for the case of the 2D hyperbola billiard. We separate
the motion into the sections of free propagation between spheres and
reflections off the spheres, and the monodromy matrix takes the
general form:
\begin{equation}
  M 
  = 
  M_{\rm prop}^{n+1 \leftarrow n} M_{\rm ref}^{n} 
    \cdots
    M_{\rm prop}^{3 \leftarrow 2} M_{\rm ref}^{2}
    M_{\rm prop}^{2 \leftarrow 1} M_{\rm ref}^{1} \: ,
  \label{eq:monod-cp}
\end{equation}
where $M_{\rm prop}^{i+1 \leftarrow i}$ describes the free propagation
from sphere $i$ to sphere $i+1$ and $M_{\rm ref}^{i}$ describes the
reflection from the sphere $i$. To explicitly calculate the matrices
one has to choose a well-defined (and convenient) coordinate system,
which is a non-trivial task in three dimensions. If we denote the
direction along the orbit by ``1'', then we have two more directions,
denoted henceforth ``2'' and ``3''. Hence there is a rotation freedom
in choosing the directions 2 and 3. For convenience of calculation of
$M_{\rm ref}$ we choose the following local convention for
coordinates: Near sphere $i$ there exists the plane ${\cal P}_{i}$
which is uniquely defined (except for normal incidence) by the
incoming segment, the outgoing segment and the normal to the sphere
$i$ at the reflection point. Direction 1 is obviously in ${\cal
  P}_{i}$. We uniquely define direction 3 to be perpendicular to
${\cal P}_{i}$ along the direction of the cross product of the
outgoing direction with the normal. Direction 2 is then uniquely
defined as $\hat{e}_2 \equiv \hat{e}_{3} \times \hat{e}_{1}$ such that
a right--handed system is formed. Obviously $\hat{e}_2$ is contained
in ${\cal P}_{i}$. The uniqueness of the local coordinate system
guarantees that the neighbourhoods of the initial and the final points
of the periodic orbits are correctly related to each other. To account
for the local coordinate systems we need to apply a rotation between
every two reflections that aligns the ``old'' system to the ``new''
one. Hence,
\begin{equation}
  M 
  = 
  M_{\rm prop}^{n+1 \leftarrow n} 
    M_{\rm rot}^{n+1 \leftarrow n}  M_{\rm ref}^{n} 
    \cdots
    M_{\rm prop}^{3 \leftarrow 2} M_{\rm rot}^{3 \leftarrow 2} 
    M_{\rm ref}^{2}
    M_{\rm prop}^{2 \leftarrow 1} M_{\rm rot}^{2 \leftarrow 1} 
    M_{\rm ref}^{1} \: .
  \label{eq:monod-crp}
\end{equation}
We should also fix the convention of the rows and columns of $M$ in
order to be able to write explicit expressions. It is chosen to be:
\begin{equation}
  \left( \begin{array}{c} \delta q_2 \\
                          \delta p_2 \\
                          \delta q_3 \\
                          \delta p_3
    \end{array} \right)_{\rm final} 
  = 
  M
  \left( \begin{array}{c} \delta q_2 \\
                          \delta p_2 \\
                          \delta q_3 \\
                          \delta p_3
    \end{array} \right)_{\rm initial} \: .
\end{equation}
A detailed calculations gives the explicit expressions for $M_{\rm
  prop}$, $M_{\rm ref}$ and $M_{\rm rot}$:
\begin{eqnarray}
  M_{\rm prop}^{i+1 \leftarrow i} 
  & = & 
  \left( \begin{array}{cccc}
    1 & L_{i+1 \leftarrow i}/p & 0 & 0                      \\
    0 & 1                      & 0 & 0                      \\
    0 & 0                      & 1 & L_{i+1 \leftarrow i}/p \\
    0 & 0                      & 0 & 1
  \end{array} \right) \: ,
  \\
  M_{\rm ref}^{i} 
  & = & 
  \left( \begin{array}{cccc}
    -1                          & 0  & 0                          & 0 \\
    -\frac{2 p}{R \cos \beta_i} & -1 & 0                          & 0 \\
    0                           & 0  & 1                          & 0 \\
    0                           & 0  & \frac{2 p \cos \beta_i}{R} & 1
  \end{array} \right) \: ,
  \\
  M_{\rm rot}^{i+1 \leftarrow i}
  & = & 
  \left( \begin{array}{cccc}
    \cos \alpha_{i+1 \leftarrow i}    & 0                               & 
      -\sin \alpha_{i+1 \leftarrow i} & 0                               \\
    0                                 &  \cos \alpha_{i+1 \leftarrow i} & 
      0                               & -\sin \alpha_{i+1 \leftarrow i} \\
    \sin \alpha_{i+1 \leftarrow i}    & 0                               & 
      \cos \alpha_{i+1 \leftarrow i}  & 0                               \\
    0                                 &  \sin \alpha_{i+1 \leftarrow i} & 
      0                               &  \cos \alpha_{i+1 \leftarrow i}
  \end{array} \right) \, .
\end{eqnarray}
In the above $p$ is the absolute value of the momentum which is a
constant, $L_{i+1 \leftarrow i}$ is the length of the orbit's segment
between spheres $i$ and $i+1$, $\beta_i$ is the reflection angle with
respect to the normal of the sphere $i$ and $\alpha_{i+1 \leftarrow
  i}$ is the angle that is needed to re-align the coordinate system
from sphere $i$ to $i+1$. Even thought the entries of $M$ are
dimensional, the eigenvalues of $M$ are dimensionless. Hence, the
eigenvalues cannot depend on $p$, which is the only variable with
dimensions of a momentum. (All other variables have either dimension
of length or are dimensionless.) Therefore, one can set $p = 1$ for
the sake of the calculations of the eigenvalues of $M$. The formulas
above for the monodromy were verified numerically for a few cases
against a direct integration of the equations of motion near a
periodic orbit of the Sinai torus. We mention the work of Sieber
\cite{Sie98} which extends the calculation of the monodromy matrix to
an arbitrary billiard in three dimensions.

\subsection{The 3D Sinai billiard case}

We next deal with the calculation of the monodromy matrix for the
periodic orbits of the desymmetrized 3D Sinai Billiard. In principle,
one can follow the same procedure as above, and calculate the
monodromy as for the Sinai torus case, this time taking into account
the presence of the symmetry planes. A reflection with a symmetry
plane is described by:
\begin{equation}
  M_{\rm ref}^{\rm plane} =
  \left( \begin{array}{cccc}
    -1 &  0 & 0 & 0 \\
     0 & -1 & 0 & 0 \\
     0 &  0 & 1 & 0 \\
     0 &  0 & 0 & 1
   \end{array} \right) \: ,
 \label{eq:monodromy-ref-plane}
\end{equation}
which is simply $M_{\rm ref}$ with $R \rightarrow \infty$. This
method, however, is computationally very cumbersome because of the
need to fold the orbit into the desymmetrized Sinai billiard. Instead,
we can use the monodromy matrix that is calculated for the unfolded
periodic orbit, because the initial and final (phase space)
neighbourhoods are the same modulo $\hat{g}$. A calculation shows,
that in order to align the axes correctly, one needs to reverse
direction 3 if $\hat{g}$ is not a pure rotation:
\begin{equation}
  M_{\hat{W}} 
  = 
  \left(
    \begin{array}{cccc}
     1 & 0 &  0              & 0               \\
     0 & 1 &  0              & 0               \\
     0 & 0 & \sigma(\hat{g}) & 0               \\
     0 & 0 &  0              & \sigma(\hat{g})
   \end{array} \right) M_{\hat{W}}^{\rm unfolded} \: ,
  \label{eq:monodromy-sb}
\end{equation}
where $\sigma(\hat{g})$ is the parity of $\hat{g}$: 
\begin{equation}
  \sigma(\hat{g}) 
  = 
  \left\{
    \begin{array}{ll}
      +1, & \mbox{$\hat{g}$ is a rotation} \\
      -1, & \mbox{$\hat{g}$ is an improper rotation
                 (rotation + inversion)}
    \end{array}
  \right. \: .
\end{equation}
The above formulas were verified numerically for a few cases by
comparing the result (\ref{eq:monodromy-sb}) to a direct integration
of the classical dynamics in the desymmetrized Sinai billiard.


\bibliographystyle{unsrt}
\bibliography{/home/harel/LATEX/general}

\end{document}